\documentclass[nonatbib]{elsarticle_biblatexcompat}

\usepackage[utf8]{inputenc}

\usepackage[T1]{fontenc}
\usepackage{lmodern}

\usepackage{float}

\usepackage[bibliography=common
]{apxproof}

\usepackage{float}
\usepackage{placeins}

\usepackage{amsmath}
\usepackage{amssymb}
\usepackage{amsthm}

\usepackage{bm}

\usepackage{mathtools}

\usepackage[dvipsnames]{xcolor}

\usepackage{adjustbox}

\usepackage[shortlabels]{enumitem}

\usepackage[save]{silence}
\usepackage[hypcap=true]{subcaption}

\usepackage{tikz}
\usetikzlibrary{automata}
\usetikzlibrary{positioning}
\usetikzlibrary{arrows.meta}
\usetikzlibrary{shapes.multipart}
\usetikzlibrary{shapes.geometric}
\usetikzlibrary{matrix}
\usetikzlibrary{backgrounds}
\usetikzlibrary{fit}
\usetikzlibrary{cd}

\usepackage{LTS_tools}

\usepackage[ruled,lined,linesnumbered,nokwfunc,noend]{algorithm2e}

\usepackage[
colorlinks=true,
linkcolor=blue,
citecolor=blue
]{hyperref}

\usepackage{cleveref}
\crefname{conjecture}{conjecture}{conjectures}

\SetNlSty{}{}{:}
\SetKwProg{Fn}{}{:}{}

\newcommand{\Assign}[3][\;]{#2 $\gets$ #3#1}
\newcommand{\Throw}[1]{\FuncSty{throw exception} \ArgSty{#1}}
\SetKwProg{uForEach}{for each}{ do}{}
\newcommand{\Parens}[1]{(#1)}
\newcommand{\settype}[1]{\ensuremath{\mathbf{Set~of~} #1}}
\newcommand{\functionLabel}[1]{%
  \crefformat{FunctionFormat#1}{##2#1##3}%
  \Crefformat{FunctionFormat#1}{##2#1##3}%
  \def\currentFunName{#1}
  \SetKwFunction{#1}{#1}
  \label[FunctionFormat#1]{fun:#1}%
}

\newcommand{\algIndentp}{%
    \SetInd{\dimexpr\skiprule+\algoskipindent}{\dimexpr\skiptext-\algoskipindent}%
    \Indp%
    }
\newcommand{\algIndentm}{%
    \Indm%
    \SetInd{\dimexpr\skiprule-\algoskipindent}{\dimexpr\skiptext+\algoskipindent}%
    }

\SetAlFnt{\scriptsize}
\SetAlgoNlRelativeSize{-1}
\SetAlCapFnt{\footnotesize}
\SetAlCapNameFnt{\footnotesize}
\SetProcNameFnt{\footnotesize}
\SetProcFnt{\footnotesize}
\SetProcArgFnt{\footnotesize}

\newcommand{\centermath}[1]{\[#1\]}

\newlength{\tabwidth}
\newcommand{\tab}[1][\tabwidth]{\hspace*{#1}}
\newenvironment{case_distinction}{\begin{enumerate}[left= 0mm .. \tabwidth,align=left,nosep,itemindent=*]}{\end{enumerate}}

\def\itemrefInternal#1:#2:#3\relax{\Cref{#1:#2}.\ref{item:#2_#3}}
\newcommand*{\itemref}[2]{\itemrefInternal#1:#2\relax}
\def\ItemrefInternal#1:#2:#3\relax{\Cref{#1:#2}.\ref{item:#2_#3}}
\newcommand*{\Itemref}[2]{\ItemrefInternal#1:#2\relax}

\newcounter{uidgencounter}
\newcommand{\itemlabelformat}[2]{%
    \crefformat{#1}{##2\cref*{#2}.\ref*{#1}##3}%
    \Crefformat{#1}{##2\Cref*{#2}.\ref*{#1}##3}%
}
\newcommand{\newItemLabel}[1]{%
    \stepcounter{uidgencounter}
    \edef\templabelname{subpointlabel\theuidgencounter:#1}
    \label{\templabelname}%
    \newcounter{\templabelname}%
    \expandafter\itemlabelformat\expandafter{\templabelname}{#1}%
    \refstepcounter{\templabelname}%
}
\newcommand*{\proofstep}[1]{{\color{blue}(*#1*)}}

\newcommand{\relationsymb}[0]{\mathcal{R}}
\newcommand{\powerset}[1]{\mathcal{P}(#1)}
\newcommand{\collectTupleSymb}[0]{\textbf{Tuple}}
\newcommand{\collectTuple}[2][]{\collectTupleSymb\withOptionalParens{#2}{#1}}

\setupcounter{spec}{s}
\setupcounter{imp}{i}

\makeatletter
\@ifpackageloaded{apxproof}{%
    \newtheoremrep{lemma}{Lemma}[section]
    \newtheoremrep{theorem}[lemma]{Theorem}
    \newtheoremrep{conjecture}[lemma]{Conjecture}
    \newtheoremrep{proposition}[lemma]{Proposition}
    \newtheoremrep{corollary}{Corollary}[lemma]
}{}

\ifcsname lemma\endcsname%
    \relax%
  \else%
    \newtheorem{lemma}{Lemma}%
  \fi%
\ifcsname definition\endcsname%
    \relax%
  \else%
    \theoremstyle{definition}%
    \newtheorem{definition}[lemma]{Definition}%
    \theoremstyle{plain}%
  \fi%
\ifcsname proposition\endcsname%
    \relax%
  \else%
    \newtheorem{proposition}[lemma]{Proposition}%
  \fi%

\theoremstyle{definition}
\newtheorem{example}[lemma]{Example}
\theoremstyle{plain}
\makeatother

\usepackage{wrapfig}

\usepackage{color}

\tikzstyle{Rightarrow}=[/tikz/commutative
diagrams/Rightarrow]
\tikzstyle{Leftrightarrow}=[/tikz/commutative
diagrams/Leftrightarrow]

\usepackage[
backend=biber,
citestyle=numeric-comp,
style=numeric
]{biblatex}
\DeclareSourcemap{
  \maps[datatype=bibtex]{
    \map{
      \step[fieldsource=doi,final]
      \step[fieldset=url,null]
    }
    \map{
      \step[fieldset=urldate,null]
    }
  }
}

\begin{document}
\begin{frontmatter}
\title{Testing Compositionality}
\author[1]{Gijs van Cuyck\texorpdfstring{\corref{cor1}}{}}
\ead{gijs.vancuyck@ru.nl}
\author[1]{Lars van Arragon}
\ead{lars.vanarragon@ru.nl}
\author[1,2]{Jan Tretmans}
\ead{jan.tretmans@ru.nl}

\affiliation[1]{organization={Radboud University, Institute iCIS},
    city={Nijmegen},
    country={The Netherlands}
}
\affiliation[2]{organization={TNO-ESI},
    city={Eindhoven},
    country={The Netherlands}
}
\tnotetext[t1]{This work is part of the project
    \textit{TiCToC - Testing in Times of Continuous Change},
    project nr 17936, part of the research program
    \textit{MasCot - Mastering Complexity},
    which is supported by the Dutch Research Council NWO.
}
\cortext[cor1]{Corresponding author}

\begin{abstract}
Compositionality supports the manipulation of large systems by working on their components.
For model-based testing, this means that large systems can be tested by modelling and testing
their components: passing tests for all components implies passing tests for the whole system. In previous work \cite{vancuyck_CompositionalityModelBasedTesting_2023a}, we defined \emph{mutual acceptance} for specification models
and proved that this property is a sufficient condition for compositionality in model-based testing. In this paper, we present three main algorithms for using mutual acceptance in practice. First, we can verify mutual acceptance on specifications, proving compositionality for all valid implementations. Second, we give a sound and exhaustive model-based testing procedure which checks mutual acceptance
on a specific black-box implementation. The result is that testing the correctness of large systems can be decomposed into testing
the component implementations for \emph{uioco-conformance} to their specifications,
and testing for \textit{environmental conformance} to the specifications of their environment. Finally, we optimise this procedure further by utilizing the constraints imposed by multiple specifications at the same time. These three algorithms together allow picking the most suitable approach for a given situation, trading in more generalizable results for faster runtime by optimising for a specific context as desired.
\end{abstract}
\end{frontmatter}

\section{Introduction}
In recent years the amount of software that is part of everyday life has skyrocketed. From the millions of lines of code powering self driving cars to the complex cyber-physical systems in the semiconductor industry, software has become more integrated and more complex year over year. As our reliance on these software systems keeps growing and their complexity keeps rising we need to ensure that even complex systems can be tested thoroughly. 

Model-Based Testing (MBT) is a state-of-the-art solution to testing, where we employ the strengths of modelling to automatically derive tests. This lowers the testing effort from creating and maintaining test suites to creating and maintaining a model of the System Under Test (SUT). Creating small, simple models is easy, but modelling quickly becomes difficult for larger more complicated systems. Larger models are also more difficult to understand, making it harder to transfer knowledge to new engineers and to check if the model itself is correct. Because modelling the entire system all at once is undesirable, theories have been developed that enable the composition of smaller models in order to deal with the complexity of larger systems. 
This leaves a question of compositionality: if we test all of the individual components with their respective models, do we then know that the composed system also works correctly? 
To answer this question, we developed a theory for compositional model-based testing called mutual acceptance \cite{vancuyck_CompositionalityModelBasedTesting_2023a}, denoted by $\mutuallyaccepts{}$. As long as models are mutually accepting, the results of tests generated for all the individual models are guaranteed to be the same as those of tests generated from the composed system model.

In this paper, we outline three different approaches to compositional testing, each with their own strengths and weaknesses. First, verification of mutual acceptance at the specification level can be performed. Secondly, compositionality can be tested using an implementation and a specification of its environment. As a third option, we can combine the testing for specification conformance with testing against the environment specification. This leads to an optimized approach at the cost of the testing results being less robust to changes in the specifications. We give an algorithm for each of these three approaches, and prove these algorithms correct. Finally, we conclude with a discussion on how compositional testing can be applied in practice.

This paper is an extended version of \cite{vancuyck_TestingCompositionality_2024}. New to this version is the third approach described describe above (\Cref{sec:combining_algorithms}).
We also added a new overview section to outline and compare our three main compositional testing approaches, and a discussion on how and when to apply them in practice.

\paragraph{Overview}
\Cref{sec:background} contains the formal preliminaries, introducing the formal context of labelled transition systems, $\uioco$-conformance, and model-based testing for $\uioco$.
We use these in \Cref{sec:problemStatement} to formalize what compositional model-based testing means, and outline the three main approaches on how we will achieve this. These three approaches: formal verification of mutual acceptance, testing for compositionality with respect to the environment, and combined functionality/environmental  testing, are then discussed in more detail in Sections \ref{sec:environmentalConformance}, \ref{sec:testing-accepting-systems}, and \ref{sec:combining_algorithms}, respectively. We then give a discussion on how and when compositional testing can be applied in practice in \Cref{sec:discussion}. \Cref{sec:related-work,sec:conclusion} contain related work, conclusions, and future work. The detailed versions of all proofs can be found in the appendix, which repeats all lemmas from the paper using the same numbering.

\section{Preliminaries}
\label{sec:background}
This section contains the theoretical background for our work.
It recapitulates definitions from earlier papers \cite{tretmans_TestGenerationInputs_1996,vancuyck_CompositionalityModelBasedTesting_2023a,vanderbijl_CompositionalTestingIoco_2004}.

\subsection{Labelled Transition Systems}

The main formalism used is that of Labelled Transition Systems (LTS); see \Cref{def:LTS}. An LTS has states and labelled transitions between states that model events. An event can be an observable input or output, or an unobservable internal event denoted by $\tau$. Formal specifications, implementations, and test cases are expressed as some kind of labelled transition system.

\begin{definition}
\label{def:LTS}
A \emph{Labelled Transition System} (LTS) is a 5-tuple $\langle Q,I,U,T,q_0 \rangle$ where: $Q$ is a non-empty, countable set of states; $I$ is a countable set of input labels; $U$ is a countable set of output labels, which is disjoint from $I$; $T\subseteq Q \times (I\cup U\cup \{\tau\})\times Q$ is a set of triples, the transition relation; and $q_0 \in Q$ is the initial state. The domain of labelled transition systems is denoted by $\LTS$.
\end{definition}

For LTS $s$, $Q_s$, $I_s$, $U_s$ and $T_s$, indicate its states, inputs, outputs, and transitions, respectively. If $s$ is clear from the context, the subscript is omitted. The shorthand $L_s$ denotes $I_s \cup U_s$. The name of an LTS is sometimes used as shorthand for its starting state.
For technical reasons, we restrict this class to strongly converging and image-finite systems. Strong convergence means that infinite sequences of $\tau$-actions are not allowed to occur. Image-finiteness means that the number of non-deterministically reachable states shall be finite.
In examples, inputs and outputs are given implicitly by prefixing inputs with $?$, and outputs with $!$. The same label can be in an input of one LTS and an output of another.
$\powerset{Q}$ denotes the power set of $Q$, i.e., the set of subsets of $Q$; $L^*$ is the set of all (finite) sequences over elements of $L$.

Reasoning about an LTS uses the concept of traces. A trace is a sequence of observable labels that can occur when walking through an LTS. Some common notation used when describing traces is repeated in \Cref{def:arrowdefs}.

\begin{samepage}
\begin{definition}
\label{def:arrowdefs}
Let $s \in \LTS$; $q, q'\in Q_s$; $\ell\in L_s$; $\sigma \in L_s^*$; $\ell_\tau\in L_s\cup\{\tau\}$; $\sigma_\tau\in (L_s\cup\{\tau\})^*$, where $\epsilon$ denotes the empty sequence of labels.
\[\begin{array}[t]{r@{~~~}c@{~~~}l}
    q\trans{\epsilon}q' & \defeq & q=q' \\
    q\trans{\ell_\tau}q' & \defeq & (q,\ell_\tau,q')\in T_s \\
    q\trans{\ell_\tau \cdot \sigma_\tau}q' & \defeq & \exists q''\in Q_s:~ q \trans{\ell_\tau} q'' \land  q''\trans{\sigma_\tau}q' \\
    q \trans{\sigma_\tau\;\;} & \defeq & \exists q'' \in Q_s :~ q \trans{\sigma_\tau} q'' \\
    q \nottrans{\sigma_\tau\;\;  } & \defeq & \nexists q'' \in Q_s :~ q \trans{\sigma_\tau} q'' \\
    q\Trans{\epsilon}q' & \defeq & \exists \varphi \in \{\tau\}^* :~ q \trans{\varphi} q'\\
    q \Trans{\sigma\cdot\ell}q' & \defeq & \exists q'',q''' \in Q_s :~
      q \Trans{\sigma} q'' \land
      q'' \trans{\ell} q''' \land
      q''' \Trans{\epsilon} q'\\
    q \Trans{\sigma} & \defeq & \exists q'' \in Q_s :~
     q \Trans{\sigma} q'' \\
    q \Nottrans{\sigma} & \defeq & \nexists q'' \in Q_s :~ q \Trans{\sigma} q'' \\
    \traces{s} & \defeq & \traces{{q_0}_s} ~~\defeq~~
    \{\,\sigma \in L^*\:|\:{q_0}_s \Trans{\sigma}\,\} \\
    \end{array}\]
\end{definition}
\end{samepage}

In order to express when an implementation is allowed to be silent while waiting for an input, and when it must produce an output, we use the concept of quiescence. Quiescence, denoted $\delta$, see \Cref{def:delta_general}, can be seen as a special output that indicates the absence of all other outputs. 

\begin{definition}
\label{def:delta_general}
Let $s \in \LTS$.
\begin{enumerate}
\item 
    A state $q_s \in Q_s$ is \emph{quiescent}, denoted by $\delta(q_s)$,
    if it does not have any outgoing output or $\tau$-transitions:
    \[
    \delta(q_s) ~~\defeq~~ \forall x \in U_s \cup \{\tau\} :~ q_s \nottrans{x}
    \]
\item \newItemLabel{def:delta_general}\label{def:delta}
    In an LTS, quiescence is indicated by a self-loop transition labelled with a special label ${\delta}$,~ $\delta \notin L_s$:
    \[
    q_s \trans{\delta} q_s ~~\iff~~ \delta(q_s)
    \]
\item 
    The extension of $s$ with $\delta$-transitions, is denoted $s_\delta$:
    \[
    s_\delta ~~\defeq~~
    \langle\,Q_s,\,I_s,\,U_s\cup\{\delta\},\,T_s\cup\{q\trans{\delta}q\,\setbar\,q\in Q_s,\,\delta(q)\},\,{q_0}_s\,\rangle
    \]
\item 
    The traces of a labelled transition system, extended with $\delta$-transitions, are called \emph{suspension traces}:
    \[
    \straces{s} ~~\defeq~~
    \{\,\sigma \in (L\cup\{\delta\})^*\:|\:s_\delta \Trans{\sigma}\,\} \\
    \]
\end{enumerate}
We use $L^\delta$ as shorthand for $L\cup\{\delta\}$;
likewise for $U^\delta$. We will not always explicitly distinguish between $s$ and $s_\delta$ if the distinction is clear from the context.
\end{definition}

In MBT, we assume that we can always give any input to an implementation. Then, to show conformance to a specification, we also assume that we can represent any real-world implementation as a specific type of LTS. This assumption is called the \emph{testability assumption} \cite{gaudel_TestingCanBe_1995}. For our testing, we assume that implementations can be modelled as \emph{Input-Enabled Transition Systems}. 
 
\begin{definition}
\label{def:iots}
$i\in \LTS$ is called an \emph{Input-Enabled Transition System} if in every reachable state, for every input, its transition relation either has a transition for that input, or reaches with just internal transitions another state that does so:
\[
\forall q \in Q_i,~ \ell \in I_i:~ q \Trans{\ell}
\]
$\IOTS$ denotes the domain of all input-enabled transition systems.
\end{definition}

To relate traces from labelled transition systems with different label sets to each other, we often have to remove (\emph{project}) or replace (\emph{substitute}) certain labels from traces. Additionally, sometimes we have to project an element out from a tuple. We define these concepts in \Cref{def:projsubst}.

\begin{definition}
\label{def:projsubst}
Let $\mathcal{L},\mathcal{L'}$ be any sets of labels;
$\sigma \in \mathcal{L}^*$;
$\ell, \ell', \ell'' \in \mathcal{L}$, with $\ell\neq\ell'$.
\begin{enumerate}
\item \newItemLabel{def:projsubst}\label{def:projection}
    \emph{Projecting} a trace to a restricted set of labels is defined as:
    \[
    \begin{array}[t]{l@{~~~}l@{~~~}l@{~~}l}
        \project{\epsilon}{\mathcal{L'}} & \defeq & \epsilon & \\
        \project{(\sigma\cdot\ell)}{\mathcal{L'}} & \defeq &
        (\project{\sigma}{\mathcal{L'}})\cdot\ell & \text{\ if\ \ } \ell \in \mathcal{L'} \\
        & & \phantom{(}\project{\sigma}{\mathcal{L'}} & \text{\ otherwise}
    \end{array}
    \]
\item\newItemLabel{def:projsubst}\label{def:substitution}
    \emph{Substitution} defines the replacement of some labels by other labels:\[
    \begin{array}[t]{l@{~~~}l@{~~~}l}
        \subst{\epsilon}{\ell}{\ell''} & \defeq & \epsilon \\
        \subst{(\sigma\cdot\ell)}{\ell}{\ell''} & \defeq & \subst{\sigma}{\ell}{\ell''}\cdot\ell''\\
        \subst{(\sigma\cdot\ell')}{\ell}{\ell''} & \defeq & \subst{\sigma}{\ell}{\ell''}\cdot\ell' \\
    \end{array}
    \]
\item \newItemLabel{def:projsubst}\label{def:project_tupple}
    Let $X, Y$ be any sets; $(x,y)\in X\times Y$. Then $\pi_1$ and $\pi_2$ define projecting the left and right element, respectively, of the tuple:
    \[
    \begin{array}[t]{l@{~~~}l@{~~~}l}
            \pi_1((x,y)) & \defeq & x \\
            \pi_2((x,y)) & \defeq & y \\
            \end{array}
    \]
\end{enumerate}
\end{definition}

\subsection{Conformance with \texorpdfstring{\uioco}{Uioco}}

A \emph{conformance relation}, also called \emph{implementation relation},  describes when an implementation is considered correct according to its specification.
We use the $\uioco$-conformance relation \cite{vanderbijl_CompositionalTestingIoco_2004}. This means that implementations are correct if their outputs, including quiescence as special output, in states reachable by $\utraces{}$ are a subset of what the specification allows. The set $\utraces{}$ is a subset of $\straces{}$ with specification traces that do not have \emph{nondeterministic underspecification}. A trace is non-deterministically underspecified if it can reach two states, where in one the next input in the trace is specified while in the other it is not. This is formalized in \Cref{def:uioco,def:outset}. 

\begin{definition}
Let $s\in\LTS$, $q_s \in Q_s$; $Q\subseteq Q_s$ and $\sigma\in {L_s^\delta}^*$.
\[\begin{array}[t]{l@{~~~}c@{~~~}l}
    q_s \after \sigma & \defeq &
    \{\ q_s' \in Q_s \ \setbar\ q_s\Trans{\sigma}q_s'\ \} \\
    \outset{q_s} & \defeq & 
    \{\ x \in U_s^\delta \ \setbar\ q_s \trans{x}\ \} \\
    \outset{Q} & \defeq &
    \bigcup\ \{\ \outset{q} \ \setbar\ q\in Q\ \} \\
    \inset{q_s} & \defeq & \{ \ \ell \in I_s \setbar q_s\Trans{\ell}\ \} \\
    \inset{Q} & \defeq & \bigcap\,\{\ \inset{q} \setbar q\in Q\ \}
\end{array}\]
\label{def:out}
\label{def:outset}
\label{def:after}
\label{def:inset}
\end{definition}

\begin{definition}
\label{def:uioco}
\label{def:straces}
Let $s\in\LTS$, $i \in \IOTS$, with $I_i = I_s$ and $U_i$ = $U_s$:
\[
\begin{array}[t]{l@{~~}l@{~~}l@{~}r@{~~}l}
    \utraces{s} & \defeq & \{\ \sigma \in \straces{s}\ \setbar\ 
    & \forall q_s \in Q_s,\:\sigma_1\cdot{\ell}\cdot\sigma_2 = \sigma,\: \ell \in I: & \\
       & & & s\Trans{\sigma_1}q_s \mbox{\emph{~~implies~~}} q_s\Trans{\ell} & \} \\
    i \uioco s & \defeq & \multicolumn{3}{l}{\forall\sigma \in \utraces{s} :~ \outset{i \after \sigma} \subseteq \outset{s \after \sigma}} \\
\end{array}
\]
\end{definition}

\subsection{Testing for \texorpdfstring{\uioco}{Uioco}}

Testing a system under test (SUT) for $\uioco$-conformance can be performed by exploring the specification on-the-fly, as described in \Cref{alg:test_uioco}  \cite{vanderbijl_CompositionalTestingIoco_2004}.
The algorithm uses an LTS specification and an SUT. The black-box SUT allows the tester to give inputs and to observe outputs or the absence of outputs (quiescence).
The algorithm keeps track of a set $X_s$ of current specification states that initially contains the initial state and the states reachable from the initial state via internal events.
The algorithm offers three nondeterministic choices, each of which leads to a valid $\uioco$-test: (\emph{A}) terminates the algorithm with a $\passState$-verdict; (\emph{B}) continues testing with giving an input to the SUT that is non-deterministically chosen from the current set of inputs $\inset{X_s}$, then the set of current specification states $X_s$ is updated to the set after the chosen input event, and the next nondeterministic pass through the algorithm is made; and (\emph{C}) observes an output or, if no real output is observed, it observes quiescence, then it calculates the next current specification states and the next nondeterministic pass through the algorithm is made, unless the set of states is empty which means that the output $\ell$ is not a specified output, $\ell \notin \outset{X_s}$, so that $\mbox{SUT} \notuioco s$ and the test terminates with the verdict $\failState$. 

\LinesNotNumbered
\begin{algorithm}
\caption{On-the-Fly Testing for $\uioco$}
\label{alg:test_uioco}
    \KwIn{$s\in\LTS$, \\
          \phantom{\textbf{Input: }}SUT allowing to give inputs and observe outputs or quiescence}
    \nl\Assign{$X_s$}{$s_0 \after \epsilon$}
    non-deterministically execute a finite number of the following cases,\\
    until the test either \passes{} or \fails{}:\\
    \textbf{\textit{(A) Stop testing:}}\\
    \Indp
        \nl the test \passes\;
    \Indm
    \textbf{\textit{(B) Give an input $\ell$ to the SUT:}}\\
    \Indp
        \nl choose $\ell \in \inset{X_s}$\;
        \nl give $\ell$ to the SUT\;
        \nl\Assign{$X_s$}{$X_s \after \ell$}
    \Indm
    \textbf{\textit{(C) Observe an output or quiescence $\ell\in U_s^\delta$ from the SUT:}}\\
    \algIndentp
        \nl\Assign{$X_s$}{$X_s \after \ell$}
        \nl\lIf{$X_s=\emptyset$}
            {the test \fails}
    \algIndentm
\end{algorithm}
\LinesNumbered

\Cref{alg:test_uioco} tests for $\uioco$-conformance.
The structure of this algorithm will be the basis for the test generation algorithms for $\eco$-testing in \Cref{sec:testing-accepting-systems}, and for combined $\uioco\!/\!\eco$-testing in \Cref{sec:combining_algorithms}.
However, when proving properties about these algorithms, the informal pseudo-code style of \Cref{alg:test_uioco} does not suffice.
To formally reason about testing, generated test cases, and their soundness and exhaustiveness, we will use a more precise, functional style to present test-generation algorithms, without side effects, where the algorithm returns formally defined test cases, and where test execution is also formalized. This style of formalization of testing is analogous to the formal approach in \cite{tretmans_TestGenerationInputs_1996,frantzen_ModelBasedTestingEnvironmental_2007} and will be used in the proofs and proof-sketch of subsequent sections. It requires the formal definition of a test case, test execution, soundness, and exhaustiveness, which are given in the next sub-section  \ref{sec:formaltesting}.
 
\subsection{Test Formalization}
\label{sec:formaltesting}

A test case (\Cref{def:testcase}) is a special, tree-structured LTS, with finite, deterministic behaviour, and with specific final states $\passState$ and $\failState$. A test case specifies a test to be executed on the SUT, by specifying the inputs that a tester will give to the SUT and the outputs, or quiescence, that the tester shall observe. When in a final state of the test case, the tester stops and a verdict is assigned, either $\passState$ or $\failState$. The inputs of a test case are the outputs of the SUT, and vice versa. Moreover, every non-final test-case state is input enabled for the outputs of the SUT. A test case has a special label $\theta$, which represents observing quiescence from the SUT.
Informally, $\theta$ can be thought of as a time-out: if no output is observed within the time-out period, quiescence is observed.

\begin{definition}
\label{def:testcase}
\label{item:testcase_no_tau}
\label{item:pass_fail_state}
Let $s\in\LTS$, and let $\theta$ be a fresh label such that $\theta \notin L_s\cup\{\tau,\delta\}$.
\begin{enumerate}
\item
The set $\mathcal{T\!\,E}_{\!s}$ of \emph{test expressions} for $s$ is the smallest set satisfying:
\[ \begin{array}[t]{lr@{~}l@{~}l}
& \passState & \in & \mathcal{T\!\,E}_{\!s} \\
& \failState & \in & \mathcal{T\!\,E}_{\!s} \\
\multicolumn{2}{l}{\mbox{for each $A \subseteq L_s\cup\{\theta\}$}:} & & \\
& \Bsum\,\{\,a\,\Bcomp\,t_a\,\setbar\,a \in A,\, t_a \in \mathcal{T\!\,E}_{\!s} \,\} & \in & \mathcal{T\!\,E}_{\!s} \\
\end{array} \]
\item
The following additional notation is used for test expressions.
For a singleton $\Bsum\!$-set, we use the notation $a\Bcomp t$,\ 
i.e, $a\Bcomp t$ $\,=\,$ $\Bsum\!\{a\Bcomp t\}$.
Moreover, for two test expressions $t_1 = \Bsum \mathcal{T}_1$ and $t_2 = \Bsum \mathcal{T}_2$, we use $t_1 \Bchoice t_2$ for the union of their $\Bsum\!$-sets, i.e.,\ $t_1 \Bchoice t_2$ $\,=\,$ $\Bsum (\mathcal{T}_1 \cup \mathcal{T}_2)$. Combining these two notations, we have $a_1\Bcomp t_1 \Bchoice a_2\Bcomp t_2$ $\,=\,$ $\Bsum\!\,\{a_1\Bcomp t_1,\,a_2\Bcomp t_2\}$.
\item
The semantics of a test expression $t \in \mathcal{T\!\,E}_{\!s}$ is
the LTS $\langle\,\mathcal{T\!\,E}_{\!s},\,U_s\cup\{\theta\},\,I_s,\,T,\,t\,\rangle$,
where $T$ is defined by the following inference rule:
\[ \begin{array}{r@{~~~}c@{~~~}l}
(a\,\Bcomp\,t_a) \in \mathcal{T} & \vdash & \Bsum\mathcal{T}\ \trans{a}\ t_a \\ 
\end{array} \]
\item
A \emph{test case} for $s$ is a test expression where each semantic state unequal to $\passState$ or $\failState$, has transitions for all outputs from the SUT, \emph{and} either one input $\ell \in I_s$, or $\theta$, i.e., if a state in the test-expression LTS has the form $\Bsum \mathcal{T}$ then
\[ \begin{array}[t]{r@{~~}l@{~}l@{~}ll}
\mbox{either} & \mathcal{T} & = &
\{\,a\,\Bcomp\,t_a\ \setbar\ a \in U_s\cup\{\ell\}\,\}, & \mbox{for some $\ell \in I_s$}, \\
\mbox{or} & \mathcal{T} & = &
\{\,a\,\Bcomp\,t_a\ \setbar\ a \in U_s\cup\{\theta\}\,\} & \\ 
\end{array} \]
\item
The domain of all test cases for $s$ is denoted as $\TTS$,
and a collection of test cases is called a \emph{test suite}: $TS\subseteq\TTS$.
\end{enumerate}
\end{definition}

Test execution consists of composing a test case with the SUT so that they can run together and synchronize. The test case is responsible for giving inputs to the SUT, and the outputs and quiescence of the SUT are observed by the test case. This parallel execution continues until a terminal state of the test case is reached, which indicates the verdict, either $\passState$ or $\failState$. 

Formally, test execution is expressed as a composition operator $\testexec$ on a test case $t \in \TTS$ and an implementation $i \in \IOTS$ modelling the SUT, which gives a composed LTS. The traces of this LTS are the possible test runs. Because of possible nondeterminism in the SUT, test execution may lead to different traces, i.e., it may be the case that testing the same SUT with the same test case leads to a different verdict. The SUT passes a test case if and only if all its test runs lead to a $\passState$-state in the test case. All this is reflected in \Cref{def:testexec_general}.

\begin{definition}
\label{def:testexec_general}
Let $t\in \TTS$, $i\in \IOTS$.
\begin{enumerate}
\item \newItemLabel{def:testexec_general}\label{def:testexec}
    \emph{Test execution} $t\testexec i\in \LTS$ is defined as follows: $Q_{t\testexec i} = Q_t\times Q_i$, $U_{t\testexec i} = U_i \cup L_t$, $I_{t\testexec i} = \emptyset $. The transition function $T_{t\testexec i}$ is defined as the minimal set satisfying the following inference rules  (where $q_t,q_t'\in Q_t$, $q_i, q_i'\in Q_i$):
    \[\begin{array}[t]{*3{l@{~~~}}c@{~~~}l}

        &q_i \trans{\ell} q_i', &\ell\in (U_i\cup \{\tau\})\setminus L_t  & \vdash& q_t \testexec q_i \trans{\ell} q_t \testexec q_i'\\

        q_t \trans{\ell} q_t', & &\ell\in L_t\setminus (L_i\cup\{\theta\}) &\vdash & q_t \testexec q_i \trans{\ell} q_t' \testexec q_i\\
        
        q_t \trans{\ell} q_t',& q_i \trans{\ell} q_i', &
        \ell \in L_i \cap L_t &
        \vdash &
        q_t \testexec q_i \trans{\ell} q_t' \testexec q_i' \\
        
        q_t \trans{\theta} q_t',& q_i\trans{\delta}q_i' && \vdash & q_t\testexec q_i \trans{\theta} q_t'\testexec q_i'\\
    \end{array}\]

\item \newItemLabel{def:testexec_general}\label{def:testrun}
    $\sigma\in L_{t\testexec i}^*$ is a \textit{test run} for $t$ on $i$ if it reaches a terminal state:
    \[\exists q_i\in Q_i:\quad t\testexec i \Trans{\sigma} \passState\testexec q_i\quad \lor\quad t\testexec i \Trans{\sigma} \failState\testexec q_i \]
    $\testruns{t\testexec i}$ represents the set of all test runs for $t$ on $i$
    
\item 
    $i \passes t$ if none of the test runs of $t$ for $i$ reaches the $\failState$ state: 
    \[i \passes t \ \defeq\ \forall \sigma\in \testruns{t\testexec i},\  \forall q_i\in Q_i:\ t\testexec i \Nottrans{\sigma} \failState\testexec q_i \]

\item 
    $i \passes TS\subseteq\TTS$ if all the tests in $TS$ pass:
    \[i \passes TS\ \defeq\ \forall t \in TS:\, i \passes t\]

\item 
    if $i$ does not pass a test case or test suite, it $\fails$.
    \end{enumerate}
\end{definition}

\begin{toappendix}
    \begin{lemmarep}
        \label{lem:L_testexec}
        take $t\in \TTS[L_i,e]$\\
        $L_{t\testexec i}= L_i \cup L_e \cup \theta$
    \end{lemmarep}
    \begin{proof}
        $L_{t\testexec i}= $\\
\proofstep{\Cref{def:testexec}}\\
$\emptyset \cup U_i \cup L_t = $\\
\proofstep{\Cref{lem:L_testcase}}\\
$ U_i \cup L_e \cup (I_i\setminus L_e) \cup \{\theta\} = $\\
$L_i \cup L_e \cup \{\theta\}$.
    \end{proof}
\end{toappendix}

We can now define what it means for a test suite to be correct in \Cref{def:complete}. A test suite is \textit{sound} with respect to a conformance relation $\mathbf{conf}\subseteq\IOTS\times\LTS$ and a specification $s$, iff it only fails for implementations that are not conforming to $s$. A test suite is \textit{exhaustive} if for every implementation that is not conforming to $s$, there is a test in the test suite that will fail for this implementation. In practice, an exhaustive test suite is usually not feasible. It would consist of an infinite number of test cases. Any practical application of testing will have to do some kind of test selection, i.e., selecting a finite test suite from the infinite, exhaustive one. It can nevertheless be useful to reason about exhaustive test suites, such that for every fault it is at least possible to select a test which will find that fault. Even randomly selecting tests from such an infinite test suite has been found as effective at detecting faults as popular finite test suites in the field of model learning \cite{garhewal_ExperimentalEvaluationConformance_2023}. 

\begin{definition}
    \label{def:complete}
    \label{def:sound}
    \label{def:exhaustive}
    Let $s \in\LTS$,\, $\mathbf{conf} \subseteq \IOTS \times \LTS$,\,  $TS\subseteq\TTS$:
    \[ \begin{array}[t]{l@{~~}l@{~~}l@{~}l@{~}c@{~}l}
        \mbox{$TS$ is \emph{sound} for $s$ and $\mathbf{conf}$} &
        \defeq & \forall i\in \IOTS: &
        i \mathrel{\mathbf{conf}} s & \implies & i \passes TS \\
        \mbox{$TS$ is \emph{exhaustive} for $s$ and $\mathbf{conf}$} &
        \defeq & \forall i\in \IOTS: &
        i \mathrel{\mathbf{conf}} s & \impliedby & i \passes TS \\
    \end{array} \]
\end{definition}

Using the above formalization of testing, \Cref{alg:test_uioco} can be rephrased in a more formal, functional style. This style is more amenable to formal reasoning and proving, so that soundness and exhaustiveness of the $\uioco$-test generation algorithm can be proven \cite{tretmans_TestGenerationInputs_1996,tretmans_ModelBasedTesting_2008,man:frantzen_SymbolicConformance_2024}.
This functional style of algorithms will be used in the proof sketches in subsequent sections.

\begin{toappendix}
    \subsection{Previous Results}
    \label{sec:old_lemmas}
    This section briefly repeats the theoretical results from \cite{vancuyck_CompositionalityModelBasedTesting_2023a}, which the lemmas and theorems in this paper build on. Several of them have been modified slightly and have a revised proof given here. For the proof of unchanged lemmas we refer back to \cite{man:vancuyck_CompositionalityModelBasedTesting_2023}. 

\begin{lemma}
\label{lem:parcomp_base_properties}
let $s,e \in \LTS$ be $\composable$, $q_s, q_s'\in Q_s, q_e, q_e'\in Q_e$

\begin{enumerate}
    \item \label{item:parcomp_base_properties_step_left}
    \[\forall \ell \in L_s \setminus L_e : q_s \parcomp q_e \xrightarrow{\ell} q_s' \parcomp q_e \iff q_s \xrightarrow{\ell} q_s'\]

    \item \label{item:parcomp_base_properties_step_right}
    \[\forall \ell \in L_e \setminus L_s : q_s \parcomp q_e \xrightarrow{\ell} q_s \parcomp q_e' \iff q_e \xrightarrow{\ell} q_e'\]

    \item \label{item:parcomp_base_properties_step_tau}
    \[q_s \parcomp q_e \xrightarrow{\tau} q_s' \parcomp q_e' \iff (q_s \xrightarrow{\tau} q_s' \land q_e = q_e')\lor (q_e \xrightarrow{\tau} q_e' \land q_s = q_s')\]

    \item \label{item:parcomp_base_properties_step_both_label}
    \begin{align*}
        \forall \ell \in (L_s \cap L_e) : q_s \parcomp q_e \xrightarrow{\ell} q_s' \parcomp q_e'\ \iff q_s \xrightarrow{\ell} q_s' \land q_e \xrightarrow{\ell} q_e'
    \end{align*}
    
    \item \label{item:parcomp_base_properties_both_delta}
    \begin{align*}
        &q_s \parcomp q_e \xrightarrow{\delta} q_s' \parcomp q_e' \land s \mutuallyaccepts{} e \;\land&& \\
        &(\exists \sigma \in \utraces{s \parcomp e}: s \parcomp e \xRightarrow{\sigma} q_s \parcomp q_e) \implies&& \\ 
        &q_s \xrightarrow{\delta} q_s' \land q_e \xrightarrow{\delta} q_e'&&
    \end{align*}

    \item \label{item:parcomp_base_properties_both_to_parcomp}
    \[q_s \xrightarrow{\delta} q_s' \land q_e \xrightarrow{\delta} q_e'  \implies q_s \parcomp q_e \xrightarrow{\delta} q_s' \parcomp q_e'\]
\end{enumerate}
\end{lemma}
\begin{proof}
    This is an adapted lemma and proof from \cite{vancuyck_CompositionalityModelBasedTesting_2023a}. This version splits the lemma up into more sub-points for better reuse in further lemmas.
~\begin{enumerate}

    \item Take $\ell\in L_s \setminus L_e$
    \begin{case_distinction}
        \item[$\Longrightarrow$:]\ \\
        $q_s \parcomp q_e \trans{\ell} q_s' \parcomp q_e'$\\
        \proofstep{Only possible through first case of definition $\parcomp_T$, \cref{def:parcomp}}\\
        $q_s \trans{\ell} q_s'$
        \item[$\Longleftarrow$:]\ \\
        $q_s \trans{\ell} q_s'$\\
        \proofstep{First case of definition $\parcomp_T$, \cref{def:parcomp}}\\
        $\forall q_e \in Q_e: q_s \parcomp q_e \trans{\ell} q_s' \parcomp q_e$
    \end{case_distinction}
    
    \item Analogous to 1.
    
    \item Analogous to 1.

    \item Analogous to 1.
    
    \item Take $\sigma \in \utraces{s\parcomp e}$\\
    
    $q_s \parcomp q_e \trans{\delta} q_s' \parcomp q_e' \land s \mutuallyaccepts{} e \land s \parcomp e \Trans{\sigma} q_s \parcomp q_e$\\
    \proofstep{\Cref{def:delta}: $\delta$}\\
    $q_s \parcomp q_e \trans{\delta} q_s' \parcomp q_e' \land s \mutuallyaccepts{} e \land s \parcomp e \Trans{\sigma} q_s \parcomp q_e \land q_s = q_s' \land q_e = q_e'$\\
    \proofstep{\Cref{def:delta}: $\delta$ + remember $q_s = q_s' \land q_e = q_e'$ as subproof, dropped here for brevity }\\
    $\forall \ell \in U_s \cup U_e \cup \{\tau\}: q_s \parcomp q_e \nottrans{\ell} \land s \mutuallyaccepts{} e \land s \parcomp e \Trans{\sigma} q_s \parcomp q_e$\\
    \proofstep{\Cref{def:parcomp}: $\parcomp$}\\
    $\forall \ell \in (U_s \setminus L_e) \cup \{\tau\}: q_s \nottrans{\ell} \land$\\
    $\forall \ell \in (U_e \setminus L_s) \cup \{\tau\}: q_e \nottrans{\ell} \land$\\
    $\forall \ell \in (U_s \cap L_e) \cup (U_e \cap L_s): q_s \nottrans{\ell} \lor\; q_e \nottrans{\ell}\land$\\
    $s \mutuallyaccepts{} e\; \land$\\
    $s \parcomp e \Trans{\sigma} q_s \parcomp q_e$\\
    \proofstep{Apply \cref{def:mutually_accepts,def:accepting}: mutually accepts and accepts}\\
    $\forall \ell \in (U_s \setminus L_e) \cup \{\tau\}: q_s \nottrans{\ell} \land$\\
    $\forall \ell \in (U_e \setminus L_s) \cup \{\tau\}: q_e \nottrans{\ell} \land$\\
    $\forall \ell \in (U_s \cap L_e) \cup (U_e \cap L_s): q_s \nottrans{\ell} \lor\; q_e \nottrans{\ell}\land$\\
    $(\forall \sigma' \in \utraces{s \parcomp e}, q_s' \in Q_s, q_e' \in Q_e: s \parcomp e \Trans{\sigma'} q_s' \parcomp q_e' \implies$\\
    $out(q_s') \cap I_e \subseteq in(q_e') \cap U_s\; \land$\\ 
    $out(q_e') \cap I_s \subseteq in(q_s') \cap U_e )\; \land $\\
    $s \parcomp e \Trans{\sigma} q_s \parcomp q_e$\\
    \proofstep{$\forall$ elimination + $\implies$ elimination}\\
     $\forall \ell \in (U_s \setminus L_e) \cup \{\tau\}: q_s \nottrans{\ell} \land$\\
    $\forall \ell \in (U_e \setminus L_s) \cup \{\tau\}: q_e \nottrans{\ell} \land$\\
    $\forall \ell \in(U_s \cap L_e) \cup (U_e \cap L_s): q_s \nottrans{\ell} \lor\; q_e \nottrans{\ell} \land$\\
    $ out(q_s) \cap I_e \subseteq in(q_e) \cap U_s \;\land$\\ 
    $ out(q_e) \cap I_s \subseteq in(q_s) \cap U_e$ \\
    \proofstep{\Cref{def:outset,def:inset}: $out$ and $in$}\\
    $\forall \ell \in (U_s \setminus L_e) \cup \{\tau\}: q_s \nottrans{\ell} \land$\\
    $\forall \ell \in (U_e \setminus L_s) \cup \{\tau\}: q_e \nottrans{\ell} \land$\\
    $\forall \ell \in(U_s \cap L_e) \cup (U_e \cap L_s): q_s \nottrans{\ell} \lor\; q_e \nottrans{\ell}\land$\\
    $\forall \ell \in U_s \cap I_e: q_s \trans{\ell} \implies  q_e \Trans{\ell} \land$\\ 
    $\forall \ell \in U_e \cap I_s: q_e \trans{\ell} \implies  q_s \Trans{\ell}$\\
    \proofstep{$q_s \nottrans{\tau} \land\; q_e\nottrans{\tau}$}\\
    $\forall \ell \in (U_s \setminus L_e) \cup \{\tau\}: q_s \nottrans{\ell} \land$\\
    $\forall \ell \in (U_e \setminus L_s) \cup \{\tau\}: q_e \nottrans{\ell} \land$\\
    $\forall \ell \in (U_s \cap L_e) \cup (U_e \cap L_s): q_s \nottrans{\ell} \lor\; q_e \nottrans{\ell}\land$\\
    $\forall \ell \in U_s \cap I_e : q_s \trans{\ell} \implies q_e \trans{\ell} \land$\\
    $\forall \ell \in U_e \cap I_s : q_e \trans{\ell} \implies q_s \trans{\ell}$\\
    \proofstep{$(A \implies B) \iff (\neg B \implies \neg A)$}\\
    $\forall \ell \in (U_s \setminus L_e) \cup \{\tau\}: q_s \nottrans{\ell} \land$\\
    $\forall \ell \in (U_e \setminus L_s) \cup \{\tau\}: q_e \nottrans{\ell} \land$\\
    $\forall \ell \in(U_s \cap L_e) \cup (U_e \cap L_s): q_s \nottrans{\ell} \lor\; q_e \nottrans{\ell}\land$\\
    $\forall \ell \in U_s \cap I_e : q_e \nottrans{\ell} \implies q_s \nottrans{\ell}\land$\\
    $\forall \ell \in U_e \cap I_s : q_s \nottrans{\ell} \implies q_e \nottrans{\ell}$\\
    \proofstep{\Cref{def:composable}: $U_s \cap U_e = \emptyset$}\\
    $\forall \ell \in (U_s \setminus L_e) \cup \{\tau\}: q_s \nottrans{\ell} \land$\\
    $\forall \ell \in (U_e \setminus L_s) \cup \{\tau\}: q_e \nottrans{\ell} \land$\\
    $\forall \ell \in (U_s \cap L_e) \cup (U_e \cap L_s): q_s \nottrans{\ell} \lor\; q_e \nottrans{\ell}\land$\\
    $\forall \ell \in U_s \cap L_e : q_e \nottrans{\ell} \implies q_s \nottrans{\ell}\land$\\
    $\forall \ell \in U_e \cap L_s : q_s \nottrans{\ell} \implies q_e \nottrans{\ell}$\\
    \proofstep{$(A \lor B) \land (A \implies B) \implies B$}\\
    $\forall \ell \in (U_s \setminus L_e) \cup \{\tau\}: q_s \nottrans{\ell} \land$\\
    $\forall \ell \in (U_e \setminus L_s) \cup \{\tau\}: q_e \nottrans{\ell} \land$\\
    $\forall \ell \in U_s \cap L_e :  q_s \nottrans{\ell}\land$\\
    $\forall \ell \in U_e \cap L_s :  q_e \nottrans{\ell}$\\
    \proofstep{$(X \setminus Y) \cup (X \cap Y) = X$}\\
    $\forall \ell \in U_s \cup \{\tau\} : q_s \nottrans{\ell} \land$\\ $\forall \ell \in U_e \cup \{\tau\} : q_e \nottrans{\ell}$\\
    \proofstep{\Cref{def:delta}: $\delta$}\\
    $q_s \trans{\delta} q_s \land q_e \trans{\delta} q_e$\\
    \proofstep{Apply subproof: $q_s = q_s' \land q_e = q_e'$}\\
    $q_s \trans{\delta} q_s' \land q_e \trans{\delta} q_e'$

    \item
    $q_s \trans{\delta} q_s' \land q_e \trans{\delta} q_e'$\\
    \proofstep{\Cref{def:delta}: $\delta$}\\
    $\forall \ell \in U_s \cup \{\tau\} : q_s \nottrans{\ell} \land\; \forall \ell \in U_e \cup \{\tau\} : q_e \nottrans{\ell} \land \; q_s = q_s' \land q_e = q_e'$\\
    \proofstep{\Cref{def:parcomp}: $\parcomp$}\\
    $\forall \ell \in U_s \cup U_e \cup \{\tau\} : q_s \parcomp q_e \nottrans{\ell} \land\; q_s = q_s' \land q_e = q_e'$\\
    \proofstep{\Cref{def:delta}: $\delta$}\\
    $q_s \parcomp q_e \trans{\delta} q_s \parcomp q_e \land q_s = q_s' \land q_e = q_e'$\\
    \proofstep{Rewrite using equalities}\\
    $q_s \parcomp q_e \trans{\delta} q_s' \parcomp q_e'$\\
\end{enumerate}
\end{proof}

\begin{lemma}
\label{prop:general_properties}
Let $s\in \LTS, \sigma,\sigma_1, \sigma_2 \in (L_s^\delta)^*$, $\ell \in L_s^\delta$, $q_s,q_s' \in Q_s$
    \begin{enumerate}
        \item \newItemLabel{prop:general_properties} \label{item:small_trans_transitive}
        The transition relation $\trans{}$ is transitive:
        \[(\exists q_s'' \in Q_s: q_s \trans{\sigma_1} q_s'' \land q_s'' \trans{\sigma_2} q_s') \iff q_s \trans{\sigma_1\cdot\sigma_2} q_s'\]

        \item  \newItemLabel{prop:general_properties}\label{item:trans_transitive}

        The transition relation $\Trans{}$ is transitive:
        \[(\exists q_s'' \in Q_s: q_s \Trans{\sigma_1} q_s'' \land q_s'' \Trans{\sigma_2} q_s') \iff q_s \Trans{\sigma_1\cdot\sigma_2} q_s'\]

        \item \newItemLabel{prop:general_properties}\label{item:utrace_prefix_closed}

        $\utraces{}$ is prefix closed:
        \[\sigma\cdot \ell \in \utraces{S} \implies \sigma \in \utraces{S}\]
    \end{enumerate}
\end{lemma}

\begin{proof}~
    \begin{enumerate}
    \item 
        We do a direct proof of the bi-implication: all proof steps are bi-implications. The proof follows by induction on $\sigma=\sigma_1\cdot\sigma_2$. Specifically, we do induction using the existence of the first label of $\sigma_1$, instead of the last label of $\sigma_2$ as usual, to match the definition of $\trans{}$\\
        \begin{case_distinction}
            \item[Base Case $\sigma=\epsilon$]\ \\
                $\exists q_s'' \in Q_s: q_s \trans{\epsilon} q_s'' \land q_s'' \trans{\epsilon} q_s'$\\
                \proofstep{\cref{def:arrowdefs}: $\trans{\epsilon}$}\\
                $\exists q_s'' \in Q_s: q_s=q_s''=q_s$\\
                \proofstep{\cref{def:arrowdefs}: $\trans{\epsilon}$}\\
                $q_s \trans{\epsilon} q_s'$
            \item[Inductive step: $\sigma=\ell'\cdot\sigma_1'\cdot\sigma_2$]\ \\
                IH: \[\forall q_s\in Q_s: (\exists q_s'' \in Q_s: q_s \trans{\sigma_1'} q_s'' \land q_s'' \trans{\sigma_2} q_s') \iff q_s \trans{\sigma_1'\cdot\sigma_2} q_s'\]
    
                $\exists q_s'' \in Q_s: q_s \trans{\ell\cdot\sigma_1'} q_s'' \land q_s'' \trans{\sigma_2} q_s'$\\
                \proofstep{\cref{def:arrowdefs}: $\trans{\ell\cdot\sigma_1'}$}\\
                $\exists q_s'',q_s''' \in Q_s: q_s \trans{\ell} q_s'''\trans{\sigma_1'} q_s'' \land q_s'' \trans{\sigma_2} q_s'$\\
                \proofstep{Apply IH}\\
                $\exists,q_s''' \in Q_s: q_s \trans{\ell} q_s'''\trans{\sigma_1'\cdot\sigma_2} q_s'$\\
                \proofstep{\cref{def:arrowdefs}: $\trans{\ell\cdot\sigma_1'\cdot\sigma_2}$}\\
                $q_s \trans{\ell\cdot\sigma_1'\cdot\sigma_2} q_s'$\\
        \end{case_distinction}
        
    \item We do a direct proof of the bi-implication: all proof steps are bi-implications. The proof follows by induction on $\sigma=\sigma_1\cdot\sigma_2$\\
    \begin{case_distinction}
        \item[Base Case $\sigma=\epsilon$]\ \\
            $\exists q_s'' \in Q_s: q_s \Trans{\epsilon} q_s'' \land q_s'' \Trans{\epsilon} q_s'$\\
            \proofstep{\cref{def:arrowdefs}: $\Trans{\epsilon}$}\\
            $\exists q_s'' \in Q_s, \sigma_\tau, \sigma_\tau' \in \{\tau\}^*: q_s \trans{\sigma_\tau} q_s'' \land q_s'' \trans{\sigma_\tau'} q_s'$\\
            \proofstep{\cref{item:small_trans_transitive}}\\
            $\exists q_s'' \in Q_s, \sigma_\tau, \sigma_\tau' \in \{\tau\}^*: q_s \trans{\sigma_\tau\cdot\sigma_\tau'} q_s'$\\
            \proofstep{\cref{def:arrowdefs}: $\Trans{\epsilon}$}\\
            $q_s \Trans{\sigma_\tau\cdot\sigma_\tau'} q_s'$
            
        \item[Inductive step: $\sigma=\sigma_1\cdot\sigma_2'\cdot\ell$]\ \\
        IH:         \[\forall q_s'\in Q_s: (\exists q_s'' \in Q_s: q_s \Trans{\sigma_1} q_s'' \land q_s'' \Trans{\sigma_2'} q_s') \iff q_s \Trans{\sigma_1\cdot\sigma_2'} q_s'\]

        $\exists q_s'' \in Q_s: q_s \Trans{\sigma_1} q_s'' \land q_s'' \Trans{\sigma_2'\cdot\ell} q_s'$\\
        \proofstep{\cref{def:arrowdefs}: $\Trans{\sigma_2'\cdot\ell}$}\\
        $\exists q_s'', q_s''',q_s'''' \in Q_s: q_s \Trans{\sigma_1} q_s'' \land q_s'' \Trans{\sigma_2'} q_s''' \trans{\ell} q_s'''' \Trans{\epsilon} q_s'$\\
        \proofstep{Apply IH}\\
        $(\exists q_s''',q_s'''' \in Q_s: q_s \Trans{\sigma_1\cdot\sigma_2'} q_s''' \trans{\ell} q_s'''' \Trans{\epsilon} q_s')$\\
        \proofstep{\cref{def:arrowdefs}: $\Trans{\sigma_1\cdot\sigma_2'\cdot\ell}$}\\        
        $q_s \Trans{\sigma_1\cdot\sigma_2'\cdot\ell} q_s'$

    \end{case_distinction}
    \item Follows directly from the definition of \utraces{}  (\cref{def:uioco}) and \cref{item:trans_transitive}
\end{enumerate}
    
\end{proof}

\begin{lemmarep}
let $s,e$ be $\composable$ $\LTS$, $q_s,q_s'\in Q_s$, $q_e,q_e' \in Q_e, \sigma \in  {L_{s\parcomp e}^{\delta}}^{*}$
\[q_s \xRightarrow{\project{\sigma}{L_s^\delta}} q_s' \land q_e \xRightarrow{\project{\sigma}{L_e^\delta}} q_e' \implies q_s \parcomp q_e \xRightarrow{\sigma} q_s' \parcomp q_e'\]
\label{lem:project_from_parcomp_light}
\end{lemmarep}
\begin{proof}
Lemma corresponds to Lemma 9 from \cite{man:vancuyck_CompositionalityModelBasedTesting_2023}. Proof not repeated here. 
\end{proof}

\begin{lemma}
let $i_s,i_e$ be $\composable$ $\IOTS$, $i_s'\in Q_{i_s}$, $i_e' \in Q_{i_e}, \sigma \in {L_{i_s\parcomp i_e}^\delta}^*$.
\centermath{i_s \parcomp i_e \xRightarrow{\sigma} i_s' \parcomp i_e' \ \iff\ 
i_s \xRightarrow{\project{\sigma}{L_{i_s}^\delta}} i_s' \:\land\:
i_e \xRightarrow{\project{\sigma}{L_{i_e}^\delta}} i_e'}
\label{lem:project_from_parcomp_IOTS}
\end{lemma}
\begin{proof}
Lemma corresponds to Lemma 2 from \cite{vancuyck_CompositionalityModelBasedTesting_2023a}. Proof not repeated here. 
\end{proof}

\begin{lemmarep}
let $s,e \in \LTS$ be $\composable$, $q_s, q_s' \in Q_s$, $q_e, q_e' \in Q_e, \sigma \in {L_{s\parcomp e}}^*$.
\[q_s \parcomp q_e \xRightarrow{\sigma} q_s' \parcomp q_e' \ \iff\ 
q_s \xRightarrow{\project{\sigma}{L_{s}}} q_s' \land
q_e \xRightarrow{\project{\sigma}{L_{e}}} q_e' \]
\label{lem:project_from_parcomp_no_delta}
\end{lemmarep}
\begin{proof}
\begin{case_distinction}
\item[($\implies$):] Proof by induction on $\sigma$. 
    \begin{case_distinction}
        \item[Base case:] $\sigma = \epsilon$. Let $\sigma_\tau,\sigma_\tau', \sigma_\tau'' \in \tau^*$\\
            $q_s \parcomp q_e \Trans{\epsilon} q_s' \parcomp q_e'$\\
            \proofstep{\Cref{def:arrowdefs}: $\Trans{\epsilon}$}\\
            $q_s \parcomp q_e \trans{\sigma_\tau} q_s' \parcomp q_e'$\\
            \proofstep{\Itemref{lem:parcomp_base_properties}{step_tau}}\\
            $q_s \trans{\sigma_\tau'} q_s' \land q_e \trans{\sigma_\tau''} q_e'$\\
            \proofstep{\Cref{def:arrowdefs}: $\Trans{\epsilon}$}\\
            $q_s \Trans{\epsilon} q_s' \land q_e \Trans{\epsilon} q_e'$\\
            \proofstep{\Cref{def:projection}: $\projectop$}\\
            $q_s \Trans{\project{\epsilon}{L_s}} q_s' \land q_e \Trans{\project{\epsilon}{L_e}} q_e'$\\
        
        \item[Induction step:] Assume the proposition holds for $\sigma'\in  \utraces{s \parcomp e}$.\\
        To proof: the proposition holds for $\sigma$, where $\sigma =\sigma' \cdot \ell$ with $\ell \in L_{s \parcomp e}$. Assume $\sigma \in  \utraces{s \parcomp e}$, otherwise the proposition trivially holds. This is divided into three cases based on $\ell$:
        \begin{case_distinction}
            \item[$\ell \in L_s \setminus L_e$:]\ \\
            $q_s \parcomp q_e \Trans{\sigma' \cdot \ell} q_s' \parcomp q_e'$\\
            \proofstep{\Cref{def:arrowdefs}: $\Trans{\sigma}$}\\
            $\exists q_s'', q_s''' \in Q_s, q_e'',q_e''' \in Q_e :
            q_s \parcomp q_e \Trans{\sigma'} q_s'' \parcomp q_e''\;\land$\\\tab
            $q_s'' \parcomp q_e'' \trans{\ell} q_s''' \parcomp q_e''' \land
            q_s''' \parcomp q_e''' \Trans{\epsilon} q_s' \parcomp q_e' $\\
            \proofstep{Apply base case}\\
            $\exists q_s'', q_s''' \in Q_s, q_e'',q_e''' \in Q_e :
            q_s \parcomp q_e \Trans{\sigma'} q_s'' \parcomp q_e''\;\land$\\\tab
            $q_s'' \parcomp q_e'' \trans{\ell} q_s''' \parcomp q_e''' \land
            q_s''' \Trans{\epsilon} q_s' \land
            q_e''' \Trans{\epsilon} q_e'$\\
            \proofstep{\Cref{def:parcomp}: $q_e'' = q_e'''$.
            From induction on structure of $T_{s\parcomp e}$ with $a \in L_s \setminus L_e$.}\\
            $\exists q_s'', q_s''' \in Q_s, q_e'' \in Q_e :
            q_s \parcomp q_e \Trans{\sigma'} q_s'' \parcomp q_e''\;\land$\\\tab
            $q_s'' \parcomp q_e'' \trans{\ell} q_s''' \parcomp q_e'' \land
            q_s''' \Trans{\epsilon} q_s' \land
            q_e'' \Trans{\epsilon} q_e'$\\
            \proofstep{\Itemref{lem:parcomp_base_properties}{step_left}}\\
            $\exists q_s'', q_s''' \in Q_s, q_e'' \in Q_e :
            q_s \parcomp q_e \Trans{\sigma'} q_s'' \parcomp q_e''\;\land$\\\tab
            $q_s''  \trans{\ell} q_s''' \land
            q_s''' \Trans{\epsilon} q_s' \land
            q_e'' \Trans{\epsilon} q_e'$\\
            \proofstep{\Cref{def:arrowdefs}: $\Trans{\ell}$}\\
            $\exists q_s'' \in Q_s, q_e'' \in Q_e :
            q_s \parcomp q_e \Trans{\sigma'} q_s'' \parcomp q_e'' \land
            q_s'' \Trans{\ell} q_s' \land
            q_e'' \Trans{\epsilon} q_e'$\\
            \proofstep{Apply IH}\\
            $\exists q_s'' \in Q_s, q_e'' \in Q_e :
            s \Trans{\project{\sigma'}{L_s}} q_s'' \land
            e \Trans{\project{\sigma'}{L_e}} q_e'' \land
            q_s'' \Trans{\ell} q_s' \land
            q_e'' \Trans{\epsilon} q_e'$\\
            \proofstep{\Cref{item:trans_transitive}}\\
            $s \Trans{\project{\sigma'}{L_s}\cdot \ell} s' \land
            e \Trans{\project{\sigma'}{L_e}} e'$\\
            \proofstep{\Cref{def:projection}: $\projectop$}\\
            $s \Trans{\project{\sigma'\cdot \ell}{L_s}} s' \land
            e \Trans{\project{\sigma'\cdot \ell}{L_e}} e'$\\
            
            \item[$\ell \in L_e \setminus L_s$:]Symmetric with the previous case.
            
            \item[$\ell\in (L_s \cap L_e)$:]\ \\
            $q_s \parcomp q_e \Trans{\sigma' \cdot \ell} q_s' \parcomp q_e'$\\
            \proofstep{\Cref{def:arrowdefs}: $\Trans{\sigma}$}\\
            $\exists q_s'', q_s''' \in Q_s, q_e'',q_e''' \in Q_e :
            q_s \parcomp q_e \Trans{\sigma'} q_s'' \parcomp q_e''\; \land$\\\tab
            $q_s'' \parcomp q_e'' \trans{\ell} q_s''' \parcomp q_e''' \land
            q_s''' \parcomp q_e''' \Trans{\epsilon} q_s' \parcomp q_e' $\\
            \proofstep{Apply base case}\\
            $\exists q_s'', q_s''' \in Q_s, q_e'',q_e''' \in Q_e :
            q_s \parcomp q_e \Trans{\sigma'} q_s'' \parcomp q_e''\; \land$\\\tab
            $q_s'' \parcomp q_e'' \trans{\ell} q_s''' \parcomp q_e''' \land
            q_s''' \Trans{\epsilon} q_s' \land
            q_e''' \Trans{\epsilon} q_e' $\\
            \proofstep{\Itemref{lem:parcomp_base_properties}{step_both_label}}\\
            $\exists q_s'', q_s''' \in Q_s, q_e'',q_e''' \in Q_e :
            q_s \parcomp q_e \Trans{\sigma'} q_s'' \parcomp q_e''\; \land$\\\tab
            $q_s'' \trans{\ell} q_s''' \land
            q_e'' \trans{\ell} q_e''' \land
            q_s''' \Trans{\epsilon} q_s' \land
            q_e''' \Trans{\epsilon} q_e' $\\
            \proofstep{\Cref{def:arrowdefs}: $\Trans{\ell}$}\\
            $\exists q_s'' \in Q_s, q_e'' \in Q_e :
            q_s \parcomp q_e \Trans{\sigma'} q_s'' \parcomp q_e'' \land
            q_s'' \Trans{\ell} q_s' \land
            q_e''\Trans{\ell} q_e' $\\
            \proofstep{Apply IH}\\
            $\exists q_s'' \in Q_s, q_e'' \in Q_e :
            s \Trans{\project{\sigma'}{L_s}} q_s'' \land e \Trans{\project{\sigma'}{L_e}} q_e'' \land
            q_s'' \Trans{\ell} q_s' \land
            q_e'' \Trans{\ell} q_e' $\\
            \proofstep{\Cref{item:trans_transitive}}\\
            $s \Trans{\project{\sigma'}{L_s}\cdot \ell} q_s' \land e \Trans{\project{\sigma'}{L_e} \cdot \ell} q_e' $\\
            \proofstep{\Cref{def:projection}: $\projectop$}\\
            $s \Trans{\project{\sigma'\cdot \ell}{L_s}} q_s' \land e \Trans{\project{\sigma'\cdot \ell}{L_e}} q_e' $\\
          
        \end{case_distinction}
    \end{case_distinction}
\item[($\impliedby$):] Covered by \cref{lem:project_from_parcomp_light}, since $\forall \sigma \in L_{s\parcomp e}, L\subseteq L_{s\parcomp e} : \project{\sigma}{L^\delta} = \project{\sigma}{L}$ Because $\delta\notin L_{s\parcomp e}$
\end{case_distinction}
\end{proof}

\begin{lemmarep}
let $s,e$ be $\composable$ $\LTS$, $s'\in Q_s$, $e' \in Q_e, \sigma \in  \utraces{s\parcomp e}$.
\centermath{s \mutuallyaccepts{} e \ \implies\ 
(\ s \parcomp e \xRightarrow{\sigma} s' \parcomp e' \ \iff\ 
s \xRightarrow{\project{\sigma}{L_s^\delta}} s' \:\land\:
e \xRightarrow{\project{\sigma}{L_e^\delta}} e'\ )}
\label{lem:project_from_parcomp}
\end{lemmarep}

\begin{proof}
    ~\begin{case_distinction}
\item[($\Longrightarrow$):] Proof by induction on $\sigma$. 
    \begin{case_distinction}
    \item[Base case:] covered by \cref{lem:project_from_parcomp_no_delta}
        
    \item[Induction step:] 
        Assume the proposition holds for $\sigma'\in  \utraces{s \parcomp e}$.\\
        To proof: the proposition holds for $\sigma$, where $\sigma =\sigma' \cdot a$ with $a \in L_{s \parcomp e}^\delta$. Assume $\sigma \in  \utraces{s \parcomp e}$, otherwise the proposition trivially holds. This is divided into four cases based on $a$:
        \begin{case_distinction}
            \item[$a \in L_s \setminus L_e$:]\ \\
                $s \parcomp e \Trans{\sigma' \cdot a} q_s \parcomp q_e$\\
                \proofstep{\Cref{def:arrowdefs}: $\Trans{\sigma}$}\\
                $\exists q_s', q_s'' \in Q_s, q_e',q_e'' \in Q_e :$\\
                \tab$s \parcomp e \Trans{\sigma'} q_s' \parcomp q_e' \land
                q_s' \parcomp q_e' \trans{a} q_s'' \parcomp q_e'' \land
                q_s'' \parcomp q_e'' \Trans{\epsilon} q_s \parcomp q_e $\\
                \proofstep{\Cref{lem:project_from_parcomp_no_delta}}\\
                $\exists q_s', q_s'' \in Q_s, q_e',q_e'' \in Q_e :$\\
                \tab$s \parcomp e \Trans{\sigma'} q_s' \parcomp q_e' \land
                q_s' \parcomp q_e' \trans{a} q_s'' \parcomp q_e'' \land
                q_s'' \Trans{\epsilon} q_s \land
                q_e'' \Trans{\epsilon} q_e$\\
                \proofstep{\Cref{def:parcomp}: $q_e' = q_e''$.
                From induction on structure of $T_{s\parcomp e}$ with $a \in L_s \setminus L_e$.}\\
                $\exists q_s', q_s'' \in Q_s, q_e'\in Q_e :$\\
                \tab$s \parcomp e \Trans{\sigma'} q_s' \parcomp q_e' \land
                q_s' \parcomp q_e' \trans{a} q_s'' \parcomp q_e' \land
                q_s'' \Trans{\epsilon} q_s \land
                q_e' \Trans{\epsilon} q_e$\\
                \proofstep{\Itemref{lem:parcomp_base_properties}{step_left}}\\
                $\exists q_s', q_s'' \in Q_s, q_e' \in Q_e :
                s \parcomp e \Trans{\sigma'} q_s' \parcomp q_e' \land
                q_s'  \trans{a} q_s'' \land
                q_s'' \Trans{\epsilon} q_s \land
                q_e' \Trans{\epsilon} q_e$\\
                \proofstep{\Cref{def:arrowdefs}: $\Trans{a}$}\\
                $\exists q_s' \in Q_s, q_e' \in Q_e :
                s \parcomp e \Trans{\sigma'} q_s' \parcomp q_e' \land
                q_s' \Trans{a} q_s \land
                q_e' \Trans{\epsilon} q_e$\\
                \proofstep{Apply IH}\\
                $\exists q_s' \in Q_s, q_e' \in Q_e :
                s \Trans{\project{\sigma'}{L_s^\delta}} q_s' \land
                e \Trans{\project{\sigma'}{L_e^\delta}} q_e' \land
                q_s' \Trans{a} q_s \land
                q_e' \Trans{\epsilon} q_e$\\
                \proofstep{\Cref{item:trans_transitive}}\\
                $s \Trans{\project{\sigma'}{L_s^\delta}\cdot a} q_s \land
                e \Trans{\project{\sigma'}{L_e^\delta}} q_e$\\
                \proofstep{\Cref{def:projection}: $\projectop$}\\
                $s \Trans{\project{\sigma'\cdot a}{L_s^\delta}} q_s \land
                e \Trans{\project{\sigma'\cdot a}{L_e^\delta}} q_e$\\
            
            \item[$a \in L_e \setminus L_s$:]Symmetric with the previous case.
            
            \item[$a\in (L_s \cap L_e)$:]\ \\
                $s \parcomp e \Trans{\sigma' \cdot a} q_s \parcomp q_e$\\
                \proofstep{\Cref{def:arrowdefs}: $\Trans{\sigma}$}\\
                $\exists q_s', q_s'' \in Q_s, q_e',q_e'' \in Q_e :$\\
                \tab$s \parcomp e \Trans{\sigma'} q_s' \parcomp q_e' \land
                q_s' \parcomp q_e' \trans{a} q_s'' \parcomp q_e'' \land
                q_s'' \parcomp q_e'' \Trans{\epsilon} q_s \parcomp q_e $\\
                \proofstep{\Cref{lem:project_from_parcomp_no_delta}}\\
                $\exists q_s', q_s'' \in Q_s, q_e',q_e'' \in Q_e :$\\
                \tab$s \parcomp e \Trans{\sigma'} q_s' \parcomp q_e' \land
                q_s' \parcomp q_e' \trans{a} q_s'' \parcomp q_e'' \land
                q_s'' \Trans{\epsilon} q_s \land
                q_e'' \Trans{\epsilon} q_e $\\
                \proofstep{\Itemref{lem:parcomp_base_properties}{step_both_label}}\\
                $\exists q_s', q_s'' \in Q_s, q_e',q_e'' \in Q_e :$\\
                \tab$s \parcomp e \Trans{\sigma'} q_s' \parcomp q_e' \land
                q_s' \trans{a} q_s'' \land
                q_e' \trans{a} q_e'' \land
                q_s'' \Trans{\epsilon} q_s \land
                q_e'' \Trans{\epsilon} q_e $\\
                \proofstep{\Cref{def:arrowdefs}: $\Trans{a}$}\\
                $\exists q_s' \in Q_s, q_e' \in Q_e :
                s \parcomp e \Trans{\sigma'} q_s' \parcomp q_e' \land
                q_s' \Trans{a} q_s \land
                q_e' \Trans{a} q_e $\\
                \proofstep{Apply IH}\\
                $\exists q_s' \in Q_s, q_e' \in Q_e :
                s \Trans{\project{\sigma'}{L_s^\delta}} q_s' \land e \Trans{\project{\sigma'}{L_e^\delta}} q_e' \land
                q_s' \Trans{a} q_s \land
                q_e' \Trans{a} q_e $\\
                \proofstep{\Cref{item:trans_transitive}}\\
                $s \Trans{\project{\sigma'}{L_s^\delta}\cdot a} q_s \land e \Trans{\project{\sigma'}{L_e^\delta} \cdot a} q_e $\\
                \proofstep{\Cref{def:projection}: $\projectop$}\\
                $s \Trans{\project{\sigma'\cdot a}{L_s^\delta}} q_s \land e \Trans{\project{\sigma'\cdot a}{L_e^\delta}} q_e $\\
            \item[$a = \delta$:]\ \\
                $s \parcomp e \Trans{\sigma' \cdot \delta} q_s \parcomp q_e$\\
                \proofstep{\Cref{def:arrowdefs}: $\Trans{\sigma}$}\\
                $\exists q_s', q_s'' \in Q_s, q_e',q_e'' \in Q_e :$\\
                \tab$s \parcomp e \Trans{\sigma'} q_s' \parcomp q_e' \land
                q_s' \parcomp q_e' \trans{\delta} q_s'' \parcomp q_e'' \land
                q_s'' \parcomp q_e'' \Trans{\epsilon} q_s \parcomp q_e $\\
                \proofstep{\Cref{lem:project_from_parcomp_no_delta}}\\
                $\exists q_s', q_s'' \in Q_s, q_e',q_e'' \in Q_e :$\\
                \tab$s \parcomp e \Trans{\sigma'} q_s' \parcomp q_e' \land
                q_s' \parcomp q_e' \trans{\delta} q_s'' \parcomp q_e'' \land
                q_s'' \Trans{\epsilon} q_s \land
                q_e'' \Trans{\epsilon} q_e $\\
                \proofstep{\Itemref{lem:parcomp_base_properties}{both_delta}}\\
                $\exists q_s', q_s'' \in Q_s, q_e',q_e'' \in Q_e :$\\
                \tab$s \parcomp e \Trans{\sigma'} q_s' \parcomp q_e' \land
                q_s' \trans{\delta} q_s'' \land
                q_e' \trans{\delta} q_e'' \land
                q_s'' \Trans{\epsilon} q_s \land
                q_e'' \Trans{\epsilon} q_e $\\
                \proofstep{\Cref{def:arrowdefs}: $\Trans{\delta}$}\\
                $\exists q_s' \in Q_s, q_e' \in Q_e :
                s \parcomp e \Trans{\sigma'} q_s' \parcomp q_e' \land
                q_s' \Trans{\delta} q_s \land
                q_e' \Trans{\delta} q_e $\\
                \proofstep{Apply IH}\\
                $\exists q_s' \in Q_s, q_e' \in Q_e :
                s \Trans{\project{\sigma'}{L_s^\delta}} q_s' \land e \Trans{\project{\sigma'}{L_e^\delta}} q_e' \land
                q_s' \Trans{\delta} q_s \land
                q_e' \Trans{\delta} q_e $\\
                \proofstep{\Cref{item:trans_transitive}}\\
                $s \Trans{\project{\sigma'}{L_s^\delta}\cdot \delta} q_s \land e \Trans{\project{\sigma'}{L_e^\delta} \cdot \delta} q_e $\\
                \proofstep{\Cref{def:projection}: $\projectop$}\\
                $s \Trans{\project{\sigma'\cdot \delta}{L_s^\delta}} q_s \land e \Trans{\project{\sigma'\cdot \delta}{L_e^\delta}} q_e $\\
          
        \end{case_distinction}
    \end{case_distinction}
    \item[($\Longleftarrow$):] Covered by \cref{lem:project_from_parcomp_light}.
\end{case_distinction}
\end{proof}

\begin{lemmarep} Let $s,e \in \LTS$ be $\composable$, $\sigma \in {L_{s\parcomp e}^\delta}^*$.
\centermath{\begin{array}[t]{ll}
s \mutuallyaccepts{} e \ \implies \\
~(\ \sigma \in \utraces{s \parcomp e} \ \iff\ \project{\sigma}{L_s^\delta} \in  \utraces{s} \;\land\; \project{\sigma}{L_e^\delta}\in  \utraces{e}\ )
\end{array}}
\label{lem:project_utraces}
\end{lemmarep}

\begin{proof}
Assume $s \mutuallyaccepts{} e$. 
\begin{case_distinction}
    \item[($\Longrightarrow$):]\ \\
    If $ \utraces{s\parcomp e} = \emptyset$, then the lemma trivially holds.\\
    Assume $\sigma \in  \utraces{s\parcomp e}$, then the proof follows by induction on $\sigma$.\\
    \begin{case_distinction}
        \item[Base case:] $\sigma = \epsilon$\\
        $\epsilon \in  \utraces{s\parcomp e}$\\
        \proofstep{\Cref{def:uioco}: $Utraces$}\\
        $\exists q_s \in Q_s, q_e \in Q_e : s \parcomp e \Trans{\epsilon} q_s \parcomp q_e$\\
        \proofstep{\Cref{lem:project_from_parcomp}}\\
        $\exists q_s \in Q_s, q_e \in Q_e : s\Trans{\project{\epsilon}{L_s^\delta}} q_s \land e\Trans{\project{\epsilon}{L_e^\delta}} q_e$\\
        \proofstep{\Cref{def:uioco}: $Utraces$}\\
        $\project{\epsilon}{L_s^\delta} \in  \utraces{s} \land \project{\epsilon}{L_e^\delta} \in  \utraces{e}$\\
        
        \item[Induction step:] Assume the proposition holds for $\sigma'$. To proof: the proposition holds for $\sigma$  where $\sigma = \sigma' \cdot a$, $a \in L_{s \parcomp e}^\delta $.\\
        We do a case distinction on $a$, splitting $L_{s \parcomp e}^\delta$ into\\
        $(I_s \setminus L_e), (I_e \setminus L_s), (I_e \cap U_s), (I_s \cap U_e), (I_s \cap I_e), (U_s \setminus L_e), (U_e \setminus L_s), \{\delta\}$. Note that $U_s \cap U_e$ is missing because it is empty (\cref{def:composable}). This fine grained distinction is required because what it means for a trace to be part of $\utraces{}$ changes based on if the last label is an input or output. Additionally, the behaviour of $\projectop$ changes based on if a label is part of $L_s$, $L_e$ or both. 
        \begin{case_distinction}
        \item[$a \in I_s \setminus L_e$:]\ \\
            $\sigma'\cdot a \in  \utraces{s\parcomp e}$\\
            \proofstep{\Cref{item:utrace_prefix_closed}}\\
            $\sigma'\in  \utraces{s\parcomp e} \land \sigma'\cdot a \in  \utraces{s\parcomp e}$\\
            \proofstep{\Cref{def:uioco}: $Utraces$}\\
            $\sigma'\in  \utraces{s\parcomp e}\; \land $\\
            $\exists q_s \in Q_s, q_e \in Q_e : s\parcomp e \Trans{\sigma'\cdot a} q_s \parcomp q_e\; \land$\\ 
            $\forall q_s' \in Q_s, q_e' \in Q_e : s \parcomp e \Trans{\sigma'} q_s' \parcomp q_e' \implies q_s' \parcomp q_e' \Trans{a}$\\
            \proofstep{\Cref{item:trans_transitive}}\\
            $\sigma'\in  \utraces{s\parcomp e} \land $\\
            $\exists q_s \in Q_s, q_e \in Q_e : s\parcomp e \Trans{\sigma'} q_s \parcomp q_e\; \land$\\
            $\forall q_s' \in Q_s, q_e' \in Q_e : s \parcomp e \Trans{\sigma'} q_s' \parcomp q_e' \implies q_s' \parcomp q_e' \Trans{a}$\\
            \proofstep{\Cref{lem:project_from_parcomp}}\\
            $\sigma'\in  \utraces{s\parcomp e}\; \land $\\
            $\exists q_e \in Q_e : e \Trans{\project{\sigma'}{L_e^\delta}} q_e\; \land$\\
            $ \forall q_s' \in Q_s, q_e' \in Q_e : s \Trans{\project{\sigma'}{L_s^\delta}} q_s' \land e \Trans{\project{\sigma'}{L_e^\delta}}  q_e' \implies q_s' \parcomp q_e' \Trans{a}$\\
            \proofstep{$\forall$ elimination}\\
            $\sigma'\in  \utraces{s\parcomp e}\; \land $\\
            $\exists q_e \in Q_e : e \Trans{\project{\sigma'}{L_e^\delta}} q_e\; \land$\\
            $ \forall q_s' \in Q_s : s \Trans{\project{\sigma'}{L_s^\delta}} q_s' \land e \Trans{\project{\sigma'}{L_e^\delta}}  q_e \implies q_s' \parcomp q_e \Trans{a} $\\
            \proofstep{($A \land B \implies C) \land B \implies (A \implies C)$}\\
            $\sigma'\in  \utraces{s\parcomp e}\; \land $\\
            $\exists q_e \in Q_e, \forall q_s' \in Q_s : s \Trans{\project{\sigma'}{L_s^\delta}} q_s' \implies q_s' \parcomp q_e \Trans{a}$\\
            \proofstep{\Cref{def:parcomp}: $\parcomp$}\\
            $\sigma'\in  \utraces{s\parcomp e}\; \land $\\
            $\forall q_s' \in Q_s : s \Trans{\project{\sigma'}{L_s^\delta}} q_s' \implies q_s' \Trans{a}$\\
            \proofstep{Apply IH}\\
            $\project{\sigma'}{L_s^\delta}\in  \utraces{s} \land \project{\sigma'}{L_e^\delta} \in  \utraces{e}\; \land$\\
            $\forall q_s' \in Q_s : s \Trans{\project{\sigma'}{L_s^\delta}} q_s' \implies q_s' \Trans{a}$\\
            \proofstep{\Cref{def:uioco}: $Utraces$}\\
            $\project{\sigma'}{L_s^\delta}\cdot a \in  \utraces{s} \land \project{\sigma'}{L_e^\delta} \in  \utraces{e}$\\
            \proofstep{\Cref{def:projection}: $\projectop$}\\
            $\project{\sigma'\cdot a}{L_s^\delta} \in  \utraces{s} \land \project{\sigma'\cdot a}{L_e^\delta} \in  \utraces{e}$
        \item[$a \in I_e \setminus L_s$:] Symmetric to previous case\\

        \item[$a \in I_s \cap U_e$:]\ \\
            $\sigma'\cdot a \in  \utraces{s\parcomp e}$\\
            \proofstep{\Cref{item:utrace_prefix_closed}}\\
            $\sigma'\in  \utraces{s\parcomp e} \land \sigma'\cdot a \in  \utraces{s\parcomp e}$\\
            \proofstep{\Cref{def:uioco}: $Utraces$}\\
            $\sigma'\in  \utraces{s\parcomp e}\; \land $\\
            $\exists q_s \in Q_s, q_e \in Q_e : s\parcomp e \Trans{\sigma'\cdot a} q_s \parcomp q_e$\\
            \proofstep{\Cref{lem:project_from_parcomp} }\\
            $\sigma'\in  \utraces{s\parcomp e}\; \land $\\
            $\exists q_e \in Q_e :  e \Trans{\project{\sigma'\cdot a}{L_e^\delta}} q_e$\\
            \proofstep{\Cref{def:mutually_accepts,def:accepting}: $\mutuallyaccepts{}$ and $\accepts{}$}\\
            $\sigma'\in  \utraces{s\parcomp e}\; \land $\\
            $\exists q_e \in Q_e :  e \Trans{\project{\sigma'\cdot a}{L_e^\delta}} q_e\; \land$\\
            $(\forall \sigma \in  \utraces{s\parcomp e}, q_s' \in Q_s, q_e' \in Q_e: s \parcomp e \Trans{\sigma} q_s' \parcomp q_e' \implies$\\
            $\outset{q_e'}\cap I_e \subseteq \inset{q_s'} \cap U_s$)\\
            \proofstep{$\forall$ elimination}\\
            $\sigma'\in  \utraces{s\parcomp e}\; \land $\\
            $\exists q_e \in Q_e :  e \Trans{\project{\sigma'\cdot a}{L_e^\delta}} q_e\; \land$\\
            $\forall q_s', \in Q_s, q_e' \in Q_e : s \parcomp e \Trans{\sigma'} q_s' \parcomp q_e' \implies \outset{q_e'} \cap I_s \subseteq \inset{q_s'} \cap U_e$\\
            \proofstep{\Cref{lem:project_from_parcomp}}\\
            $\sigma'\in  \utraces{s\parcomp e}\; \land $\\
            $\exists q_e \in Q_e :  e \Trans{\project{\sigma'\cdot a}{L_e^\delta}} q_e\; \land$\\
            $\forall q_s', \in Q_s, q_e' \in Q_e : s \Trans{\project{\sigma'}{L_s^\delta}} q_s' \land e \Trans{\project{\sigma'}{L_e^\delta}} q_e' \implies \outset{q_e'} \cap I_s \subseteq \inset{q_s'} \cap U_e$\\
            \proofstep{\Cref{def:outset,def:inset}: $\inset{}$ and $\outset{}$, and $a \in I_s \cap U_e$}\\
            $\sigma'\in  \utraces{s\parcomp e}\; \land $\\
            $\exists q_e \in Q_e :  e \Trans{\project{\sigma'\cdot a}{L_e^\delta}} q_e\; \land$\\
            $\forall q_s', \in Q_s, q_e' \in Q_e : s \Trans{\project{\sigma'}{L_s^\delta}} q_s' \land e \Trans{\project{\sigma'}{L_e^\delta}} q_e' \implies (q_e' \trans{a} \implies q_s' \Trans{a})$\\
            \proofstep{\Cref{def:projection}: $\projectop$}\\
            $\sigma'\in  \utraces{s\parcomp e}\; \land $\\
            $\exists q_e \in Q_e :  e \Trans{\project{\sigma'}{L_e^\delta}\cdot a} q_e\; \land$\\
            $\forall q_s', \in Q_s, q_e' \in Q_e : s \Trans{\project{\sigma'}{L_s^\delta}} q_s' \land e \Trans{\project{\sigma'}{L_e^\delta}} q_e' \implies (q_e' \trans{a} \implies q_s' \Trans{a})$\\
            \proofstep{\Cref{def:arrowdefs}: $e \Trans{\project{\sigma'}{L_e^\delta}\cdot a} q$}\\
            $\sigma'\in  \utraces{s\parcomp e}\; \land $\\
            $\exists q_e, q_e' \in Q_e :  e \Trans{\project{\sigma'}{L_e^\delta}\cdot a} q_e \land e \Trans{\project{\sigma'}{L_e^\delta}} q_e' \land  q_e' \trans{a}\; \land$\\
            $\forall q_s', \in Q_s, q_e' \in Q_e : s \Trans{\project{\sigma'}{L_s^\delta}} q_s' \land e \Trans{\project{\sigma'}{L_e^\delta}} q_e' \implies (q_e' \trans{a} \implies q_s' \Trans{a})$\\
            \proofstep{$\forall$ elimination}\\
            $\sigma'\in  \utraces{s\parcomp e}\; \land $\\
            $\exists q_e, q_e' \in Q_e : e \Trans{\project{\sigma'}{L_e^\delta}\cdot a} q_s \land e \Trans{\project{\sigma'}{L_e^\delta}} q_e' \land  q_e' \trans{a}\;\land$\\
            $\forall q_s', \in Q_s : s \Trans{\project{\sigma'}{L_s^\delta}} q_s' \land e \Trans{\project{\sigma'}{L_e^\delta}} q_e' \implies (q_e' \trans{a} \implies q_s' \Trans{a})$\\
            \proofstep{($A \land B \implies C) \land B \implies (A \implies C)$}\\
            $\sigma'\in  \utraces{s\parcomp e}\; \land $\\
            $\exists q_e, q_e' \in Q_e : e \Trans{\project{\sigma'}{L_e^\delta}\cdot a} q_e \land q_e' \trans{a}\; \land$\\
            $\forall q_s', \in Q_s: s \Trans{\project{\sigma'}{L_s^\delta}} q_s' \implies (q_e' \trans{a} \implies q_s' \Trans{a})$\\
            \proofstep{($A \implies B) \land A \implies B$}\\
            $\sigma'\in  \utraces{s\parcomp e}\; \land $\\
            $\exists q_e \in Q_e : e \Trans{\project{\sigma'}{L_e^\delta}\cdot a} q_e\; \land$\\
            $\forall q_s', \in Q_s : s \Trans{\project{\sigma'}{L_s^\delta}} q_s' \implies q_s' \Trans{a}$\\
            \proofstep{Apply IH}\\
            $\project{\sigma'}{L_s^\delta}\in  \utraces{s} \land \project{\sigma'}{L_e^\delta} \in  \utraces{e} \land$\\
            $\exists q_e \in Q_e : e \Trans{\project{\sigma'}{L_e^\delta}\cdot a} q_e\; \land$\\
            $\forall q_s', \in Q_s : s \Trans{\project{\sigma'}{L_s^\delta}} q_s' \implies q_s' \Trans{a}$\\
            \proofstep{\Cref{def:uioco}: $Utraces$}\\
            $\project{\sigma'}{L_s^\delta}\cdot a \in  \utraces{s} \land \project{\sigma'}{L_e^\delta}\cdot a \in  \utraces{e}$\\
            \proofstep{\Cref{def:projection}: $\projectop$}\\
            $\project{\sigma'\cdot a}{L_s^\delta} \in  \utraces{s} \land \project{\sigma'\cdot a}{L_e^\delta} \in  \utraces{e}$
            
        \item[$a \in I_e \cap U_s$:] Symmetric to previous case.\\

        \item[$a \in I_e \cap I_s$:]\ \\
            $\sigma'\cdot a \in  \utraces{s\parcomp e}$\\
            \proofstep{\Cref{item:utrace_prefix_closed}}\\
            $\sigma'\in  \utraces{s\parcomp e} \land \sigma'\cdot a \in  \utraces{s\parcomp e}$\\
            \proofstep{\Cref{def:uioco}: $Utraces$}\\
            $\sigma'\in  \utraces{s\parcomp e}\; \land $\\
            $\exists q_s \in Q_s, q_e \in Q_e : s\parcomp e \Trans{\sigma'\cdot a} q_s \parcomp q_e\; \land$\\ 
            $\forall q_s' \in Q_s, q_e' \in Q_e : s \parcomp e \Trans{\sigma'} q_s' \parcomp q_e' \implies q_s' \parcomp q_e' \Trans{a}$\\
            \proofstep{\Cref{item:trans_transitive}}\\
            $\sigma'\in  \utraces{s\parcomp e} \land $\\
            $\exists q_s \in Q_s, q_e \in Q_e : s\parcomp e \Trans{\sigma'} q_s \parcomp q_e\; \land$\\
            $\forall q_s' \in Q_s, q_e' \in Q_e : s \parcomp e \Trans{\sigma'} q_s' \parcomp q_e' \implies q_s' \parcomp q_e' \Trans{a}$\\
            \proofstep{\Cref{lem:project_from_parcomp}}\\
            $\sigma'\in  \utraces{s\parcomp e}\; \land $\\
            $\exists q_s \in Q_s, q_e \in Q_e : s \Trans{\project{\sigma'}{L_s^\delta}} q_s \land e \Trans{\project{\sigma'}{L_e^\delta}} q_e\; \land$\\
            $ \forall q_s' \in Q_s, q_e' \in Q_e : s \Trans{\project{\sigma'}{L_s^\delta}} q_s' \land e \Trans{\project{\sigma'}{L_e^\delta}}  q_e' \implies q_s' \parcomp q_e' \Trans{a}$\\
            \proofstep{$\forall$ elimination X2}\\
            $\sigma'\in  \utraces{s\parcomp e}\; \land $\\
            $\exists q_s \in Q_s, q_e \in Q_e  : s \Trans{\project{\sigma'}{L_s^\delta}} q_s \land e \Trans{\project{\sigma'}{L_e^\delta}} q_e\; \land$\\
            $ \forall q_s' \in Q_s : s \Trans{\project{\sigma'}{L_s^\delta}} q_s' \land e \Trans{\project{\sigma'}{L_e^\delta}}  q_e \implies q_s' \parcomp q_e \Trans{a}\;\land$\\
            $ \forall q_e' \in Q_e : s \Trans{\project{\sigma'}{L_s^\delta}} q_s \land e \Trans{\project{\sigma'}{L_e^\delta}}  q_e' \implies q_s \parcomp q_e' \Trans{a}$\\
            \proofstep{($A \land B \implies C) \land B \implies (A \implies C)$}\\
            $\sigma'\in  \utraces{s\parcomp e}\; \land $\\
            $\exists q_e \in Q_e, \forall q_s' \in Q_s : s \Trans{\project{\sigma'}{L_s^\delta}} q_s' \implies q_s' \parcomp q_e \Trans{a}\;\land$\\
            $\exists q_s \in Q_s, \forall q_e' \in Q_e : e \Trans{\project{\sigma'}{L_e^\delta}}  q_e' \implies q_s \parcomp q_e' \Trans{a}$\\
            \proofstep{\Cref{def:parcomp}: $\parcomp$}\\
            $\sigma'\in  \utraces{s\parcomp e}\; \land $\\
            $\forall q_s' \in Q_s : s \Trans{\project{\sigma'}{L_s^\delta}} q_s' \implies q_s' \Trans{a}\;\land$\\
            $\forall q_e' \in Q_e : e \Trans{\project{\sigma'}{L_e^\delta}}  q_e' \implies q_e' \Trans{a}$\\
            \proofstep{Apply IH}\\
            $\project{\sigma'}{L_s^\delta}\in  \utraces{s} \land \project{\sigma'}{L_e^\delta} \in  \utraces{e}\; \land$\\
            $\forall q_s' \in Q_s : s \Trans{\project{\sigma'}{L_s^\delta}} q_s' \implies q_s' \Trans{a}\;\land$\\
            $\forall q_e' \in Q_e : e \Trans{\project{\sigma'}{L_e^\delta}}  q_e' \implies q_e' \Trans{a}$\\\
            \proofstep{\Cref{def:uioco}: $Utraces$}\\
            $\project{\sigma'}{L_s^\delta}\cdot a \in  \utraces{s} \land \project{\sigma'}{L_e^\delta}\cdot a \in  \utraces{e}$\\
            \proofstep{\Cref{def:projection}: $\projectop$}\\
            $\project{\sigma'\cdot a}{L_s^\delta} \in  \utraces{s} \land \project{\sigma'\cdot a}{L_e^\delta} \in  \utraces{e}$
            
        \item[$a \in U_s \setminus L_e$:]\ \\
            $\sigma'\cdot a \in  \utraces{s\parcomp e}$\\
            \proofstep{\Cref{item:utrace_prefix_closed}}\\
            $\sigma'\in  \utraces{s\parcomp e} \land \sigma'\cdot a \in  \utraces{s\parcomp e}$\\
            \proofstep{\Cref{def:uioco}: $Utraces$}\\
            $\sigma'\in  \utraces{s\parcomp e} \land \sigma'\cdot a \in  \utraces{s\parcomp e}\; \land $\\
            $\exists q_s \in Q_s, q_e \in Q_e : s\parcomp e \Trans{\sigma'\cdot a} q_s \parcomp q_e$\\
            \proofstep{\Cref{lem:project_from_parcomp} }\\
            $\sigma'\in  \utraces{s\parcomp e}\; \land $\\
            $\exists q_s \in Q_s : s \Trans{\project{\sigma'\cdot a}{L_s^\delta}} q_s$\\
            \proofstep{\Cref{def:projection}: $\projectop$}\\
            $\sigma'\in  \utraces{s\parcomp e}\; \land $\\
            $\exists q_s \in Q_s: s \Trans{\project{\sigma'}{L_s^\delta}\cdot a} q_s$\\
            \proofstep{Apply IH}\\
            $\project{\sigma'}{L_s^\delta}\in  \utraces{s} \land \project{\sigma'}{L_e^\delta} \in  \utraces{e}\; \land$\\
            $\exists q_s \in Q_s : s \Trans{\project{\sigma'}{L_s^\delta}\cdot a} q_s$\\
            \proofstep{\Cref{def:uioco}: $Utraces$}\\
            $\project{\sigma'}{L_s^\delta}\cdot a \in  \utraces{s} \land \project{\sigma'}{L_e^\delta}\in  \utraces{e}$\\
            \proofstep{\Cref{def:projection}: $\projectop$}\\
            $\project{\sigma'\cdot a}{L_s^\delta} \in  \utraces{s} \land \project{\sigma'\cdot a}{L_e^\delta} \in  \utraces{e}$
            
        \item[$a \in U_e \setminus L_s$:]Symmetric to previous case.\\
            
        \item [$a = \delta$:]\ \\
            $\sigma'\cdot \delta \in  \utraces{s\parcomp e}$\\
            \proofstep{\Cref{item:utrace_prefix_closed}}\\
            $\sigma'\in  \utraces{s\parcomp e} \land \sigma'\cdot \delta \in  \utraces{s\parcomp e}$\\
            \proofstep{\Cref{def:uioco}: $Utraces$}\\
            $\sigma'\in  \utraces{s\parcomp e} \land \sigma'\cdot \delta \in  \utraces{s\parcomp e} \;\land $\\
            $\exists q_s \in Q_s, q_e \in Q_e : s\parcomp e \Trans{\sigma'\cdot \delta} q_s \parcomp q_e$\\
            \proofstep{\Cref{lem:project_from_parcomp} }\\
            $\sigma'\in  \utraces{s\parcomp e}\; \land $\\
            $\exists q_s \in Q_s, q_e \in Q_e : s \Trans{\project{\sigma'\cdot \delta}{L_s^\delta}} q_s \land e \Trans{\project{\sigma'\cdot \delta}{L_e^\delta}} q_e$\\
            \proofstep{\Cref{def:projection}: $\projectop$}\\
            $\sigma'\in  \utraces{s\parcomp e}\; \land $\\
            $\exists q_s \in Q_s, q_e \in Q_e : s \Trans{\project{\sigma'}{L_s^\delta}\cdot \delta} q_s \land e \Trans{\project{\sigma'}{L_e^\delta}\cdot \delta} q_e$\\
            \proofstep{Apply IH}\\
            $\project{\sigma'}{L_s^\delta}\in  \utraces{s} \land \project{\sigma'}{L_e^\delta} \in  \utraces{e}\; \land$\\
            $\exists q_s \in Q_s, q_e \in Q_e : s \Trans{\project{\sigma'}{L_s^\delta}\cdot \delta} q_e \land e \Trans{\project{\sigma'}{L_e^\delta}\cdot \delta} q_e$\\
            \proofstep{\Cref{def:uioco}: $Utraces$}\\
            $\project{\sigma'}{L_s^\delta}\cdot \delta \in  \utraces{s} \land \project{\sigma'}{L_e^\delta}\cdot \delta\in  \utraces{e}$\\
            \proofstep{\Cref{def:projection}: $\projectop$}\\
            $\project{\sigma'\cdot \delta}{L_s^\delta} \in  \utraces{s} \land \project{\sigma'\cdot \delta}{L_e^\delta} \in  \utraces{e}$
        \end{case_distinction}
    \end{case_distinction}
    \item[($\Longleftarrow$):]\ \\
     Assume $\project{\sigma}{L_s^\delta} \in  \utraces{s}\; \land\; \project{\sigma}{L_e^\delta} \in  \utraces{e}$, then the proof follows by induction on $\sigma$.\\
    \begin{case_distinction}
        \item[Base case:] $\sigma = \epsilon$\\
        
        $\project{\epsilon}{L_s^\delta} \in  \utraces{s} \land \project{\epsilon}{L_e^\delta} \in  \utraces{e}$\\
        \proofstep{\Cref{def:uioco}: $Utraces$}\\
        $\exists q_s \in Q_s, q_e \in Q_e : s\Trans{\project{\epsilon}{L_s^\delta}} q_s \land e\Trans{\project{\epsilon}{L_e^\delta}} q_e$\\
        \proofstep{\Cref{lem:project_from_parcomp_light}}\\
        $\exists q_s \in Q_s, q_e \in Q_e : s \parcomp e \Trans{\epsilon} q_s \parcomp q_e$\\
        \proofstep{\Cref{def:uioco}: $Utraces$}\\
        $\epsilon \in  \utraces{s\parcomp e}$\\
        
        \item[Induction step:]\ \\
        IH: the proposition holds for $\sigma'$. To proof: the proposition holds for $\sigma$ where $\sigma = \sigma' \cdot a$, $a \in L_{s \parcomp e}^\delta $.
        We do a case distinction on $a$, splitting $L_{s \parcomp e}^\delta$ into\\
        $(I_s \setminus L_e), (I_e \setminus L_s), ((I_s \cap U_e)\cup (I_e \cap U_s)\cup\{\delta\}), (I_s \cap I_e), (U_s \setminus L_e), (U_e \setminus L_s)$. Note that $U_s \cap U_e$ is missing because it is empty (\cref{def:composable}).:
        \begin{case_distinction}
            \item[$a \in I_s \setminus L_e$:]\ \\
            $\project{\sigma'\cdot a}{L_s^\delta} \in  \utraces{s} \land \project{\sigma'\cdot a}{L_e^\delta} \in  \utraces{e}$\\
            \proofstep{\Cref{def:projection}: $\projectop$}\\
            $\project{\sigma'}{L_s^\delta}\cdot a \in  \utraces{s} \land \project{\sigma'}{L_e^\delta} \in  \utraces{e}$\\
            \proofstep{\Cref{def:uioco}: $Utraces$}\\
            $\project{\sigma'}{L_s^\delta}\cdot a\in  \utraces{s} \land \project{\sigma'}{L_e^\delta} \in  \utraces{e}\; \land$\\
            $\exists q_s \in Q_s, q_e \in Q_e: s \Trans{\project{\sigma'}{L_s^\delta}\cdot a} q_s\land e \Trans{\project{\sigma'}{L_e^\delta}} q_e$\\
            $\forall q_s' \in Q_s : s \Trans{\project{\sigma'}{L_s^\delta}} q_s' \implies q_s' \Trans{a}$\\
            \proofstep{\Cref{item:utrace_prefix_closed}}\\
            $\project{\sigma'}{L_s^\delta}\in  \utraces{s} \land \project{\sigma'}{L_e^\delta} \in  \utraces{e}\; \land$\\
            $\exists q_s \in Q_s, q_e \in Q_e: s \Trans{\project{\sigma'}{L_s^\delta}\cdot a} q_s\land e \Trans{\project{\sigma'}{L_e^\delta}} q_e$\\
            $\forall q_s' \in Q_s : s \Trans{\project{\sigma'}{L_s^\delta}} q_s' \implies q_s' \Trans{a}$\\
            \proofstep{Apply IH}\\
            $\sigma'\in  \utraces{s\parcomp e}\; \land $\\
            $\exists q_s \in Q_s, q_e \in Q_e: s \Trans{\project{\sigma'}{L_s^\delta}\cdot a} q_s\land e \Trans{\project{\sigma'}{L_e^\delta}} q_e$\\
            $\forall q_s' \in Q_s : s \Trans{\project{\sigma'}{L_s^\delta}} q_s' \implies q_s' \Trans{a}$\\
            \proofstep{\Cref{def:projection}: $\projectop$}\\
            $\sigma'\in  \utraces{s\parcomp e}\; \land $\\
            $\exists q_s \in Q_s, q_e \in Q_e : s\Trans{\project{\sigma'\cdot a}{L_s^\delta}} q_s \land e \Trans{\project{\sigma'\cdot a}{L_e^\delta}} q_e\; \land$\\
            $\forall q_s' \in Q_s: s \Trans{\project{\sigma'}{L_s^\delta}} q_s' \implies q_s'  \Trans{\project{a}{L_s^\delta}}$\\
            \proofstep{$(A \implies C) \implies (A \land B \implies C)$}\\
            $\sigma'\in  \utraces{s\parcomp e}\; \land $\\
            $\exists q_s \in Q_s, q_e \in Q_e : s\Trans{\project{\sigma'\cdot a}{L_s^\delta}} q_s \land e \Trans{\project{\sigma'\cdot a}{L_e^\delta}} q_e\; \land$\\
            $\forall q_s' \in Q_s, q_e' \in Q_e : s \Trans{\project{\sigma'}{L_s^\delta}} q_s' \land  e \Trans{\project{\sigma'}{L_e^\delta}} q_e' \implies q_s'  \Trans{\project{a}{L_s^\delta}}$\\
            \proofstep{\Cref{def:arrowdefs}: $\forall q: q \Trans{\epsilon}$}\\
            $\sigma'\in  \utraces{s\parcomp e}\; \land $\\
            $\exists q_s \in Q_s, q_e \in Q_e : s\Trans{\project{\sigma'\cdot a}{L_s^\delta}} q_s \land e \Trans{\project{\sigma'\cdot a}{L_e^\delta}} q_e\; \land$\\
            $\forall q_s' \in Q_s, q_e' \in Q_e : s \Trans{\project{\sigma'}{L_s^\delta}} q_s' \land  e \Trans{\project{\sigma'}{L_e^\delta}} q_e' \implies q_s'  \Trans{\project{a}{L_s^\delta}} \land\; q_e' \Trans{\epsilon}$\\
            \proofstep{\Cref{def:projection}: $\projectop$}\\
            $\sigma'\in  \utraces{s\parcomp e}\; \land $\\
            $\exists q_s \in Q_s, q_e \in Q_e : s\Trans{\project{\sigma'\cdot a}{L_s^\delta}} q_s \land e \Trans{\project{\sigma'\cdot a}{L_e^\delta}} q_e\; \land$\\
            $\forall q_s' \in Q_s, q_e' \in Q_e : s \Trans{\project{\sigma'}{L_s^\delta}} q_s' \land  \Trans{\project{\sigma'}{L_e^\delta}} q_e' \implies q_s'  \Trans{\project{a}{L_s^\delta}} \land\; q_e' \Trans{\project{ a}{L_e^\delta}}$\\
            \proofstep{\Cref{lem:project_from_parcomp}}\\
            $\sigma'\in  \utraces{s\parcomp e}\; \land $\\
            $\exists q_s \in Q_s, q_e \in Q_e : s\parcomp e \Trans{\sigma'\cdot a} q_s \parcomp q_e\; \land$\\ 
            $\forall q_s' \in Q_s, q_e' \in Q_e : s \parcomp e \Trans{\sigma'} q_s' \parcomp q_e' \implies q_s' \parcomp q_e' \Trans{a}$\\
            \proofstep{\Cref{def:uioco}: $Utraces$}\\
            $\sigma'\cdot a \in  \utraces{s\parcomp e}$\\

            \item[$a \in I_e \setminus L_s$:] Symmetric to previous case\\
            
            \item[$a \in (I_s \cap U_e)\cup (I_e \cap U_s)\cup\{\delta\}$:]\ \\
            $\project{\sigma'\cdot a}{L_s^\delta} \in  \utraces{s} \land \project{\sigma'\cdot a}{L_e^\delta} \in  \utraces{e}$\\
            \proofstep{\Cref{def:uioco}: $Utraces$}\\
            $\project{\sigma'\cdot a}{L_s^\delta}\in  \utraces{s} \land \project{\sigma'\cdot a}{L_e^\delta}\in  \utraces{e} \land$\\
            $\exists q_s \in Q_s, q_e \in Q_e: s \Trans{\project{\sigma'\cdot a}{L_s^\delta}} q_s\land e \Trans{\project{\sigma'\cdot a}{L_e^\delta}} q_e$\\
            \proofstep{\Cref{def:projection}: $\projectop$}\\
            $\project{\sigma'}{L_s^\delta}\cdot a\in  \utraces{s} \land \project{\sigma'}{L_e^\delta}\cdot a\in  \utraces{e} \land$\\
            $\exists q_s \in Q_s, q_e \in Q_e: s \Trans{\project{\sigma'\cdot a}{L_s^\delta}} q_s\land e \Trans{\project{\sigma'\cdot a}{L_e^\delta}} q_e$\\
            \proofstep{\Cref{item:utrace_prefix_closed}}\\
            $\project{\sigma'}{L_s^\delta}\in  \utraces{s} \land \project{\sigma'}{L_e^\delta} \in  \utraces{e} \land$\\
            $\exists q_s \in Q_s, q_e \in Q_e: s \Trans{\project{\sigma'\cdot a}{L_s^\delta}} q_s\land e \Trans{\project{\sigma'\cdot a}{L_e^\delta}} q_e$\\
            \proofstep{Apply IH}\\
            $\sigma'\in  \utraces{s\parcomp e} \land $\\
            $\exists q_s \in Q_s, q_e \in Q_e: s \Trans{\project{\sigma'\cdot a}{L_s^\delta}} q_s\land e \Trans{\project{\sigma'\cdot a}{L_e^\delta}} q_e$\\
            \proofstep{\Cref{lem:project_from_parcomp_light}}\\
            $\sigma'\in  \utraces{s\parcomp e} \land $\\
            $\exists q_s \in Q_s, q_e \in Q_e : s\parcomp e \Trans{\sigma'\cdot a} q_s \parcomp q_e$\\ 
            \proofstep{\Cref{def:uioco}: $Utraces + a \notin I_{s\parcomp e}$}\\
            $\sigma'\cdot a \in  \utraces{s\parcomp e}$\\

            \item[$a \in I_s \cap I_e$:]\ \\
            $\project{\sigma'\cdot a}{L_s^\delta} \in  \utraces{s} \land \project{\sigma'\cdot a}{L_e^\delta} \in  \utraces{e}$\\
            \proofstep{\Cref{def:projection}: $\projectop$}\\
            $\project{\sigma'}{L_s^\delta}\cdot a \in  \utraces{s} \land \project{\sigma'}{L_e^\delta}\cdot a \in  \utraces{e}$\\
            \proofstep{\Cref{def:uioco}: $Utraces$}\\
            $\project{\sigma'}{L_s^\delta}\cdot a\in  \utraces{s} \land \project{\sigma'}{L_e^\delta}\cdot a \in  \utraces{e}\; \land$\\
            $\exists q_s \in Q_s, q_e \in Q_e: s \Trans{\project{\sigma'}{L_s^\delta}\cdot a} q_s\land e \Trans{\project{\sigma'}{L_e^\delta}\cdot a} q_e\;\land$\\
            $\forall q_s' \in Q_s : s \Trans{\project{\sigma'}{L_s^\delta}} q_s' \implies q_s' \Trans{a} \;\land$\\
            $\forall q_e' \in Q_e : e \Trans{\project{\sigma'}{L_s^\delta}} q_e' \implies q_e' \Trans{a}$\\
            \proofstep{\Cref{item:utrace_prefix_closed}}\\
            $\project{\sigma'}{L_s^\delta}\in  \utraces{s} \land \project{\sigma'}{L_e^\delta} \in  \utraces{e}\; \land$\\
            $\exists q_s \in Q_s, q_e \in Q_e: s \Trans{\project{\sigma'}{L_s^\delta}\cdot a} q_s\land e \Trans{\project{\sigma'}{L_e^\delta}\cdot a} q_e\;\land$\\
            $\forall q_s' \in Q_s : s \Trans{\project{\sigma'}{L_s^\delta}} q_s' \implies q_s' \Trans{a} \;\land$\\
            $\forall q_e' \in Q_e : e \Trans{\project{\sigma'}{L_s^\delta}} q_e' \implies q_e' \Trans{a}$\\
            \proofstep{Apply IH}\\
            $\sigma'\in  \utraces{s\parcomp e}\; \land $\\
            $\exists q_s \in Q_s, q_e \in Q_e: s \Trans{\project{\sigma'}{L_s^\delta}\cdot a} q_s\land e \Trans{\project{\sigma'}{L_e^\delta}\cdot a} q_e\;\land$\\
            $\forall q_s' \in Q_s : s \Trans{\project{\sigma'}{L_s^\delta}} q_s' \implies q_s' \Trans{a} \;\land$\\
            $\forall q_e' \in Q_e : e \Trans{\project{\sigma'}{L_s^\delta}} q_e' \implies q_e' \Trans{a}$\\
            \proofstep{$(A \implies C) \land (B \implies D) \implies (A \land B \implies C \land D)$}\\
            $\sigma'\in  \utraces{s\parcomp e}\; \land $\\
             $\exists q_s \in Q_s, q_e \in Q_e: s \Trans{\project{\sigma'}{L_s^\delta}\cdot a} q_s\land e \Trans{\project{\sigma'}{L_e^\delta}\cdot a} q_e\;\land$\\
            $\forall q_s' \in Q_s, q_e' \in Q_e : s \Trans{\project{\sigma'}{L_s^\delta}} q_s' \land  e \Trans{\project{\sigma'}{L_e^\delta}} q_e' \implies q_s'  \Trans{a}\land\; q_e' \Trans{a}$\\
            \proofstep{\Cref{def:projection}: $\projectop$}\\
            $\sigma'\in  \utraces{s\parcomp e}\; \land $\\
             $\exists q_s \in Q_s, q_e \in Q_e: s \Trans{\project{\sigma'\cdot a}{L_s^\delta}} q_s\land e \Trans{\project{\sigma'\cdot a}{L_e^\delta}} q_e\;\land$\\
            $\forall q_s' \in Q_s, q_e' \in Q_e : s \Trans{\project{\sigma'}{L_s^\delta}} q_s' \land  e \Trans{\project{\sigma'}{L_e^\delta}} q_e' \implies q_s'  \Trans{\project{a}{L_s^\delta}}\land q_e' \Trans{\project{a}{L_e^\delta}}$\\
            \proofstep{\Cref{lem:project_from_parcomp}}\\
            $\sigma'\in  \utraces{s\parcomp e}\; \land $\\
            $\exists q_s \in Q_s, q_e \in Q_e : s\parcomp e \Trans{\sigma'\cdot a} q_s \parcomp q_e\; \land$\\ 
            $\forall q_s' \in Q_s, q_e' \in Q_e : s \parcomp e \Trans{\sigma'} q_s' \parcomp q_e' \implies q_s'  \parcomp q_e' \Trans{a}$\\
            \proofstep{\Cref{def:uioco}: $Utraces$}\\
            $\sigma'\cdot a \in  \utraces{s\parcomp e}$\\
            
            \item[$a \in (U_s \setminus L_e)$:]\ \\
            $\project{\sigma'\cdot a}{L_s^\delta} \in  \utraces{s} \land \project{\sigma'\cdot a}{L_e^\delta} \in  \utraces{e}$\\
            \proofstep{\Cref{def:uioco}: $Utraces$}\\
            $\project{\sigma'\cdot a}{L_s^\delta} \in  \utraces{s} \land \project{\sigma'\cdot a}{L_e^\delta} \in  \utraces{e}\;\land$\\
            $\exists q_s \in Q_s, q_e \in Q_e : s \Trans{\project{\sigma'\cdot a}{L_s^\delta}} q_s \land e \Trans{\project{\sigma'\cdot a}{L_e^\delta}} q_e$\\
            \proofstep{\Cref{def:projection}: $\projectop$}\\
            $\project{\sigma'}{L_s^\delta}\cdot a \in  \utraces{s} \land \project{\sigma'}{L_e^\delta}\in  \utraces{e}\;\land$\\
            $\exists q_s \in Q_s, q_e \in Q_e : s \Trans{\project{\sigma'\cdot a}{L_s^\delta}} q_s \land e \Trans{\project{\sigma'\cdot a}{L_e^\delta}} q_e$\\
            \proofstep{\Cref{item:utrace_prefix_closed}}\\
            $\project{\sigma'}{L_s^\delta} \in  \utraces{s} \land \project{\sigma'}{L_e^\delta} \in  \utraces{e}\; \land$\\
            $\exists q_s \in Q_s, q_e \in Q_e : s \Trans{\project{\sigma'\cdot a}{L_s^\delta}} q_s \land e \Trans{\project{\sigma'\cdot a}{L_e^\delta}} q_e$\\
            \proofstep{Apply IH}\\
            $\sigma'\in  \utraces{s\parcomp e}\; \land $\\
            $\exists q_s \in Q_s, q_e \in Q_e : s \Trans{\project{\sigma'\cdot a}{L_s^\delta}} q_s \land e \Trans{\project{\sigma'\cdot a}{L_e^\delta}} q_e$\\
            \proofstep{\Cref{lem:project_from_parcomp_light} }\\
            $\sigma'\in  \utraces{s\parcomp e}\; \land $\\
            $\exists q_s \in Q_s, q_e \in Q_e : s\parcomp e \Trans{\sigma'\cdot a} q_s \parcomp q_e$\\
            \proofstep{\Cref{def:uioco}: $Utraces$}\\
            $\sigma'\cdot a \in  \utraces{s\parcomp e}$\\
            
            \item[$a \in (U_e \setminus L_s)$:]Symmetric to previous case.\\
        \end{case_distinction}
    \end{case_distinction}
\end{case_distinction}
\end{proof}

\end{toappendix}

\section{Compositional Testing}
\label{sec:problemStatement}
\FloatBarrier
To manage the size and complexity of modern software systems, a widely adopted strategy is to decompose the system into smaller, well-defined units known as components. This modular approach enables focused development and analysis at the component level, thereby improving manageability and scalability.
The overall system is constructed by composing these components. This approach offers several advantages: components are smaller and simpler to understand, they can be developed and tested independently, replaced more easily, and faults can be traced and localized more effectively.
A key to successful decomposition is \emph{compositionality}: the principle that properties and results established for individual components are preserved when the components are combined into a larger system.
In testing, compositionality implies that components can be tested in isolation, and the results of these tests remain valid when the components are integrated into the complete system.
In this section, we formalize our approach to compositional model-based testing using labelled transition systems (\Cref{def:LTS}), the implementation relation $\uioco$ (\Cref{def:uioco}), and parallel composition of LTS for component-based system construction (\Cref{def:parcomp}).

\subsection{Composition for Labelled Transition Systems}

Components are combined using the parallel composition operator on LTS. This operator models that all components are executed in parallel, executing independently from each other, while synchronizing on shared labels. For a well-defined and associative parallel composition, we require that the two component systems are \emph{composable}, i.e., they shall have disjunct sets of outputs. In practice, this is not a restriction, as non-composable systems can easily be made composable by renaming shared output labels.

In \Cref{fig:composition_label_division}, the communication structure is visualized for two components: a system component $s$ and an environment component $e$.
Shared labels, input for the one and output for the other component, are communicated between the two components and show as output to the outside world (arrows 4 and 5). Labels that only occur in one of the two components, communicate with the outside world, they are invisible to the other component, and are therefore called non-interacting (arrows 1, 2, 6 and 7). Shared input labels come from the outside world and are given to both components synchronously (arrow 3). The requirement of being composable forbids shared outputs. Inputs to the composed system are provided via arrows 1, 3, and 7, while outputs occur on arrows 2, 4, 5, and 6. The absence of any outputs is called quiescence (\Cref{def:delta_general}), and is denoted as $\delta$ (arrow 8). Quiescence for a composed system is more complicated than on the component level and can occur in two ways: either all components are quiescent, or there are components that are trying to communicate an output, but the other component is not ready to receive the corresponding input. These communication errors are the main reason why MBT is not always compositional, as quiescence of the system does not imply the quiescence of all components. We denote quiescence of a specific component $s$ as $\delta_s$ where this is relevant.

Composition is formalized for this two-component setup, which is straightforwardly extended to multiple components in a hierarchical way, treating each parallel composition of two components as a single new component on a higher level. We can chain this in any order, as parallel composition is both commutative and associative \cite{vancuyck_CompositionalityModelBasedTesting_2023a}.

\begin{definition}
\label{def:composable}
$s, e\in \LTS$ are \emph{composable} iff their respective output sets $U_s$ and $U_e$ are disjoint:~ $U_s \cap U_e = \emptyset$.
\end{definition}

\begin{definition}
\label{def:parcomp}
\emph{Parallel composition} $\parcomp$ on two $\composable$ labelled transition systems $s$ and $e$ is defined as ~ $s\parcomp e ~\defeq~ \langle Q,I,U,T,q_0 \rangle$,~ where:\\
$Q = \{\;q_s\parcomp q_e \:\setbar\: q_s\in Q_s, q_e\in Q_e\;\}$;\ 
$I\hspace{3pt} = (I_s \setminus U_e) \cup (I_e \setminus U_s)$;\ 
$U = U_s\cup U_e$;\ 
$q_0 = {q_0}_s\parcomp {q_0}_e$;\ 
$T$ is the smallest set satisfying the following inference rules (where $q_s,q_s' \in Q_s, q_e,q_e' \in Q_e$):
    \[\begin{array}[t]{l@{~}r@{~~~~}c@{~~~~}l}
        q_s \trans{\ell} q_s' &
        \ell \in (L_s \cup \{\tau\})\setminus L_e &
        \vdash &
        q_s \parcomp q_e \trans{\ell} q_s' \parcomp q_e \\
        q_e \trans{\ell} q_e' &
        \ell \in (L_e \cup \{\tau\})\setminus L_s &
        \vdash &
        q_s \parcomp q_e \trans{\ell} q_s \parcomp q_e' \\
        q_s \trans{\ell} q_s',~ q_e \trans{\ell} q_e' &
        \ell \in L_s \cap L_e &
        \vdash &
        q_s \parcomp q_e \trans{\ell} q_s' \parcomp q_e' \\
    \end{array}\]
\end{definition}

\begin{proposition}
\label{prop:assoc}
Let $s,p,e\in\LTS$, then
$s \parcomp e \equiv e \parcomp s$, ~and~
$(s \parcomp e) \parcomp p \equiv s \parcomp (e \parcomp p)$, \\
\emph{where $\equiv$ is LTS isomorphism}  \emph{\cite{vancuyck_CompositionalityModelBasedTesting_2023a}}.
\end{proposition}

\begin{figure}
    \centering
    \begin{tikzpicture}[on grid]
        \newlength{\arrowlength}
        \setlength{\arrowlength}{3.5cm}
        \setlength{\nodedistance}{\arrowlength}
        \tikzset{component/.style={rectangle,rounded corners, draw, inner xsep=15,inner ysep=10}}
        \tikzset{labelarrow/.style={thick,-{Latex[length=2.5mm,width=1.8mm]}}}
        \node[component] (s) {\Huge $s$};
        \node[component,right= \nodedistance of s] (e) {\Huge $e$};
        \node[coordinate] (sin) at (s.160) {};
        \node[coordinate, left = 0.6\arrowlength  of sin] (sinstart) {};
        \node[coordinate] (sout) at (s.200) {};
        \node[coordinate, left = 0.6\arrowlength  of sout] (soutend) {};
        \node[coordinate] (soutsync) at (s.20) {};
        \node[coordinate] (sinsync) at (s.340) {};
        \node[coordinate] (eout) at (e.20) {};
        \node[coordinate] (ein) at (e.340) {};
        \node[coordinate, right = 0.6\arrowlength  of ein] (einstart) {};
        \node[coordinate, right = 0.6\arrowlength  of eout] (eoutend) {};
        \node[coordinate] (eoutsync) at (e.200) {};
        \node[coordinate] (einsync) at (e.160) {};
        \node[coordinate, above right = .3\nodedistance and .5\nodedistance of s] (globalsyncinmerge) {};
        \node[coordinate, above = 0.5cm of globalsyncinmerge] (globalsyncin) {}; 
        \node[coordinate, below right = .3\nodedistance and .5\nodedistance of s] (deltamerge) {};
        \node[coordinate, below = 0.5cm of deltamerge] (delta) {}; 

        \path[labelarrow,font=\large] 
            (sinstart) edge []   node [above] {$1:I_s\setminus L_e$} (sin)
            (sout) edge []   node [below] {$2:U_s\setminus I_e$} (soutend)
            (einstart) edge []   node [below] {$7:I_e\setminus L_s$} (ein)
            (eout) edge []   node [above] {$6:U_e\setminus I_s$} (eoutend)
            (soutsync) edge []   node [above] {$4:U_s\cap I_e$} (einsync)
            (eoutsync) edge []   node [below] {$5:U_e\cap I_s$} (sinsync)
            (globalsyncinmerge) edge [] (s.north)
            (globalsyncinmerge) edge [] (e.north)
            (deltamerge) edge [] node [below=0.3cm] {$8: \delta$} (delta);
        \path[thick,-,font=\large]
            (s.south) edge [] node [below left] {$\delta_s$} (deltamerge)
            (e.south) edge [] node [below right] {$\delta_e$} (deltamerge)
            (globalsyncin) edge []   node [above=.3cm] {$3:I_s\cap I_s$} (globalsyncinmerge);
    \end{tikzpicture}
    \caption{Labels of a composed system}
    \label{fig:composition_label_division}
\end{figure}

\subsection{Compositional Testing for Labelled Transition Systems}

Compositionality in model-based testing with LTS means that we test component implementations with respect to their specifications, and that test results transfer to the composed system, i.e., the parallel composition of the components. For specifications $s,e\in\LTS$ and their implementations $i_s,i_e\in\IOTS$, respectively, this means that 
\begin{equation}
\label{eq:compos}
i_s \uioco s\ \land\ i_e \uioco e\ \implies\ i_s \parcomp i_e \uioco s \parcomp e
\end{equation}
It is known that in general compositionality according to \Cref{eq:compos} does not hold, 
but that it does hold when both specifications are input enabled $s,e\in\IOTS$ \cite{vanderbijl_CompositionalTestingIoco_2004}.
In this paper, We will give more precise conditions under which compositionality holds and, in addition, we will give procedures for these deciding these conditions, either by formal verification or by testing.

Three approaches with conditions of decreasing strength are proposed,
which are schematically depicted in 
\Cref{subfig:modelchecking_aproach,subfig:testing_aproach,subfig:cioco},
and which are elaborated in \Cref{sec:environmentalConformance,sec:testing-accepting-systems,sec:combining_algorithms}.
For all three of these approaches the desired outcome is that the composition of the implementations is $\uioco{}$-conforming to the composition of the specifications, as depicted in \Cref{subfig:comp_mbt_conclusion}. 

The first approach works on the level of specifications and uses the concept of \emph{mutual acceptance} introduced in \cite{vancuyck_CompositionalityModelBasedTesting_2023}:
if two specifications are mutually accepting then compositionality according to \Cref{eq:compos} holds.
We will give an algorithm to verify whether two specifications are mutually accepting. This is depicted in \Cref{subfig:modelchecking_aproach} and discussed in more detail in \Cref{sec:environmentalConformance}.
The advantage of formal verification is that it is complete and can therefore prove that two specifications are indeed mutually accepting, as opposed to testing, which would require an infinite amount of tests to be complete. The disadvantage is that it might not scale due the infamous state-space explosion problem.

The second approach uses testing as depicted in \Cref{subfig:testing_aproach} and elaborated in \Cref{sec:testing-accepting-systems}. Instead of verifying mutual acceptance between the specifications, we test for environmental conformance  \cite{frantzen_ModelBasedTestingEnvironmental_2007} between an implementation and a specification of its environment. We use the existing $\eco$ relation instead of $\mutuallyaccepts{}$ because $\eco$ already has some algorithms available. To also bring these algorithms to $\mutuallyaccepts{}$, and as a means to validate our definition for $\mutuallyaccepts{}$, we prove that both relations are actually equivalent on common domains in \Cref{sec:environmentalConformance}.

The third approach combines testing for $\uioco$ and $\eco$.
In the second approach each implementation is separately tested for $\uioco$ with respect to its specification and for $\eco$ with respect to its environment, which implies four test procedures. There is however a large overlap between the $\uioco$ and $\eco$ tests, which could be optimized by running them together. Both specifications put constraints on the implementation, and by using multiple specifications at once more of those constraints can be used to prune away redundant tests. For example, if $s$ states that the output $!x$ is not allowed and the test should fail, than any tests from $e$ for the behaviour after $!x$ can be removed without altering the test outcome.
This means that $i_s$ is tested for $\uioco\!$-conformance to $s$ only in the context of $e$, and the other way around, which means that each conformance individually might not hold.

We describe this idea in more detail in \Cref{sec:combining_algorithms},
and it is depicted in \Cref{subfig:cioco}, which shows that this approach involves arrows touching more than two entities (models or implementations). All three of these approaches have their own strengths and weaknesses, and are most useful in different contexts. We explore this further in \Cref{sec:discussion}.

\begin{figure}[htb]
\captionsetup{justification=centering}
\tikzset{uioco/.style={Black}}
\tikzset{highlight/.style={RoyalBlue}}
\tikzset{mutaccepts/.style={highlight}}
\tikzset{eco/.style={highlight}}
\tikzset{imp/.style={rounded corners=0}}
        \centering
        \begin{subfigure}{0.4\linewidth}
        \adjustbox{scale=1.5,center}{
            \begin{tikzcd}[sep=large,cells={nodes={draw=black,rounded corners,minimum size=15pt}}]
            s 
                \arrow[r, mutaccepts, Leftrightarrow, dashed, "\mutuallyaccepts{}" scale=1.5] & 
            e \\%
            |[imp]| i_s 
                \arrow[u, uioco, Leftrightarrow, "\uioco"] & 
            |[imp]| i_e 
                \arrow[u, uioco, swap, Leftrightarrow, "\uioco"] 
            \end{tikzcd}
        }
        \caption{Formal verification approach\\(\Cref{sec:environmentalConformance})}
        \label{subfig:modelchecking_aproach}
    \end{subfigure}
    \hspace{1cm}
    \begin{subfigure}{0.4\linewidth}
        \adjustbox{scale=1.5,center}{
            \begin{tikzcd}[sep=large,cells={nodes={draw=black,rounded corners,minimum size=15pt}}]
            s & 
            e \\%
            |[imp]| i_s 
                \arrow[u, uioco, Leftrightarrow,"\uioco"] 
                \arrow[ur, eco, Leftrightarrow, dashed, pos=0.25, sloped, "\eco"] & 
            |[imp]| i_e 
                \arrow[u, uioco, swap, Leftrightarrow, "\uioco"] 
                \arrow[ul, eco, Leftrightarrow, dashed, pos=0.25, sloped, "\eco"]
            \end{tikzcd}
        }
        \caption{Testing approach\\(\Cref{sec:testing-accepting-systems})}
        \label{subfig:testing_aproach}
    \end{subfigure}\\\vspace{5pt}
    \begin{subfigure}{0.4\linewidth}
        \adjustbox{scale=1.5,center}{
            \begin{tikzcd}[sep=large,cells={nodes={draw=black,rounded corners,minimum size=15pt}}]
            |[alias=s]| s 
                \arrow[phantom, d, "" {coordinate, pos=0.6, name=s_merge}]
                \arrow[from=s_merge, to=e, highlight, Rightarrow]
                & 
            |[alias=e]|e
                \arrow[phantom, d, "" {coordinate, pos=0.6, name=e_merge}]
                \arrow[from=e_merge, to=s, highlight, Rightarrow]
                \\%
            |[imp]| i_s 
                \arrow[u, highlight, pos=0.4, Leftrightarrow, "\cioco{}{~s}{e}" {description,xshift=-0.5cm}] & 
            |[imp]| i_e 
                \arrow[u, highlight, pos=0.4, Leftrightarrow, "\cioco{}{~e}{s}" {description,xshift=0.5cm}]
            \end{tikzcd}
        }
        \caption{Setting specific testing\\(\Cref{sec:combining_algorithms})}
        \label{subfig:cioco}
    \end{subfigure}
    \hspace{1cm}
    \begin{subfigure}{0.4\linewidth}
        \adjustbox{scale=1.5,center}{
            \begin{tikzcd}[
            row sep=large, column sep=normal,
            cells={nodes={draw=black,rounded corners,minimum size=15pt}},
            /tikz/execute at end picture={
                \node (largetop) [rectangle, highlight, rounded corners, draw, fit=(s) (e)] {};
                \node (largebot) [rectangle, highlight, imp , draw, fit=(is) (ie)] {};
                \draw[Leftrightarrow, uioco, highlight] (largebot) -- node [right] {$\scriptstyle\uioco$} (largetop);}
            ]
            |[alias=s]| s 
                \arrow[r, phantom, sloped, "\parcomp"] & 
            |[alias=e]| e \\%
            |[alias=is, imp]| i_s  
                \arrow[r, phantom, sloped, "\parcomp"] &
            |[alias=ie ,imp ]| i_e 
            \end{tikzcd}
        }
        \caption{Desired testing result}
        \label{subfig:comp_mbt_conclusion}
    \end{subfigure}
\caption{Compositional model-based testing overview:\\ three methods (a, b, c) to achieve a correct system (d)}
\label{fig:compMBTsetup}
\end{figure}

\FloatBarrier
\section{Mutually Accepting Specifications}
\label{sec:eco-verificaction}

In this section we discuss the first of three ways to do compositional model based testing: we check for compositionality at the specification level, and then use this to prove that all valid implementations are also compositional.

While parallel composition does not preserve correctness under $\uioco{}$ in general, it does preserve this correctness for specifications that are mutually accepting (\Cref{def:mutually_accepts}), which is the main result of our previous work \cite{vancuyck_CompositionalityModelBasedTesting_2023a}, repeated in \Cref{lem:compositional_testing_mutaccepts}. 
If $s \accepts{} e$, then whenever $e$ produces any output $\ell$ intended for $s$ (arrow 5 of \Cref{fig:composition_label_division}), $s$ must accept $\ell$ as input.

\begin{definition}
Let $s, e\in\LTS$ be composable. \\
Then $s$ \textbf{accepts}($\accepts{}$) / \textbf{mutually accepts}($\mutuallyaccepts{}$) $e$, respectively, then
\[\begin{array}[t]{l@{~~~}l@{~~~}l}
    s \accepts{} e & \defeq &
    \forall \sigma \in \utraces{s\parcomp e},\ q_s \in Q_s,\ q_e \in Q_e: \\
    & & ~~~~~~~ s \parcomp e \Trans{\sigma} q_s \parcomp q_e \ \implies\ 
    \outset{q_e}\cap I_s \:\subseteq\: \inset{q_s}\\
    s \mutuallyaccepts{} e ~~ & \defeq & ~~ s \accepts{} e \ \land\ e \accepts{} s
\end{array}\]
\label{def:accepting}
\label{def:mutually_accepts}
\end{definition}

\begin{theoremrep}
Let  $s,e \in\LTS$ be composable, $i_s, i_e\in \IOTS$, then\\
    \[s \mutuallyaccepts{} e \ \land\  
    i_s \uioco s \ \land\  i_e \uioco e \ \implies
    \ i_s \parcomp i_e \uioco s \parcomp e\]
\label{lem:compositional_testing_mutaccepts}
\end{theoremrep}

\begin{proof}
Theorem corresponds to Theorem 1 from \cite{vancuyck_CompositionalityModelBasedTesting_2023a}. Proof not repeated here.
\end{proof}

We give an alternative characterization for $\mutuallyaccepts{}$ in \Cref{sec:characterization}. We then prove this characterization equal to the original definition in \Cref{sec:eco_mutaccepts_proofsketch}, and end by giving an algorithm using this new characterization in \Cref{sec:eco_modelcheck_alg}
\label{sec:environmentalConformance}

\subsection{Characterization of Mutual Acceptance}
\label{sec:characterization}
One way to do compositional model based testing is to check if the specifications of all the components correctly interact with each other, as depicted in \Cref{subfig:modelchecking_aproach}. By doing this check once at the specification level, the results can be reused for all possible combinations of implementations. In previous work we defined such a relation called \emph{mutual acceptance} (\mutuallyaccepts{}, \Cref{def:mutually_accepts}). Mutually accepting systems are useful for model-based testing, because they preserve correctness: $\uioco$-correct components imply the $\uioco$-correctness of their parallel composition. However, the definition does not intuitively lead to an algorithm. This is because $\mutuallyaccepts{}$ is a statement over all $\utraces{}$ of the parallel composition. Since this is an infinite set, it is non-trivial to iterate over them. It is also unnecessary, as many $\utraces{}$ will reach the same states and thus have the same behaviour. We therefore give an alternative characterization of $\mutuallyaccepts{}$ which ranges over the reachable states instead of over traces. We do this by defining a type of \emph{bisimulation}, which is a property over a relation between two transition systems (\Cref{def:eco_rel}). If a relation with this property exists, we call the the two systems \textit{accepting}$-conform$ ($\aco$, \Cref{def:eco_conformance}).

\begin{samepage}
\begin{definition}
    \label{def:eco_rel}\ \\
    $\acceptrel \subseteq \powerset{Q_s} \times \powerset{Q_e}$ is an $\acceptsim$ for $s$ and $e \in \LTS$ iff: 
    \begin{enumerate}
        \item $(s \after \epsilon, e \after \epsilon) \in \acceptrel$ \label{item:eco_rel_base}
        \item  $\forall(X_s,X_e)\in \acceptrel$: 
        \begin{enumerate}
            \item $\forall \ell \in \outset{X_e} \cap I_s : \ell \in \inset{X_s} \land (X_s \after \ell, X_e \after \ell) \in \acceptrel$ \label{item:eco_rel_sync_out_env}
            \item  $\forall \ell \in \outset{X_s} \cap I_e : \ell \in \inset{X_e}\land (X_s \after \ell, X_e \after \ell) \in \acceptrel$ \label{item:eco_rel_sync_out_imp}
            \item $\forall \ell \in (\outset{X_e}\cup \inset{X_e})\setminus L_s^\delta: (X_s, X_e \after \ell) \in \acceptrel$ \label{item:eco_rel_internal_env}
            \item $\forall \ell \in (\outset{X_s}\cup \inset{X_s})\setminus L_e^\delta: (X_s \after \ell, X_e) \in \acceptrel$ \label{item:eco_rel_internal_imp}
            \item $\forall \ell \in (\inset{X_s}\cap \inset{X_e}):(X_s \after \ell, X_e \after \ell) \in \acceptrel$ \label{item:eco_rel_sync_in}
            \item $ \delta \in\outset{X_s} \land \delta \in \outset{X_e} \implies (X_s \after \delta, X_e \after \delta) \in \acceptrel$ \label{item:eco_rel_delta}
        \end{enumerate}
    \end{enumerate}
    
\end{definition}
\end{samepage}

\begin{definition}
    Let $s,e \in\LTS$. Then
    $s$ is $\textit{accepting}$-conform to $e$, denoted $s \aco e$, iff there exists an $\acceptsim$ for $s$ and $e$.
     \label{def:eco_conformance}
\end{definition}

An $\acceptsim$ relates the sets of states the components of a composed system can be in. It then adds the requirement that synchronised outputs generated by one component have to be defined as inputs in all corresponding reachable states of the other component, similar to mutual acceptance. 

Points \ref{item:eco_rel_sync_out_env} and \ref{item:eco_rel_sync_out_imp} cover the synchronised outputs. If they happen, both components change states. If an output is communicated, the receiving component must define how to handle it as an input. 
Points \ref{item:eco_rel_internal_env} and \ref{item:eco_rel_internal_imp} cover the non-interacting labels which only happen in one component. Point \ref{item:eco_rel_sync_in} covers synchronised inputs that are not outputs of the other component. These come from an unspecified third component that is part of the new environment after composition. These are only relevant if they are enabled in both $s$ and $e$, because otherwise they are not part of the \utraces{} of the composed system. The last point, \ref{item:eco_rel_delta}, covers quiescence. This is relevant because if the system displays quiescence, this is observable and it reduces the uncertainty over which state the system is in. Quiescence is only observable in a composed system if all components are quiescent, so cases where only one component is quiescent can be ignored here. 

\begin{lemmarep}
    \label{lem:eco_symetrical}
    $\aco$ \textup{is} symmetric, \textup{but not} reflexive \textup{or} transitive.
\end{lemmarep}
\begin{proof}
     \begin{case_distinction}

\item[\emph{Symmetric}:] 
    $\forall s, e \in \LTS: s \aco e \iff e \aco s$.\\
    The proof follows from the structure of $\acceptrel$. Swapping out $s$ and $e$ in \Cref{def:eco_rel} will result in the same definition with the left and right elements of $\acceptrel$ swapped. \Itemref{def:eco_rel}{base}, \itemref{def:eco_rel}{sync_in}, and \itemref{def:eco_rel}{delta} are all symetrical definitions, so swapping $s$ and $e$ around will result in exactly the same statement. \Itemref{def:eco_rel}{sync_out_env} and \itemref{def:eco_rel}{sync_out_imp}, and \itemref{def:eco_rel}{internal_env} and \itemref{def:eco_rel}{internal_imp} are pairwise symmetrical with each other. So swapping out $s$ and $e$ here will result in the corresponding other statement from the pair.
\item[\emph{Not reflexive}:] $\neg\forall s\in \LTS: s \eco s$. This follows trivially from \Cref{def:composable}: only an specification with no outputs, which is not a very useful specification, would be composable with itself. Non composable specifications are also automatically not $\aco$, as $\aco$ is only defined for composable specifications.
\item[\emph{Not transitive}:] 
    $\neg\forall s, e, e'\in \LTS: s \aco e' \land e' \aco e \implies s \aco e$. 
    Any communication between $s$ and $e$ is not checked in either $s\aco e'$ or in $e' \aco e$. This means that $s$ can contain communication errors, but if they are on a different interface that is not visible to $e$ then they are not verified unless $s\aco e$ is checked directly. 
\end{case_distinction}

\end{proof}

An $\acceptsim$ describes that the interfaces of two specifications fit well together, not that they are equivalent in some way. As such it is symmetric , but neither reflexive nor transitive (\Cref{lem:eco_symetrical}). Composing a component with itself would violate the property of composability. Transitivity does not work, because $\aco$ only talks about one interface at a time, usually between $s$ and $e$. Checking that communication can correctly flow trough a middleware component ($s \aco e'$ and $e' \aco e$) does not automatically imply that communication trough a more direct interface also works ($s \aco e$).

\begin{example}
An \acceptsim{} $\acceptrel$ for the two specifications in \Cref{fig:ecosym_example} is $\{(\{1\},\{A\}),(\{2\},\{B,C\})\}$. It contains $(\Cref{spec:ecosym_s} \after \epsilon, \Cref{spec:ecosym_e} \after \epsilon)$ as per point \ref{item:eco_rel_base} of \Cref{def:eco_rel}. Then it adds $(\Cref{spec:ecosym_s} \after \ltslabel{a}, \Cref{spec:ecosym_e} \after \ltslabel{a})$ per point \ref{item:eco_rel_sync_out_imp}. Finally, $(\Cref{spec:ecosym_s} \after \ltslabel{a}\cdot\ltslabel{b}, \Cref{spec:ecosym_e} \after \ltslabel{a}\cdot\ltslabel{b})$ should also be added as per point \ref{item:eco_rel_sync_out_env}, but since this is $(\{1\},\{A\})$ again which is already in the relation, we are done. This expresses among other things, that when running \Cref{spec:ecosym_s} and \Cref{spec:ecosym_e} in parallel, we can reach a configuration where \Cref{spec:ecosym_s} is in state 2, and \Cref{spec:ecosym_e} is in either state $B$ or $C$. For such a system to function correctly, it is therefore required that state 2 defines all communicated outputs from either state $B$ or $C$ as an input (and the other way around). State 2 does not need to define the outputs of state $A$ as inputs, as there is no element $(X_{\Cref{spec:ecosym_s}},X_{\Cref{spec:ecosym_e}})\in\acceptrel$ such that $2\in X_{\Cref{spec:ecosym_s}} $ and $A\in X_{\Cref{spec:ecosym_e}}$.
\end{example}

\begin{figure}
    \centering
    \begin{subfigure}[b]{.5\linewidth}
    \centering
    \newspeclabel\label{spec:ecosym_s}
    \begin{tikzpicture}[LTS]
    \node[state,initial] (1) {1};
    \node[state, below = of 1] (2) {2};

    \path[->] 
        (1) edge [bend left]    node [auto] {\ltslabel{!a}} (2)
        (2) edge [bend left]    node [auto] {\ltslabel{?b}} (1)
            edge [loop left]    node [auto] {$\delta$} (2)
        ;

    \end{tikzpicture}%
    \caption{\currentspec,  with I=\{\ltslabel{b}\} and U=\{\ltslabel{a}\}}
    \end{subfigure}%
    \begin{subfigure}[b]{.5\linewidth}
    \centering
    \newspeclabel\label{spec:ecosym_e}
    \begin{tikzpicture}[LTS]
    \node[state,initial] (A) {A};
    \node[state, below right = .7\nodedistance and .4\nodedistance of A] (B) {B};
    \node[state, below left = .7\nodedistance and .4\nodedistance  of A] (C) {C};

    \path[->] 
        (A) edge [bend left]    node [auto] {\ltslabel{?a}} (B)
            edge [loop above]   node [auto] {$\delta$} (A)
        (B) edge [bend left]    node [auto] {$\tau$} (C)
        (C) edge [bend left]    node [auto] {\ltslabel{!b}} (A)
        ;
    
    \end{tikzpicture}
    \caption{\currentspec,  with I=\{\ltslabel{a}\} and U=\{\ltslabel{b}\}}
    \end{subfigure}

    \begin{subfigure}{\linewidth}
        \centering
        \[\acceptrel: \{(\{1\},\{A\}),(\{2\},\{B,C\})\}\]
        \caption{An $\acceptsim$ for \Cref{spec:ecosym_s} and \Cref{spec:ecosym_e}}
    \end{subfigure}
    
    \caption{$\acceptsim{}$ example}
    \label{fig:ecosym_example}
\end{figure}

\Cref{def:eco_conformance,def:eco_rel} are based on the $\ecosim$ from \cite{man:frantzen_SymbolicConformance_2024}, which we adapt several ways. We expand the scope with non-interacting labels, as well as synchronized inputs. We also take quiescence into account, which is required to make the link between $\aco$-conformance of components and $\uioco{}$-conformance of the composed system. Finally, we relate two specifications, while the original $\ecosim$ was only defined between an implementation and a specification. Since implementations are just input-enabled specifications in our setting, this is a strict increase in scope.

The alternative characterization for $\mutuallyaccepts{}$ given in \Cref{def:eco_rel} is equivalent to the original definition, as described in \Cref{lem:eco_equals_mutaccepts}. This is because the behaviour after a trace $\sigma\in \straces{s}$, i.e. whether $\sigma\cdot\ell$ is possible or not, is entirely described by the set of states that could be reached with that trace ($s \after \sigma$). It does not matter which trace was used, only the possible end states influence the next possible step.  We give a more detailed proof sketch in \Cref{sec:eco_mutaccepts_proofsketch}, and then use the fact that we can range over sets of states instead of traces to develop an algorithm for checking $\mutuallyaccepts{}$ in \Cref{sec:eco_modelcheck_alg}.

\begin{theoremrep}
    \label{lem:eco_equals_mutaccepts}
     Let $s,e\in \LTS$ be $\composable$, then
    \[s \aco e \iff s \mutuallyaccepts{} e\]
\end{theoremrep}

\begin{proof}
    Proof follows directly from \Cref{lem:mutaccepts_implies_eco,lem:eco_implies_mutaccepts}
\end{proof}

\subsection{Proof Sketch}
\label{sec:eco_mutaccepts_proofsketch}

 In order to prove the existence of an $\acceptsim{}$ we will first construct a specific relation $\uiocorel$, and then prove that $\uiocorel$ is an $\acceptsim$. The relation we use for this is shown in \Cref{def:uioco_rel}. It is almost equivalent to \Cref{def:eco_rel}, but without the extra conditions imposed by rules \ref{item:eco_rel_sync_out_env} and \ref{item:eco_rel_sync_out_imp}. Because of this, there is exactly one $\uiocorel$ for every $s,e\in \LTS$, and it represents the sets of states that can be reached with the $\utraces{}$ of their parallel composition. We write $\uiocorel$ for $\uiocorel(s,e)$ when $s$ and $e$ are clear from context. We split the  equivalence proof for \Cref{lem:eco_equals_mutaccepts} up into two implications and give a proof sketch for each side separately below. For a more detailed proof per lemma, see the appendix.

\begin{samepage}
\setcounter{equation}{0}
\begin{definition}
    \label{def:uioco_rel}
    Let $s,e \in\LTS$, $X_s \subseteq Q_s$, $X_e \subseteq Q_e$, $\ell \in L_s \cup L_e \cup \{\delta\}$.\\Then
    $\uiocorel(s,e) \subseteq \powerset{Q_s} \times \powerset{Q_e}$, is the smallest set satisfying the following inference rules:
    \small
    \begin{alignat}{2}
        &&&\vdash (s \after \epsilon, e \after \epsilon) \in \uiocorel \label{item:uioco_rel_base} \\
        &(X_s,X_e)\in\uiocorel, \quad &\ell \in \outset{X_e} \cap I_s  &\vdash (X_s \after \ell, X_e \after \ell) \in \uiocorel \label{item:uioco_rel_sync_out_env}\\
        &(X_s,X_e)\in\uiocorel, \quad &\ell \in \outset{X_s} \cap I_e &\vdash (X_s \after \ell, X_e \after \ell) \in \uiocorel \label{item:uioco_rel_sync_out_imp}\\
        &(X_s,X_e)\in\uiocorel, \quad &\ell \in (\outset{X_e}\cup \inset{X_e})\setminus L_s^\delta &\vdash (X_s, X_e \after \ell) \in \uiocorel\label{item:uioco_rel_internal_env}\\
        &(X_s,X_e)\in\uiocorel, \quad &\ell \in (\outset{X_s}\cup \inset{X_s})\setminus L_e^\delta &\vdash (X_s \after \ell, X_e) \in \uiocorel\label{item:uioco_rel_internal_imp}\\
        &(X_s,X_e)\in\uiocorel, \quad &\ell \in (\inset{X_s}\cap \inset{X_e}) &\vdash (X_s \after \ell, X_e \after \ell) \in \uiocorel \label{item:uioco_rel_sync_in}\\
        &(X_s,X_e)\in\uiocorel, \quad &\delta \in\outset{X_s} \land \delta \in \outset{X_e} &\vdash (X_s \after \delta, X_e \after \delta) \in \uiocorel \label{item:uioco_rel_delta}
    \end{alignat}
\end{definition}
\end{samepage}

\begin{toappendix}
    \begin{lemma}
        Let $s,e \in \LTS$, $X_s\subseteq Q_s$, $X_e\subseteq Q_e$ :
        \[(X_s,X_e)\in \uiocorel(s,e) \implies X_s \after \epsilon = X_s \land X_e \after \epsilon = X_e\]
    \label{lem:uioco_rel_after_epsilon}
    \end{lemma}
    \begin{proof}
        By induction on the structure of $\uiocorel$ (\Cref{def:uioco_rel}) we can easily see that all $(X_s,X_e)\in \uiocorel$ are of the form $(s\after \sigma,e\after \sigma')$, for some $\sigma \in L_s^{\delta*},\sigma'\in L_e^{\delta*}$. This combined with the fact that $(X \after \sigma) \after \epsilon = X \after \sigma$ completes the proof.
    \end{proof}
\end{toappendix}

\begin{lemmarep}
    \label{lem:uiocosim_minimal}
   Let $s,e \in\LTS$ and $\acceptrel$ be an $\acceptsim$ for $s$  and $e$, then
    \[\uiocorel \subseteq \acceptrel\]
\end{lemmarep}
\begin{proof}
    Proof follows by induction on the structure of $\uiocorel$. Case names follow the names of \Cref{def:uioco_rel}. For the recursive cases, the IH assumes the lemma is true for the elements on the left side of the $\vdash$ of the current case of \Cref{def:uioco_rel}.
\begin{case_distinction}
    \item[Case \ref{item:uioco_rel_base}:]\ \\
        $(s \after \epsilon, e\after \epsilon) \in \uiocorel$\\
        \proofstep{\Itemref{def:eco_rel}{base}}\\
        $(s \after \epsilon, e\after \epsilon) \in \acceptrel$
    \item[Case \ref{item:uioco_rel_sync_out_env}:]\ \\
        $\ell \in \outset{X_e}\cap I_s \land(X_s,X_e)\in\uiocorel\land (X_s \after \ell, X_e \after \ell) \in \uiocorel$\\
        \proofstep{Apply IH}\\
        $\ell \in \outset{X_e}\cap I_s \land(X_s,X_e)\in\acceptrel$\\
        \proofstep{\Itemref{def:eco_rel}{sync_out_env}}\\
        $(X_s \after \ell, X_e \after \ell) \in \acceptrel$
    \item[Case \ref{item:uioco_rel_sync_out_imp}:]\ \\
        $\ell \in \outset{X_s}\cap I_e \land(X_s,X_e)\in\uiocorel\land (X_s \after \ell, X_e \after \ell) \in \uiocorel$\\
        \proofstep{Apply IH}\\
        $\ell \in \outset{X_s}\cap I_e \land(X_s,X_e)\in\acceptrel$\\
        \proofstep{\Itemref{def:eco_rel}{sync_out_imp}}\\
        $(X_s \after \ell, X_e \after \ell) \in \acceptrel$
    \item[Case \ref{item:uioco_rel_internal_env}:]\ \\
        $\ell \in (\outset{X_e}\cup \inset{X_e})\setminus L_s^\delta \land(X_s,X_e)\in\uiocorel\land (X_s, X_e \after \ell) \in \uiocorel$\\
        \proofstep{Apply IH}\\
        $\ell \in (\outset{X_e}\cup \inset{X_e})\setminus L_s^\delta \land(X_s,X_e)\in\acceptrel$\\
        \proofstep{\Itemref{def:eco_rel}{internal_env}}\\
        $(X_s, X_e \after \ell) \in \acceptrel$
    \item[Case \ref{item:uioco_rel_internal_imp}:]\ \\
        $\ell \in (\outset{X_s}\cup \inset{X_s})\setminus L_e^\delta \land(X_s,X_e)\in\uiocorel\land (X_s \after \ell, X_e) \in \uiocorel$\\
        \proofstep{Apply IH}\\
        $\ell \in (\outset{X_s}\cup \inset{X_s})\setminus L_e^\delta \land(X_s,X_e)\in\acceptrel$\\
        \proofstep{\Itemref{def:eco_rel}{internal_imp}}\\
        $(X_s \after \ell, X_e) \in \acceptrel$
    \item[Case \ref{item:uioco_rel_sync_in}:]\ \\
        $\ell \in (\inset{X_s}\cap \inset{X_e}) \land(X_s,X_e)\in\uiocorel\land (X_s \after \ell, X_e \after \ell) \in \uiocorel$\\
        \proofstep{Apply IH}\\
        $\ell \in (\inset{X_s}\cap \inset{X_e}) \land(X_s,X_e)\in\acceptrel$\\
        \proofstep{\Itemref{def:eco_rel}{sync_in}}\\
        $(X_s \after \ell, X_e \after \ell) \in \acceptrel$
    \item[Case \ref{item:uioco_rel_delta}:]\ \\
        $\delta \in \outset{X_s} \land \delta \in \outset{X_e} \land(X_s,X_e)\in\uiocorel \land (X_s \after \delta, X_e \after \delta) \in \uiocorel$\\
        \proofstep{Apply IH}\\
        $\delta \in \outset{X_s} \land \delta \in \outset{X_e} \land(X_s,X_e)\in\acceptrel$\\
        \proofstep{\Itemref{def:eco_rel}{delta}}\\
        $(X_s \after \delta, X_e \after \delta) \in \acceptrel$
\end{case_distinction}

\end{proof}

\begin{corollaryrep}
\label{lem:constructive_ecosim}
Let $s,e \in\LTS$, then
    \[s \aco e \iff \uiocorel \textit{ is an } \acceptsim \textit{ for } s \textit{ and } e\]
\end{corollaryrep}

\begin{proof}
    \begin{case_distinction}
    \item[$\implies$:] 
    $s\aco e$ gives $\exists \acceptrel$. Then \Cref{lem:uiocosim_minimal} states that $\uiocorel = \acceptrel$.
    \item[$\impliedby$:] Follows directly from \Cref{def:eco_conformance}.
\end{case_distinction}
\end{proof}

\subsubsection{Accepting Conformance Implies Mutual Acceptance}
For states within an $\acceptsim$ $\acceptrel$, quiescence is preserved over parallel composition, which is not the case in general. Quiescence at the system level could be caused by communication errors, in which case not all individual components are quiescent. These communication errors are excluded by $\aco$, and since $\utraces{}$ without $\delta$ are preserved under parallel composition, this generalises to all $\utraces{}$ in \Cref{lem:project_from_parcomp_eco}. However, just showing that the traces are preserved under composition is not enough on its own. We must also show that the states within an $\acceptsim$ are exactly those reachable by $\utraces{}$ (\Cref{lem:R_utraces_bi-implication}).

\begin{toappendix}
\begin{lemmarep}
     Let $s,e\in\LTS$ be $\composable$, $s\aco e$, $X_s\subseteq Q_s$, $X_e\subseteq Q_e$, $(X_s,X_e)\in \uiocorel$, $q_{xs}\in X_s$, $q_{xe}\in X_e$, $q_s\in Q_s$, $q_e\in  Q_e$: 
     \[q_{xs} \parcomp q_{xe} \Trans{\delta} q_s \parcomp q_e \iff q_{xs} \Trans{\delta} q_s \land q_{x_e} \Trans{\delta} q_e\]
     \label{lem:parcomp_delta_eco_rel}
\end{lemmarep}
\begin{proof}
    \begin{case_distinction}
        \item[$\implies$:]\ \\
            $q_{xs} \parcomp q_{xe} \Trans{\delta} q_s \parcomp q_e$\\
            \proofstep{\Cref{def:arrowdefs}: $\Trans{\delta}$}\\
            $\exists q_{xs}', q_s'\in Q_s, q_{xe}', q_e'\in Q_e:$\\
            $ q_{xs} \parcomp q_{xe} \Trans{\epsilon} q_{xs}' \parcomp q_{xe}' \land  q_{xs}' \parcomp q_{xe}'  \trans{\delta} q_s' \parcomp q_e' \land q_s' \parcomp q_e' \Trans{\epsilon} q_s \parcomp q_e$\\
            \proofstep{\Cref{def:delta}: $\delta$}\\
            $\exists q_{xs}'\in Q_s, q_{xe}'\in Q_e:$\\
            $ q_{xs} \parcomp q_{xe} \Trans{\epsilon} q_{xs}' \parcomp q_{xe}' \land  q_{xs}' \parcomp q_{xe}'  \trans{\delta} q_{xs}' \parcomp q_{xe}' \land q_{xs}' \parcomp q_{xe}' \Trans{\epsilon} q_s \parcomp q_e$\\
            \proofstep{\Cref{lem:uioco_rel_after_epsilon}}\\
            $\exists q_{xs}'\in X_s, q_{xe}'\in X_e:$\\
            $ q_{xs} \parcomp q_{xe} \Trans{\epsilon} q_{xs}' \parcomp q_{xe}' \land  q_{xs}' \parcomp q_{xe}'  \trans{\delta} q_{xs}' \parcomp q_{xe}' \land q_{xs}' \parcomp q_{xe}' \Trans{\epsilon} q_s \parcomp q_e$\\
            \proofstep{\Cref{def:delta}: $\delta$}\\
            $\exists q_{xs}'\in X_s, q_{xe}'\in X_e: q_{xs} \parcomp q_{xe} \Trans{\epsilon} q_{xs}' \parcomp q_{xe}'  \land q_{xs}' \parcomp q_{xe}' \Trans{\epsilon} q_s \parcomp q_e$\\
            $\forall \ell \in U_s \cup U_e \cup \{\tau\}: q_{xs}' \parcomp q_{xe}' \nottrans{\ell}$\\
            \proofstep{\Cref{def:parcomp}: $\parcomp$}\\
            $\exists q_{xs}'\in X_s, q_{xe}'\in X_e: q_{xs} \parcomp q_{xe} \Trans{\epsilon} q_{xs}' \parcomp q_{xe}'  \land q_{xs}' \parcomp q_{xe}' \Trans{\epsilon} q_s \parcomp q_e$\\
            $\forall \ell \in (U_s \setminus L_e) \cup \{\tau\}: q_{xs}' \nottrans{\ell} \land$\\
            $\forall \ell \in (U_e \setminus L_s) \cup \{\tau\}: q_{xe}' \nottrans{\ell} \land$\\
            $\forall \ell \in (U_s \cap L_e) \cup (U_e \cap L_s): q_{xs}' \nottrans{\ell} \lor\; q_{xe}' \nottrans{\ell}$\\
            \proofstep{\Itemref{def:eco_rel}{sync_out_env} and \itemref{def:eco_rel}{sync_out_imp}: $\acceptrel$, combined with \Cref{lem:constructive_ecosim}}\\
            $\exists q_{xs}'\in X_s, q_{xe}'\in X_e: q_{xs} \parcomp q_{xe} \Trans{\epsilon} q_{xs}' \parcomp q_{xe}'  \land q_{xs}' \parcomp q_{xe}' \Trans{\epsilon} q_s \parcomp q_e$\\
            $\forall \ell \in (U_s \setminus L_e) \cup \{\tau\}: q_{xs}' \nottrans{\ell} \land$\\
            $\forall \ell \in (U_e \setminus L_s) \cup \{\tau\}: q_{xe}' \nottrans{\ell} \land$\\
            $\forall \ell \in (U_s \cap L_e) \cup (U_e \cap L_s): q_{xs}' \nottrans{\ell} \lor\; q_{xe}' \nottrans{\ell}\land$\\
            $\forall \ell \in \outset{X_s}\cap I_e: \ell \in \inset{X_e}$\\
            $\forall \ell \in \outset{X_e}\cap I_s: \ell \in \inset{X_s}$\\
            \proofstep{\Cref{def:out,def:inset}: $out$ and $in$}\\
            $\exists q_{xs}'\in X_s, q_{xe}'\in X_e: q_{xs} \parcomp q_{xe} \Trans{\epsilon} q_{xs}' \parcomp q_{xe}'  \land q_{xs}' \parcomp q_{xe}' \Trans{\epsilon} q_s \parcomp q_e$\\
            $\forall \ell \in (U_s \setminus L_e) \cup \{\tau\}: q_{xs}' \nottrans{\ell} \land$\\
            $\forall \ell \in (U_s \setminus L_e) \cup \{\tau\}: q_{xs}' \nottrans{\ell} \land$\\
            $\forall \ell \in (U_e \setminus L_s) \cup \{\tau\}: q_{xe}' \nottrans{\ell} \land$\\
            $\forall \ell \in(U_s \cap L_e) \cup (U_e \cap L_s): q_{xs}' \nottrans{\ell} \lor\; q_{xe}' \nottrans{\ell}\land$\\
            $\forall \ell \in U_s \cap I_e: q_{xs}' \trans{\ell} \implies  q_{xe}' \Trans{\ell} \land$\\ 
            $\forall \ell \in U_e \cap I_s: q_{xe}' \trans{\ell} \implies  q_{xs}' \Trans{\ell}$\\
            \proofstep{$q_{xs}' \nottrans{\tau} \land\; q_{xe}'\nottrans{\tau}$}\\
            $\exists q_{xs}'\in X_s, q_{xe}'\in X_e: q_{xs} \parcomp q_{xe} \Trans{\epsilon} q_{xs}' \parcomp q_{xe}'  \land q_{xs}' \parcomp q_{xe}' \Trans{\epsilon} q_s \parcomp q_e$\\
            $\forall \ell \in (U_s \setminus L_e) \cup \{\tau\}: q_{xs}' \nottrans{\ell} \land$\\
            $\forall \ell \in (U_s \setminus L_e) \cup \{\tau\}: q_{xs}' \nottrans{\ell} \land$\\
            $\forall \ell \in (U_e \setminus L_s) \cup \{\tau\}: q_{xe}' \nottrans{\ell} \land$\\
            $\forall \ell \in (U_s \cap L_e) \cup (U_e \cap L_s): q_{xs}' \nottrans{\ell} \lor\; q_{xe}' \nottrans{\ell}\land$\\
            $\forall \ell \in U_s \cap I_e : q_{xs}' \trans{\ell} \implies q_{xe}' \trans{\ell} \land$\\
            $\forall \ell \in U_e \cap I_s : q_{xe}' \trans{\ell} \implies q_{xs}' \trans{\ell}$\\
            \proofstep{$(A \implies B) \iff (\neg B \implies \neg A)$}\\
            $\exists q_{xs}'\in X_s, q_{xe}'\in X_e: q_{xs} \parcomp q_{xe} \Trans{\epsilon} q_{xs}' \parcomp q_{xe}'  \land q_{xs}' \parcomp q_{xe}' \Trans{\epsilon} q_s \parcomp q_e$\\
            $\forall \ell \in (U_s \setminus L_e) \cup \{\tau\}: q_{xs}' \nottrans{\ell} \land$\\
            $\forall \ell \in (U_e \setminus L_s) \cup \{\tau\}: q_{xe}' \nottrans{\ell} \land$\\
            $\forall \ell \in(U_s \cap L_e) \cup (U_e \cap L_s): q_{xs}' \nottrans{\ell} \lor\; q_{xe}' \nottrans{\ell}\land$\\
            $\forall \ell \in U_s \cap I_e : q_{xe}' \nottrans{\ell} \implies q_{xs}' \nottrans{\ell}\land$\\
            $\forall \ell \in U_e \cap I_s : q_{xs}' \nottrans{\ell} \implies q_{xe}' \nottrans{\ell}$\\
            \proofstep{\Cref{def:composable}: $U_s \cap U_e = \emptyset$}\\
            $\exists q_{xs}'\in X_s, q_{xe}'\in X_e: q_{xs} \parcomp q_{xe} \Trans{\epsilon} q_{xs}' \parcomp q_{xe}'  \land q_{xs}' \parcomp q_{xe}' \Trans{\epsilon} q_s \parcomp q_e$\\
            $\forall \ell \in (U_s \setminus L_e) \cup \{\tau\}: q_{xs}' \nottrans{\ell} \land$\\
            $\forall \ell \in (U_e \setminus L_s) \cup \{\tau\}: q_{xe}' \nottrans{\ell} \land$\\
            $\forall \ell \in (U_s \cap L_e) \cup (U_e \cap L_s): q_{xs}' \nottrans{\ell} \lor\; q_{xe}' \nottrans{\ell}\land$\\
            $\forall \ell \in U_s \cap L_e : q_{xe}' \nottrans{\ell} \implies q_{xs}' \nottrans{\ell}\land$\\
            $\forall \ell \in U_e \cap L_s : q_{xs}' \nottrans{\ell} \implies q_{xe}' \nottrans{\ell}$\\
            \proofstep{$(A \lor B) \land (A \implies B) \implies B$}\\
            $\exists q_{xs}'\in X_s, q_{xe}'\in X_e: q_{xs} \parcomp q_{xe} \Trans{\epsilon} q_{xs}' \parcomp q_{xe}'  \land q_{xs}' \parcomp q_{xe}' \Trans{\epsilon} q_s \parcomp q_e$\\
            $\forall \ell \in (U_s \setminus L_e) \cup \{\tau\}: q_{xs}' \nottrans{\ell} \land$\\
            $\forall \ell \in (U_e \setminus L_s) \cup \{\tau\}: q_{xe}' \nottrans{\ell} \land$\\
            $\forall \ell \in U_s \cap L_e :  q_{xs}' \nottrans{\ell}\land$\\
            $\forall \ell \in U_e \cap L_s :  q_{xe}' \nottrans{\ell}$\\
            \proofstep{$(X \setminus Y) \cup (X \cap Y) = X$}\\
            $\exists q_{xs}'\in X_s, q_{xe}'\in X_e: q_{xs} \parcomp q_{xe} \Trans{\epsilon} q_{xs}' \parcomp q_{xe}'  \land q_{xs}' \parcomp q_{xe}' \Trans{\epsilon} q_s \parcomp q_e$\\
            $\forall \ell \in U_s \cup \{\tau\} : q_{xs}' \nottrans{\ell} \land$\\ $\forall \ell \in U_e \cup \{\tau\} : q_{xe}' \nottrans{\ell}$\\
            \proofstep{\Cref{def:delta}: $\delta$}\\
            $\exists q_{xs}'\in X_s, q_{xe}'\in X_e: q_{xs} \parcomp q_{xe} \Trans{\epsilon} q_{xs}' \parcomp q_{xe}'  \land q_{xs}' \parcomp q_{xe}' \Trans{\epsilon} q_s \parcomp q_e$\\
            $q_{xs}' \trans{\delta} q_{xs}' \land q_{xe}' \trans{\delta} q_{xe}'$\\
            \proofstep{\Cref{lem:project_from_parcomp_light} and \Cref{def:arrowdefs,}: $\Trans{\delta}$}\\
            $q_{xs} \Trans{\delta} q_s \land q_{xe} \Trans{\delta} q_e$
        \item[$\impliedby$:] Covered by \Cref{lem:project_from_parcomp_light}
    \end{case_distinction}
\end{proof}
\end{toappendix}

\begin{toappendix}
    \begin{lemma}
    Let $s,e \in \LTS$ be $\composable$, $s\aco e$, $\sigma \in \utraces{s\parcomp e}$, $ q_s\in Q_s$, $q_e\in Q_e:$
    \[s \parcomp e \Trans{\sigma} q_s\parcomp q_e \implies (s \after \project{\sigma}{L_s^\delta}, e \after \project{\sigma}{L_e^\delta}) \in \uiocorel \land s \Trans{ \project{\sigma}{L_s^\delta}} q_s \land e \Trans{ \project{\sigma}{L_e^\delta}} q_e\]
    \label{lem:once_in_R_always_in_R}
    \end{lemma}
    \begin{proof}
        Proof by induction on $\sigma$, with further case distinction on the last label of $\sigma$. Due to \cref{lem:constructive_ecosim}, $\uiocorel$ is also an $\acceptsim$ and is used as such directly in the proof.
\begin{case_distinction}
    \item[Base case: $\sigma=\epsilon$.]
        Immediate from \itemref{def:eco_rel}{base} and \cref{lem:project_from_parcomp_no_delta}
    \item[Inductive step: $\sigma=\sigma'\cdot\ell$.]\ \\
        Induction Hypothesis:
        \begin{align*}
            &\forall q_s'\in Q_s, q_e'\in Q_e: s \parcomp e \Trans{\sigma'} q_s'\parcomp q_e' \implies\\
            &\tab(s \after \project{\sigma'}{L_s^\delta}, e \after \project{\sigma'}{L_e^\delta}) \in \uiocorel \;\land\\
            &\tab s \Trans{ \project{\sigma'}{L_s^\delta}} q_s' \land e \Trans{ \project{\sigma'}{L_e^\delta}} q_e'
        \end{align*}
    \begin{case_distinction}
        \item[$\sigma=\sigma'\cdot\ell$, $(\ell \in U_s\cap I_e):$]\ \\
            $s \parcomp e \Trans{\sigma\cdot\ell} q_s\parcomp q_e $\\
            \proofstep{\Cref{item:trans_transitive}}\\
            $\exists q_s'\in Q_s, q_e'\in Q_e:  s \parcomp e \Trans{\sigma'} q_s'\parcomp q_e' \land q_s'\parcomp q_e' \Trans{\ell} q_s\parcomp q_e$\\
            \proofstep{Apply IH}\\
            $(s \after \project{\sigma'}{L_s^\delta}, e \after \project{\sigma'}{L_e^\delta}) \in \uiocorel \;\land$\\
            $\exists q_s'\in Q_s, q_e'\in Q_e: s \Trans{ \project{\sigma'}{L_s^\delta}} q_s' \land e \Trans{ \project{\sigma'}{L_e^\delta}} q_e'\land q_s'\parcomp q_e' \Trans{\ell} q_s\parcomp q_e$\\
            \proofstep{\Cref{lem:project_from_parcomp_no_delta} and \cref{def:projection}: $\projectop$}\\
            $(s \after \project{\sigma'}{L_s^\delta}, e \after \project{\sigma'}{L_e^\delta}) \in \uiocorel \;\land$\\
            $\exists q_s'\in Q_s, q_e'\in Q_e: s \Trans{ \project{\sigma'}{L_s^\delta}} q_s' \land e \Trans{ \project{\sigma'}{L_e^\delta}} q_e'\land q_s' \Trans{\ell} q_s\land q_e' \Trans{\ell} q_e$\\
            \proofstep{\Cref{item:trans_transitive} and \cref{def:projection}: $\projectop$}\\
            $(s \after \project{\sigma'}{L_s^\delta}, e \after \project{\sigma'}{L_e^\delta}) \in \uiocorel \;\land$\\
            $\exists q_s'\in Q_s, q_e'\in Q_e: s \Trans{ \project{\sigma'}{L_s^\delta}} q_s' \land q_s'\Trans{\ell}\;\land$\\
            $s \Trans{ \project{\sigma'\cdot\ell}{L_s^\delta}} q_s \land e \Trans{ \project{\sigma'\cdot\ell}{L_e^\delta}} q_e$\\
            \proofstep{\Cref{def:after,def:outset}: $\after$ and $\outset{}$}\\
            $(s \after \project{\sigma'}{L_s^\delta}, e \after \project{\sigma'}{L_e^\delta}) \in \uiocorel \;\land$\\
            $\exists q_s'\in Q_s, q_e'\in Q_e: \ell \in \outset{s \after \project{\sigma'}{L_s^\delta}} \;\land$\\
            $s \Trans{ \project{\sigma'\cdot\ell}{L_s^\delta}} q_s \land e \Trans{ \project{\sigma'\cdot\ell}{L_e^\delta}} q_e$\\
            \proofstep{\Itemref{def:eco_rel}{sync_out_imp}: $\uiocorel$}\\
            $(s \after \project{\sigma'}{L_s^\delta} \after \ell, e \after \project{\sigma'}{L_e^\delta} \after \ell) \in \uiocorel\;\land$\\
            $s \Trans{ \project{\sigma'\cdot\ell}{L_s^\delta}} q_s \land e \Trans{ \project{\sigma'\cdot\ell}{L_e^\delta}} q_e$\\
            \proofstep{\Cref{def:after,def:projection}: $\after$ and $\projectop$}\\
            $(s \after \project{\sigma'\cdot\ell}{L_s^\delta}, e \after \project{\sigma'\cdot\ell}{L_e^\delta}) \in \uiocorel\;\land$\\
            $s \Trans{ \project{\sigma'\cdot\ell}{L_s^\delta}} q_s \land e \Trans{ \project{\sigma'\cdot\ell}{L_e^\delta}} q_e$
    
        \item[$\sigma=\sigma'\cdot\ell$, $(\ell \in U_e\cap I_s):$] Symmetric to previous case
        \item[$\sigma=\sigma'\cdot\ell$, $(\ell \in U_s \setminus L_s^\delta):$]\ \\
            $s \parcomp e \Trans{\sigma\cdot\ell} q_s\parcomp q_e $\\
            \proofstep{\Cref{item:trans_transitive}}\\
            $\exists q_s'\in Q_s, q_e'\in Q_e:  s \parcomp e \Trans{\sigma'} q_s'\parcomp q_e' \land q_s'\parcomp q_e' \Trans{\ell} q_s \parcomp q_e$\\
            \proofstep{Apply IH}\\
            $(s \after \project{\sigma'}{L_s^\delta}, e \after \project{\sigma'}{L_e^\delta}) \in \uiocorel \;\land$\\
            $\exists q_s'\in Q_s, q_e'\in Q_e: s \Trans{ \project{\sigma'}{L_s^\delta}} q_s' \land e \Trans{ \project{\sigma'}{L_e^\delta}} q_e'\land q_s'\parcomp q_e' \Trans{\ell}q_s\parcomp q_e$\\
            \proofstep{\Cref{lem:project_from_parcomp_no_delta} and \cref{def:projection}: $\projectop$}\\
            $(s \after \project{\sigma'}{L_s^\delta}, e \after \project{\sigma'}{L_e^\delta}) \in \uiocorel \;\land$\\
            $\exists q_s'\in Q_s, q_e'\in Q_e: s \Trans{ \project{\sigma'}{L_s^\delta}} q_s' \land e \Trans{ \project{\sigma'}{L_e^\delta}} q_e'\land q_s' \Trans{\ell} q_s\land q_e' \Trans{\epsilon} q_e$\\
            \proofstep{\Cref{item:trans_transitive} and \cref{def:projection}: $\projectop$}\\
            $(s \after \project{\sigma'}{L_s^\delta}, e \after \project{\sigma'}{L_e^\delta}) \in \uiocorel \;\land$\\
            $\exists q_s'\in Q_s: s \Trans{ \project{\sigma'}{L_s^\delta}} q_s' \land q_s'\Trans{\ell}\;\land$\\
            $s \Trans{ \project{\sigma'\cdot\ell}{L_s^\delta}} q_s \land e \Trans{ \project{\sigma'\cdot\ell}{L_e^\delta}} q_e$\\
            \proofstep{\Cref{def:after,def:outset}: $\after$ and $\outset{}$}\\
            $(s \after \project{\sigma'}{L_s^\delta}, e \after \project{\sigma'}{L_e^\delta}) \in \uiocorel \;\land$\\
            $\ell \in \outset{s \after \project{\sigma'}{L_s^\delta}} \;\land$\\
            $s \Trans{ \project{\sigma'\cdot\ell}{L_s^\delta}} q_s \land e \Trans{ \project{\sigma'\cdot\ell}{L_e^\delta}} q_e$\\
            \proofstep{\Itemref{def:eco_rel}{internal_imp}: $\uiocorel$}\\
            $(s \after \project{\sigma'}{L_s^\delta} \after \ell, e \after \project{\sigma'}{L_e^\delta} \after \ell) \in \uiocorel\;\land$\\
            $s \Trans{ \project{\sigma'\cdot\ell}{L_s^\delta}} q_s \land e \Trans{ \project{\sigma'\cdot\ell}{L_e^\delta}} q_e$\\
            \proofstep{\Cref{def:after,def:projection}: $\after$ and $\projectop$}\\
            $(s \after \project{\sigma'\cdot\ell}{L_s^\delta}, e \after \project{\sigma'\cdot\ell}{L_e^\delta}) \in \uiocorel\;\land$\\
            $s \Trans{ \project{\sigma'\cdot\ell}{L_s^\delta}} q_s \land e \Trans{ \project{\sigma'\cdot\ell}{L_e^\delta}} q_e$
        \item[$\sigma=\sigma'\cdot\ell$, $(\ell \in U_e \setminus L_e^\delta):$]Symmetric to previous case
        \item[$\sigma=\sigma'\cdot\ell$, $(\ell \in I_e \setminus L_s^\delta):$]\ \\
            $s \parcomp e \Trans{\sigma\cdot\ell} q_s\parcomp q_e \land  \sigma'\cdot\ell\in \utraces{s\parcomp e}$\\
            \proofstep{\Cref{item:trans_transitive}}\\
            $\exists q_s'\in Q_s, q_e'\in Q_e:  s \parcomp e \Trans{\sigma'} q_s'\parcomp q_e' \land q_s'\parcomp q_e' \Trans{\ell} q_s\parcomp q_e\;\land$\\
            $\sigma'\cdot\ell\in \utraces{s\parcomp e}$\\
            \proofstep{Apply IH}\\
            $(s \after \project{\sigma'}{L_s^\delta}, e \after \project{\sigma'}{L_e^\delta}) \in \uiocorel \;\land$\\
            $\exists q_s'\in Q_s, q_e'\in Q_e: s \Trans{ \project{\sigma'}{L_s^\delta}} q_s' \land e \Trans{ \project{\sigma'}{L_e^\delta}} q_e' \land q_s'\parcomp q_e' \Trans{\ell} q_s\parcomp q_e \; \land $\\
            $\sigma'\cdot\ell\in \utraces{s\parcomp e}$\\
            \proofstep{\Cref{lem:project_from_parcomp_no_delta} and \cref{def:projection}: $\projectop$}\\
            $(s \after \project{\sigma'}{L_s^\delta}, e \after \project{\sigma'}{L_e^\delta}) \in \uiocorel \;\land$\\
            $\exists q_s'\in Q_s, q_e'\in Q_e: s \Trans{ \project{\sigma'}{L_s^\delta}} q_s' \land e \Trans{ \project{\sigma'}{L_e^\delta}} q_e'\land q_s' \Trans{\epsilon} q_s\land q_e' \Trans{\ell} q_e \; \land $\\
            $\sigma'\cdot\ell\in \utraces{s\parcomp e}$\\
            \proofstep{\Cref{item:trans_transitive} and \cref{def:projection}: $\projectop$}\\
            $(s \after \project{\sigma'}{L_s^\delta}, e \after \project{\sigma'}{L_e^\delta}) \in \uiocorel \;\land$\\
            $\exists q_s'\in Q_s: s \Trans{ \project{\sigma'}{L_s^\delta}} q_s' \;\land$\\
            $\sigma'\cdot\ell\in \utraces{s\parcomp e}\;\land$\\
            $s \Trans{ \project{\sigma'\cdot\ell}{L_s^\delta}} q_s \land e \Trans{ \project{\sigma'\cdot\ell}{L_e^\delta}} q_e$\\
            \proofstep{\Cref{def:uioco,def:inset}: $\inset{}$ and $\utraces{}$}\\
            $(s \after \project{\sigma'}{L_s^\delta}, e \after \project{\sigma'}{L_e^\delta}) \in \uiocorel \;\land$\\
            $\exists q_s'\in Q_s: s \Trans{ \project{\sigma'}{L_s^\delta}} q_s'\; \land $\\
            $\ell\in\inset{s\parcomp e \after \sigma'}\;\land$\\
            $s \Trans{ \project{\sigma'\cdot\ell}{L_s^\delta}} q_s \land e \Trans{ \project{\sigma'\cdot\ell}{L_e^\delta}} q_e$\\
            \proofstep{\Cref{lem:project_from_parcomp_light}}\\
            $(s \after \project{\sigma'}{L_s^\delta}, e \after \project{\sigma'}{L_e^\delta}) \in \uiocorel \;\land$\\
            $\exists q_s'\in Q_s, \forall q_e''\in e \after \project{\sigma'}{L_e^\delta}: s \parcomp e \Trans {\sigma'} q_s'\parcomp q_e'' \; \land $\\
            $\ell\in\inset{s\parcomp e \after \sigma'}\;\land$\\
            $s \Trans{ \project{\sigma'\cdot\ell}{L_s^\delta}} q_s \land e \Trans{ \project{\sigma'\cdot\ell}{L_e^\delta}} q_e$\\
            \proofstep{\Cref{def:inset}: $\inset{s\parcomp e \after \sigma'}$}\\
            $(s \after \project{\sigma'}{L_s^\delta}, e \after \project{\sigma'}{L_e^\delta}) \in \uiocorel \;\land$\\
            $\exists q_s'\in Q_s, \forall q_e''\in e \after \project{\sigma'}{L_e^\delta}: q_s'\parcomp q_e'' \Trans{\ell}\;\land$\\
            $s \Trans{ \project{\sigma'\cdot\ell}{L_s^\delta}} q_s \land e \Trans{ \project{\sigma'\cdot\ell}{L_e^\delta}} q_e$\\
            \proofstep{\Cref{lem:project_from_parcomp_no_delta}}\\
            $(s \after \project{\sigma'}{L_s^\delta}, e \after \project{\sigma'}{L_e^\delta}) \in \uiocorel \;\land$\\
            $\forall q_e''\in e \after \project{\sigma'}{L_e^\delta}: q_e'' \Trans{\ell}\;\land$\\
            $s \Trans{ \project{\sigma'\cdot\ell}{L_s^\delta}} q_s \land e \Trans{ \project{\sigma'\cdot\ell}{L_e^\delta}} q_e$\\
            \proofstep{\Cref{def:inset}: $\inset{e \after \project{\sigma'}{L_e^\delta}}$}\\
            $(s \after \project{\sigma'}{L_s^\delta}, e \after \project{\sigma'}{L_e^\delta}) \in \uiocorel \land \ell\in\inset{e \after \project{\sigma'}{L_e^\delta}}\;\land$\\
            $s \Trans{ \project{\sigma'\cdot\ell}{L_s^\delta}} q_s \land e \Trans{ \project{\sigma'\cdot\ell}{L_e^\delta}} q_e$\\
            \proofstep{\Itemref{def:eco_rel}{internal_env}: $\uiocorel$}\\
            $(s \after \project{\sigma'}{L_s^\delta}, e \after \project{\sigma'}{L_e^\delta} \after \ell) \in \uiocorel\;\land$\\
            $s \Trans{ \project{\sigma'\cdot\ell}{L_s^\delta}} q_s \land e \Trans{ \project{\sigma'\cdot\ell}{L_e^\delta}} q_e$\\
            \proofstep{\Cref{def:after,def:projection}: $\after$ and $\projectop$}\\
            $(s \after \project{\sigma'\cdot\ell}{L_s^\delta}, e \after \project{\sigma'\cdot\ell}{L_e^\delta}) \in \uiocorel\;\land$\\
            $s \Trans{ \project{\sigma'\cdot\ell}{L_s^\delta}} q_s \land e \Trans{ \project{\sigma'\cdot\ell}{L_e^\delta}} q_e$\\
           
        \item[$\sigma=\sigma'\cdot\ell$, $(\ell \in I_s \setminus L_e^\delta):$]Symmetric to previous case
        \item[$\sigma=\sigma'\cdot\ell$, $(\ell \in I_e\cap  I_s):$]\ \\
             $s \parcomp e \Trans{\sigma\cdot\ell} q_s\parcomp q_e \land \sigma'\cdot\ell\in \utraces{s\parcomp e} $\\
            \proofstep{\Cref{item:trans_transitive}}\\
            $\exists q_s'\in Q_s, q_e'\in Q_e:  s \parcomp e \Trans{\sigma'} q_s'\parcomp q_e' \land q_s'\parcomp q_e' \Trans{\ell} q_s \parcomp q_e\;\land$\\
            $\sigma'\cdot\ell\in \utraces{s\parcomp e}$\\
            \proofstep{Apply IH}\\
            $(s \after \project{\sigma'}{L_s^\delta}, e \after \project{\sigma'}{L_e^\delta}) \in \uiocorel \;\land$\\
            $\exists q_s'\in Q_s: s \Trans{ \project{\sigma'}{L_s^\delta}} q_s'\land e \Trans{ \project{\sigma'}{L_e^\delta}} q_e'\land q_s'\parcomp q_e' \Trans{\ell} q_s \parcomp q_e\;\land$\\
            $\sigma'\cdot\ell\in \utraces{s\parcomp e}$\\
            \proofstep{\Cref{lem:project_from_parcomp_no_delta} and \cref{def:projection}: $\projectop$}\\
            $(s \after \project{\sigma'}{L_s^\delta}, e \after \project{\sigma'}{L_e^\delta}) \in \uiocorel \;\land$\\
            $\exists q_s'\in Q_s, q_e'\in Q_e: s \Trans{ \project{\sigma'}{L_s^\delta}} q_s' \land e \Trans{ \project{\sigma'}{L_e^\delta}} q_e'\land q_s' \Trans{\ell} q_s\land q_e' \Trans{\ell} q_e \; \land $\\
            $\sigma'\cdot\ell\in \utraces{s\parcomp e}$\\
            \proofstep{\Cref{item:trans_transitive} and \cref{def:projection}: $\projectop$}\\
            $(s \after \project{\sigma'}{L_s^\delta}, e \after \project{\sigma'}{L_e^\delta}) \in \uiocorel \;\land$\\
            $\sigma'\cdot\ell\in \utraces{s\parcomp e}\;\land$\\
            $s \Trans{ \project{\sigma'\cdot\ell}{L_s^\delta}} q_s \land e \Trans{ \project{\sigma'\cdot\ell}{L_e^\delta}} q_e$\\
            \proofstep{\Cref{def:uioco,def:inset}: $\utraces{}$ and $\inset{}$}\\
            $(s \after \project{\sigma'}{L_s^\delta}, e \after \project{\sigma'}{L_e^\delta}) \in \uiocorel \;\land$\\
            $\ell\in\inset{s\parcomp e \after \sigma'}\;\land$\\
            $s \Trans{ \project{\sigma'\cdot\ell}{L_s^\delta}} q_s \land e \Trans{ \project{\sigma'\cdot\ell}{L_e^\delta}} q_e$\\
            \proofstep{\Cref{lem:project_from_parcomp_light}}\\
            $(s \after \project{\sigma'}{L_s^\delta}, e \after \project{\sigma'}{L_e^\delta}) \in \uiocorel \;\land$\\
            $\ell\in\inset{s\parcomp e \after \sigma'}\;\land$\\
            $\forall q_s''\in s \after \project{\sigma'}{L_s^\delta}, q_e''\in e \after \project{\sigma'}{L_e^\delta}: q_s''\parcomp q_e''\in s\parcomp e \after \sigma'\;\land$\\
            $s \Trans{ \project{\sigma'\cdot\ell}{L_s^\delta}} q_s \land e \Trans{ \project{\sigma'\cdot\ell}{L_e^\delta}} q_e$\\
            \proofstep{\Cref{def:inset}: $\inset{s\parcomp e \after \sigma'}$}\\
            $(s \after \project{\sigma'}{L_s^\delta}, e \after \project{\sigma'}{L_e^\delta}) \in \uiocorel \;\land$\\
            $\forall q_s''\in s \after \project{\sigma'}{L_s^\delta}, q_e''\in e \after \project{\sigma'}{L_e^\delta}: q_s''\parcomp q_e'' \Trans{\ell}\;\land$\\
            $s \Trans{ \project{\sigma'\cdot\ell}{L_s^\delta}} q_s \land e \Trans{ \project{\sigma'\cdot\ell}{L_e^\delta}} q_e$\\
            \proofstep{\Cref{lem:project_from_parcomp_no_delta} and \cref{def:projection}: $\projectop$ }\\
            $(s \after \project{\sigma'}{L_s^\delta}, e \after \project{\sigma'}{L_e^\delta}) \in \uiocorel \;\land$\\
            $\forall q_s''\in s \after \project{\sigma'}{L_s^\delta}, q_e''\in e \after \project{\sigma'}{L_e^\delta}: q_s'' \Trans{\ell} \land\; q_e'' \Trans{\ell}\;\land$\\
            $s \Trans{ \project{\sigma'\cdot\ell}{L_s^\delta}} q_s \land e \Trans{ \project{\sigma'\cdot\ell}{L_e^\delta}} q_e$\\
            \proofstep{\Cref{def:inset,def:parcomp}: $\inset{}$ and $\parcomp$}\\
            $(s \after \project{\sigma'}{L_s^\delta}, e \after \project{\sigma'}{L_e^\delta}) \in \uiocorel\; \land$\\
            $\ell\in\inset{e \after \project{\sigma'}{L_e^\delta}} \land \ell\in\inset{s \after \project{\sigma'}{L_s^\delta}}\;\land$\\
            $s \Trans{ \project{\sigma'\cdot\ell}{L_s^\delta}} q_s \land e \Trans{ \project{\sigma'\cdot\ell}{L_e^\delta}} q_e$\\
            \proofstep{\Itemref{def:eco_rel}{sync_in}:
            $\uiocorel$}\\
            $(s \after \project{\sigma'}{L_s^\delta} \after \ell, e \after \project{\sigma'}{L_e^\delta} \after \ell) \in \uiocorel\;\land$\\
            $s \Trans{ \project{\sigma'\cdot\ell}{L_s^\delta}} q_s \land e \Trans{ \project{\sigma'\cdot\ell}{L_e^\delta}} q_e$\\
            \proofstep{\Cref{def:after,def:projection}: $\after$ and $\projectop$}\\
            $(s \after \project{\sigma'\cdot\ell}{L_s^\delta}, e \after \project{\sigma'\cdot\ell}{L_e^\delta}) \in \uiocorel\;\land$\\
            $s \Trans{ \project{\sigma'\cdot\ell}{L_s^\delta}} q_s \land e \Trans{ \project{\sigma'\cdot\ell}{L_e^\delta}} q_e$\\
        \item[$\sigma=\sigma'\cdot\ell$, $(\ell \in U_e\cap U_s):$] Impossible due to \cref{def:composable}
        \item[$\sigma=\sigma'\cdot\delta:$]\ \\
            $s \parcomp e \Trans{\sigma\cdot\delta} q_s\parcomp q_e $\\
            \proofstep{\Cref{item:trans_transitive}}\\
            $\exists q_s'\in Q_s, q_e'\in Q_e:  s \parcomp e \Trans{\sigma'} q_s'\parcomp q_e' \land q_s'\parcomp q_e' \Trans{\delta} q_s\parcomp q_e$\\
            \proofstep{Apply IH}\\
            $(s \after \project{\sigma'}{L_s^\delta}, e \after \project{\sigma'}{L_e^\delta}) \in \uiocorel \;\land$\\
            $\exists q_s'\in Q_s, q_e'\in Q_e: s \Trans{ \project{\sigma'}{L_s^\delta}} q_s' \land e \Trans{ \project{\sigma'}{L_e^\delta}} q_e'\land q_s'\parcomp q_e' \Trans{\delta} q_s\parcomp q_e$\\
            \proofstep{\Cref{def:after}: $\after$}\\
            $(s \after \project{\sigma'}{L_s^\delta}, e \after \project{\sigma'}{L_e^\delta}) \in \uiocorel \;\land$\\
            $\exists q_s'\in Q_s, q_e'\in Q_e: s \Trans{ \project{\sigma'}{L_s^\delta}} q_s' \land e \Trans{ \project{\sigma'}{L_e^\delta}} q_e'$\\$
            q_s'\in s \after \project{\sigma'}{L_s^\delta} \land q_e'\in e \after \project{\sigma'}{L_e^\delta} \land q_s'\parcomp q_e' \Trans{\delta} q_s\parcomp q_e$\\
            \proofstep{\Cref{lem:parcomp_delta_eco_rel}}\\
            $(s \after \project{\sigma'}{L_s^\delta}, e \after \project{\sigma'}{L_e^\delta}) \in \uiocorel \;\land$\\
            $\exists q_s'\in Q_s, q_e'\in Q_e: s \Trans{ \project{\sigma'}{L_s^\delta}} q_s' \land e \Trans{ \project{\sigma'}{L_e^\delta}} q_e'$\\$
            q_s'\in s \after \project{\sigma'}{L_s^\delta} \land q_e'\in e \after \project{\sigma'}{L_e^\delta} \land q_s' \Trans{\delta} q_s \land q_e' \Trans{\delta} q_e $\\
            \proofstep{\Itemref{def:eco_rel}{delta} and \cref{def:outset}: $\outset{}$}\\
            $\exists q_s'\in Q_s, q_e'\in Q_e: s \Trans{ \project{\sigma'}{L_s^\delta}} q_s' \land e \Trans{ \project{\sigma'}{L_e^\delta}} q_e' \land q_s' \Trans{\delta} q_s \land q_e' \Trans{\delta} q_e \;\land$\\
            $(s \after \project{\sigma'}{L_s^\delta} \after \delta, e \after \project{\sigma'}{L_e^\delta} \after \delta) \in \uiocorel$\\
            \proofstep{\Cref{item:trans_transitive}}\\
            $s \Trans{ \project{\sigma'}{L_s^\delta}\cdot\delta} q_s \land e \Trans{ \project{\sigma'}{L_e^\delta}\cdot \delta} q_e\;\land$\\
            $(s \after \project{\sigma'}{L_s^\delta} \after \delta, e \after \project{\sigma'}{L_e^\delta} \after \delta) \in \uiocorel$\\
            \proofstep{\Cref{def:after,def:projection}: $\after$ and $\projectop$}\\
            $s \Trans{ \project{\sigma'\cdot\delta}{L_s^\delta}} q_s \land e \Trans{ \project{\sigma'\cdot\delta}{L_e^\delta}} q_e\;\land$\\
            $(s \after \project{\sigma'\cdot\delta}{L_s^\delta}, e \after \project{\sigma'\cdot\delta}{L_e^\delta}) \in \uiocorel$
    \end{case_distinction}
\end{case_distinction}
    \end{proof}
\end{toappendix}

\begin{lemmarep}
Let $s,e \in\LTS$ be composable, $\sigma \in \utraces{s\parcomp e}$, $ q_s\in Q_s$, $q_e\in Q_e$, then 
\[s\aco e \implies (s \parcomp e \Trans{\sigma} q_s\parcomp q_e \iff s \Trans{ \project{\sigma}{L_s^\delta}} q_s \land e \Trans{ \project{\sigma}{L_e^\delta}} q_e)\]
\label{lem:project_from_parcomp_eco}
\end{lemmarep}
\begin{proof}
\ \\
    \begin{case_distinction}
    \item[$\implies$:]  The proof follows directly from \Cref{lem:once_in_R_always_in_R}
    \item[$\impliedby$:] Follows directly from \Cref{lem:project_from_parcomp_light}
\end{case_distinction}
\end{proof}

\begin{toappendix}
    \begin{lemma}
    Let $s,e \in \LTS$ be $\composable$, $s \aco e$, 
    $\sigma \in \utraces{s\parcomp e}$, $q_s,q_s'\in Q_s$, $q_e,q_e'\in Q_e:$
    \[s\parcomp e \Trans{\sigma} q_s\parcomp q_e \land s\parcomp e \Trans{\sigma} q_s' \parcomp q_e' \iff s\parcomp e \Trans{\sigma} q_s\parcomp q_e' \land s\parcomp e \Trans{\sigma} q_s' \parcomp q_e \]
    \label{lem:trans_complete_eco_rel}
    \end{lemma}
    \begin{proof} 
        Proof follows almost directly from \Cref{lem:once_in_R_always_in_R}.The proofs for the $\implies$ and $\impliedby$ case are symmetric with each other, So we will only prove the $\implies$ case here:\\
$s\parcomp e \Trans{\sigma} q_s\parcomp q_e \land s\parcomp e \Trans{\sigma} q_s' \parcomp q_e'\land s \eco e \land \sigma\in \utraces{s\parcomp e}$\\
\proofstep{\Cref{lem:once_in_R_always_in_R}}\\
$ s \Trans{\project{\sigma}{L_s^\delta}} q_s  \land e \Trans{\project{\sigma}{L_e^\delta}} q_e \land s \Trans{\project{\sigma}{L_s^\delta}} q_s'  \land e \Trans{\project{\sigma}{L_e^\delta}} q_e' $\\
\proofstep{\Cref{lem:project_from_parcomp_light}}\\
$s\parcomp e \Trans{\sigma} q_s\parcomp q_e' \land s\parcomp e \Trans{\sigma} q_s' \parcomp q_e$
    \end{proof}
    
    \begin{definition}
    Let $X$ be a set of tuples, then:
        \[\collectTuple{X} \defeq (\{x \setbar (x,y)\in X\}, \{y \setbar (x,y) \in X\})\]
        \label{def:collect_tuple}
    \end{definition}

    \begin{lemmarep}
        Let $s,e\in\LTS$ be $\composable$, $s\aco e$, $q_s\in Q_s$, $q_e\in Q_e$, $X_s\subseteq Q_s$, $X_e\subseteq Q_e$, $\sigma \in \utraces{s\parcomp e}$, $\collectTuple{s \parcomp e \after \sigma} = (X_s,X_e):$
        \[ q_s \in X_s \land q_e \in X_e \iff s\parcomp e \Trans{\sigma} q_s \parcomp q_e\]%
        \label{lem:tupple_trace_reachable_eco_rel}%
    \end{lemmarep}
    \begin{proof}
        \begin{case_distinction}
    \item[$\implies:$]\ \\
        $\collectTuple{s\parcomp e \after \sigma} = (X_s,X_e) \land q_s\in X_s \land q_e\in X_e$\\
        \proofstep{\Cref{def:collect_tuple,def:after}: $\collectTupleSymb$ and $\after$}\\
        $\exists q_s'\in Q_s, q_e'\in Q_e: s\parcomp e \Trans{\sigma} q_s'\parcomp q_e \land s\parcomp e \Trans{\sigma} q_s\parcomp q_e'$\\
        \proofstep{\Cref{lem:trans_complete_eco_rel}}\\
        $s\parcomp e \Trans{\sigma} q_s \parcomp q_e$
    \item[$\impliedby:$]\ \\
        $s\parcomp e \Trans{\sigma} q_s \parcomp q_e$\\
        \proofstep{\Cref{def:after}: $\after$}\\
        $q_s \parcomp q_e \in s\parcomp e \after \sigma$\\
        \proofstep{\Cref{def:collect_tuple}: $\collectTupleSymb$ and $\collectTuple{q_s \parcomp q_e \after \sigma} = (X_s,X_e)$}\\
        $q_s\in X_s \land q_e\in X_e$
\end{case_distinction}
    \end{proof}

    \begin{lemmarep}
         Let $s,e\in \LTS$ be $\composable$, $s\aco e$, $X_e\subseteq Q_e, X_s\subseteq Q_s$.
        \[(X_s,X_e)\in \uiocorel \iff \exists \sigma \in \utraces{s\parcomp e}: \collectTuple{s\parcomp e \after \sigma} = (X_s,X_e)\]
        \label{lem:eco_rel_iff_tuple_utraces}
    \end{lemmarep}
    \begin{proof}
        \begin{case_distinction}
        \item[$\implies$:] 
        Proof by induction on the structure of $\uiocorel$. Case names follow naming of \cref{def:uioco_rel}. For the recursive cases the induction hypothesis (IH) is: 
        \[\exists \sigma'\in \utraces{s\parcomp e}: \collectTuple{s\parcomp e \after \sigma'} = (X_s',X_e')\]
        Where $(X_s', X_e')$ comes from the left side of the $\vdash$ of the corresponding case of \cref{def:uioco_rel}. 
        To prove the set equality, we pick a $\sigma$ and split each goal further into three parts: A proof that the given $\sigma$ is part of $\utraces{s\parcomp e}$, and the two subset statements corresponding to the set equality. For brevity the definitions of $\collectTupleSymb$ and $\after$ are implicitly expanded here. I.e. to prove $\exists\sigma\in \utraces{s\parcomp e}: \collectTuple{s\parcomp e \after \sigma} = (X_s,X_e)$, we fix a $\sigma$ (containing $\sigma'$) and then prove:
        \begin{enumerate}
            \item $\sigma\in \utraces{s\parcomp e}$
            \item $s\parcomp e \Trans{\sigma} q_s\parcomp q_e$ for arbitrary $q_s\in X_s, q_e\in X_e$ ($\supseteq$)
            \item $q_s \in X_s$ and $q_e \in X_e$, for arbitrary $q_s \parcomp q_e \in s\parcomp e \after \sigma$ ($\subseteq$).
        \end{enumerate}
        \begin{case_distinction}
            \item[case \ref{item:uioco_rel_base} ($X_s= s \after \epsilon,X_e=e \after \epsilon$):] Take $\sigma = \epsilon$\\
                \begin{case_distinction}
                    \item[$\utraces{}$:] $\epsilon$ is always trivially part of $\utraces{}$
                    \item[$\supseteq$:]
                    $q_s \in s \after \epsilon \land q_e \in e \after \epsilon$\\
                    \proofstep{\Cref{def:after}: $\after$}\\
                    $s \Trans{\epsilon} q_s \land e \Trans{\epsilon} q_e $\\
                    \proofstep{\Cref{lem:project_from_parcomp_no_delta}}\\
                    $s\parcomp e \Trans{\epsilon} q_s \parcomp q_e$\\
                    \proofstep{\Cref{def:uioco}: $\utraces{}$}\\
                    $s\parcomp e \Trans{\epsilon} q_s \parcomp q_e \land \epsilon \in \utraces{s\parcomp e}$
                    
                    \item[$\subseteq$:]
                    $s\parcomp e \Trans{\epsilon} q_s \parcomp q_e$\\
                    \proofstep{\Cref{lem:project_from_parcomp_no_delta}}\\
                    $s \Trans{\epsilon} q_s \land e \Trans{\epsilon} q_e $\\
                    \proofstep{\Cref{def:after,def:uioco}: $\utraces{}$ and $\after$}\\
                    $q_s \in s \after \epsilon \land q_e \in e \after \epsilon \land \epsilon \in \utraces{s\parcomp e}$
                \end{case_distinction}
                
            \item[case \ref{item:uioco_rel_sync_out_env} ($X_s= X_s'\after \ell, X_e=X_e'\after \ell$):] Take $\sigma =\sigma'\cdot\ell$
            \begin{case_distinction}
                \item[$\utraces{}$:] We have $\sigma'\in \utraces{s\parcomp e}$ (IH) and $s\parcomp e \Trans{\sigma'} q_s'\parcomp q_e'$ for all $q_s'\in X_s'$, $q_e'\in X_e'$ (IH and \cref{lem:tupple_trace_reachable_eco_rel}). Combine with $\ell \in \outset{X_e'} \cap I_s \land \ell \in \inset{X_s'}$ from \itemref{def:eco_rel}{sync_out_env} and \cref{lem:constructive_ecosim}, we get $s\parcomp e \Trans{\sigma'\cdot\ell}$. Since $\ell\in U_s \land \ell\in I_e$, we have $\ell\in U_{s\parcomp e}$ (\cref{def:parcomp}), and therefore $\sigma\cdot\ell\in\utraces{s\parcomp e}$ (\cref{def:uioco}).
                \item[$\supseteq$:]
                    $q_s \in X_s' \after \ell \land q_e \in X_e' \after \ell$\\
                    \proofstep{\Cref{def:after}: $\after$}\\
                    $\exists q_s' \in X_s', q_e'\in X_e': q_s'\Trans{\ell}q_s \land q_e' \Trans{\ell} q_e$\\
                    \proofstep{\Cref{lem:project_from_parcomp_light,def:projection}: $\projectop$}\\
                    $\exists q_s'\in X_s', q_e'\in X_e': q_s' \parcomp q_e' \Trans{\ell}q_s \parcomp q_e$\\
                    \proofstep{Apply IH}\\
                    $\collectTuple{s\parcomp e \after \sigma'} = (X_s',X_e')$\\
                    $\exists q_s'\in X_s', q_e'\in X_e': q_s' \parcomp q_e' \Trans{\ell}q_s \parcomp q_e$\\
                    \proofstep{\Cref{lem:tupple_trace_reachable_eco_rel}}\\
                    $\exists q_s'\in X_s', q_e'\in X_e': s\parcomp e \Trans{\sigma'} q_s'\parcomp q_e' \land q_s' \parcomp q_e' \Trans{\ell}q_s \parcomp q_e $\\
                    \proofstep{\Cref{item:trans_transitive}}\\
                    $s\parcomp e \Trans{\sigma'\cdot \ell} q_s \parcomp q_e $
                \item[$\subseteq$:] 
                    $s\parcomp e \Trans{\sigma'\cdot\ell} q_s \parcomp q_e$\\
                    \proofstep{\Cref{item:trans_transitive}}\\
                    $\exists q_s'\in Q_s, q_e'\in Q_e: s \parcomp e \Trans{\sigma'} q_s'\parcomp q_e' \land q_s'\parcomp q_e' \Trans{\ell}q_s \parcomp q_e$\\
                    \proofstep{Apply IH}\\
                    $\collectTuple{s\parcomp e \after \sigma'} = (X_s',X_e')$\\
                    $\exists q_s'\in Q_s, q_e'\in Q_e: s \parcomp e \Trans{\sigma'} q_s'\parcomp q_e' \land q_s'\parcomp q_e' \Trans{\ell}q_s \parcomp q_e$\\
                    \proofstep{\Cref{def:collect_tuple,def:after}: $\collectTupleSymb$ and $\after$}\\
                    $\exists q_s'\in X_s', q_e'\in X_e': q_s'\parcomp q_e' \Trans{\ell}q_s \parcomp q_e$\\
                    \proofstep{\Cref{lem:project_from_parcomp_no_delta,def:projection}: $\projectop$}\\
                    $\exists q_s'\in X_s', q_e'\in X_e': q_s' \Trans{\ell} q_s \land  q_e' \Trans{\ell} q_e$\\
                    \proofstep{\Cref{def:after}: $\after$}\\
                    $q_s \in X_s' \after \ell \land q_e \in X_e' \after \ell$
            \end{case_distinction}
                
            \item[case \ref{item:uioco_rel_sync_out_imp}:] Symmetric to previous case
            \item[case \ref{item:uioco_rel_internal_env} ($X_s= X_s', X_e=X_e'\after \ell$):]Take $\sigma =\sigma'\cdot\ell$
            \begin{case_distinction}
                 \item[$\utraces{}$:] We have $\sigma'\in \utraces{s\parcomp e}$ (IH) and $s\parcomp e \Trans{\sigma'} q_s'\parcomp q_e'$ for all $q_s'\in X_s'$, $q_e'\in X_e'$ (IH and \cref{lem:tupple_trace_reachable_eco_rel}). Combine with $\ell \in (\outset{X_e'}\cup \inset{X_e'})\setminus L_s^\delta$ from \itemref{def:uioco_rel}{internal_env}, we get $s\parcomp e \Trans{\sigma'\cdot\ell}$. If $\ell \in U_s$ then $\ell\in U_{s\parcomp e}$ and we have $\sigma\cdot\ell\in\utraces{s\parcomp e}$ (\cref{def:uioco}). 
                 
                 If $\ell \in I_s$ then $\ell \in I_{s\parcomp e}$. We then have to prove that $\ell\in \inset{s\parcomp e \after \sigma'}$. Since inputs cannot be part of $\outset{X_e'}$, we know $\ell\in \inset{X_e'}$. From the IH we know $q_s'\parcomp q_e' \in s\parcomp e \after \sigma' \implies q_e'\in X_e'$. Combined with the fact that $\ell\notin L_s^\delta$ this then gives $\ell\in \inset{s\parcomp e \after \sigma'}$.
            
                \item[$\supseteq$:]
                    $q_s \in X_s' \land q_e \in X_e' \after \ell$\\
                    \proofstep{\Cref{def:after}: $\after$}\\
                    $ q_s \in X_s'\land \exists q_e'\in X_e':  q_e' \Trans{\ell} q_e$\\
                    \proofstep{\Cref{lem:project_from_parcomp_light,def:projection}: $\projectop$}\\
                    $ q_s \in X_s'\land \exists q_e'\in X_e':  q_s\parcomp q_e' \Trans{\ell} q_s \parcomp q_e \land$\\
                    \proofstep{Apply IH}\\
                    $\collectTuple{s\parcomp e \after \sigma'} = (X_s',X_e')\;\land$\\
                    $q_s \in X_s'\land \exists q_e'\in X_e':  q_s\parcomp q_e' \Trans{\ell} q_s \parcomp q_e$\\
                    \proofstep{\Cref{lem:tupple_trace_reachable_eco_rel}}\\
                    $\exists q_e'\in X_e': s\parcomp e \Trans{\sigma'} q_s\parcomp q_e' \land   q_s\parcomp q_e' \Trans{\ell} q_s \parcomp q_e$\\
                    \proofstep{\Cref{item:trans_transitive}}\\
                    $s\parcomp e \Trans{\sigma'\cdot\ell} q_s \parcomp q_e$
                \item [$\subseteq$:]
                    $s\parcomp e \Trans{\sigma'\cdot\ell} q_s \parcomp q_e$\\
                    \proofstep{\Cref{item:trans_transitive}}\\
                    $\exists q_s'\in Q_s, q_e'\in Q_e: s \parcomp e \Trans{\sigma'} q_s'\parcomp q_e' \land q_s'\parcomp q_e' \Trans{\ell}q_s \parcomp q_e$\\
                    \proofstep{Apply IH}\\
                    $\collectTuple{s\parcomp e \after \sigma'} = (X_s',X_e')$\\
                    $\exists q_s'\in Q_s, q_e'\in Q_e: s \parcomp e \Trans{\sigma'} q_s'\parcomp q_e' \land q_s'\parcomp q_e' \Trans{\ell}q_s \parcomp q_e$\\
                    \proofstep{\Cref{def:collect_tuple,def:after}: $\collectTupleSymb$ and $\after$}\\
                    $\exists q_s'\in X_s', q_e'\in X_e': q_s'\parcomp q_e' \Trans{\ell}q_s \parcomp q_e$\\
                    \proofstep{\Cref{lem:project_from_parcomp_no_delta,def:projection}: $\projectop$}\\
                    $\exists q_s'\in X_s', q_e'\in X_e': q_s' \Trans{\epsilon} q_s \land  q_e' \Trans{\ell} q_e$\\
                    \proofstep{\Cref{def:after,def:parcomp,def:arrowdefs} and IH: $\after$, $\parcomp$, and $\Trans{\epsilon}$}\\
                    $q_s \in X_s' \land \exists, q_e'\in X_e':q_e' \Trans{\ell} q_e$\\
                    \proofstep{\Cref{def:after}: $\after$}\\
                    $q_s \in X_s'\land q_e \in X_e' \after \ell$
            \end{case_distinction}
             
            \item[case \ref{item:uioco_rel_internal_imp}:] Symmetrical to previous case
            \item[case \ref{item:uioco_rel_sync_in} ($X_s= X_s' \after \ell, X_e=X_e'\after \ell$):]Take $\sigma =\sigma'\cdot\ell$
            \begin{case_distinction}
                 \item[$\utraces{}$:] We have $\sigma'\in \utraces{s\parcomp e}$ (IH) and $s\parcomp e \Trans{\sigma'} q_s'\parcomp q_e'$ for all $q_s'\in X_s'$, $q_e'\in X_e'$ (IH and \cref{lem:tupple_trace_reachable_eco_rel}). Combine with $\ell \in  (\inset{X_e'}\cap \inset{X_s'})$ from \itemref{def:uioco_rel}{sync_in}, we get $s\parcomp e \Trans{\sigma'\cdot\ell}$. 
                 
                 Since $\ell\in I_s \land \ell\in I_e$, we have $\ell\in I_{s\parcomp e}$ (\cref{def:parcomp}), and therefore we need to prove  $\ell\in \inset{s\parcomp e \after \sigma'}$. From the IH we know $q_s'\parcomp q_e' \in s\parcomp e \after \sigma' \implies q_s' \in X_e' \land q_e'\in X_e'$. Combined with $\ell \in  (\inset{X_e'}\cap \inset{X_s'})$ this gives $\ell\in \inset{s\parcomp e \after \sigma'}$.
                 \item [$\supseteq$:]
                    $q_s \in X_s' \after \ell \land q_e \in X_e' \after \ell$\\
                    \proofstep{\Cref{def:after}: $\after$}\\
                    $\exists q_s' \in X_s', q_e'\in X_e': q_s'\Trans{\ell}q_s \land q_e' \Trans{\ell} q_e$\\
                    \proofstep{\Cref{lem:project_from_parcomp_light,def:projection}: $\projectop$}\\
                    $\exists q_s'\in X_s', q_e'\in X_e': q_s' \parcomp q_e' \Trans{\ell}q_s \parcomp q_e$\\
                    \proofstep{Apply IH}\\
                    $\collectTuple{s\parcomp e \after \sigma'} = (X_s',X_e')$\\
                    $\exists q_s'\in X_s', q_e'\in X_e': q_s' \parcomp q_e' \Trans{\ell}q_s \parcomp q_e$\\
                    \proofstep{\Cref{lem:tupple_trace_reachable_eco_rel}}\\
                    $\exists q_s'\in X_s', q_e'\in X_e': s\parcomp e \Trans{\sigma'} q_s'\parcomp q_e' \land q_s' \parcomp q_e' \Trans{\ell}q_s \parcomp q_e $\\
                    \proofstep{\Cref{item:trans_transitive}}\\
                    $s\parcomp e \Trans{\sigma'\cdot \ell} q_s \parcomp q_e $
                 \item [$\subseteq$:]
                    $s\parcomp e \Trans{\sigma'\cdot\ell} q_s \parcomp q_e$\\
                    \proofstep{\Cref{item:trans_transitive}}\\
                    $\exists q_s'\in Q_s, q_e'\in Q_e: s \parcomp e \Trans{\sigma'} q_s'\parcomp q_e' \land q_s'\parcomp q_e' \Trans{\ell}q_s \parcomp q_e$\\
                    \proofstep{Apply IH}\\
                    $\collectTuple{s\parcomp e \after \sigma'} = (X_s',X_e')$\\
                    $\exists q_s'\in Q_s, q_e'\in Q_e: s \parcomp e \Trans{\sigma'} q_s'\parcomp q_e' \land q_s'\parcomp q_e' \Trans{\ell}q_s \parcomp q_e$\\
                    \proofstep{\Cref{def:collect_tuple,def:after}: $\collectTupleSymb$ and $\after$}\\
                    $\exists q_s'\in X_s', q_e'\in X_e': q_s'\parcomp q_e' \Trans{\ell}q_s \parcomp q_e$\\
                    \proofstep{\Cref{lem:project_from_parcomp_no_delta,def:projection}: $\projectop$}\\
                    $\exists q_s'\in X_s', q_e'\in X_e': q_s' \Trans{\ell} q_s \land  q_e' \Trans{\ell} q_e$\\
                    \proofstep{\Cref{def:after}: $\after$}\\
                    $q_s \in X_s' \after \ell \land q_e \in X_e' \after \ell$
            \end{case_distinction}
            \item[case \ref{item:uioco_rel_delta}($X_s= X_s' \after \delta, X_e=X_e'\after \delta$):]Take $\sigma =\sigma'\cdot\delta$
            \begin{case_distinction}
                 \item[$\utraces{}$:] We have $\sigma'\in \utraces{s\parcomp e}$ (IH) and $s\parcomp e \Trans{\sigma'} q_s'\parcomp q_e'$ for all $q_s'\in X_s'$, $q_e'\in X_e'$ (IH and \cref{lem:tupple_trace_reachable_eco_rel}). $\delta \in\outset{X_e'} \land \delta \in \outset{X_s'}$ from \itemref{def:eco_rel}{delta} plus \cref{lem:constructive_ecosim} gives that both $X_e'$ and $X_s'$ contain at least one quiescent state ($p_{s\delta}$ and $p_{e\delta}$). since $p_{s\delta}\in X_s'$ and $p_{e\delta} \in X_e'$, we have $s\parcomp e \Trans{\sigma'}p_{s\delta} \parcomp p_{e\delta}$. Since outputs in composed states come from one of their component states, $p_{s\delta} \parcomp p_{e\delta}$ is itself also quiescent by definition. This gives $s\parcomp e \Trans{\sigma'\cdot\delta}$, which then leads to $\sigma' \cdot \delta\in \utraces{s\parcomp e}$.

                 \item [$\supseteq$:]
                    $q_s \in X_s' \after \delta \land q_e \in X_e' \after \delta$\\
                    \proofstep{\Cref{def:after,def:delta}: $\after$ and $\delta$}\\
                    $\exists q_s' \in X_s', q_e'\in X_e': q_s'\Trans{\delta} q_s \land q_e' \Trans{\delta} q_e$\\
                    \proofstep{\Cref{lem:project_from_parcomp_light,def:projection}: $\projectop$}\\
                    $\exists q_s'\in X_s', q_e'\in X_e': q_s' \parcomp q_e' \Trans{\delta}q_s \parcomp q_e$\\
                    \proofstep{Apply IH}\\
                    $\collectTuple{s\parcomp e \after \sigma'} = (X_s',X_e')$\\
                    $\exists q_s'\in X_s', q_e'\in X_e': q_s' \parcomp q_e' \Trans{\delta}q_s \parcomp q_e$\\
                    \proofstep{\Cref{lem:tupple_trace_reachable_eco_rel}}\\
                    $\exists q_s'\in X_s', q_e'\in X_e': s\parcomp e \Trans{\sigma'} q_s'\parcomp q_e' \land q_s' \parcomp q_e' \Trans{\delta}q_s \parcomp q_e $\\
                    \proofstep{\Cref{item:trans_transitive}}\\
                    $s\parcomp e \Trans{\sigma'\cdot \delta} q_s \parcomp q_e $
                 \item [$\subseteq$:]
                    $s\parcomp e \Trans{\sigma'\cdot\delta} q_s \parcomp q_e$\\
                    \proofstep{\Cref{def:arrowdefs}: $\Trans{\sigma'\cdot\delta}$}\\
                    $\exists q_s',q_s''\in Q_s, q_e',q_e''\in Q_e:$\\
                    $ s \parcomp e \Trans{\sigma'} q_s'\parcomp q_e' \land q_s'\parcomp q_e' \trans{\delta} q_s''\parcomp q_e'' \land q_s''\parcomp q_e'' \Trans{\epsilon}  q_s \parcomp q_e$\\
                    \proofstep{\Cref{def:delta}: $\trans{\delta}$}\\
                    $\exists q_s'\in Q_s, q_e'\in Q_e: s \parcomp e \Trans{\sigma'} q_s'\parcomp q_e' \land q_s'\parcomp q_e' \trans{\delta} q_s'\parcomp q_e' \land q_s'\parcomp q_e' \Trans{\epsilon}  q_s \parcomp q_e$\\
                    \proofstep{Apply IH}\\
                    $\collectTuple{s\parcomp e \after \sigma'} = (X_s',X_e')$\\
                    $\exists q_s'\in Q_s, q_e'\in Q_e: s \parcomp e \Trans{\sigma'} q_s'\parcomp q_e' \land q_s'\parcomp q_e' \trans{\delta} q_s'\parcomp q_e' \land q_s'\parcomp q_e' \Trans{\epsilon}  q_s \parcomp q_e$\\
                    \proofstep{\Cref{def:collect_tuple,def:after}: $\collectTupleSymb$ and $\after$}\\
                    $\exists q_s'\in X_s', q_e'\in X_e':  q_s'\parcomp q_e' \trans{\delta} q_s'\parcomp q_e' \land q_s'\parcomp q_e' \Trans{\epsilon}  q_s \parcomp q_e$\\
                    \proofstep{\Cref{lem:uiocosim_minimal}}\\
                    $(X_s',X_e')\in \acceptrel$\\
                    $\exists q_s'\in X_s', q_e'\in X_e':  q_s'\parcomp q_e' \trans{\delta} q_s'\parcomp q_e' \land q_s'\parcomp q_e' \Trans{\epsilon}  q_s \parcomp q_e$\\
                    \proofstep{\Cref{def:eco_rel}.\ref{item:eco_rel_sync_out_env} and \ref{def:eco_rel}.\ref{item:eco_rel_sync_out_imp} ($q_s'\in X_s'$, $q_e'\in X_e'$, $(X_e',X_s')\in \acceptrel$)}\\
                    $\exists q_s'\in X_s', q_e'\in X_e':  q_s'\parcomp q_e' \trans{\delta} q_s'\parcomp q_e' \land q_s'\parcomp q_e' \Trans{\epsilon}  q_s \parcomp q_e\; \land $\\
                    $\forall\ell\in out(q_s')\cap I_e: \ell\in \inset{q_e'}$\\
                    $\forall\ell\in out(q_e')\cap I_s: \ell\in \inset{q_s'}$\\
                    \proofstep{\Cref{def:parcomp}: $q_s'\parcomp q_e'\trans{\delta} q_s'\parcomp q_e'$}\\
                    $\exists q_s'\in X_s', q_e'\in X_e': q_s'\parcomp q_e' \Trans{\epsilon}  q_s \parcomp q_e\; \land $\\
                    $\forall\ell\in \outset{q_s'}\cap I_e: \ell\in \inset{q_e'}\;\land$\\
                    $\forall\ell\in \outset{q_e'}\cap I_s: \ell\in \inset{q_s'}\;\land$\\
                    $\forall\ell\in \outset{q_s'}\cap I_e \land \ell\in \inset{q_e'}: q_s\nottrans{\delta}\;\land$\\
                    $\forall\ell\in \outset{q_e'}\cap I_s \land \ell\in \inset{q_s'}: q_e'\nottrans{\delta}\; \land$\\
                    $\forall \ell \in (\outset{q_s'}\setminus I_e) \cup \{\tau\}: q_s'\nottrans{\ell}\;\land$\\
                    $\forall \ell \in (\outset{q_e'}\setminus I_s) \cup \{\tau\}: q_e'\nottrans{\ell}\;\land$\\
                    \proofstep{\Cref{def:delta}: $\delta$}\\
                    $\exists q_s'\in X_s', q_e'\in X_e':q_s'\parcomp q_e' \Trans{\epsilon}  q_s \parcomp q_e \land q_s'\trans{\delta}q_s'\land q_e'\trans{\delta} q_e'$\\
                    \proofstep{\Itemref{lem:parcomp_base_properties}{step_tau} and \cref{def:arrowdefs}: $\Trans{\epsilon}$}\\
                    $\exists q_s'\in X_s', q_e'\in X_e': q_s'\Trans{\epsilon} q_s \land  q_e' \Trans{\epsilon}  q_e\land q_s'\trans{\delta}q_s'\land q_e'\trans{\delta} q_e'$\\
                    \proofstep{\Cref{def:arrowdefs}: $\Trans{\delta}$}\\
                    $\exists q_s'\in X_s', q_e'\in X_e': q_s'\Trans{\delta} q_s \land  q_e' \Trans{\delta}  q_e$\\
                    \proofstep{\Cref{def:after}: $\after$}\\
                    $q_s \in X_s' \after \delta \land q_e \in X_e' \after \delta$
                   
            \end{case_distinction}
        \end{case_distinction}
        \item [$\impliedby$:]
        To make induction easier to follow, we will prove:
            \[\forall \sigma \in \utraces{s\parcomp e}: \collectTuple{s\parcomp e \after \sigma} = (X_s,X_e) \implies (X_s,X_e)\in \uiocorel\]
        which is equivalent to the original statement according to \cref{lem:exists_forall_if_no_goal}.
        The proof follows by induction on $\sigma$. Base case $\sigma=\epsilon$, and inductive step $\sigma = \sigma'\cdot\ell$, for $\ell\in L_{s\parcomp e}^\delta$. Further case distinction on $\ell$ where required. IH:
        \[\forall X_s'\subseteq Q_x, X_e'\subseteq Q_e: \collectTuple{s\parcomp e \after \sigma'} = (X_s',X_e') \implies (X_s',X_e')\in \uiocorel\]
        Since $\collectTuple{}$ is a function, for every set $X$ there is always exactly one tuple of sets $(Y,Y')$ for which $\collectTuple{X}=(Y,Y')$ is true. The IH therefore rewrites to the easier to use statement:
        \[\exists X_s'\subseteq Q_x, X_e'\subseteq Q_e: \collectTuple{s\parcomp e \after \sigma'} = (X_s',X_e') \land (X_s',X_e')\in \uiocorel\]
        \begin{case_distinction}
            \item[$\sigma=\epsilon$:]\ \\
                $\collectTuple{s\parcomp e \after \epsilon} = (X_s,X_e)$\\
                \proofstep{\Cref{def:after}: $\after$}\\
                $\collectTuple{\{(q_s,q_e) \setbar s \parcomp e \Trans{\epsilon} q_s \parcomp q_e \}} = (X_s,X_e)$\\
                \proofstep{\Cref{lem:project_from_parcomp_no_delta}}\\
                $\collectTuple{\{(q_s,q_e) \setbar s \Trans{\epsilon} q_s \land e \Trans{\epsilon} q_e\}} = (X_s,X_e)$\\
                \proofstep{\Cref{def:collect_tuple}: $\collectTupleSymb$}\\
                $X_s = \{q_s \setbar s \Trans{\epsilon} q_s \land e \Trans{\epsilon} q_e\} \land X_e = \{q_e \setbar s \Trans{\epsilon} q_s \land e \Trans{\epsilon} q_e\}$\\
                \proofstep{\Cref{def:after,def:LTS,def:arrowdefs}: $s\after \epsilon \neq \emptyset$}\\
                $X_s=s\after \epsilon \land X_e = e \after \epsilon$\\
                \proofstep{\Itemref{def:uioco_rel}{base}: $\uiocorel$}\\
                $(X_s,X_e)\in\uiocorel$
            \item[$\sigma = \sigma'\cdot\ell$ ($\ell\in L_s\setminus L_e$):]\ \\
                $\collectTuple{s\parcomp e \after \sigma'\cdot\ell} = (X_s,X_e)$\\
                \proofstep{\Cref{def:after}: $\after$}\\
                $\collectTuple{\{(q_s,q_e) \setbar s \parcomp e \Trans{\sigma'\cdot\ell} q_s \parcomp q_e \}} = (X_s,X_e)$\\
                \proofstep{\Cref{item:trans_transitive}}\\
                $\collectTuple{\{(q_s,q_e) \setbar \exists q_s',\in Q_s, q_e'\in Q_e :s \parcomp e \Trans{\sigma'} q_s'\parcomp q_e'\land  q_s'\parcomp q_e' \Trans{\ell} q_s \parcomp q_e \}} = (X_s,X_e)$\\
                \proofstep{Apply IH}\\
                $\collectTuple{s\parcomp e \after \sigma'} = (X_s',X_e')\land(X_s',X_e')\in \uiocorel\;\land$\\
                $\collectTuple{\{(q_s,q_e) \setbar \exists q_s',\in Q_s, q_e'\in Q_e :s \parcomp e \Trans{\sigma'} q_s'\parcomp q_e'\land  q_s'\parcomp q_e' \Trans{\ell} q_s \parcomp q_e \}} = (X_s,X_e)$\\
                \proofstep{\Cref{lem:tupple_trace_reachable_eco_rel})}\\
                $\collectTuple{s\parcomp e \after \sigma'} = (X_s',X_e')\land(X_s',X_e')\in \uiocorel\;\land$\\
                $\collectTuple{\{(q_s,q_e) \setbar \exists q_s'\in X_s', q_e' \in X_e':  q_s'\parcomp q_e' \Trans{\ell} q_s \parcomp q_e \}} = (X_s,X_e)$\\
                \proofstep{\Cref{lem:project_from_parcomp_no_delta}}\\
                $\collectTuple{s\parcomp e \after \sigma'} = (X_s',X_e')\land(X_s',X_e')\in \uiocorel\;\land$\\
                $\collectTuple{\{(q_s,q_e) \setbar \exists q_s'\in X_s', q_e' \in X_e':  q_s' \Trans{\ell} q_s  \land q_e'\Trans{\epsilon} q_e \}} = (X_s,X_e)$\\
                \proofstep{\Cref{def:outset}: $\outset{X_s'}$}\\
                $\collectTuple{s\parcomp e \after \sigma'} = (X_s',X_e')\land(X_s',X_e')\in \uiocorel\;\land$\\
                $\collectTuple{\{(q_s,q_e) \setbar \exists q_s'\in X_s', q_e' \in X_e':  q_s' \Trans{\ell} q_s  \land q_e'\Trans{\epsilon} q_e \}} = (X_s,X_e)\;\land$\\
                $\ell\in U_s \implies \ell\in \outset{X_s'}$\\
                \proofstep{\Cref{def:inset,def:uioco}: $\inset{X_s'}$ and $\sigma'\cdot\ell\in\utraces{s\parcomp e}$}\\
                $\collectTuple{s\parcomp e \after \sigma'} = (X_s',X_e')\land(X_s',X_e')\in \uiocorel\;\land$\\
                $\collectTuple{\{(q_s,q_e) \setbar \exists q_s'\in X_s', q_e' \in X_e':  q_s' \Trans{\ell} q_s  \land q_e'\Trans{\epsilon} q_e \}} = (X_s,X_e)\;\land$\\
                $\ell\in (\outset{X_s'} \cup \inset{X_s'})$\\
                \proofstep{\Cref{def:after}: $\after$}\\
                $\collectTuple{s\parcomp e \after \sigma'} = (X_s',X_e')\land(X_s',X_e')\in \uiocorel\;\land$\\
                $\collectTuple{\{(q_s,q_e) \setbar q_s \in X_s'\after\ell \land q_e\in X_e'\after\epsilon \}} = (X_s,X_e)\;\land$\\
                $\ell\in (\outset{X_s'} \cup \inset{X_s'})$\\
                \proofstep{\Cref{def:collect_tuple,def:LTS,def:uioco_rel}: $\collectTupleSymb$ and $X_s'\after\ell\neq \emptyset \land X_e' \neq \emptyset$}\\
                $\collectTuple{s\parcomp e \after \sigma'} = (X_s',X_e')\land(X_s',X_e')\in \uiocorel\;\land$\\
                $ X_s =  X_s'\after\ell \land X_e= X_e'\after\epsilon \land \ell\in (\outset{X_s'} \cup \inset{X_s'})$\\
                \proofstep{\Cref{lem:uioco_rel_after_epsilon}}\\
                $\collectTuple{s\parcomp e \after \sigma'} = (X_s',X_e')\land(X_s',X_e')\in \uiocorel\;\land$\\
                $ X_s =  X_s'\after\ell \land X_e= X_e'\land \ell\in (\outset{X_s'} \cup \inset{X_s'})$\\
                \proofstep{\Itemref{def:uioco_rel}{internal_imp}: $\uiocorel$}\\
                $(X_s,X_e)\in\uiocorel$
            \item[$\sigma = \sigma'\cdot\ell$ ($\ell\in L_e\setminus L_s$):] symmetric to previous case\\
            \item[$\sigma = \sigma'\cdot\ell$ ($\ell\in U_s\cap I_e$):]\ \\
                $\collectTuple{s\parcomp e \after \sigma'\cdot\ell} = (X_s,X_e)$\\
                \proofstep{\Cref{def:after}: $\after$}\\
                $\collectTuple{\{(q_s,q_e) \setbar s \parcomp e \Trans{\sigma'\cdot\ell} q_s \parcomp q_e \}} = (X_s,X_e)$\\
                \proofstep{\Cref{item:trans_transitive}}\\
                $\collectTuple{\{(q_s,q_e) \setbar \exists q_s',\in Q_s, q_e'\in Q_e :s \parcomp e \Trans{\sigma'} q_s'\parcomp q_e'\land  q_s'\parcomp q_e' \Trans{\ell} q_s \parcomp q_e \}} = (X_s,X_e)$\\
                \proofstep{Apply IH}\\
                $\collectTuple{s\parcomp e \after \sigma'} = (X_s',X_e')\land(X_s',X_e')\in \uiocorel\;\land$\\
                $\collectTuple{\{(q_s,q_e) \setbar \exists q_s',\in Q_s, q_e'\in Q_e :s \parcomp e \Trans{\sigma'} q_s'\parcomp q_e'\land  q_s'\parcomp q_e' \Trans{\ell} q_s \parcomp q_e \}} = (X_s,X_e)$\\
                \proofstep{\Cref{lem:tupple_trace_reachable_eco_rel}}\\
                $\collectTuple{s\parcomp e \after \sigma'} = (X_s',X_e')\land(X_s',X_e')\in \uiocorel\;\land$\\
                $\collectTuple{\{(q_s,q_e) \setbar \exists q_s'\in X_s', q_e' \in X_e':  q_s'\parcomp q_e' \Trans{\ell} q_s \parcomp q_e \}} = (X_s,X_e)$\\
                \proofstep{\Cref{lem:project_from_parcomp_no_delta}}\\
                $\collectTuple{s\parcomp e \after \sigma'} = (X_s',X_e')\land(X_s',X_e')\in \uiocorel\;\land$\\
                $\collectTuple{\{(q_s,q_e) \setbar \exists q_s'\in X_s', q_e' \in X_e':  q_s' \Trans{\ell} q_s  \land q_e'\Trans{\ell} q_e \}} = (X_s,X_e)$\\
                \proofstep{\Cref{def:outset}: $\outset{X_s'}$}\\
                $\collectTuple{s\parcomp e \after \sigma'} = (X_s',X_e')\land(X_s',X_e')\in \uiocorel \land \ell\in\outset{X_s'}\;\land$\\
                $\collectTuple{\{(q_s,q_e) \setbar \exists q_s'\in X_s', q_e' \in X_e':  q_s' \Trans{\ell} q_s  \land q_e'\Trans{\ell} q_e \}} = (X_s,X_e)$\\
                \proofstep{\Cref{def:after}: $\after$}\\
                $\collectTuple{s\parcomp e \after \sigma'} = (X_s',X_e')\land(X_s',X_e')\in \uiocorel\land \ell\in\outset{X_s'}\;\land$\\
                $\collectTuple{\{(q_s,q_e) \setbar q_s \in X_s'\after\ell \land q_e\in X_e'\after\ell \}} = (X_s,X_e)$\\
                \proofstep{\Cref{def:collect_tuple,def:LTS,def:uioco_rel}: $\collectTupleSymb$ and $X_s'\after\ell\neq \emptyset \land X_e' \after \ell \neq \emptyset$}\\
                $\collectTuple{s\parcomp e \after \sigma'} = (X_s',X_e')\land(X_s',X_e')\in \uiocorel\land \ell\in\outset{X_s'}\;\land$\\
                $ X_s =  X_s'\after\ell \land X_e= X_e'\after\ell $\\
                \proofstep{\Itemref{def:uioco_rel}{sync_out_imp}: $\uiocorel$}\\
                $(X_s,X_e)\in\uiocorel$
            \item[$\sigma = \sigma'\cdot\ell$ ($\ell\in I_s\cap U_e$):] Symmetric to previous case\\
            \item[$\sigma = \sigma'\cdot\ell$ ($\ell\in I_s\cap I_e$):] Symmetric to previous case\\
            \item[$\sigma = \sigma'\cdot\ell$ ($\ell\in U_s\cap U_e$):] Case not possible (\cref{def:composable})\\
            \item[$\sigma = \sigma'\cdot\delta$:]\ \\
                $\collectTuple{s\parcomp e \after \sigma'\cdot\delta} = (X_s,X_e)$\\
                \proofstep{\Cref{def:after}: $\after$}\\
                $\collectTuple{\{(q_s,q_e) \setbar s \parcomp e \Trans{\sigma'\cdot\delta} q_s \parcomp q_e \}} = (X_s,X_e)$\\
                \proofstep{\Cref{item:trans_transitive}}\\
                $\collectTuple{\{(q_s,q_e) \setbar \exists q_s',\in Q_s, q_e'\in Q_e :s \parcomp e \Trans{\sigma'} q_s'\parcomp q_e'\land  q_s'\parcomp q_e' \Trans{\delta} q_s \parcomp q_e \}} = (X_s,X_e)$\\
                \proofstep{Apply IH}\\
                $\collectTuple{s\parcomp e \after \sigma'} = (X_s',X_e')\land(X_s',X_e')\in \uiocorel\;\land$\\
                $\collectTuple{\{(q_s,q_e) \setbar \exists q_s',\in Q_s, q_e'\in Q_e :s \parcomp e \Trans{\sigma'} q_s'\parcomp q_e'\land  q_s'\parcomp q_e' \Trans{\delta} q_s \parcomp q_e \}} = (X_s,X_e)$\\
                \proofstep{\Cref{lem:tupple_trace_reachable_eco_rel})}\\
                $\collectTuple{s\parcomp e \after \sigma'} = (X_s',X_e')\land(X_s',X_e')\in \uiocorel\;\land$\\
                $\collectTuple{\{(q_s,q_e) \setbar \exists q_s'\in X_s', q_e' \in X_e':  q_s'\parcomp q_e' \Trans{\delta} q_s \parcomp q_e \}} = (X_s,X_e)$\\
                \proofstep{\Cref{lem:parcomp_delta_eco_rel}}\\
                $\collectTuple{s\parcomp e \after \sigma'} = (X_s',X_e')\land(X_s',X_e')\in \uiocorel\;\land$\\
                $\collectTuple{\{(q_s,q_e) \setbar \exists q_s'\in X_s', q_e' \in X_e':  q_s' \Trans{\delta} q_s  \land q_e'\Trans{\delta} q_e \}} = (X_s,X_e)$\\
                \proofstep{\Cref{def:outset}: $\outset{X_s'}$ and $\outset{X_e'}$}\\
                $\collectTuple{s\parcomp e \after \sigma'} = (X_s',X_e')\land(X_s',X_e')\in \uiocorel\;\land$\\
                $\collectTuple{\{(q_s,q_e) \setbar \exists q_s'\in X_s', q_e' \in X_e':  q_s' \Trans{\delta} q_s  \land q_e'\Trans{\delta} q_e \}} = (X_s,X_e)\;\land$\\
                $\delta\in \outset{X_e'}\land \delta\in \outset{X_s'}$\\
                \proofstep{\Cref{def:after}: $\after$}\\
                $\collectTuple{s\parcomp e \after \sigma'} = (X_s',X_e')\land(X_s',X_e')\in \uiocorel\;\land$\\
                $\collectTuple{\{(q_s,q_e) \setbar q_s \in X_s'\after\delta \land q_e\in X_e'\after\delta \}} = (X_s,X_e)\;\land$\\
                $\delta\in \outset{X_e'}\land \delta\in \outset{X_s'}$\\
                \proofstep{\Cref{def:collect_tuple,def:LTS,def:uioco_rel}: $\collectTupleSymb$ and $X_s'\after\delta\neq \emptyset \land X_e' \after \delta \neq \emptyset$}\\
                $(X_s',X_e')\in \uiocorel\;\land$\\
                $ X_s =  X_s'\after\delta \land X_e= X_e'\after\delta \land \delta\in \outset{X_e'}\land \delta\in \outset{X_s'}$\\
                \proofstep{\Itemref{def:uioco_rel}{delta}: $\uiocorel$}\\
                $(X_s,X_e)\in\uiocorel$
            
        \end{case_distinction}
    \end{case_distinction}
    \end{proof}
\end{toappendix}

\begin{lemmarep}
    Let $s,e \in\LTS$ be composable, $s\aco e$, $q_s\in Q_s, q_e\in Q_e$, then
    \[\exists (X_s,X_e)\in \uiocorel: q_s \in X_s \land q_e \in X_e \iff \exists \sigma \in \utraces{s\parcomp e}: s\parcomp e \Trans{\sigma} q_s \parcomp q_e\]
    \label{lem:R_utraces_bi-implication}
\end{lemmarep}
\begin{proof}
\ \\
    \begin{case_distinction}
        \item[$\implies$:]\ \\
        $\exists (X_s, X_e) \in \uiocorel: q_s\in X_s \land q_e\in X_e$\\
        \proofstep{\Cref{lem:eco_rel_iff_tuple_utraces}}\\
        $\exists (X_s, X_e) \in \uiocorel: q_s\in X_s \land q_e\in X_e\;\land$\\
        $\exists \sigma \in \utraces{s\parcomp e}: \collectTuple{s\parcomp e \after \sigma} = (X_s,X_e)$\\
        \proofstep{\Cref{def:collect_tuple,def:after}: $\collectTupleSymb$ and $\after$}\\
        $\exists (X_s, X_e) \in \uiocorel: q_s\in X_s \land q_e\in X_e\;\land$\\
        $\exists \sigma \in \utraces{s\parcomp e}: X_s = \{q_s'\setbar s\parcomp e \Trans{\sigma} q_s'\parcomp q_e'\} \land X_e = \{q_e'\setbar s\parcomp e \Trans{\sigma} q_s'\parcomp q_e'\}$\\
        \proofstep{\Cref{lem:project_from_parcomp_eco}}\\
        $\exists (X_s, X_e) \in \uiocorel: q_s\in X_s \land q_e\in X_e\;\land$\\
        $\exists \sigma \in \utraces{s\parcomp e}: X_s = \{q_s'\setbar s \Trans{\project{\sigma}{L_s^\delta}} q_s'\} \land X_e = \{q_e'\setbar e \Trans{\project{\sigma}{L_e^\delta}} q_e'\}$\\
        \proofstep{$q_s\in X_s$ and $q_e\in X_e$}\\
        $\exists \sigma \in \utraces{s\parcomp e}: s \Trans{\project{\sigma}{L_s^\delta}} q_s \land e \Trans{\project{\sigma}{L_e^\delta}} q_e$\\
        \proofstep{\Cref{lem:project_from_parcomp_light}}\\
        $\exists \sigma \in \utraces{s\parcomp e}: s \parcomp e \Trans{\sigma} q_s \parcomp q_e $\\
    \item [$\impliedby$:]\ \\
        $\exists \sigma \in \utraces{s\parcomp e}: s \parcomp e \Trans{\sigma} q_s \parcomp q_e $\\
        \proofstep{\Cref{lem:once_in_R_always_in_R}}\\
        $\exists \sigma \in \utraces{s\parcomp e}: $\\
        \tab$(s \after \project{\sigma}{L_s^\delta}, e \after \project{\sigma}{L_e^\delta}) \in \uiocorel \land s \Trans{ \project{\sigma}{L_s^\delta}} q_s \land e \Trans{ \project{\sigma}{L_e^\delta}} q_e$\\
        \proofstep{\Cref{def:after}: $\after$}\\
        $\exists \sigma \in \utraces{s\parcomp e}: (s \after \project{\sigma}{L_s^\delta}, e \after \project{\sigma}{L_e^\delta}) \in \uiocorel\;\land$\\
        $q_s \in s  \after \project{\sigma}{L_s^\delta}  \land q_e \in e \after \project{\sigma}{L_e^\delta}$\\
        \proofstep{$\exists$-introduction}\\
        $\exists (X_s,X_e)\in \uiocorel: q_s \in X_s \land q_e \in X_e$
\end{case_distinction}
\end{proof}

Using these two results, we can now prove that the existence of an $\acceptsim$ for $s$ and $e$ implies certain properties over the states reachable by $\utraces{}$ in the parallel composition of $s$ and $e$. 
Most notably, we can conclude from \itemref{def:eco_rel}{sync_out_env} and \itemref{def:eco_rel}{sync_out_imp} that synchronised outputs of one component are always accepted by the other component, which coincides with the definition of $\mutuallyaccepts{}$. This means that establishing the existence of an $\acceptsim$ is enough to apply the theoretical results previously obtained for $\mutuallyaccepts{}$ in \cite{vancuyck_CompositionalityModelBasedTesting_2023a}, the main result being the possibility of compositional testing as expressed in \Cref{lem:compositional_testing_eco}.

\begin{toappendix} 
    \begin{lemmarep}
    Let $s,e\in \LTS$ be $\composable$.
    \[s \aco e \implies s \accepts{} e\]
    \label{lem:eco_implies_accepts}
    \end{lemmarep}
    \begin{proof}
    \ \\
                Take $\sigma\in \utraces{s\parcomp e}$, $q_s\in Q_s$, $q_e\in Q_e$, such that $s\parcomp e \Trans{\sigma} q_s \parcomp q_e$.\\
        Take $\ell \in \outset{q_e} \cap I_s$. To Prove: $\ell \in \inset{q_s}$\\\\
        
        \noindent
        $s \eco e \land s\parcomp e \Trans{\sigma} q_s \parcomp q_e \land \ell \in \outset{q_e} \cap I_s$\\
        \proofstep{\Cref{lem:once_in_R_always_in_R}}\\
        $s\parcomp e \Trans{\sigma} q_s \parcomp q_e\land \ell \in \outset{q_e} \cap I_s\;\land$\\
        $(s \after \project{\sigma}{L_s^\delta}, e \after \project{\sigma}{L_e^\delta}) \in \uiocorel \land s \Trans{ \project{\sigma}{L_s^\delta}} q_s \land e \Trans{ \project{\sigma}{L_e^\delta}} q_e$\\
        \proofstep{\Itemref{def:eco_rel}{sync_out_env} and \Cref{lem:uiocosim_minimal}: $\acceptrel$}\\
        $\ell\in \inset{s \after \project{\sigma}{L_s^\delta}} \land s \Trans{ \project{\sigma}{L_s^\delta}} q_s$\\
        \proofstep{\Cref{def:inset,def:after}: $\inset{}$ and $\after$}\\
        $\ell \in \inset{q_s}$

    \end{proof}
\end{toappendix}

\begin{theoremrep}
Let $s,e \in\LTS$ be composable, then
\[s \aco e \implies s \mutuallyaccepts{} e\]
\label{lem:eco_implies_mutaccepts}
\end{theoremrep}

\begin{proof}
    \ \\
    $s\eco e$\\
\proofstep{\Cref{lem:eco_symetrical}}\\
$s\eco e\land e \eco s$\\
\proofstep{\Cref{lem:eco_implies_accepts}}\\
$s\accepts{} e\land e \accepts{} s$\\
\proofstep{\Cref{def:mutually_accepts} ($\mutuallyaccepts{}$)}\\
$s\mutuallyaccepts{} e$
\end{proof}

\begin{theoremrep}
    Let $s,e \in\LTS$ be composable, $i_s,i_e \in \IOTS$, then
    \[s \aco e \land i_s \uioco s \land i_e \uioco e \implies i_s \parcomp i_e \uioco s \parcomp e\]
\label{lem:compositional_testing_eco}
\end{theoremrep}

\begin{proof}
    Follows directly from \Cref{lem:compositional_testing_mutaccepts,lem:eco_implies_mutaccepts}
\end{proof}

\subsubsection{Mutual Acceptance Implies Accepting Conformance}

The previous section shows that $\aco$ is at least as strong as $\mutuallyaccepts{}$. In this section we show that the other direction also holds, and that $\aco$ and $\mutuallyaccepts{}$ are just two different ways of writing down the same relation. This means $\aco$ is not just sufficient, but also necessary with respect to mutual acceptance, and that \aco{} and \mutuallyaccepts{} can be used interchangeably wherever this is convenient.

To prove this, we show that under the assumption of $s\mutuallyaccepts{} e$, \Cref{def:eco_rel,def:uioco_rel} coincide (\Cref{lem:uiocosim_implies_ecosim}). The proof follows the same structure and lemmas as the previous section, but steps relying on \itemref{def:eco_rel}{sync_out_env} and \itemref{def:eco_rel}{sync_out_imp} are replaced with steps relying on $s\mutuallyaccepts{} e$ instead. The equivalent lemma for \Cref{lem:project_from_parcomp_eco} is already a result from a previous paper (Lemma 3 of \cite{vancuyck_CompositionalityModelBasedTesting_2023a}) and is therefore not repeated here.

\Cref{lem:mutaccepts_implies_eco} combined with \Cref{lem:eco_implies_mutaccepts} then gives the main result of this section: $s \aco e$ and $s \mutuallyaccepts{} e$ are equivalent.

\begin{toappendix}
    \begin{lemmarep}
    Let $s,e \in \LTS$ be $\composable$, $s \mutuallyaccepts{} e$, 
    $\sigma \in \utraces{s\parcomp e}$, $q_s,q_s'\in Q_s$, $q_e,q_e'\in Q_e:$
    \[s\parcomp e \Trans{\sigma} q_s\parcomp q_e \land s\parcomp e \Trans{\sigma} q_s' \parcomp q_e' \iff s\parcomp e \Trans{\sigma} q_s\parcomp q_e' \land s\parcomp e \Trans{\sigma} q_s' \parcomp q_e \]
    \label{lem:trans_complete_uioco_rel}
    \end{lemmarep}
    \begin{proof}
        The $\implies$ and $\impliedby$ case are symmetric with each other up to alpha renaming, So we will only prove the $\implies$ case here.\\

\noindent
$s\parcomp e \Trans{\sigma} q_s\parcomp q_e \land s\parcomp e \Trans{\sigma} q_s' \parcomp q_e' \land s\eco e \land \sigma \in \utraces{s\parcomp e}$\\
\proofstep{\Cref{lem:project_from_parcomp}}\\
$s\Trans{\project{\sigma}{L_s^\delta}} q_s \land e\Trans{\project{\sigma}{L_e^\delta}} q_e \land s\Trans{\project{\sigma}{L_s^\delta}} q_s' \land e\Trans{\project{\sigma}{L_e^\delta}} q_e'$\\
\proofstep{\Cref{lem:project_from_parcomp_light}}\\
$s\parcomp e \Trans{\sigma} q_s\parcomp q_e' \land s\parcomp e \Trans{\sigma} q_s' \parcomp q_e$\\
    \end{proof}

    \begin{lemmarep}
        \label{lem:tupple_trace_reachable_uioco_rel}
        Let $s,e\in\LTS$ be $\composable$, $s\mutuallyaccepts{} e$,
        $(X_s,X_e)\in \uiocorel$\\
        $q_s\in Q_s$, $q_e\in Q_e$, $\sigma \in \utraces{s\parcomp e}$, $\collectTuple{s \parcomp e \after \sigma} = (X_s,X_e):$
        \[ q_s \in X_s \land q_e \in X_e \iff s\parcomp e \Trans{\sigma} q_s \parcomp q_e\]
        
    \end{lemmarep}
    \begin{proof}
        \begin{case_distinction}
    \item[$\implies:$]\ \\
        $\collectTuple{s\parcomp e \after \sigma} = (X_s,X_e) \land q_s\in X_s \land q_e\in X_e \land (X_s,X_e)\in \uiocorel$\\
        \proofstep{\Cref{def:collect_tuple,def:after}: $\collectTupleSymb$ and $\after$}\\
        $\exists q_s'\in Q_s, q_e'\in Q_e: s\parcomp e \Trans{\sigma} q_s'\parcomp q_e \land s\parcomp e \Trans{\sigma} q_s\parcomp q_e'$\\
        \proofstep{\Cref{lem:trans_complete_uioco_rel}}\\
        $s\parcomp e \Trans{\sigma} q_s \parcomp q_e$
    \item[$\impliedby:$]\ \\
        $s\parcomp e \Trans{\sigma} q_s \parcomp q_e$\\
        \proofstep{\Cref{def:after}: $\after$}\\
        $q_s \parcomp q_e \in s\parcomp e \after \sigma$\\
        \proofstep{\Cref{def:collect_tuple}: $\collectTupleSymb$ and $\collectTuple{q_s \parcomp q_e \after \sigma} = (X_s,X_e)$}\\
        $q_s\in X_s \land q_e\in X_e$
\end{case_distinction}
    \end{proof}

    \begin{lemmarep}
         \label{lem:uioco_rel_iff_tuple_utraces}
         Let $s,e\in \LTS$ be $\composable$, $s\mutuallyaccepts{} e$, $X_e\subseteq Q_e, X_s\subseteq Q_s$
        \[(X_s,X_e)\in \uiocorel \iff \exists \sigma \in \utraces{s\parcomp e}: \collectTuple{s\parcomp e \after \sigma} = (X_s,X_e)\]
       
    \end{lemmarep}
    
    \begin{proof}
        \begin{case_distinction}
        \item[$\implies$:] 
        Proof by induction on the structure of $\uiocorel$. Case names follow naming of \cref{def:uioco_rel}. For the recursive cases the induction hypothesis (IH) is: 
        \[\exists \sigma'\in \utraces{s\parcomp e}: \collectTuple{s\parcomp e \after \sigma'} = (X_s',X_e') \]
        Where $(X_s',X_e')$ comes from the left side of the $\vdash$ of the relevant case of \cref{def:uioco_rel}. 
        
        To prove the set equality, we split each goal further into three parts: A proof that the given $\sigma$ is part of $\utraces{s\parcomp e}$, and the two subset statements corresponding to the set equality. For brevity the definitions of $\collectTupleSymb$ and $\after$ are implicitly expanded here. I.e. to prove $\exists\sigma\in \utraces{s\parcomp e}: \collectTuple{s\parcomp e \after \sigma} = (X_s,X_e)$, we fix a $\sigma$ (containing $\sigma'$ from the IH) and then prove:
        \begin{enumerate}
            \item $\sigma\in \utraces{s\parcomp e}$
            \item $s\parcomp e \Trans{\sigma} q_s\parcomp q_e$ for arbitrary $q_s\in X_s, q_e\in X_e$ ($\supseteq$)
            \item $q_s \in X_s$ and $q_e \in X_e$, for arbitrary $q_s \parcomp q_e \in s\parcomp e \after \sigma$ ($\subseteq$).
        \end{enumerate}
        \begin{case_distinction}
            \item[case \ref{item:uioco_rel_base} ($X_s= s \after \epsilon,X_e=e \after \epsilon$):] Take $\sigma = \epsilon$\\
                \begin{case_distinction}
                    \item[$\utraces{}$:] $\epsilon$ is always trivially part of $\utraces{}$
                    \item[$\supseteq$:]
                    $q_s \in s \after \epsilon \land q_e \in e \after \epsilon$\\
                    \proofstep{\Cref{def:after}: $\after$}\\
                    $s \Trans{\epsilon} q_s \land e \Trans{\epsilon} q_e $\\
                    \proofstep{\Cref{lem:project_from_parcomp_no_delta}}\\
                    $s\parcomp e \Trans{\epsilon} q_s \parcomp q_e$\\
                    
                    \item[$\subseteq$:]
                    $s\parcomp e \Trans{\epsilon} q_s \parcomp q_e$\\
                    \proofstep{\Cref{lem:project_from_parcomp_no_delta}}\\
                    $s \Trans{\epsilon} q_s \land e \Trans{\epsilon} q_e $\\
                    \proofstep{\Cref{def:after}: $\utraces{}$ and $\after$}\\
                    $q_s \in s \after \epsilon \land q_e \in e \after \epsilon$
                \end{case_distinction}
                
            \item[case \ref{item:uioco_rel_sync_out_env} ($X_s= X_s'\after \ell, X_e=X_e'\after \ell$):] Take $\sigma =\sigma'\cdot\ell$
            \begin{case_distinction}
                \item[$\utraces{}$:] We have $\sigma'\in \utraces{s\parcomp e}$ (IH) and $s\parcomp e \Trans{\sigma'} q_s'\parcomp q_e'$ for all $q_s'\in X_s'$, $q_e'\in X_e'$ (IH and \cref{lem:tupple_trace_reachable_uioco_rel}). We also have $\ell \in \outset{X_e'} \cap I_s$ from case $\ref{item:uioco_rel_sync_out_env}$, which then gives $\ell \in \inset{q_s'}$ for all $q_s'\in X_s'$ from \cref{def:mutually_accepts}. This gives $s\parcomp e \Trans{\sigma'\cdot\ell}$. Since $\ell\in U_s \land \ell\in I_e$, we have $\ell\in U_{s\parcomp e}$ (\cref{def:parcomp}), and therefore $\sigma\cdot\ell\in\utraces{s\parcomp e}$ (\cref{def:uioco}).
                \item[$\supseteq$:]
                    $q_s \in X_s' \after \ell \land q_e \in X_e' \after \ell$\\
                    \proofstep{\Cref{def:after}: $\after$}\\
                    $\exists q_s' \in X_s', q_e'\in X_e': q_s'\Trans{\ell}q_s \land q_e' \Trans{\ell} q_e$\\
                    \proofstep{\Cref{lem:project_from_parcomp_light,def:projection}: $\projectop$}\\
                    $\exists q_s'\in X_s', q_e'\in X_e': q_s' \parcomp q_e' \Trans{\ell}q_s \parcomp q_e$\\
                    \proofstep{Apply IH}\\
                    $\collectTuple{s\parcomp e \after \sigma'} = (X_s',X_e')$\\
                    $\exists q_s'\in X_s', q_e'\in X_e': q_s' \parcomp q_e' \Trans{\ell}q_s \parcomp q_e$\\
                    \proofstep{\Cref{lem:tupple_trace_reachable_uioco_rel}}\\
                    $\exists q_s'\in X_s', q_e'\in X_e': s\parcomp e \Trans{\sigma'} q_s'\parcomp q_e' \land q_s' \parcomp q_e' \Trans{\ell}q_s \parcomp q_e $\\
                    \proofstep{\Cref{item:trans_transitive}}\\
                    $s\parcomp e \Trans{\sigma'\cdot \ell} q_s \parcomp q_e $
                \item[$\subseteq$:] 
                    $s\parcomp e \Trans{\sigma'\cdot\ell} q_s \parcomp q_e$\\
                    \proofstep{\Cref{item:trans_transitive}}\\
                    $\exists q_s'\in Q_s, q_e'\in Q_e: s \parcomp e \Trans{\sigma'} q_s'\parcomp q_e' \land q_s'\parcomp q_e' \Trans{\ell}q_s \parcomp q_e$\\
                    \proofstep{Apply IH}\\
                    $\collectTuple{s\parcomp e \after \sigma'} = (X_s',X_e')$\\
                    $\exists q_s'\in Q_s, q_e'\in Q_e: s \parcomp e \Trans{\sigma'} q_s'\parcomp q_e' \land q_s'\parcomp q_e' \Trans{\ell}q_s \parcomp q_e$\\
                    \proofstep{\Cref{def:collect_tuple,def:after}: $\collectTupleSymb$ and $\after$}\\
                    $\exists q_s'\in X_s', q_e'\in X_e': q_s'\parcomp q_e' \Trans{\ell}q_s \parcomp q_e$\\
                    \proofstep{\Cref{lem:project_from_parcomp_no_delta,def:projection}: $\projectop$}\\
                    $\exists q_s'\in X_s', q_e'\in X_e': q_s' \Trans{\ell} q_s \land  q_e' \Trans{\ell} q_e$\\
                    \proofstep{\Cref{def:after}: $\after$}\\
                    $q_s \in X_s' \after \ell \land q_e \in X_e' \after \ell$
            \end{case_distinction}
                
            \item[case \ref{item:uioco_rel_sync_out_imp}:] Symmetric to previous case
            \item[case \ref{item:uioco_rel_internal_env} ($X_s= X_s', X_e=X_e'\after \ell$):]Take $\sigma =\sigma'\cdot\ell$
            \begin{case_distinction}
                 \item[$\utraces{}$:]
                 We have $\sigma'\in \utraces{s\parcomp e}$ (IH) and $s\parcomp e \Trans{\sigma'} q_s'\parcomp q_e'$ for all $q_s'\in X_s'$, $q_e'\in X_e'$ (IH and \cref{lem:tupple_trace_reachable_uioco_rel}). Combine with $\ell \in (\outset{X_e'}\cup \inset{X_e'})\setminus L_s^\delta$ from case $\ref{item:uioco_rel_internal_env}$, we get $s\parcomp e \Trans{\sigma'\cdot\ell}$. If $\ell \in U_s$ then $\ell\in U_{s\parcomp e}$ and we have $\sigma\cdot\ell\in\utraces{s\parcomp e}$ (\cref{def:uioco}). 
                 
                 If $\ell \in I_s$ then $\ell \in I_{s\parcomp e}$. We then have to prove that $\ell\in \inset{s\parcomp e \after \sigma'}$. Since inputs cannot be part of $\outset{X_e'}$, we know $\ell\in \inset{X_e'}$. From the IH we know $q_s'\parcomp q_e' \in s\parcomp e \after \sigma' \implies q_e'\in X_e'$. Combined with the fact that $\ell\notin L_s^\delta$ this then gives $\ell\in \inset{s\parcomp e \after \sigma'}$.
            
                \item[$\supseteq$:]
                    $q_s \in X_s' \land q_e \in X_e' \after \ell$\\
                    \proofstep{\Cref{def:after}: $\after$}\\
                    $ q_s \in X_s'\land \exists q_e'\in X_e':  q_e' \Trans{\ell} q_e$\\
                    \proofstep{\Cref{lem:project_from_parcomp_light,def:projection}: $\projectop$}\\
                    $ q_s \in X_s'\land \exists q_e'\in X_e':  q_s\parcomp q_e' \Trans{\ell} q_s \parcomp q_e \land$\\
                    \proofstep{Apply IH}\\
                    $\collectTuple{s\parcomp e \after \sigma'} = (X_s',X_e')\;\land$\\
                    $q_s \in X_s'\land \exists q_e'\in X_e':  q_s\parcomp q_e' \Trans{\ell} q_s \parcomp q_e$\\
                    \proofstep{\Cref{lem:tupple_trace_reachable_uioco_rel}}\\
                    $\exists q_e'\in X_e': s\parcomp e \Trans{\sigma'} q_s\parcomp q_e' \land   q_s\parcomp q_e' \Trans{\ell} q_s \parcomp q_e$\\
                    \proofstep{\Cref{item:trans_transitive}}\\
                    $s\parcomp e \Trans{\sigma'\cdot\ell} q_s \parcomp q_e$
                \item [$\subseteq$:]
                    $s\parcomp e \Trans{\sigma'\cdot\ell} q_s \parcomp q_e$\\
                    \proofstep{\Cref{item:trans_transitive}}\\
                    $\exists q_s'\in Q_s, q_e'\in Q_e: s \parcomp e \Trans{\sigma'} q_s'\parcomp q_e' \land q_s'\parcomp q_e' \Trans{\ell}q_s \parcomp q_e$\\
                    \proofstep{Apply IH}\\
                    $\collectTuple{s\parcomp e \after \sigma'} = (X_s',X_e')$\\
                    $\exists q_s'\in Q_s, q_e'\in Q_e: s \parcomp e \Trans{\sigma'} q_s'\parcomp q_e' \land q_s'\parcomp q_e' \Trans{\ell}q_s \parcomp q_e$\\
                    \proofstep{\Cref{def:collect_tuple,def:after}: $\collectTupleSymb$ and $\after$}\\
                    $\exists q_s'\in X_s', q_e'\in X_e': q_s'\parcomp q_e' \Trans{\ell}q_s \parcomp q_e$\\
                    \proofstep{\Cref{lem:project_from_parcomp_no_delta,def:projection}: $\projectop$}\\
                    $\exists q_s'\in X_s', q_e'\in X_e': q_s' \Trans{\epsilon} q_s \land  q_e' \Trans{\ell} q_e$\\
                    \proofstep{\Cref{def:after,def:arrowdefs}($\after$ and $\Trans{\epsilon}$)}\\
                    $\exists q_e'\in X_e': q_s \in X_s' \after \epsilon \land  q_e' \Trans{\ell} q_e$\\
                    \proofstep{\Cref{lem:uioco_rel_after_epsilon}}\\
                    $\exists q_e'\in X_e': q_s \in X_s'\land q_e' \Trans{\ell} q_e$\\
                    \proofstep{\Cref{def:after}: $\after$}\\
                    $q_s \in X_s'\land q_e \in X_e' \after \ell$
            \end{case_distinction}
             
            \item[case \ref{item:uioco_rel_internal_imp}:] Symmetrical to previous case
            \item[case \ref{item:uioco_rel_sync_in} ($X_s= X_s' \after \ell, X_e=X_e'\after \ell$):]Take $\sigma =\sigma'\cdot\ell$
            \begin{case_distinction}
                 \item[$\utraces{}$:] We have $\sigma'\in \utraces{s\parcomp e}$ (IH) and $s\parcomp e \Trans{\sigma'} q_s'\parcomp q_e'$ for all $q_s'\in X_s'$, $q_e'\in X_e'$ (IH and \cref{lem:tupple_trace_reachable_uioco_rel}). Combine with $\ell \in  (\inset{X_e'}\cap \inset{X_s'})$ from case $\ref{item:uioco_rel_sync_in}$, we get $s\parcomp e \Trans{\sigma'\cdot\ell}$. 
                 
                 Since $\ell\in I_s \land \ell\in I_e$, we have $\ell\in I_{s\parcomp e}$ (\cref{def:parcomp}), and therefore we need to prove  $\ell\in \inset{s\parcomp e \after \sigma'}$. From the IH we know $q_s'\parcomp q_e' \in s\parcomp e \after \sigma' \implies q_s' \in X_e' \land q_e'\in X_e'$. Combined with $\ell \in  (\inset{X_e'}\cap \inset{X_s'})$ this gives $\ell\in \inset{s\parcomp e \after \sigma'}$.
                 \item [$\supseteq$:]
                    $q_s \in X_s' \after \ell \land q_e \in X_e' \after \ell$\\
                    \proofstep{\Cref{def:after}: $\after$}\\
                    $\exists q_s' \in X_s', q_e'\in X_e': q_s'\Trans{\ell}q_s \land q_e' \Trans{\ell} q_e$\\
                    \proofstep{\Cref{lem:project_from_parcomp_light,def:projection}: $\projectop$}\\
                    $\exists q_s'\in X_s', q_e'\in X_e': q_s' \parcomp q_e' \Trans{\ell}q_s \parcomp q_e$\\
                    \proofstep{Apply IH}\\
                    $\collectTuple{s\parcomp e \after \sigma'} = (X_s',X_e')$\\
                    $\exists q_s'\in X_s', q_e'\in X_e': q_s' \parcomp q_e' \Trans{\ell}q_s \parcomp q_e$\\
                    \proofstep{\Cref{lem:tupple_trace_reachable_uioco_rel}}\\
                    $\exists q_s'\in X_s', q_e'\in X_e': s\parcomp e \Trans{\sigma'} q_s'\parcomp q_e' \land q_s' \parcomp q_e' \Trans{\ell}q_s \parcomp q_e $\\
                    \proofstep{\Cref{item:trans_transitive}}\\
                    $s\parcomp e \Trans{\sigma'\cdot \ell} q_s \parcomp q_e $
                 \item [$\subseteq$:]
                    $s\parcomp e \Trans{\sigma'\cdot\ell} q_s \parcomp q_e$\\
                    \proofstep{\Cref{item:trans_transitive}}\\
                    $\exists q_s'\in Q_s, q_e'\in Q_e: s \parcomp e \Trans{\sigma'} q_s'\parcomp q_e' \land q_s'\parcomp q_e' \Trans{\ell}q_s \parcomp q_e$\\
                    \proofstep{Apply IH}\\
                    $\collectTuple{s\parcomp e \after \sigma'} = (X_s',X_e')$\\
                    $\exists q_s'\in Q_s, q_e'\in Q_e: s \parcomp e \Trans{\sigma'} q_s'\parcomp q_e' \land q_s'\parcomp q_e' \Trans{\ell}q_s \parcomp q_e$\\
                    \proofstep{\Cref{def:collect_tuple,def:after}: $\collectTupleSymb$ and $\after$}\\
                    $\exists q_s'\in X_s', q_e'\in X_e': q_s'\parcomp q_e' \Trans{\ell}q_s \parcomp q_e$\\
                    \proofstep{\Cref{lem:project_from_parcomp_no_delta,def:projection}: $\projectop$}\\
                    $\exists q_s'\in X_s', q_e'\in X_e': q_s' \Trans{\ell} q_s \land  q_e' \Trans{\ell} q_e$\\
                    \proofstep{\Cref{def:after}: $\after$}\\
                    $q_s \in X_s' \after \ell \land q_e \in X_e' \after \ell$
            \end{case_distinction}
            \item[case \ref{item:uioco_rel_delta}($X_s= X_s' \after \delta, X_e=X_e'\after \delta$):]Take $\sigma =\sigma'\cdot\delta$
            \begin{case_distinction}
                 \item[$\utraces{}$:] We have $\sigma'\in \utraces{s\parcomp e}$ (IH) and $s\parcomp e \Trans{\sigma'} q_s'\parcomp q_e'$ for all $q_s'\in X_s'$, $q_e'\in X_e'$ (IH and \cref{lem:tupple_trace_reachable_uioco_rel}). $\delta \in\outset{X_e'} \land \delta \in \outset{X_s'}$ from case $\ref{item:uioco_rel_delta}$ gives that both $X_e'$ and $X_s'$ contain at least one quiescent state ($p_{s\delta}$ and $p_{e\delta}$). since $p_{s\delta}\in X_s'$ and $p_{e\delta} \in X_e'$, we have $s\parcomp e \Trans{\sigma'}p_{s\delta} \parcomp p_{e\delta}$. Since outputs in composed states come from one of their component states, $p_{s\delta} \parcomp p_{e\delta}$ is itself also quiescent by definition. This gives $s\parcomp e \Trans{\sigma'\cdot\delta}$, which then leads to $\sigma' \cdot \delta\in \utraces{s\parcomp e}$.

                 \item [$\supseteq$:]
                    $q_s \in X_s' \after \delta \land q_e \in X_e' \after \delta$\\
                    \proofstep{\Cref{def:after,def:delta}: $\after$ and $\delta$}\\
                    $\exists q_s' \in X_s', q_e'\in X_e': q_s'\Trans{\delta} q_s \land q_e' \Trans{\delta} q_e$\\
                    \proofstep{\Cref{lem:project_from_parcomp_light,def:projection}: $\projectop$}\\
                    $\exists q_s'\in X_s', q_e'\in X_e': q_s' \parcomp q_e' \Trans{\delta}q_s \parcomp q_e$\\
                    \proofstep{Apply IH}\\
                    $\collectTuple{s\parcomp e \after \sigma'} = (X_s',X_e')$\\
                    $\exists q_s'\in X_s', q_e'\in X_e': q_s' \parcomp q_e' \Trans{\delta}q_s \parcomp q_e$\\
                    \proofstep{\Cref{lem:tupple_trace_reachable_uioco_rel}}\\
                    $\exists q_s'\in X_s', q_e'\in X_e': s\parcomp e \Trans{\sigma'} q_s'\parcomp q_e' \land q_s' \parcomp q_e' \Trans{\delta}q_s \parcomp q_e $\\
                    \proofstep{\Cref{item:trans_transitive}}\\
                    $s\parcomp e \Trans{\sigma'\cdot \delta} q_s \parcomp q_e $
                 \item [$\subseteq$:]
                    $s\eco e \land s\parcomp e \Trans{\sigma'\cdot\delta} q_s \parcomp q_e$\\
                    \proofstep{\Cref{lem:project_from_parcomp}}\\
                    $s\Trans{\project{\sigma'\cdot\delta}{L_s^\delta}} q_s \land e\Trans{\project{\sigma'\cdot\delta}{L_e^\delta}} q_e $\\
                    \proofstep{\Cref{def:projection}: $\projectop$}\\
                    $s\Trans{\project{\sigma'}{L_s^\delta}\cdot\delta} q_s \land e\Trans{\project{\sigma'}{L_e^\delta}\cdot\delta} q_e $\\
                    \proofstep{\Cref{item:trans_transitive}}\\
                    $\exists q_s'\in Q_s, q_e'\in Q_e: s\Trans{\project{\sigma'}{L_s^\delta}} q_s' \land q_s' \Trans{\delta} q_s \land e\Trans{\project{\sigma'}{L_e^\delta}} q_e' \land q_e' \Trans{\delta} q_e $\\
                    \proofstep{\Cref{lem:project_from_parcomp_light}}\\
                    $\exists q_s'\in Q_s, q_e'\in Q_e: s \parcomp e \Trans{\sigma'} q_s' \parcomp q_e' \land q_s' \Trans{\delta} q_s \land q_e' \Trans{\delta} q_e $\\
                    \proofstep{Apply IH}\\
                    $\collectTuple{s\parcomp e \after \sigma'} = (X_s',X_e')$\\
                    $\exists q_s'\in Q_s, q_e'\in Q_e: s \parcomp e \Trans{\sigma'} q_s' \parcomp q_e' \land q_s' \Trans{\delta} q_s \land q_e' \Trans{\delta} q_e $\\
                    \proofstep{\Cref{def:collect_tuple,def:after}: $\collectTupleSymb$ and $\after$}\\
                    $\exists q_s'\in X_s', q_e'\in X_e':  q_s' \Trans{\delta} q_s \land q_e' \Trans{\delta} q_e $\\
                    \proofstep{\Cref{def:after}: $\after$}\\
                    $q_s \in X_s' \after \delta \land q_e \in X_e' \after \delta$
                   
            \end{case_distinction}
        \end{case_distinction}
        \item [$\impliedby$:] 
        To make induction easier to follow, we will prove:
            \[\forall \sigma \in \utraces{s\parcomp e}: \collectTuple{s\parcomp e \after \sigma} = (X_s,X_e) \implies (X_s,X_e)\in \uiocorel\]
        which is equivalent to the original statement according to \cref{lem:exists_forall_if_no_goal}.
        The proof follows by induction on $\sigma$. Base case $\sigma=\epsilon$, and inductive step $\sigma = \sigma'\cdot\ell$, for $\ell\in L_{s\parcomp e}^\delta$ with further case distinction on $\ell$ where required. IH:
        \[\forall X_s'\subseteq Q_x, X_e'\subseteq Q_e: \collectTuple{s\parcomp e \after \sigma'} = (X_s',X_e') \implies (X_s',X_e')\in \uiocorel\]
        Since $\collectTuple{}$ is a function, for every set $X$ there is always exactly one tuple of sets $(Y,Y')$ for which $\collectTuple{X}=(Y,Y')$ is true. The IH therefore rewrites to the easier to use statement:
        \[\exists X_s'\subseteq Q_x, X_e'\subseteq Q_e: \collectTuple{s\parcomp e \after \sigma'} = (X_s',X_e') \land (X_s',X_e')\in \uiocorel\]
        \begin{case_distinction}
            \item[$\sigma=\epsilon$:]\ \\
                $\collectTuple{s\parcomp e \after \epsilon} = (X_s,X_e)$\\
                \proofstep{\Cref{def:after}: $\after$}\\
                $\collectTuple{\{(q_s,q_e) \setbar s \parcomp e \Trans{\epsilon} q_s \parcomp q_e \}} = (X_s,X_e)$\\
                \proofstep{\Cref{lem:project_from_parcomp_no_delta}}\\
                $\collectTuple{\{(q_s,q_e) \setbar s \Trans{\epsilon} q_s \land e \Trans{\epsilon} q_e\}} = (X_s,X_e)$\\
                \proofstep{\Cref{def:collect_tuple}  ($\collectTupleSymb$)}\\
                $X_s = \{q_s \setbar s \Trans{\epsilon} q_s \land e \Trans{\epsilon} q_e\} \land X_e = \{q_e \setbar s \Trans{\epsilon} q_s \land e \Trans{\epsilon} q_e\}$\\
                \proofstep{\Cref{def:after,def:LTS,def:arrowdefs}: $s\after \epsilon \neq \emptyset$}\\
                $X_s=s\after \epsilon \land X_e = e \after \epsilon$\\
                \proofstep{\Itemref{def:uioco_rel}{base}: $\uiocorel$}\\
                $(X_s,X_e)\in\uiocorel$
            \item[$\sigma = \sigma'\cdot\ell$ ($\ell\in L_s\setminus L_e$):]
                $\collectTuple{s\parcomp e \after \sigma'\cdot\ell} = (X_s,X_e)$\\
                \proofstep{\Cref{def:after}: $\after$}\\
                $\collectTuple{\{(q_s,q_e) \setbar s \parcomp e \Trans{\sigma'\cdot\ell} q_s \parcomp q_e \}} = (X_s,X_e)$\\
                \proofstep{\Cref{item:trans_transitive}}\\
                $\collectTuple{\{(q_s,q_e) \setbar \exists q_s',\in Q_s, q_e'\in Q_e :s \parcomp e \Trans{\sigma'} q_s'\parcomp q_e'\land  q_s'\parcomp q_e' \Trans{\ell} q_s \parcomp q_e \}} = (X_s,X_e)$\\
                \proofstep{Apply IH}\\
                $\collectTuple{s\parcomp e \after \sigma'} = (X_s',X_e')\land(X_s',X_e')\in \uiocorel\;\land$\\
                $\collectTuple{\{(q_s,q_e) \setbar \exists q_s',\in Q_s, q_e'\in Q_e :s \parcomp e \Trans{\sigma'} q_s'\parcomp q_e'\land  q_s'\parcomp q_e' \Trans{\ell} q_s \parcomp q_e \}} = (X_s,X_e)$\\
                \proofstep{\Cref{lem:tupple_trace_reachable_uioco_rel})}\\
                $\collectTuple{s\parcomp e \after \sigma'} = (X_s',X_e')\land(X_s',X_e')\in \uiocorel\;\land$\\
                $\collectTuple{\{(q_s,q_e) \setbar \exists q_s'\in X_s', q_e' \in X_e':  q_s'\parcomp q_e' \Trans{\ell} q_s \parcomp q_e \}} = (X_s,X_e)$\\
                \proofstep{\Cref{lem:project_from_parcomp_no_delta}}\\
                $\collectTuple{s\parcomp e \after \sigma'} = (X_s',X_e')\land(X_s',X_e')\in \uiocorel\;\land$\\
                $\collectTuple{\{(q_s,q_e) \setbar \exists q_s'\in X_s', q_e' \in X_e':  q_s' \Trans{\ell} q_s  \land q_e'\Trans{\epsilon} q_e \}} = (X_s,X_e)$\\
                \proofstep{\Cref{def:outset}: $\outset{X_s'}$ and $\sigma'\cdot\ell\in\utraces{s\parcomp e}$ }\\
                $\collectTuple{s\parcomp e \after \sigma'} = (X_s',X_e')\land(X_s',X_e')\in \uiocorel\;\land$\\
                $\collectTuple{\{(q_s,q_e) \setbar \exists q_s'\in X_s', q_e' \in X_e':  q_s' \Trans{\ell} q_s  \land q_e'\Trans{\epsilon} q_e \}} = (X_s,X_e)\;\land$\\
                $\ell\in U_s \implies \ell\in \outset{X_s'}$\\
                \proofstep{\Cref{def:inset,def:uioco}: $\inset{X_s'}$ and $\sigma'\cdot\ell\in\utraces{s\parcomp e}$}\\
                $\collectTuple{s\parcomp e \after \sigma'} = (X_s',X_e')\land(X_s',X_e')\in \uiocorel\;\land$\\
                $\collectTuple{\{(q_s,q_e) \setbar \exists q_s'\in X_s', q_e' \in X_e':  q_s' \Trans{\ell} q_s  \land q_e'\Trans{\epsilon} q_e \}} = (X_s,X_e)\;\land$\\
                $\ell\in (\outset{X_s'} \cup \inset{X_s'})$\\
                \proofstep{\Cref{def:after}: $\after$}\\
                $\collectTuple{s\parcomp e \after \sigma'} = (X_s',X_e')\land(X_s',X_e')\in \uiocorel\;\land$\\
                $\collectTuple{\{(q_s,q_e) \setbar q_s \in X_s'\after\ell \land q_e\in X_e'\after\epsilon \}} = (X_s,X_e)\;\land$\\
                $\ell\in (\outset{X_s'} \cup \inset{X_s'})$\\
                \proofstep{\Cref{def:collect_tuple,def:LTS,def:uioco_rel}: $\collectTupleSymb$ and $X_s'\after\ell\neq \emptyset \land X_e' \neq \emptyset$}\\
                $\collectTuple{s\parcomp e \after \sigma'} = (X_s',X_e')\land(X_s',X_e')\in \uiocorel\;\land$\\
                $ X_s =  X_s'\after\ell \land X_e= X_e'\after\epsilon \land \ell\in (\outset{X_s'} \cup \inset{X_s'})$\\
                \proofstep{\Cref{lem:uioco_rel_after_epsilon}}\\
                $\collectTuple{s\parcomp e \after \sigma'} = (X_s',X_e')\land(X_s',X_e')\in \uiocorel\;\land$\\
                $ X_s =  X_s'\after\ell \land X_e= X_e'\land \ell\in (\outset{X_s'} \cup \inset{X_s'})$\\
                \proofstep{\Itemref{def:uioco_rel}{internal_imp}: $\uiocorel$}\\
                $(X_s,X_e)\in\uiocorel$
            \item[$\sigma = \sigma'\cdot\ell$ ($\ell\in L_e\setminus L_s$):] symmetric to previous case\\
            \item[$\sigma = \sigma'\cdot\ell$ ($\ell\in U_s\cap I_e$):]\ \\
                $\collectTuple{s\parcomp e \after \sigma'\cdot\ell} = (X_s,X_e)$\\
                \proofstep{\Cref{def:after}: $\after$}\\
                $\collectTuple{\{(q_s,q_e) \setbar s \parcomp e \Trans{\sigma'\cdot\ell} q_s \parcomp q_e \}} = (X_s,X_e)$\\
                \proofstep{\Cref{item:trans_transitive}}\\
                $\collectTuple{\{(q_s,q_e) \setbar \exists q_s',\in Q_s, q_e'\in Q_e :s \parcomp e \Trans{\sigma'} q_s'\parcomp q_e'\land  q_s'\parcomp q_e' \Trans{\ell} q_s \parcomp q_e \}} = (X_s,X_e)$\\
                \proofstep{Apply IH}\\
                $\collectTuple{s\parcomp e \after \sigma'} = (X_s',X_e')\land(X_s',X_e')\in \uiocorel\;\land$\\
                $\collectTuple{\{(q_s,q_e) \setbar \exists q_s',\in Q_s, q_e'\in Q_e :s \parcomp e \Trans{\sigma'} q_s'\parcomp q_e'\land  q_s'\parcomp q_e' \Trans{\ell} q_s \parcomp q_e \}} = (X_s,X_e)$\\
                \proofstep{\Cref{lem:tupple_trace_reachable_uioco_rel})}\\
                $\collectTuple{s\parcomp e \after \sigma'} = (X_s',X_e')\land(X_s',X_e')\in \uiocorel\;\land$\\
                $\collectTuple{\{(q_s,q_e) \setbar \exists q_s'\in X_s', q_e' \in X_e':  q_s'\parcomp q_e' \Trans{\ell} q_s \parcomp q_e \}} = (X_s,X_e)$\\
                \proofstep{\Cref{lem:project_from_parcomp_no_delta}}\\
                $\collectTuple{s\parcomp e \after \sigma'} = (X_s',X_e')\land(X_s',X_e')\in \uiocorel\;\land$\\
                $\collectTuple{\{(q_s,q_e) \setbar \exists q_s'\in X_s', q_e' \in X_e':  q_s' \Trans{\ell} q_s  \land q_e'\Trans{\ell} q_e \}} = (X_s,X_e)$\\
                \proofstep{\Cref{def:outset}: $\outset{X_s'}$ and $\sigma'\cdot\ell\in\utraces{s\parcomp e}$}\\
                $\collectTuple{s\parcomp e \after \sigma'} = (X_s',X_e')\land(X_s',X_e')\in \uiocorel \land \ell\in\outset{X_s'}\;\land$\\
                $\collectTuple{\{(q_s,q_e) \setbar \exists q_s'\in X_s', q_e' \in X_e':  q_s' \Trans{\ell} q_s  \land q_e'\Trans{\ell} q_e \}} = (X_s,X_e)$\\
                \proofstep{\Cref{def:after}: $\after$}\\
                $\collectTuple{s\parcomp e \after \sigma'} = (X_s',X_e')\land(X_s',X_e')\in \uiocorel\land \ell\in\outset{X_s'}\;\land$\\
                $\collectTuple{\{(q_s,q_e) \setbar q_s \in X_s'\after\ell \land q_e\in X_e'\after\ell \}} = (X_s,X_e)$\\
                \proofstep{\Cref{def:collect_tuple,def:LTS,def:uioco_rel}: $\collectTupleSymb$ and $X_s'\after\ell\neq \emptyset \land X_e' \after \ell \neq \emptyset$}\\
                $\collectTuple{s\parcomp e \after \sigma'} = (X_s',X_e')\land(X_s',X_e')\in \uiocorel\land \ell\in\outset{X_s'}\;\land$\\
                $ X_s =  X_s'\after\ell \land X_e= X_e'\after\ell $\\
                \proofstep{\Itemref{def:uioco_rel}{sync_out_imp}: $\uiocorel$}\\
                $(X_s,X_e)\in\uiocorel$
            \item[$\sigma = \sigma'\cdot\ell$ ($\ell\in I_s\cap U_e$):] Symmetric to previous case\\
            \item[$\sigma = \sigma'\cdot\ell$ ($\ell\in I_s\cap I_e$):]\ \\ 
                $\collectTuple{s\parcomp e \after \sigma'\cdot\ell} = (X_s,X_e)$\\
                \proofstep{\Cref{def:after}: $\after$}\\
                $\collectTuple{\{(q_s,q_e) \setbar s \parcomp e \Trans{\sigma'\cdot\ell} q_s \parcomp q_e \}} = (X_s,X_e)$\\
                \proofstep{\Cref{item:trans_transitive}}\\
                $\collectTuple{\{(q_s,q_e) \setbar \exists q_s',\in Q_s, q_e'\in Q_e :s \parcomp e \Trans{\sigma'} q_s'\parcomp q_e'\land  q_s'\parcomp q_e' \Trans{\ell} q_s \parcomp q_e \}} = (X_s,X_e)$\\
                \proofstep{Apply IH}\\
                $\collectTuple{s\parcomp e \after \sigma'} = (X_s',X_e')\land(X_s',X_e')\in \uiocorel\;\land$\\
                $\collectTuple{\{(q_s,q_e) \setbar \exists q_s',\in Q_s, q_e'\in Q_e :s \parcomp e \Trans{\sigma'} q_s'\parcomp q_e'\land  q_s'\parcomp q_e' \Trans{\ell} q_s \parcomp q_e \}} = (X_s,X_e)$\\
                \proofstep{\Cref{lem:tupple_trace_reachable_uioco_rel})}\\
                $\collectTuple{s\parcomp e \after \sigma'} = (X_s',X_e')\land(X_s',X_e')\in \uiocorel\;\land$\\
                $\collectTuple{\{(q_s,q_e) \setbar \exists q_s'\in X_s', q_e' \in X_e':  q_s'\parcomp q_e' \Trans{\ell} q_s \parcomp q_e \}} = (X_s,X_e)$\\
                \proofstep{\Cref{lem:project_from_parcomp_no_delta}}\\
                $\collectTuple{s\parcomp e \after \sigma'} = (X_s',X_e')\land(X_s',X_e')\in \uiocorel\;\land$\\
                $\collectTuple{\{(q_s,q_e) \setbar \exists q_s'\in X_s', q_e' \in X_e':  q_s' \Trans{\ell} q_s  \land q_e'\Trans{\ell} q_e \}} = (X_s,X_e)$\\
                \proofstep{\Cref{lem:trans_complete_uioco_rel} and $\sigma'\cdot\ell\in\utraces{s\parcomp e}$}\\
                $\collectTuple{s\parcomp e \after \sigma'} = (X_s',X_e')\land(X_s',X_e')\in \uiocorel\; \land$\\
                $\ell\in\inset{X_s'}\land \ell \in \inset{X_e'}\;\land$\\
                $\collectTuple{\{(q_s,q_e) \setbar \exists q_s'\in X_s', q_e' \in X_e':  q_s' \Trans{\ell} q_s  \land q_e'\Trans{\ell} q_e \}} = (X_s,X_e)$\\
                \proofstep{\Cref{def:after}: $\after$}\\
                $\collectTuple{s\parcomp e \after \sigma'} = (X_s',X_e')\land(X_s',X_e')\in \uiocorel\;\land$\\
                $\ell\in\inset{X_s'}\land \ell \in \inset{X_e'}\;\land$\\
                $\collectTuple{\{(q_s,q_e) \setbar q_s \in X_s'\after\ell \land q_e\in X_e'\after\ell \}} = (X_s,X_e)$\\
                \proofstep{\Cref{def:collect_tuple,def:LTS,def:uioco_rel}: $\collectTupleSymb$ and $X_s'\after\ell\neq \emptyset \land X_e' \after \ell \neq \emptyset$}\\
                $\collectTuple{s\parcomp e \after \sigma'} = (X_s',X_e')\land(X_s',X_e')\in \uiocorel\;\land$\\
                $\ell\in\inset{X_s'}\land \ell \in \inset{X_e'}\;\land$\\
                $ X_s =  X_s'\after\ell \land X_e= X_e'\after\ell $\\
                \proofstep{\Itemref{def:uioco_rel}{sync_in}: $\uiocorel$}\\
                $(X_s,X_e)\in\uiocorel$
            \item[$\sigma = \sigma'\cdot\ell$ ($\ell\in U_s\cap U_e$):] Case not possible (\cref{def:composable})\\
            \item[$\sigma = \sigma'\cdot\delta$:]\ \\
                $\sigma'\cdot \delta \in \utraces{s\parcomp e}\;\land$\\
                $\collectTuple{s\parcomp e \after \sigma'\cdot\delta} = (X_s,X_e)$\\
                \proofstep{\Cref{def:after}: $\after$}\\
                $\sigma'\cdot \delta \in \utraces{s\parcomp e}\;\land$\\
                $\collectTuple{\{(q_s,q_e) \setbar s \parcomp e \Trans{\sigma'\cdot\delta} q_s \parcomp q_e \}} = (X_s,X_e)$\\
                \proofstep{\Cref{lem:project_from_parcomp}}\\
                $\sigma'\cdot \delta \in \utraces{s\parcomp e}\;\land$\\
                $\collectTuple{\{(q_s,q_e) \setbar s \Trans{\project{\sigma'\cdot\delta}{L_s^\delta}} q_s \land e \Trans{\project{\sigma'\cdot\delta}{L_e^\delta}} q_e\}} = (X_s,X_e)$\\
                \proofstep{\Cref{item:trans_transitive} and \cref{def:projection}: $\projectop$}\\
                $\sigma'\cdot \delta \in \utraces{s\parcomp e}\;\land$\\
                $\collectTupleSymb(\{(q_s,q_e) \setbar \exists q_s',\in Q_s, q_e'\in Q_e :s \Trans{\project{\sigma'}{L_s^\delta}} q_s' \land q_s'\Trans{\delta} q_s\;\land$\\
                \tab$e \Trans{\project{\sigma'}{L_e^\delta}} q_e'\land q_e'\Trans{\delta} q_e\}) = (X_s,X_e)$\\
                \proofstep{\Cref{lem:project_from_parcomp}}\\
                $\sigma'\cdot \delta \in \utraces{s\parcomp e}\;\land$\\
                $\collectTupleSymb(\{(q_s,q_e) \setbar \exists q_s',\in Q_s, q_e'\in Q_e :s \parcomp e \Trans{\sigma'} q_s' \parcomp q_e' \land q_s'\Trans{\delta} q_s \;\land$\\
                \tab$ q_e'\Trans{\delta} q_e\}) = (X_s,X_e)$\\
                \proofstep{Apply IH}\\
                $\sigma'\cdot \delta \in \utraces{s\parcomp e}\;\land$\\
                $\collectTuple{s\parcomp e \after \sigma'} = (X_s',X_e')\land(X_s',X_e')\in \uiocorel\;\land$\\
                $\collectTupleSymb(\{(q_s,q_e) \setbar \exists q_s',\in Q_s, q_e'\in Q_e :s \parcomp e \Trans{\sigma'} q_s' \parcomp q_e' \land q_s'\Trans{\delta} q_s \;\land$\\
                \tab$ q_e'\Trans{\delta} q_e\}) = (X_s,X_e)$\\
                \proofstep{\Cref{lem:tupple_trace_reachable_uioco_rel})}\\
                $\sigma'\cdot \delta \in \utraces{s\parcomp e}\;\land$\\
                $\collectTuple{s\parcomp e \after \sigma'} = (X_s',X_e')\land(X_s',X_e')\in \uiocorel\;\land$\\
                $\collectTuple{\{(q_s,q_e) \setbar \exists q_s'\in X_s', q_e' \in X_e':   q_s'\Trans{\delta} q_s \land q_e'\Trans{\delta} q_e \}} = (X_s,X_e)$\\
                \proofstep{\Cref{def:uioco}: $\utraces{}$}\\
                $s \parcomp e \Trans{\sigma'\cdot\delta}\;\land$\\
                $\collectTuple{s\parcomp e \after \sigma'} = (X_s',X_e')\land(X_s',X_e')\in \uiocorel\;\land$\\
                $\collectTuple{\{(q_s,q_e) \setbar \exists q_s'\in X_s', q_e' \in X_e':   q_s'\Trans{\delta} q_s \land q_e'\Trans{\delta} q_e \}} = (X_s,X_e)$\\
                \proofstep{\Cref{item:trans_transitive} and \cref{def:projection}: $\projectop$}\\
                $\exists q_s''\in Q_s, q_e''\in Q_e: s \parcomp e \Trans{\sigma'} q_s'' \parcomp q_e'' \land q_s'' \parcomp q_e'' \Trans{\delta} \;\land$\\
                $\collectTuple{s\parcomp e \after \sigma'} = (X_s',X_e')\land(X_s',X_e')\in \uiocorel\;\land$\\
                $\collectTuple{\{(q_s,q_e) \setbar \exists q_s'\in X_s', q_e' \in X_e':   q_s'\Trans{\delta} q_s \land q_e'\Trans{\delta} q_e \}} = (X_s,X_e)$\\
                \proofstep{\Cref{lem:tupple_trace_reachable_uioco_rel}}\\
                $\exists q_s''\in X_s', q_e''\in X_e': q_s'' \parcomp q_e'' \Trans{\delta} \;\land$\\
                $\collectTuple{s\parcomp e \after \sigma'} = (X_s',X_e')\land(X_s',X_e')\in \uiocorel\;\land$\\
                $\collectTuple{\{(q_s,q_e) \setbar \exists q_s'\in X_s', q_e' \in X_e':   q_s'\Trans{\delta} q_s \land q_e'\Trans{\delta} q_e \}} = (X_s,X_e)$\\
                \proofstep{\Cref{lem:project_from_parcomp_light} and \cref{def:projection}: $\projectop$}\\
                $\exists q_s''\in X_s', q_e''\in X_e': q_s'' \Trans{\delta} \land q_e''\Trans{\delta}\;\land$\\
                $\collectTuple{s\parcomp e \after \sigma'} = (X_s',X_e')\land(X_s',X_e')\in \uiocorel\;\land$\\
                $\collectTuple{\{(q_s,q_e) \setbar \exists q_s'\in X_s', q_e' \in X_e':   q_s'\Trans{\delta} q_s \land q_e'\Trans{\delta} q_e \}} = (X_s,X_e)$\\
                \proofstep{\Cref{def:outset}: $\outset{X_s'}$ and $\outset{X_e'}$ }\\
                $\collectTuple{s\parcomp e \after \sigma'} = (X_s',X_e')\land(X_s',X_e')\in \uiocorel\;\land$\\
                $\collectTuple{\{(q_s,q_e) \setbar \exists q_s'\in X_s', q_e' \in X_e':  q_s' \Trans{\delta} q_s  \land q_e'\Trans{\delta} q_e \}} = (X_s,X_e)\;\land$\\
                $\delta\in \outset{X_e'}\land \delta\in \outset{X_s'}$\\
                \proofstep{\Cref{def:after}: $\after$}\\
                $\collectTuple{s\parcomp e \after \sigma'} = (X_s',X_e')\land(X_s',X_e')\in \uiocorel\;\land$\\
                $\collectTuple{\{(q_s,q_e) \setbar q_s \in X_s'\after\delta \land q_e\in X_e'\after\delta \}} = (X_s,X_e)\;\land$\\
                $\delta\in \outset{X_e'}\land \delta\in \outset{X_s'}$\\
                \proofstep{\Cref{def:collect_tuple,def:LTS,def:uioco_rel}: $\collectTupleSymb$ and $X_s'\after\delta\neq \emptyset \land X_e' \after \delta \neq \emptyset$}\\
                $(X_s',X_e')\in \uiocorel\;\land$\\
                $ X_s =  X_s'\after\delta \land X_e= X_e'\after\delta \land \delta\in \outset{X_e'}\land \delta\in \outset{X_s'}$\\
                \proofstep{\Itemref{def:uioco_rel}{delta}: $\uiocorel$}\\
                $(X_s,X_e)\in\uiocorel$
            
        \end{case_distinction}
    \end{case_distinction}
    \end{proof}
\end{toappendix}

\begin{lemmarep}
    \label{lem:uioco_rel_utraces_bi-implication}
    Let $s\mutuallyaccepts{} e$, $q_e\in Q_e,q_s\in Q_s$.
    \[\exists (X_s,X_e)\in \uiocorel: q_s \in X_s \land q_e \in X_e \implies \exists \sigma \in \utraces{s\parcomp e}: s\parcomp e \Trans{\sigma} q_s \parcomp q_e\]
\end{lemmarep}
\begin{proof}
\ \\
    $\exists (X_s, X_e) \in \uiocorel: q_s\in X_s \land q_e\in X_e$\\
\proofstep{\Cref{lem:uioco_rel_iff_tuple_utraces}}\\
$\exists (X_s, X_e) \in \uiocorel: q_s\in X_s \land q_e\in X_e\;\land$\\
$\exists \sigma \in \utraces{s\parcomp e}: \collectTuple{s\parcomp e \after \sigma} = (X_s,X_e)$\\
\proofstep{\Cref{def:collect_tuple,def:after}: $\collectTupleSymb$ and $\after$}\\
$\exists (X_s, X_e) \in \uiocorel: q_s\in X_s \land q_e\in X_e\;\land$\\
$\exists \sigma \in \utraces{s\parcomp e}: X_s = \{q_s'\setbar s\parcomp e \Trans{\sigma} q_s'\parcomp q_e'\} \land X_e = \{q_e'\setbar s\parcomp e \Trans{\sigma} q_s'\parcomp q_e'\}$\\
\proofstep{\Cref{lem:project_from_parcomp}}\\
$\exists (X_s, X_e) \in \uiocorel: q_s\in X_s \land q_e\in X_e\;\land$\\
$\exists \sigma \in \utraces{s\parcomp e}: X_s = \{q_s'\setbar s \Trans{\project{\sigma}{L_s^\delta}} q_s'\} \land X_e = \{q_e'\setbar e \Trans{\project{\sigma}{L_e^\delta}} q_e'\}$\\
\proofstep{$q_s\in X_s$ and $q_e\in X_e$}\\
$\exists \sigma \in \utraces{s\parcomp e}: s \Trans{\project{\sigma}{L_s^\delta}} q_s \land e \Trans{\project{\sigma}{L_e^\delta}} q_e$\\
\proofstep{\Cref{lem:project_from_parcomp_light}}\\
$\exists \sigma \in \utraces{s\parcomp e}: s \parcomp e \Trans{\sigma} q_s \parcomp q_e $\\
\end{proof}

\begin{lemmarep}
    \label{lem:uiocosim_implies_ecosim}
    Let $s,e \in\LTS$ be composable, then
    \[s\mutuallyaccepts{} e \implies \uiocorel \text{ is an }\acceptsim\]
\end{lemmarep}

\begin{proof}
   Proof by induction on the definition of $\uiocosim$ (\Cref{def:uioco_rel}). Most cases are identical between the definition of $\acceptsim$ and $\uiocosim$. The two cases that are not identical (respectively \itemref{def:eco_rel}{sync_out_env} and \itemref{def:eco_rel}{sync_out_imp}, and \itemref{def:uioco_rel}{sync_out_env} and \itemref{def:uioco_rel}{sync_out_imp}) are symmetrical. We therefore only give the proof here for case \itemref{def:eco_rel}{sync_out_env}: Take $(X_s,X_e)\in\uiocorel$, $\ell\in\outset{X_e}\cap I_s$. To prove: $\ell\in \inset{X_s}$.\\\\
 
\noindent
$s\mutuallyaccepts{} e \land (X_s,X_e)\in\uiocorel \land \ell\in \outset{X_e}\cap I_s$\\
\proofstep{\Cref{lem:uioco_rel_iff_tuple_utraces}}\\
$s\mutuallyaccepts{} e \land (X_s,X_e)\in\uiocorel \land \ell\in \outset{X_e}\cap I_s\;\land$\\
$\exists \sigma \in \utraces{s\parcomp e}: \collectTuple{s\parcomp e \after \sigma} = (X_s,X_e)$\\
\proofstep{\Cref{def:out}: $\outset{X_e}$}\\
$s\mutuallyaccepts{} e \land (X_s,X_e)\in\uiocorel \land \ell\in \outset{X_e}\cap I_s\;\land$\\
$\exists \sigma \in \utraces{s\parcomp e}, q_e\in X_e: \collectTuple{s\parcomp e \after \sigma} = (X_s,X_e) \land q_e\trans{\ell}$\\
\proofstep{$\sigma\in\utraces{s\parcomp e}$}\\
$s\mutuallyaccepts{} e \land (X_s,X_e)\in\uiocorel \land \ell\in \outset{X_e}\cap I_s\;\land$\\
$\exists \sigma \in \utraces{s\parcomp e}, q_s\in X_s, q_e \in X_e: \collectTuple{s\parcomp e \after \sigma} = (X_s,X_e)\land q_e\trans{\ell}$\\
\proofstep{\Cref{lem:uioco_rel_utraces_bi-implication}}\\
$s\mutuallyaccepts{} e \land \ell\in \outset{X_e}\cap I_s\;\land$\\
$\exists q_e \in X_e: q_e\trans{\ell}\;\land$\\
$\forall q_s\in X_s, \exists \sigma \in \utraces{s\parcomp e}:  s\parcomp e \Trans{\sigma} q_s \parcomp q_e$\\
\proofstep{\Cref{def:mutually_accepts}: $\mutuallyaccepts{}$}\\
$\ell\in \outset{X_e}\cap I_s\;\land$\\
$\exists q_e \in X_e: q_e\trans{\ell}\;\land$\\
$\forall q_s\in X_s: \outset{q_e}\cap I_s \:\subseteq\: \inset{q_s} \cap U_e$\\
\proofstep{definition $\subseteq$ and \Cref{def:inset,def:out}: $\inset{}$ and  $\outset{}$}\\
$\forall q_s\in X_s: \ell\in\inset{q_s}$\\
\proofstep{\Cref{def:inset}: $\inset{}$}\\
$\ell\in\inset{X_s}$
\end{proof}

\begin{lemmarep}
    \label{lem:mutaccepts_implies_eco}
    Let $s,e\in \LTS$ be $\composable$, then
    \[s \aco e \impliedby s \mutuallyaccepts{} e\]
\end{lemmarep}

\begin{proof}
    Proof follows directly from \Cref{lem:uiocosim_implies_ecosim}
\end{proof}

\subsection{Algorithm}
\label{sec:eco_modelcheck_alg}
The previous sections described an alternative characterization for $\mutuallyaccepts{}$ ($\aco$), and proved this characterization equal to the original in \Cref{def:mutually_accepts}. In this section we will make use of this by giving an algorithm for checking $\mutuallyaccepts{}$ between two specifications.

\begin{function}
\functionLabel{DecideEco}
\caption{\currentFunName{} (\ProcArgFnt$\Parens{X_s,X_e}:\powerset{Q_s}\times \powerset{Q_e}$)}
\KwData{$\uiocorel: \settype{(\powerset{Q_s}, \powerset{Q_e})}$}
    \eIf{%
        \label{line:main_check}%
        $\outset{X_s} \cap I_e \subseteq \inset{X_e} \land \outset{X_e} \cap I_s \subseteq \inset{X_s}$}
        {\Assign{$\uiocorel$}{$\uiocorel \cup \{(X_s,X_e)\}$}
        \Assign{$\mathcal{A}$}{$(\outset{X_s}\cap I_e)\cup (\outset{X_e}\cap I_s)\cup (\inset{X_s}\cap \inset{X_e})$}\label{line:decideeco_synclabels}
        \If{$\delta \in \outset{X_e}\land \delta \in \outset{X_s}$}
            {\Assign{$\mathcal{A}$}{$\mathcal{A}\cup \{\delta\}$}}\label{line:decideeco_delta}
        \ForEach{$\ell \in \mathcal{A}$}
            {\Assign{$next$}{$(X_s \after \ell,X_e \after \ell)$}
            \If{$next\notin \uiocorel$}
                {\DecideEco{$next$}}
            }\label{line:decideeco_syncloop}
        \ForEach{$\ell \in (\outset{X_s}\cup \inset{X_s})\setminus L_e^\delta$}
            {\Assign{$next$}{$(X_s \after \ell,X_e)$}\label{line:decideeco_internal_imp}
            \If{$next\notin \uiocorel$}
                {\DecideEco{$next$}}
            }
        \ForEach{$\ell \in (\outset{X_e}\cup \inset{X_e})\setminus L_s^\delta$}
            {\Assign{$next$}{$(X_s,X_e \after \ell)$}\label{line:decideeco_internal_env}
            \If{$next\notin \uiocorel$}
                {\DecideEco{$next$}}
            }
        }
        {\Throw{$s \noteco e$}}
\end{function}

\begin{algorithm}
    \caption{Deciding eco}
    \label{alg:decideEco}
    \KwIn{$s,e: \LTS$ with finite number of states and labels}
    \Assign{$\uiocorel$}{$\emptyset$}
    \DecideEco{$(s_0 \after \epsilon, e_0 \after \epsilon)$} \label{line:dececo_base}
\end{algorithm}

Because \aco{} comes down to the existence of an $\acceptsim$, the algorithm shown in \Cref{alg:decideEco} and \DecideEco works by constructing the concrete relation $R_U$, while checking that this relation is indeed an \acceptsim{}. All the points in \Cref{def:eco_rel} are handled explicitly by \DecideEco: Points \ref{item:eco_rel_sync_out_env}, \ref{item:eco_rel_sync_out_imp} and \ref{item:eco_rel_sync_in} at \cref{line:decideeco_synclabels}, Point \ref{item:eco_rel_delta} at \cref{line:decideeco_delta}, and Point \ref{item:eco_rel_internal_env} and \ref{item:eco_rel_internal_imp} at \cref{line:decideeco_internal_imp,line:decideeco_internal_env}, respectively. The initialization of Point \ref{item:eco_rel_base} is handled 
in \cref{line:dececo_base} of \Cref{alg:decideEco}. The important parts are the two main restraints imposed by \Cref{def:eco_rel} in Points \ref{item:eco_rel_sync_out_imp} and \ref{item:eco_rel_sync_out_env}: communicated outputs have to be enabled in every possible receiving state. This is handled in \cref{line:main_check} of \DecideEco. The correctness of the algorithm follows from \Cref{lem:constructive_ecosim}. 

For simplicity \Cref{alg:decideEco} does not give a counter-example and stops at the first found problem. A counter-example could however be generated by also keeping track of the sequence of labels that was used to reach a certain set of states, at the cost of increased memory consumption.

Since constructing $\uiocorel$ means exploring the part of the state space that is reachable by $\utraces{}$, the algorithm will only terminate if this state space is finite. The number of enabled labels in a given state $q$: $\outset{q} \cup \inset{q}$, must also be finite. We will show how to get around these restrictions in the next section.

\section{Testing Mutual Acceptance}
\label{sec:testing-accepting-systems}

\Cref{alg:decideEco} can be used to verify $\mutuallyaccepts{}$, but due to the state space explosion problem this is not always feasible. We can however still use testing to get some idea on the conformance between an implementation and a specification of its environment.

In this section, we develop such an approach by building on the results obtained for $\mutuallyaccepts{}$ so far. We reuse the naming from \cite{frantzen_ModelBasedTestingEnvironmental_2007} and call this compositionality based conformance relation \emph{environmental conformance} ($\eco$). We use almost the same definition as $\mutuallyaccepts{}$, but restrict the scope to relate an implementation and a specification of its environment, instead of two specifications (\Cref{def:eco}). We use a black-box testing approach, where we assume the implementation is represented by an \textit{IOTS}, but we do not have access to it.
In \Cref{sec:eco_testing} we give an algorithm for this approach, and prove this algorithm correct in \Cref{sec:algorithm_correctness}.
Finally, in \Cref{sec:component_based_testing_eco} we show that because the $\eco$-based testing approach does not have access to the implementation model, it results in a weaker relation than $\mutuallyaccepts{}$. This means that using $\eco$ will allow more specifications than when using $\mutuallyaccepts{}$, but despite this $\eco$ is still strong enough to allow for compositional testing.

\begin{definition}
Let $i\in \IOTS$ and $e\in\LTS$ be composable.
\[i\eco e \defeq e \accepts{} i\]
\label{def:eco}
\end{definition}

\begin{corollary}
    Let $i\in \IOTS$ and $e\in\LTS$ be composable.
    \[i \emph{~\eco~} e \iff i \mutuallyaccepts{} e\]
\end{corollary}

\subsection{On-the-Fly Testing}
\label{sec:eco_testing}

\Cref{alg:test_eco} can be used to test whether an implementation \eco{}-conforms to the specification of its environment. Intuitively \Cref{alg:test_eco} is very similar to the algorithm used for testing $\uioco$-conformance shown in the preliminaries (\Cref{alg:test_uioco}) and consists of five possible options which repeat non-deterministically until the algorithm terminates:
\begin{enumerate}[label=(\Alph*):]
    \item We stop testing and the test passes.
    \item We give a valid input to the SUT, which is either an output from the environment, or a shared input with the environment.
    \item We can observe an output from the SUT, and check if the environment can handle this output.
    \item We can simulate non-interacting behaviour of the environment
    \item We can generate non-interacting inputs for the SUT. This can be optimised further if a specification of $s$ is available, by restricting to specified inputs instead of random ones, which we discuss in more detail in \Cref{sec:combining_algorithms}. 
\end{enumerate}

We need more cases here than in \Cref{alg:test_uioco}, because we treat synchronized labels differently from non-interacting labels, but otherwise follow the same structure. We claim the algorithm is sound and complete. This means that it is possible to generate a test that will fail, if and only if the SUT and its environment are not $\eco$-conformant. We will further formalise and prove this claim in \Cref{sec:algorithm_correctness}.

\LinesNotNumbered
\begin{algorithm}
\caption{On-the-Fly Testing for $\eco$}
\label{alg:test_eco}
    \KwIn{A SUT, on which we can perform inputs and observe outputs (and quiescence)}
    \KwIn{$e\in\LTS$; $I_s, U_s$: labels of SUT}
    \nl\Assign{$X_e$}{$e_0 \after \epsilon$}
    non-deterministically execute a finite number of the following cases,\\
    until the test either Passes or Fails:\\
    \textbf{\textit{(A) Stop testing:}}\\
    \Indp
        \nl the test Passes\;
    \Indm
    \textbf{\textit{(B) Perform an input $\ell$ on the SUT:}}\\
    \Indp
        \nl choose $\ell \in (\outset{X_e}\cup \inset{X_e})\cap I_s$\tcp*{Only send inputs allowed by e}    \label{line:test_eco_output_selection}
        \nl send $\ell$ to the SUT\;
        \nl\Assign{$X_e$}{$X_e \after \ell$}
    \Indm
    \textbf{\textit{(C) Observe an output or quiescence  $\ell\in U_s^\delta$ from the SUT:}}\\
    \algIndentp
    \nl\If{$\ell \in I_e$}
        {
            \nl\lIf{$\ell \notin \inset{X_e}$}
                {the test Fails}
            \Else
            {
                \nl\Assign{$X_e$}{$X_e \after \ell$}
            }
        }\ElseIf{$\ell=\delta\land \delta\in\outset{X_e}$}
        {
            \nl\Assign[]{$X_e$}{$X_e \after \ell$}\tcp*{Quiescence is only observable if e is also quiescent.}
        }
        \nl\Else{\tcp{do nothing}}
    \algIndentm
    \SetInd{\dimexpr\skiprule-\algoskipindent}{\dimexpr\skiptext+\algoskipindent}
    \textbf{\textit{(D) Simulate non-interacting behaviour in the environment}}\\
    \Indp
        \nl choose $\ell \in (\outset{X_e} \cup \inset{X_e})\setminus L_s^\delta$ \; 
        \nl \Assign{$X_e$}{$X_e \after \ell$}
    \Indm
    \textbf{\textit{(E) Generate non-interacting behaviour of the SUT}}\\
    \Indp
        \nl choose $\ell \in I_s \setminus L_e$ \; 
        \nl send $\ell$ to the SUT\;
    \Indm

\end{algorithm}
\LinesNumbered

\subsection{Component Based Testing}
\label{sec:component_based_testing_eco}
\FloatBarrier
\FloatBarrier
Using the algorithm introduced in the previous section, we can test for environmental conformance between an implementation and a specification of its environment ($i_s \eco e$). This however still leaves the question if it is also useful. Does black-box testing using an implementation give us enough information to draw the same conclusions as when the specifications were mutually accepting? Environmental conformance for an implementation $i_s$ and its environment specification $e$ does not imply mutual acceptance between the specification $s$ and its environment: $i_s \eco e \land i_s \uioco s \notimplies s \mutuallyaccepts{} e$. The reason for this is that $\uioco$ does not require that every output from the specification is also part of the implementation. This means it is possible that $s$ allows for communication errors to occur, while because of implementation choices none of these possible errors are present in $i_s$. Since we are usually more interested in the compositionality of the implementations than that of the specifications, testing for environmental conformance is still useful:
it gives enough information to guarantee correctness of compositional testing, which is formulated in \Cref{lem:test_eco_implies_uioco}, and visualised in \Cref{subfig:testing_aproach}. Each arrow from \Cref{subfig:testing_aproach} represents one of the conditions of \Cref{lem:test_eco_implies_uioco}. All arrows touch exactly two things, which could be specifications or implementations. If anything changes, the connected arrows need to be rerun, but the unconnected arrows stay the same and therefore can be skipped, as they can reuse the results from last run. We briefly sketch the proof for \Cref{lem:test_eco_implies_uioco} in \Cref{sec:test_eco_implies_uioco_proofsketch}.

The $\eco$ based testing approach is weaker than the one for $\mutuallyaccepts{}$ outlined in \Cref{sec:environmentalConformance}, because it allows more specifications: $\mutuallyaccepts{}$ would fail iff the given specifications allow a communication error to occur, while $\eco$ only fails if such an error occurs with the current implementation. In this way $\mutuallyaccepts{}$ states that all implementations of $s$ and $e$ are compositional, while $\eco$ only says something about the one we are currently testing.

\FloatBarrier

\begin{theoremrep}
    \label{lem:test_eco_implies_uioco}
      Let $s,e\in \LTS$ be $\composable$, $i_s, i_e \in \IOTS$, then
    \[i_s \eco{} e \land i_e \eco{} s \land i_s \uioco s \land i_e \uioco e \implies i_s \parcomp i_e \uioco s \parcomp e\]
\end{theoremrep}

\begin{proof}
    Take $\sigma \in  \utraces{s\parcomp e}, \ell \in \outset{i_s \parcomp i_e \after \sigma}$. To prove: $\ell \in \outset{s\parcomp e \after \sigma}$\\
We do a case distinction on $\ell$, splitting $U_{s \parcomp e}^\delta$ into\\
$(U_s \setminus L_e)$, $(U_e \setminus L_s)$, $(U_s \cap I_e)$, $(U_e \cap I_s)$, and $\{\delta\}$. Note that $U_s \cap U_e$ is missing because it is empty (\cref{def:composable}).\\
Since $i_s \uioco s$ implies $L_{i_s} = L_s$ (\cref{def:uioco}), this proof will only use $L_s$ and $L_e$ to increase readability.\\ 

\begin{case_distinction}
\item[$\ell\in (U_s \setminus L_e)$:]\ \\
    $i_s \eco{} e\land i_e \eco{} s \land i_s \uioco s \land i_e \uioco e \land \sigma \in  \utraces{s\parcomp e}\;\land$\\
    $ \ell \in \outset{i_s \parcomp i_e \after \sigma}$\\
    \proofstep{\Cref{def:outset,def:after}: $out$ and $\after$}\\
    $i_s \eco{} e\land i_e \eco{} s \land i_s \uioco s \land i_e \uioco e \land \sigma \in  \utraces{s\parcomp e}\;\land$\\
    $i_s \parcomp i_e \Trans{\sigma\cdot \ell}$\\
    \proofstep{\Cref{item:trans_transitive}}\\
    $i_s \eco{} e\land i_e \eco{} s \land i_s \uioco s \land i_e \uioco e \land \sigma \in  \utraces{s\parcomp e}\;\land$\\
    $i_s \parcomp i_e \Trans{\sigma\cdot \ell}\land\;i_s \parcomp i_e \Trans{\sigma}$\\
    \proofstep{\Cref{lem:test_project_utraces}}\\
    $i_s \uioco s \land \project{\sigma}{L_s^\delta} \in  \utraces{s}\land \project{\sigma}{L_e^\delta} \in  \utraces{e}\;\land$\\
    $i_s \parcomp i_e \Trans{\sigma\cdot \ell}$\\
    \proofstep{\Cref{lem:project_from_parcomp_IOTS}}\\
    $i_s \uioco s\land \project{\sigma}{L_s^\delta} \in  \utraces{s}\land \project{\sigma}{L_e^\delta} \in  \utraces{e}\;\land$\\
    $i_s \Trans{\project{\sigma\cdot \ell}{L_s^\delta}} $\\
    \proofstep{\Cref{def:projection}: $\projectop$}\\
    $i_s \uioco s \land \project{\sigma}{L_s^\delta} \in  \utraces{s}\land \project{\sigma}{L_e^\delta} \in  \utraces{e}\;\land$\\
    $i_s \Trans{\project{\sigma}{L_s^\delta}\cdot \ell}$\\
    \proofstep{\Cref{def:uioco}: $\utraces{}$}\\
    $i_s \uioco s \land \project{\sigma}{L_s^\delta} \in  \utraces{s}\;\land$\\
    $i_s \Trans{\project{\sigma}{L_s^\delta}\cdot \ell} \land\; e \Trans{\project{\sigma}{L_e^\delta}}$\\
    \proofstep{\Cref{def:uioco}: $\uioco{}$}\\
    $s \Trans{\project{\sigma}{L_s^\delta}\cdot \ell}\land\; e \Trans{\project{\sigma}{L_e^\delta}}$\\
    \proofstep{\Cref{def:projection}: $\projectop$}\\
    $s \Trans{\project{\sigma\cdot\ell}{L_s^\delta}}\land\; e \Trans{\project{\sigma\cdot\ell}{L_e^\delta}}$\\
    \proofstep{\Cref{lem:project_from_parcomp_light}}\\
    $s \parcomp e \Trans{\sigma\cdot\ell}$\\
    \proofstep{\Cref{def:arrowdefs,def:outset}: $\Trans{}$ and $\outset{}$}\\
    $\ell \in \outset{s\parcomp e \after \sigma}$\\
\item[$\ell\in (U_e \setminus L_s)$:]Symmetrical to previous case
\item[$\ell\in (U_s \cap I_e)$:]\ \\
    $i_s \eco{} e\land i_e \eco{} s \land i_s \uioco s \land i_e \uioco e \land \sigma \in  \utraces{s\parcomp e}\;\land$\\
    $ \ell \in \outset{i_s \parcomp i_e \after \sigma}$\\
    \proofstep{\Cref{def:outset,def:after}: $out$ and $\after$}\\
    $i_s \eco{} e\land i_e \eco{} s \land i_s \uioco s \land i_e \uioco e \land \sigma \in  \utraces{s\parcomp e}\;\land$\\
    $i_s \parcomp i_e \Trans{\sigma\cdot \ell}$\\
    \proofstep{\Cref{item:trans_transitive}}\\
    $i_s \eco{} e\land i_e \eco{} s \land i_s \uioco s \land i_e \uioco e \land \sigma \in  \utraces{s\parcomp e}\;\land$\\
    $i_s \parcomp i_e \Trans{\sigma\cdot \ell}\land\;i_s \parcomp i_e \Trans{\sigma}$\\
    \proofstep{\Cref{lem:test_project_utraces}}\\
    $i_s \eco{} e\land i_s \uioco s \land \project{\sigma}{L_s^\delta} \in  \utraces{s}\land \project{\sigma}{L_e^\delta} \in  \utraces{e}\;\land$\\
    $i_s \parcomp i_e \Trans{\sigma\cdot \ell}$\\
    \proofstep{\Cref{lem:project_from_parcomp_IOTS}}\\
    $i_s \eco{} e\land i_s \uioco s \land \project{\sigma}{L_s^\delta} \in  \utraces{s}\land \project{\sigma}{L_e^\delta} \in  \utraces{e}\;\land$\\
    $i_s \Trans{\project{\sigma\cdot \ell}{L_s^\delta}}$\\
    \proofstep{\Cref{lem:straces_equals_utraces_iots}}\\
    $i_s \eco{} e\land i_s \uioco s \land \project{\sigma}{L_s^\delta} \in  \utraces{s}\land \project{\sigma}{L_e^\delta} \in  \utraces{e}\;\land$\\
    $i_s \Trans{\project{\sigma\cdot \ell}{L_s^\delta}}\land \project{\sigma\cdot \ell}{L_s^\delta}\in \utraces{i_s} $\\
    \proofstep{\Cref{def:uioco}: $\utraces{}$}\\
    $i_s \eco{} e\land i_s \uioco s \land \project{\sigma}{L_s^\delta} \in  \utraces{s}\land \project{\sigma}{L_e^\delta} \in  \utraces{e}\;\land$\\
    $i_s \Trans{\project{\sigma\cdot \ell}{L_s^\delta}}\land \project{\sigma\cdot \ell}{L_s^\delta}\in \utraces{i_s} \land e \Trans{\project{\sigma}{L_e^\delta}}$\\
    \proofstep{\Cref{lem:combine_utrace_imp}}\\
    $i_s \eco{} e\land i_s \uioco s \land \project{\sigma}{L_s^\delta} \in  \utraces{s}\land \sigma \in  \utraces{s_i\parcomp e}\;\land$\\
    $i_s \Trans{\project{\sigma\cdot \ell}{L_s^\delta}}\land\; e \Trans{\project{\sigma}{L_e^\delta}}$\\
    \proofstep{\Cref{def:projection}: $\projectop$}\\
    $i_s \eco{} e\land i_s \uioco s \land \project{\sigma}{L_s^\delta} \in  \utraces{s}\land \sigma \in  \utraces{s_i\parcomp e}\;\land$\\
    $i_s \Trans{\project{\sigma}{L_{i_s}^\delta}\cdot \ell}\land\; e \Trans{\project{\sigma}{L_e^\delta}}$\\
    \proofstep{\Cref{def:uioco}: $\uioco{}$}\\
    $i_s \eco{} e \land \sigma \in  \utraces{s_i\parcomp e} \;\land$\\
    $i_s \Trans{\project{\sigma}{L_{i_s}^\delta}\cdot \ell}\land\; s \Trans{\project{\sigma}{L_{i_s}^\delta}\cdot \ell}\land\; e \Trans{\project{\sigma}{L_e^\delta}}$\\
    \proofstep{\Cref{def:mutually_accepts,def:accepting,lem:eco_equals_mutaccepts}: $\mutuallyaccepts{}$ + ($\mutuallyaccepts{} \iff \eco{}$)}\\
    $i_s \Trans{\project{\sigma}{L_{i_s}^\delta}\cdot \ell}\land\; s \Trans{\project{\sigma}{L_{i_s}^\delta}\cdot \ell}\land\; e \Trans{\project{\sigma}{L_e^\delta}}\;\land$\\
    $\forall q_{i_s} \in Q_{i_s}, q_e \in Q_e: i_s \parcomp e \Trans{\sigma} q_s \parcomp q_e \implies \outset{q_{i_s}}\cap I_e \subseteq \inset{q_e} \cap U_s$\\
    \proofstep{\Cref{lem:project_from_parcomp_light}}\\
    $i_s \Trans{\project{\sigma}{L_{i_s}^\delta}\cdot \ell}\land\; s \Trans{\project{\sigma}{L_{i_s}^\delta}\cdot \ell}\land\; e \Trans{\project{\sigma}{L_e^\delta}}\;\land$\\
    $\forall q_{i_s} \in Q_{i_s}, q_e \in Q_e:$\\
    \tab$i_s \Trans{\project{\sigma}{L_{i_s^\delta}}} q_{i_s} \land e \Trans{\project{\sigma'}{L_e^\delta}} q_e \implies \outset{q_{i_s}}\cap I_e \subseteq \inset{q_e} \cap U_s$\\
    \proofstep{\Cref{def:arrowdefs,def:outset}: $\Trans{}$ and $\outset{}$}\\
    $\exists q_{i_s}' \in Q_{i_s}, q_e'\in Q_e: i_s \Trans{\project{\sigma}{L_{i_s}^\delta}} q_{i_s}' \land \ell\in \outset{q_{i_s}'} \land s \Trans{\project{\sigma}{L_{i_s}^\delta}\cdot \ell}\land\; e \Trans{\project{\sigma}{L_e^\delta}} q_e'\;\land$\\
    $\forall q_{i_s} \in Q_{i_s}, q_e \in Q_e:$\\
    \tab$i_s \Trans{\project{\sigma}{L_{i_s^\delta}}} q_{i_s} \land e \Trans{\project{\sigma'}{L_e^\delta}} q_e \implies \outset{q_{i_s}}\cap I_e \subseteq \inset{q_e} \cap U_s$\\
    \proofstep{$\forall$ elimination}\\
    $\exists q_{i_s}' \in Q_{i_s}, q_e'\in Q_e: i_s \Trans{\project{\sigma}{L_{i_s}^\delta}} q_{i_s}' \land \ell\in \outset{q_{i_s}'} \land s \Trans{\project{\sigma}{L_{i_s}^\delta}\cdot \ell}\land\; e \Trans{\project{\sigma}{L_e^\delta}} q_e'\;\land$\\
    $i_s \Trans{\project{\sigma}{L_{i_s^\delta}}} q_{i_s}' \land e \Trans{\project{\sigma'}{L_e^\delta}} q_e' \implies \outset{q_{i_s'}}\cap I_e \subseteq \inset{q_e'} \cap U_s$\\
    \proofstep{$A\land B\land (A\land B \implies C) \implies c$}\\
    $\exists q_{i_s}' \in Q_{i_s}, q_e'\in Q_e:  \ell\in \outset{q_{i_s}'} \land s \Trans{\project{\sigma}{L_{i_s}^\delta}\cdot \ell}\land\; e \Trans{\project{\sigma}{L_e^\delta}} q_e'\;\land$\\
    $\outset{q_{i_s'}}\cap I_e \subseteq \inset{q_e'} \cap U_s$\\
    \proofstep{definition $\subseteq$}\\
    $\exists q_e'\in Q_e:  \ell\in \inset{q_e'} \land s \Trans{\project{\sigma}{L_{i_s}^\delta}\cdot \ell}\land\; e \Trans{\project{\sigma}{L_e^\delta}} q_e'$\\
    \proofstep{\Cref{def:inset,def:arrowdefs}: $\inset{}$ and $\Trans{}$}\\
    $s \Trans{\project{\sigma}{L_{i_s}^\delta}\cdot \ell}\land\; e \Trans{\project{\sigma}{L_e^\delta}\cdot\ell}$\\
    \proofstep{\Cref{def:projection}: $\projectop$}\\
    $s \Trans{\project{\sigma\cdot \ell}{L_{i_s}^\delta}}\land\; e \Trans{\project{\sigma\cdot \ell}{L_e^\delta}}$\\
    \proofstep{\Cref{lem:project_from_parcomp_light}}\\
    $s \parcomp e  \Trans{\sigma\cdot \ell}$\\
    \proofstep{\Cref{def:arrowdefs,def:outset}: $\Trans{}$ and $\outset{}$}\\
    $\ell \in \outset{s\parcomp e \after \sigma}$\\
\item[$\ell\in (U_e \cap I_s)$:]Symmetrical to previous case
\item[$\ell = \delta$:]\ \\
    $i_s \eco{} e\land i_e \eco{} s \land i_s \uioco s \land i_e \uioco e \land \sigma \in  \utraces{s\parcomp e}\;\land$\\
    $ \delta \in \outset{i_s \parcomp i_e \after \sigma}$\\
    \proofstep{\Cref{def:outset,def:after}: $out$ and $\after$}\\
    $i_s \eco{} e\land i_e \eco{} s \land i_s \uioco s \land i_e \uioco e \land \sigma \in  \utraces{s\parcomp e}\;\land$\\
    $i_s \parcomp i_e \Trans{\sigma\cdot \delta}$\\
    \proofstep{\Cref{item:trans_transitive}}\\
    $i_s \eco{} e\land i_e \eco{} s \land i_s \uioco s \land i_e \uioco e \land \sigma \in  \utraces{s\parcomp e}\;\land$\\
    $i_s \parcomp i_e \Trans{\sigma\cdot \delta}\land\;i_s \parcomp i_e \Trans{\sigma}$\\
    \proofstep{\Cref{lem:test_project_utraces}}\\
    $i_s \uioco s \land i_e \uioco e \land \project{\sigma}{L_s^\delta} \in  \utraces{s}\land \project{\sigma}{L_e^\delta} \in  \utraces{e}\;\land$\\
    $i_s \parcomp i_e \Trans{\sigma\cdot \delta}$\\
    \proofstep{\Cref{lem:project_from_parcomp_IOTS}}\\
    $i_s \uioco s \land i_e \uioco e \land \project{\sigma}{L_s^\delta} \in  \utraces{s}\land \project{\sigma}{L_e^\delta} \in  \utraces{e}\;\land$\\
    $i_s \Trans{\project{\sigma\cdot \delta}{L_s^\delta}} \land i_e \Trans{\project{\sigma\cdot \delta}{L_e^\delta}} $\\
    \proofstep{\Cref{def:projection}: $\projectop$}\\
    $i_s \uioco s \land i_e \uioco e \land \project{\sigma}{L_s^\delta} \in  \utraces{s}\land \project{\sigma}{L_e^\delta} \in  \utraces{e}\;\land$\\
    $i_s \Trans{\project{\sigma}{L_s^\delta}\cdot \delta} \land\; i_e \Trans{\project{\sigma}{L_e^\delta}\cdot \delta}$\\
    \proofstep{\Cref{def:uioco}: $\uioco{}$}\\
    $s \Trans{\project{\sigma}{L_s^\delta}\cdot \delta}\land\; e \Trans{\project{\sigma}{L_e^\delta}\cdot\delta}$\\
    \proofstep{\Cref{def:projection}: $\projectop$}\\
    $s \Trans{\project{\sigma\cdot\delta}{L_s^\delta}}\land\; e \Trans{\project{\sigma\cdot\delta}{L_e^\delta}}$\\
    \proofstep{\Cref{lem:project_from_parcomp_light}}\\
    $s \parcomp e \Trans{\sigma\cdot\delta}$\\
    \proofstep{\Cref{def:arrowdefs,def:outset}: $\Trans{}$ and $\outset{}$}\\
    $\delta \in \outset{s\parcomp e \after \sigma}$\\
\end{case_distinction}
\end{proof}

\subsubsection{Proof Sketch}
We give a brief sketch of the proof of \Cref{lem:test_eco_implies_uioco}, working backwards from the goal.
The main challenge lies in relating the $\utraces{}$ of the various transition systems involved. We want to prove something about the $\utraces{}$ of $s\parcomp e$, but all of the assumptions give us information about the $\utraces{}$ of $s$ and $e$, $i_s \parcomp e$ and $i_e \parcomp s$. We already handled translating traces between all of these systems in \cite{vancuyck_CompositionalityModelBasedTesting_2023a}, but we still need to prove that being a \textit{Utrace} is preserved over these transformations. In \Cref{lem:test_project_utraces} we express that we can split the $\utraces{}$ of $s\parcomp e$ into a `left' and a `right' part, and both of these parts are still $\utraces{}$ of $s$ and $e$ respectively. This requires a lot of assumptions, where the most notable one is that it is not enough that the trace in question is just part of the specification, it must also be part of the implementation. To also make use of the $i_s \eco e$ and $i_e \eco s$ results, we need \Cref{lem:combine_utrace_imp} to show that combining \textit{Utraces} is also possible, as long as one of the two transition systems is input enabled. The basis for \Cref{lem:test_project_utraces,lem:combine_utrace_imp} lies in that the traces of $s$ and $e$ only reach states that are also reachable by the traces of $s\parcomp e$, as expressed in \Cref{lem:traces_after_projection_subset}. The reason why \Cref{lem:traces_after_projection_subset} talks about complicated subset constructions instead of just stating set equality lies in the fact that it is very hard to preserve quiescence over projection. The presence of communication errors can create queisence in the full system, while at least one of the individual components is actively trying to communicate and therefore not quiescent. This possibiliy is excluded by the combined $\eco$ and $\uioco$ assumptions of \Cref{lem:test_project_utraces}, and the fact that an input enabled system is always ready to receive communications in  \Cref{lem:traces_after_projection_imp_superset}.
 
\label{sec:test_eco_implies_uioco_proofsketch}
\begin{toappendix}
    \begin{lemmarep}
    Let $X$ and $Y$ be arbitrary sets. Let $ARG$ and $GOAL$ be predicates over $X \times Y$ and $Y$ respectively. Then $\forall x\in X:$
    \[\begin{array}[t]{ll}
    \Big(\big(\exists y \in Y: ARG(x,y)\big) \implies GOAL(x)\Big) \iff\\
    \big(\forall y \in Y: ARG(x,y) \implies GOAL(x)\big)
    \end{array}\]
    \label{lem:exists_forall_if_no_goal}
    \end{lemmarep}
    
    \begin{proof}
        \begin{case_distinction}
    \item[$\implies$:]\ \\
        Assume $\big(\exists y \in Y: ARG(x,y)\big) \implies GOAL(x)$. Take an arbitrary $y'\in Y$. To prove: $ARG(x,y') \implies GOAL(x)$. The goal is an implication, so we assume the argument. since $ARG(x,y')$ implies $\exists y \in Y: ARG(x,y)$, We immediately get $GOAL(x)$ which proves the statement.
    \item[$\impliedby$:]\ \\
        Assume $\forall y \in Y: ARG(x,y) \implies GOAL(x)$, and $\exists y \in Y: ARG(x,y)$. To prove: $GOAL(x)$. Using the $\exists$ elimination gives $ARG(x,y')$ for some fresh $y'\in Y$. Filling this in in the other assumption gives $GOAL(x)$, which proves the statement.
    
\end{case_distinction}
    \end{proof}
\end{toappendix}

\begin{lemmarep}
    \label{lem:traces_after_projection_subset}
    Let $s,e \in \LTS$ be $\composable$, $\sigma\in L_{s\parcomp e}^{\delta*}$, $s \Trans{\project{\sigma}{L_s^\delta}}$, $e \Trans{\project{\sigma}{L_e^\delta}}$.
    \[\begin{array}[t]{ll}
        s \after (\project{\sigma}{L_s^\delta}) \subseteq \{ q_s \setbar \exists q_e \in Q_e : q_s \parcomp q_e \in s\parcomp e \after \sigma\}\; \land\\
        e \after (\project{\sigma}{L_e^\delta}) \subseteq \{ q_e \setbar \exists q_s \in Q_s: q_s \parcomp q_e \in s\parcomp e \after \sigma\}
    \end{array}\]
\end{lemmarep}

\begin{proof}
        The lemma is symmetrical, because the definition of $\parcomp$ is symmetrical. For brevity, we only show half the proof\\ 
    (for $s \after (\project{\sigma}{L_s^\delta}) \subseteq \{ q_s \setbar \exists q_e \in Q_e: q_s \parcomp q_e \in s\parcomp e \after \sigma\}$).\\
    This proof can be repeated to also obtain the other half of the lemma. We assume some $q_s\in s \after (\project{\sigma}{L_s^\delta})$. To prove: $\exists q_e \in Q_e : q_s \parcomp q_e \in s \parcomp e \after \sigma$.\\

    \noindent
    $q_s\in s \after (\project{\sigma}{L_s^\delta}) \land e \Trans{\project{\sigma}{L_e^\delta}}$\\
    \proofstep{\Cref{def:after}: $\after$}\\
    $s \Trans{\project{\sigma}{L_s^\delta}} q_s \land e \Trans{\project{\sigma}{L_e^\delta}}$\\
    \proofstep{\Cref{lem:project_from_parcomp_light}}\\
    $s \parcomp e \Trans{\sigma} q_s \parcomp q_e$\\
    \proofstep{\Cref{def:after}: $\after$}\\
    $q_s \parcomp q_e \in s \parcomp e \after \sigma$
\end{proof}

\begin{lemmarep}
    \label{lem:traces_after_projection_imp_superset}
    Let $s \in \LTS, i_e\in \IOTS$ be $\composable$, $\sigma\in L_{s\parcomp i_e}^{\delta*}$.
    \[
        s \after (\project{\sigma}{L_s^\delta}) \supseteq \{ q_s \setbar \exists q_{i_e}\in Q_{i_e}: q_s \parcomp q_{i_e} \in s\parcomp i_e \after \sigma\}
    \]
\end{lemmarep}

\begin{proof}
     Assume some $q_s\parcomp q_{i_e}\in s \parcomp i_e \after \sigma$. To prove: $q_s \in s \after  \project{\sigma}{L_s^\delta}$. The proof follows from induction over $\sigma$. \\

\
 \begin{case_distinction}
    \item[Base case:] $\sigma = \epsilon$\\
        $q_s \parcomp q_{i_e} \in s \parcomp i_e \after \epsilon$\\
        \proofstep{\Cref{def:after}: $\after$}\\
        $s \parcomp i_e \Trans{\epsilon} q_s \parcomp q_{i_e}$\\
        \proofstep{\Cref{lem:project_from_parcomp_no_delta}}\\
        $s \Trans{\project{\epsilon}{L_s}} q_s$\\
        \proofstep{\Cref{def:projection}:$\projectop$, ($\delta\notin\epsilon$)}\\
        $s \Trans{\project{\epsilon}{L_s^\delta}} q_s$\\
        \proofstep{\Cref{def:after}: $\after$}\\
        $q_s\in s \after (\project{\epsilon}{L_s^\delta})$\\

    \item[Induction step:] 
        Assume the proposition holds for $\sigma'$. To prove: the proposition holds for $\sigma$  where $\sigma = \sigma' \cdot \ell$, $\ell \in L_{s \parcomp i_e}^\delta $.\\
        We do a case distinction on whether $\ell$ is $\delta$ or not.
        \begin{case_distinction}
            \item[$\ell \neq \delta$:]\ \\
                $q_s \parcomp q_{i_e} \in s \parcomp i_e \after \sigma'\cdot\ell$\\
                \proofstep{\Cref{def:after}: $\after$}\\
                $s \parcomp i_e \Trans{\sigma'\cdot\ell} q_s \parcomp q_{i_e}$\\
                \proofstep{\Cref{item:trans_transitive}}\\
                $\exists q_s'\in Q_s, q_{i_e}' \in Q_{i_e}: s \parcomp i_e \Trans{\sigma'} q_s'\parcomp q_{i_e}' \land q_s' \parcomp q_{i_e}'\Trans{\ell} q_s \parcomp q_{i_e}$\\
                \proofstep{Apply IH, and \cref{def:after}: $\after$}\\
                $\exists q_s'\in Q_s, q_{i_e}' \in Q_{i_e}: s  \Trans{\project{\sigma'}{L_s^\delta}} q_s' \land q_s' \parcomp q_{i_e}'\Trans{\ell} q_s \parcomp q_{i_e}$\\
                \proofstep{\Cref{lem:project_from_parcomp_no_delta}}\\
                $\exists q_s'\in Q_s: s  \Trans{\project{\sigma'}{L_s^\delta}} q_s' \land q_s' \Trans{\project{\ell}{L_s}} q_s$\\
                \proofstep{\Cref{item:trans_transitive}}\\
                $s  \Trans{\project{\sigma'}{L_s^\delta} \cdot\project{\ell}{L_s}} q_s$\\
                \proofstep{\Cref{def:projection}:$\projectop$ and $\ell\neq\delta$}\\
                $s  \Trans{\project{\sigma'\cdot\ell}{L_s^\delta}} q_s$\\
                \proofstep{\Cref{def:after}: $\after$}\\
                $q_s\in s \after (\project{\sigma'\cdot\ell}{L_s^\delta})$\\
            \item[$\ell = \delta$:]\ \\
                $q_s \parcomp q_{i_e} \in s \parcomp i_e \after \sigma'\cdot\delta$\\
                \proofstep{\Cref{def:after}: $\after$}\\
                $s \parcomp i_e \Trans{\sigma'\cdot\delta} q_s \parcomp q_{i_e}$\\
                \proofstep{\Cref{def:arrowdefs}: $\Trans{\sigma'\cdot\delta}$}\\
                $\exists q_s',q_s''\in Q_s, q_{i_e}',q_{i_e}'' \in Q_{i_e}:$\\
                \tab$s \parcomp i_e \Trans{\sigma'} q_s'\parcomp q_{i_e}' \land q_s' \parcomp q_{i_e}'\trans{\delta} q_s'' \parcomp q_{i_e}''\land q_s''\parcomp q_{i_e}'' \Trans{\epsilon} q_s \parcomp q_{i_e}$\\
                \proofstep{Apply IH, and \cref{def:after}: $\after$}\\
                $\exists q_s',q_s''\in Q_s, q_{i_e}',q_{i_e}'' \in Q_{i_e}:$\\
                \tab$s \Trans{\project{\sigma'}{L_s^\delta}} q_s' \land q_s' \parcomp q_{i_e}'\trans{\delta} q_s'' \parcomp q_{i_e}''\land q_s''\parcomp q_{i_e}'' \Trans{\epsilon} q_s \parcomp q_{i_e}$\\
                \proofstep{\Cref{lem:project_from_parcomp_no_delta}}\\
                $\exists q_s',q_s''\in Q_s, q_{i_e}',q_{i_e}'' \in Q_{i_e}: s \Trans{\project{\sigma'}{L_s^\delta}} q_s' \land q_s' \parcomp q_{i_e}'\trans{\delta} q_s'' \parcomp q_{i_e}''\land q_s'' \Trans{\project{\epsilon}{L_s^\delta}} q_s$\\
                \proofstep{\Cref{def:delta}: $\delta$}\\
                $\exists q_s'\in Q_s, q_{i_e}' \in Q_{i_e}: s \Trans{\project{\sigma'}{L_s^\delta}} q_s'\land q_s' \Trans{\project{\epsilon}{L_s^\delta}} q_s\;\land$\\
                $\forall a \in U_{s\parcomp e_i}: q_s'\parcomp q_{i_e}' \nottrans{a}$\\
                \proofstep{\Cref{def:parcomp}: $\parcomp$}\\
                $\exists q_s'\in Q_s, q_{i_e}' \in Q_{i_e}: s \Trans{\project{\sigma'}{L_s^\delta}} q_s'\land q_s' \Trans{\project{\epsilon}{L_s^\delta}} q_s\;\land$\\
                $\forall a \in (U_s \setminus L_{i_e}) \cup \{\tau\}: q_s' \nottrans{a} \land$\\
                $\forall a \in (U_{i_e} \setminus L_s) \cup \{\tau\}: q_{i_e}' \nottrans{a} \land$\\
                $\forall a \in (U_s \cap I_{i_e}): q_s' \nottrans{a} \lor\; q_{i_e}' \nottrans{a}$\\
                \proofstep{\Cref{def:iots}: $\IOTS$}\\
                $\exists q_s'\in Q_s, q_{i_e}' \in Q_{i_e}: s \Trans{\project{\sigma'}{L_s^\delta}} q_s'\land q_s' \Trans{\project{\epsilon}{L_s^\delta}} q_s\;\land$\\
                $\forall a \in (U_s \setminus L_{i_e}) \cup \{\tau\}: q_s' \nottrans{a} \land$\\
                $\forall a \in (U_{i_e} \setminus L_s) \cup \{\tau\}: q_{i_e}' \nottrans{a} \land$\\
                $\forall a \in (U_s \cap I_{i_e}): (q_s' \nottrans{a} \lor\; q_{i_e}' \nottrans{a}) \;\land$\\
                $\forall q_{i_e}'' \in Q_{i_e}, a \in I_{i_e}: q_{i_e}''\Trans{a}$\\
                \proofstep{$\forall$ elimination}\\
                $\exists q_s'\in Q_s, q_{i_e}' \in Q_{i_e}: s \Trans{\project{\sigma'}{L_s^\delta}} q_s'\land q_s' \Trans{\project{\epsilon}{L_s^\delta}} q_s\;\land$\\
                $\forall a \in (U_s \setminus L_{i_e}) \cup \{\tau\}: q_s' \nottrans{a} \land$\\
                $\forall a \in (U_{i_e} \setminus L_s) \cup \{\tau\}: q_{i_e}' \nottrans{a} \land$\\
                $\forall a \in (U_s \cap I_{i_e}): (q_s' \nottrans{a} \lor\; q_{i_e}' \nottrans{a})\land q_{i_e}'\Trans{a}$\\
                \proofstep{\Cref{def:arrowdefs}: $\Trans{a}$ and $q_{i_e}'\nottrans{\tau}$}\\
                $\exists q_s'\in Q_s, q_{i_e}' \in Q_{i_e}: s \Trans{\project{\sigma'}{L_s^\delta}} q_s'\land q_s' \Trans{\project{\epsilon}{L_s^\delta}} q_s\;\land$\\
                $\forall a \in (U_s \setminus L_{i_e}) \cup \{\tau\}: q_s' \nottrans{a} \land$\\
                $\forall a \in (U_s \cap I_{i_e}): (q_s' \nottrans{a} \lor\; q_{i_e}' \nottrans{a})\land q_{i_e}'\trans{a}$\\
                \proofstep{$(\neg A \lor \neg B)\land B = \neg A$}\\
                $\exists q_s'\in Q_s: s \Trans{\project{\sigma'}{L_s^\delta}} q_s'\land q_s' \Trans{\project{\epsilon}{L_s^\delta}} q_s\;\land$\\
                $\forall a \in (U_s \setminus L_{i_e}) \cup \{\tau\}: q_s' \nottrans{a} \land$\\
                $\forall a \in (U_s \cap I_{i_e}): q_s' \nottrans{a}$\\
                \proofstep{\Cref{def:composable}: $\composable$}\\
                $\exists q_s'\in Q_s: s \Trans{\project{\sigma'}{L_s^\delta}} q_s'\land q_s' \Trans{\project{\epsilon}{L_s^\delta}} q_s\;\land$\\
                $\forall a \in (U_s \setminus L_{i_e}) \cup \{\tau\}: q_s' \nottrans{a} \land$\\
                $\forall a \in (U_s \cap L_{i_e}): q_s' \nottrans{a}$\\
                \proofstep{\Cref{def:delta}: $\delta$}\\
                $\exists q_s'\in Q_s: s \Trans{\project{\sigma'}{L_s^\delta}} q_s'\land q_s'  \Trans{\project{\epsilon}{L_s^\delta}} q_s \land q_s'\trans{\delta}q_s'$\\
                \proofstep{\Cref{def:arrowdefs}:$\Trans{}$}\\
                $s \Trans{(\project{\sigma'}{L_s^\delta})\cdot\delta\cdot(\project{\epsilon}{L_s^\delta})} q_s$\\
                \proofstep{\Cref{def:projection}:$\projectop$}\\
                $s  \Trans{\project{\sigma'\cdot\delta}{L_s^\delta}} q_s$\\
                \proofstep{\Cref{def:after}: $\after$}\\
                $q_s\in s \after (\project{\sigma'\cdot\delta}{L_s^\delta})$\\

        \end{case_distinction}

 \end{case_distinction}

\end{proof}

\begin{lemmarep}
    \label{lem:combine_utrace_imp}
    Let $s \in \LTS$ and $i_e \in \IOTS$  be $\composable$. 
    \[
        \project{\sigma}{L_s^\delta} \in \utraces{s} \land \project{\sigma}{L_{i_e}^\delta} \in \utraces{i_e} \implies
        \sigma \in \utraces{s\parcomp i_e} 
    \]
\end{lemmarep}

\begin{proof}
    Proof by induction on $\sigma$.
\begin{case_distinction}
\item[Base case: $\sigma = \epsilon$] Covered directly by \cref{lem:project_from_parcomp_light}\\

\item[Induction step:] 
    Assume the proposition holds for $\sigma'$. To prove: the proposition holds for $\sigma$ where $\sigma = \sigma' \cdot \ell$, $\ell \in L_{s \parcomp i_e}^\delta $.
    We do a case distinction on $\ell$, splitting $L_{s \parcomp i_e}^\delta$ into
    $(I_s \setminus L_{i_e}), (I_{i_e} \setminus L_s), (I_s \cap U_{i_e}), (U_s \setminus L_{i_e}), (U_{i_e} \setminus L_s)$ and $(I_{i_e} \cap U_s) \cup (I_s \cap I_{i_e}) \cup \{\delta\}$.\\
    Note that $U_s \cap U_{i_e}$ is missing because it is empty (\cref{def:composable}).
    \begin{case_distinction}
        \item[$\ell\in (I_s \setminus L_{i_e})$: ]\ \\
            $\project{\sigma'\cdot\ell}{L_s^\delta}\in \utraces{s} \land \project{\sigma'\cdot\ell}{L_{i_e}^\delta}\in \utraces{i_e}$\\
            \proofstep{\Cref{def:projection}: $\projectop$}\\
            $\project{\sigma'}{L_s^\delta}\cdot\ell\in \utraces{s} \land \project{\sigma'}{L_{i_e}^\delta}\in \utraces{i_e}$\\
            \proofstep{\Cref{def:uioco}: $\utraces{}$}\\
            $\project{\sigma'}{L_s^\delta}\cdot\ell\in \utraces{s} \land \project{\sigma'}{L_{i_e}^\delta}\in \utraces{i_e}\;\land$\\
            $\forall q_s \in Q_s : s\Trans{\project{\sigma'}{L_s^\delta}} q_s \implies q_s\Trans{\ell}$\\
            \proofstep{\Cref{item:utrace_prefix_closed}}\\
            $\project{\sigma'}{L_s^\delta}\in \utraces{s} \land \project{\sigma'}{L_{i_e}^\delta}\in \utraces{i_e}\;\land$\\
            $\forall q_s \in Q_s : s\Trans{\project{\sigma'}{L_s^\delta}} q_s \implies q_s\Trans{\ell}$\\
            \proofstep{Apply IH}\\
            $\sigma'\in \utraces{s\parcomp i_e}\; \land$\\
            $\forall q_s \in Q_s : s\Trans{\project{\sigma'}{L_s^\delta}} q_s \implies q_s\Trans{\ell}$\\
            \proofstep{\Cref{def:after}: $\after$}\\
            $\sigma'\in \utraces{s\parcomp i_e}\; \land$\\
            $\forall q_s \in s \after \project{\sigma'}{L_s^\delta}: q_s\Trans{\ell}$\\
            \proofstep{\Cref{lem:traces_after_projection_imp_superset}}\\
            $\sigma'\in \utraces{s\parcomp i_e}\; \land$\\
            $\forall q_s \in \pi_1(s\parcomp i_e \after \sigma'): q_s\Trans{\ell}$\\
            \proofstep{\Cref{def:after}: $\after$}\\
            $\sigma'\in \utraces{s\parcomp i_e}\; \land$\\
            $\forall q_s \in \{q_s \setbar \exists q_{i_e}\in Q_{i_e}: s\parcomp i_e \Trans{\sigma'} q_s \parcomp q_{i_e} \}: q_s\Trans{\ell}$\\
            \proofstep{Rewrite set notation}\\
            $\sigma'\in \utraces{s\parcomp i_e}\; \land$\\
            $\forall q_s \in Q_s: (\exists q_{i_e} \in Q_{i_e} : s \parcomp i_e \Trans{\sigma'} q_s \parcomp q_{i_e}) \implies q_s\Trans{\ell}$\\
            \proofstep{\Cref{lem:exists_forall_if_no_goal}}\\
            $\sigma'\in \utraces{s\parcomp i_e}\; \land$\\
            $\forall q_s \in Q_s, q_{i_e} \in Q_{i_e} : s \parcomp i_e \Trans{\sigma'} q_s \parcomp q_{i_e} \implies q_s\Trans{\ell}$\\
            \proofstep{\Cref{def:arrowdefs}: $\Trans{\epsilon}$}\\
            $\sigma'\in \utraces{s\parcomp i_e}\; \land$\\
            $\forall q_s \in Q_s, q_{i_e} \in Q_{i_e} : s \parcomp i_e \Trans{\sigma'} q_s \parcomp q_{i_e} \implies (q_s\Trans{\ell} \land\; q_{i_e}\Trans{\epsilon})$\\
            \proofstep{\Cref{def:projection}: $\projectop$}\\
            $\sigma'\in \utraces{s\parcomp i_e}\; \land$\\
            $\forall q_s \in Q_s, q_{i_e} \in Q_{i_e} : s \parcomp i_e \Trans{\sigma'} q_s \parcomp q_{i_e} \implies (q_s\Trans{\project{\ell}{L_s^\delta}} \land\; q_{i_e}\Trans{\project{\ell}{L_e^\delta}})$\\
            \proofstep{\Cref{lem:project_from_parcomp_light}}\\
            $\sigma'\in \utraces{s\parcomp i_e}\; \land$\\
            $\forall q_s \in Q_s, q_{i_e} \in Q_{i_e} : s \parcomp i_e \Trans{\sigma'} q_s \parcomp q_{i_e} \implies q_s \parcomp q_{i_e} \Trans{\ell}$\\
            \proofstep{\Cref{def:uioco}: $\utraces{}$}\\
            $\sigma'\cdot\ell\in \utraces{s\parcomp i_e}$\\
        \item[$\ell\in (I_{i_e} \setminus L_s)$: ]\ \\
            $\project{\sigma'\cdot\ell}{L_s^\delta}\in \utraces{s} \land \project{\sigma'\cdot\ell}{L_{i_e}^\delta}\in \utraces{i_e}$\\
            \proofstep{\Cref{def:uioco}: $\utraces{}$}\\
            $\project{\sigma'\cdot\ell}{L_s^\delta}\in \utraces{s} \land \project{\sigma'\cdot\ell}{L_{i_e}^\delta}\in \utraces{i_e}\;\land$\\
            $\exists q_s\in Q_s, q_{i_e} \in Q_{i_e}: s \Trans{\project{\sigma'\cdot\ell}{L_s^\delta}} q_s \land e \Trans{\project{\sigma'\cdot\ell}{L_{i_e}^\delta}} q_{i_e}$\\
            \proofstep{\Cref{lem:project_from_parcomp_light}}\\
            $\project{\sigma'\cdot\ell}{L_s^\delta}\in \utraces{s} \land \project{\sigma'\cdot\ell}{L_{i_e}^\delta}\in \utraces{i_e}\;\land$\\
            $\exists q_s\in Q_s, q_{i_e} \in Q_{i_e}: s \parcomp e \Trans{\sigma'\cdot\ell} q_s \parcomp q_{i_e}$\\
            \proofstep{\Cref{def:projection}: $\projectop$}\\
            $\project{\sigma'}{L_s^\delta}\in \utraces{s} \land \project{\sigma'}{L_{i_e}^\delta}\cdot\ell\in \utraces{i_e}\;\land$\\
            $\exists q_s\in Q_s, q_{i_e} \in Q_{i_e}: s \parcomp e \Trans{\sigma'\cdot\ell} q_s \parcomp q_{i_e}$\\
            \proofstep{\Cref{item:utrace_prefix_closed}}\\
            $\project{\sigma'}{L_s^\delta}\in \utraces{s} \land \project{\sigma'}{L_{i_e}^\delta}\in \utraces{i_e}\;\land$\\
            $\exists q_s\in Q_s, q_{i_e} \in Q_{i_e}: s \parcomp e \Trans{\sigma'\cdot\ell} q_s \parcomp q_{i_e}$\\
            \proofstep{Apply IH}\\
            $\sigma'\in \utraces{s\parcomp i_e}\; \land$\\
            $\exists q_s\in Q_s, q_{i_e} \in Q_{i_e}: s \parcomp e \Trans{\sigma'\cdot\ell} q_s \parcomp q_{i_e}$\\
            \proofstep{\Cref{def:iots}: $\IOTS$}\\
            $\sigma'\in \utraces{s\parcomp i_e}\; \land$\\
            $\exists q_s\in Q_s, q_{i_e} \in Q_{i_e}: s \parcomp e \Trans{\sigma'\cdot\ell} q_s \parcomp q_{i_e}\;\land$\\
            $\forall q_{i_e}' \in Q_{i_e}, \ell'\in I_{i_e} : q_{i_e}'\Trans{\ell'}$\\
            \proofstep{\Cref{def:arrowdefs}: $\Trans{\epsilon}$}\\
            $\sigma'\in \utraces{s\parcomp i_e}\; \land$\\
            $\exists q_s\in Q_s, q_{i_e} \in Q_{i_e}: s \parcomp e \Trans{\sigma'\cdot\ell} q_s \parcomp q_{i_e}\;\land$\\
            $\forall q_{i_e}' \in Q_{i_e}, \ell'\in I_{i_e} : q_{i_e}'\Trans{\ell'}\;\land$\\
            $\forall q_s' \in Q_s : q_s'\Trans{\epsilon}$\\
            \proofstep{\Cref{def:projection}: $\projectop$}\\
            $\sigma'\in \utraces{s\parcomp i_e}\; \land$\\
            $\exists q_s\in Q_s, q_{i_e} \in Q_{i_e}: s \parcomp e \Trans{\sigma'\cdot\ell} q_s \parcomp q_{i_e}\;\land$\\
            $\forall q_{i_e}' \in Q_{i_e}, \ell'\in I_{i_e} : q_{i_e}'\Trans{\project{\ell'}{L_{i_e}^\delta}}\;\land$\\
            $\forall q_s' \in Q_s : q_s'\Trans{\project{\epsilon}{L_s^\delta}}$\\
            \proofstep{\Cref{lem:project_from_parcomp_light}}\\
            $\sigma'\in \utraces{s\parcomp i_e}\; \land$\\
            $\exists q_s\in Q_s, q_{i_e} \in Q_{i_e}: s \parcomp e \Trans{\sigma'\cdot\ell} q_s \parcomp q_{i_e}\;\land$\\
            $\forall q_{i_e}' \in Q_{i_e},  q_s' \in Q_s, \ell'\in I_{i_e} : q_s' \parcomp q_{i_e}'\Trans{\ell'}$\\
            \proofstep{\Cref{def:uioco}: $\utraces{}$}\\
            $\sigma'\cdot\ell\in \utraces{s\parcomp i_e}$\\
        \item[$\ell \in (I_s \cup I_{i_e})$:]\ \\
            $\project{\sigma'\cdot\ell}{L_s^\delta}\in \utraces{s} \land \project{\sigma'\cdot\ell}{L_{i_e}^\delta}\in \utraces{i_e}$\\
            \proofstep{\Cref{def:projection}: $\projectop$}\\
            $\project{\sigma'}{L_s^\delta}\cdot\ell\in \utraces{s} \land \project{\sigma'}{L_{i_e}^\delta}\cdot\ell\in \utraces{i_e}$\\
            \proofstep{\Cref{def:uioco}: $\utraces{}$}\\
            $\project{\sigma'}{L_s^\delta}\cdot\ell\in \utraces{s} \land \project{\sigma'}{L_{i_e}^\delta}\cdot\ell\in \utraces{i_e}\;\land$\\
            $\forall q_s \in Q_s : s\Trans{\project{\sigma'}{L_s^\delta}} q_s \implies q_s\Trans{\ell}$\\
            \proofstep{\Cref{item:utrace_prefix_closed}}\\
            $\project{\sigma'}{L_s^\delta}\in \utraces{s} \land \project{\sigma'}{L_{i_e}^\delta}\in \utraces{i_e}\;\land$\\
            $\forall q_s \in Q_s : s\Trans{\project{\sigma'}{L_s^\delta}} q_s \implies q_s\Trans{\ell}$\\
            \proofstep{Apply IH}\\
            $\sigma'\in \utraces{s\parcomp i_e}\; \land$\\
            $\forall q_s \in Q_s : s\Trans{\project{\sigma'}{L_s^\delta}} q_s \implies q_s\Trans{\ell}$\\
            \proofstep{\Cref{def:after}: $\after$}\\
            $\sigma'\in \utraces{s\parcomp i_e}\; \land$\\
            $\forall q_s \in s \after \project{\sigma'}{L_s^\delta}: q_s\Trans{\ell}$\\
            \proofstep{\Cref{lem:traces_after_projection_imp_superset}}\\
            $\sigma'\in \utraces{s\parcomp i_e}\; \land$\\
            $\forall q_s \in \pi_1(s\parcomp i_e \after \sigma'): q_s\Trans{\ell}$\\
            \proofstep{\Cref{def:after}: $\after$}\\
            $\sigma'\in \utraces{s\parcomp i_e}\; \land$\\
            $\forall q_s \in \{q_s \setbar \exists q_{i_e}\in Q_{i_e}: s\parcomp i_e \Trans{\sigma'} q_s \parcomp q_{i_e} \}: q_s\Trans{\ell}$\\
            \proofstep{Rewrite set notation}\\
            $\sigma'\in \utraces{s\parcomp i_e}\; \land$\\
            $\forall q_s \in Q_s: (\exists q_{i_e} \in Q_{i_e} : s \parcomp i_e \Trans{\sigma'} q_s \parcomp q_{i_e}) \implies q_s\Trans{\ell}$\\
            \proofstep{\Cref{lem:exists_forall_if_no_goal}}\\
            $\sigma'\in \utraces{s\parcomp i_e}\; \land$\\
            $\forall q_s \in Q_s, q_{i_e} \in Q_{i_e} : s \parcomp i_e \Trans{\sigma'} q_s \parcomp q_{i_e} \implies q_s\Trans{\ell}$\\
            \proofstep{\Cref{def:iots}: $\IOTS$}\\
            $\sigma'\in \utraces{s\parcomp i_e}\; \land$\\
            $\forall q_s \in Q_s, q_{i_e} \in Q_{i_e} : s \parcomp i_e \Trans{\sigma'} q_s \parcomp q_{i_e} \implies (q_s\Trans{\ell}\;\land\; q_{i_e}\Trans{\ell})$\\
            \proofstep{\Cref{def:projection}: $\projectop$}\\
            $\sigma'\in \utraces{s\parcomp i_e}\; \land$\\
            $\forall q_s \in Q_s, q_{i_e} \in Q_{i_e} : s \parcomp i_e \Trans{\sigma'} q_s \parcomp q_{i_e} \implies (q_s\Trans{\project{\ell}{L_s^\delta}} \land\; q_{i_e}\Trans{\project{\ell}{L_e^\delta}})$\\
            \proofstep{\Cref{lem:project_from_parcomp_light}}\\
            $\sigma'\in \utraces{s\parcomp i_e}\; \land$\\
            $\forall q_s \in Q_s, q_{i_e} \in Q_{i_e} : s \parcomp i_e \Trans{\sigma'} q_s \parcomp q_{i_e} \implies q_s \parcomp q_{i_e} \Trans{\ell}$\\
            \proofstep{\Cref{def:uioco}: $\utraces{}$}\\
            $\sigma'\cdot\ell\in \utraces{s\parcomp i_e}$\\
            
        \item[$\ell\in (U_s \setminus L_{i_e})$: ]\ \\
            $\project{\sigma'\cdot\ell}{L_s^\delta}\in \utraces{s} \land \project{\sigma'\cdot\ell}{L_{i_e}^\delta}\in \utraces{i_e}$\\
            \proofstep{\Cref{def:uioco}: $\utraces{}$}\\
            $\project{\sigma'\cdot\ell}{L_s^\delta}\in \utraces{s} \land \project{\sigma'\cdot\ell}{L_{i_e}^\delta}\in \utraces{i_e}\;\land$\\
            $\exists q_s\in Q_s, q_{i_e} \in Q_{i_e}: s \Trans{\project{\sigma'\cdot\ell}{L_s^\delta}} q_s \land e \Trans{\project{\sigma'\cdot\ell}{L_{i_e}^\delta}} q_{i_e}$\\
            \proofstep{\Cref{lem:project_from_parcomp_light}}\\
            $\project{\sigma'\cdot\ell}{L_s^\delta}\in \utraces{s} \land \project{\sigma'\cdot\ell}{L_{i_e}^\delta}\in \utraces{i_e}\;\land$\\
            $\exists q_s\in Q_s, q_{i_e} \in Q_{i_e}: s \parcomp e \Trans{\sigma'\cdot\ell} q_s \parcomp q_{i_e}$\\
            \proofstep{\Cref{def:projection}: $\projectop$}\\
            $\project{\sigma'}{L_s^\delta}\cdot\ell\in \utraces{s} \land \project{\sigma'}{L_{i_e}^\delta}\in \utraces{i_e}\;\land$\\
            $\exists q_s\in Q_s, q_{i_e} \in Q_{i_e}: s \parcomp e \Trans{\sigma'\cdot\ell} q_s \parcomp q_{i_e}$\\
            \proofstep{\Cref{item:utrace_prefix_closed}}\\
            $\project{\sigma'}{L_s^\delta}\in \utraces{s} \land \project{\sigma'}{L_{i_e}^\delta}\in \utraces{i_e}\;\land$\\
            $\exists q_s\in Q_s, q_{i_e} \in Q_{i_e}: s \parcomp e \Trans{\sigma'\cdot\ell} q_s \parcomp q_{i_e}$\\
            \proofstep{Apply IH}\\
            $\sigma'\in \utraces{s\parcomp i_e}\; \land$\\
            $\exists q_s\in Q_s, q_{i_e} \in Q_{i_e}: s \parcomp e \Trans{\sigma'\cdot\ell} q_s \parcomp q_{i_e}$\\
            \proofstep{\Cref{def:uioco}: $\utraces{}$}\\
            $\sigma'\cdot\ell\in \utraces{s\parcomp i_e}$\\
        \item[$\ell\in (U_{i_e} \setminus L_s)$: ] Symmetrical to previous case\\

        \item[$\ell \in (I_s \cap U_{i_e}) \cup (I_{i_e} \cap U_s) \cup \{\delta\}$:]\ \\
            $\project{\sigma'\cdot\ell}{L_s^\delta}\in \utraces{s} \land \project{\sigma'\cdot\ell}{L_{i_e}^\delta}\in \utraces{i_e}$\\
            \proofstep{\Cref{def:uioco}: $\utraces{}$}\\
            $\project{\sigma'\cdot\ell}{L_s^\delta}\in \utraces{s} \land \project{\sigma'\cdot\ell}{L_{i_e}^\delta}\in \utraces{i_e}\;\land$\\
            $\exists q_s\in Q_s, q_{i_e} \in Q_{i_e}: s \Trans{\project{\sigma'\cdot\ell}{L_s^\delta}} q_s \land e \Trans{\project{\sigma'\cdot\ell}{L_{i_e}^\delta}} q_{i_e}$\\
            \proofstep{\Cref{lem:project_from_parcomp_light}}\\
            $\project{\sigma'\cdot\ell}{L_s^\delta}\in \utraces{s} \land \project{\sigma'\cdot\ell}{L_{i_e}^\delta}\in \utraces{i_e}\;\land$\\
            $\exists q_s\in Q_s, q_{i_e} \in Q_{i_e}: s \parcomp e \Trans{\sigma'\cdot\ell} q_s \parcomp q_{i_e}$\\
            \proofstep{\Cref{def:projection}: $\projectop$}\\
            $\project{\sigma'}{L_s^\delta}\cdot\ell\in \utraces{s} \land \project{\sigma'}{L_{i_e}^\delta}\cdot \ell\in \utraces{i_e}\;\land$\\
            $\exists q_s\in Q_s, q_{i_e} \in Q_{i_e}: s \parcomp e \Trans{\sigma'\cdot\ell} q_s \parcomp q_{i_e}$\\
            \proofstep{\Cref{item:utrace_prefix_closed}}\\
            $\project{\sigma'}{L_s^\delta}\in \utraces{s} \land \project{\sigma'}{L_{i_e}^\delta}\in \utraces{i_e}\;\land$\\
            $\exists q_s\in Q_s, q_{i_e} \in Q_{i_e}: s \parcomp e \Trans{\sigma'\cdot\ell} q_s \parcomp q_{i_e}$\\
            \proofstep{Apply IH}\\
            $\sigma'\in \utraces{s\parcomp i_e}\; \land$\\
            $\exists q_s\in Q_s, q_{i_e} \in Q_{i_e}: s \parcomp e \Trans{\sigma'\cdot\ell} q_s \parcomp q_{i_e}$\\
            \proofstep{\Cref{def:uioco}: $\utraces{}$}\\
            $\sigma'\cdot\ell\in \utraces{s\parcomp i_e}$\\

    \end{case_distinction}

\end{case_distinction}
\end{proof}

\begin{toappendix} 
    \begin{lemmarep}
        Let $i \in \IOTS$, $\sigma\in L_i^{\delta*}$.
        \[i \Trans{\sigma} \implies \sigma \in \utraces{i}\]
        \label{lem:straces_equals_utraces_iots}
    \end{lemmarep}
    
    \begin{proof}
        Proof follows directly from \Cref{def:iots} and \Cref{def:uioco}. $\utraces{}$ require all used inputs to be defined in any state you could non-deterministically be in when giving the input. Input enabled systems have every input enabled in every state, so they satisfy this property by default.
    \end{proof}
\end{toappendix}

\begin{lemmarep}
    \label{lem:test_project_utraces}
    Let $s,e \in \LTS$ be $\composable$, $i_s, i_e \in \IOTS$ , $\sigma \in  L_{s\parcomp e}^\delta$.
    \[\begin{array}[t]{ll}
       i_s \eco{} e \land i_e \eco{} s \land i_s \uioco s \land i_e \uioco e  \land i_s \parcomp i_e \Trans{\sigma} \land\; \sigma \in \utraces{s\parcomp e}  \implies\\
        \tab \project{\sigma}{L_s^\delta} \in \utraces{s} \land \project{\sigma}{L_e^\delta} \in \utraces{e}
    \end{array}\]
\end{lemmarep}

\begin{proof}
    Proof by induction on $\sigma$. Since $i_s \uioco s$ implies $L_{i_s} = L_s$ (\cref{def:uioco}), this proof will only use $L_s$ and $L_e$ to increase readability.
\begin{case_distinction}
    \item[Base case:] $\sigma = \epsilon$\\
        Nothing to prove, as $\epsilon$ is part of the $\utraces{}$ of every $LTS$.
    \item[Induction step:] 
        Assume the proposition holds for $\sigma'$. To prove: the proposition holds for $\sigma$  where $\sigma = \sigma' \cdot \ell$, $\ell \in L_{s \parcomp e}^\delta $.\\
        We do a case distinction on $\ell$, splitting $L_{s \parcomp e}^\delta$ into\\
        $(I_s \setminus L_e), (I_e \setminus L_s), (I_e \cap U_s), (I_s \cap U_e), (I_s \cap I_e), (U_s \setminus L_e), (U_e \setminus L_s), \{\delta\}$. Note that $U_s \cap U_e$ is missing because it is empty (\cref{def:composable}). This fine grained distinction is required because what it means for a trace to be part of $\utraces{}$ changes based on if the last label is an input or output. Additionally, the behaviour of $\projectop$ changes based on if a label is part of $L_s$, $L_e$ or both. 
        \begin{case_distinction}
            \item[$\ell \in I_s \setminus L_e$:]\ \\
            $\sigma'\cdot\ell\in\utraces{s\parcomp e}$\\
            \proofstep{\Cref{item:utrace_prefix_closed}}\\
            $\sigma'\in\utraces{s\parcomp e} \land \sigma'\cdot\ell\in\utraces{s\parcomp e} $\\
            \proofstep{\Cref{def:uioco}: $\utraces{}$}\\
            $\sigma'\in\utraces{s\parcomp e}\; \land$\\
            $\forall q_s \in Q_s, q_e \in Q_e: s \parcomp e \Trans{\sigma'} q_s \parcomp q_e \implies q_s \parcomp q_e \Trans{\ell}$\\
            \proofstep{Apply IH}\\
            $\project{\sigma'}{L_s^\delta}\in\utraces{s}\land \project{\sigma'}{L_e^\delta}\in\utraces{e}  \; \land$\\
            $\forall q_s \in Q_s, q_e \in Q_e: s \parcomp e \Trans{\sigma'} q_s \parcomp q_e \implies q_s \parcomp q_e \Trans{\ell}$\\
            \proofstep{\Cref{lem:traces_after_projection_subset}}\\
            $\project{\sigma'}{L_s^\delta}\in\utraces{s}\land \project{\sigma'}{L_e^\delta}\in\utraces{e}  \; \land$\\
            $\forall q_s \in Q_s, q_e \in Q_e: s \parcomp e \Trans{\sigma'} q_s \parcomp q_e \implies q_s \parcomp q_e \Trans{\ell}\;\land$\\
            $s \after (\project{\sigma'}{L_s^\delta}) \subseteq \pi_1(s\parcomp e \after \sigma')$\\
            \proofstep{\Cref{def:after}: $\after$ and definition $\subseteq$}\\
            $\project{\sigma'}{L_s^\delta}\in\utraces{s}\land \project{\sigma'}{L_e^\delta}\in\utraces{e}  \; \land$\\
            $\forall q_s \in Q_s, q_e \in Q_e: s \parcomp e \Trans{\sigma'} q_s \parcomp q_e \implies q_s \parcomp q_e \Trans{\ell}\;\land$\\
            $\forall q_s' \in Q_s : s \Trans{\project{\sigma'}{L_s^\delta}} q_s' \implies (\exists q_e'\in Q_e: s \parcomp e \Trans{\sigma'} q_s' \parcomp q_e')$\\
            \proofstep{$\forall$ elimination and $(A \implies B) \land (B \implies C) \implies (A \implies C)$}\\
            $\project{\sigma'}{L_s^\delta}\in\utraces{s}\land \project{\sigma'}{L_e^\delta}\in\utraces{e}  \; \land$\\
            $\forall q_s' \in Q_s:  s \Trans{ \project{\sigma'}{L_s^\delta}} q_s' \implies (\exists q_e' \in Q_e: q_s' \parcomp q_e' \Trans{\ell})$\\
            \proofstep{\Cref{lem:project_from_parcomp_light}}\\
            $\project{\sigma'}{L_s^\delta}\in\utraces{s}\land \project{\sigma'}{L_e^\delta}\in\utraces{e}  \; \land$\\
            $\forall q_s' \in Q_s:  s \Trans{ \project{\sigma'}{L_s^\delta}} q_s' \implies q_s' \Trans{\ell}$\\
            \proofstep{\Cref{def:uioco}: $\utraces{}$}\\
            $\project{\sigma'}{L_e^\delta}\in\utraces{e}  \; \land$\\
            $\project{\sigma'}{L_s^\delta}\cdot\ell \in \utraces{s}$\\
            \proofstep{\Cref{def:projection}: $\projectop$}\\
            $\project{\sigma'\cdot\ell}{L_e^\delta}\in\utraces{e}  \; \land$\\
            $\project{\sigma'\cdot\ell}{L_s^\delta} \in \utraces{s}$
        \item[$\ell \in I_e \setminus L_s$:] Symmetrical to previous case
        \item[$\ell \in I_s \cap I_e$:]\ \\
            $\sigma'\cdot\ell\in\utraces{s\parcomp e}$\\
            \proofstep{\Cref{item:utrace_prefix_closed}}\\
            $\sigma'\in\utraces{s\parcomp e} \land \sigma'\cdot\ell\in\utraces{s\parcomp e} $\\
            \proofstep{\Cref{def:uioco}: $\utraces{}$}\\
            $\sigma'\in\utraces{s\parcomp e}\; \land$\\
            $\forall q_s \in Q_s, q_e \in Q_e: s \parcomp e \Trans{\sigma'} q_s \parcomp q_e \implies q_s \parcomp q_e \Trans{\ell}$\\
            \proofstep{Apply IH}\\
            $\project{\sigma'}{L_s^\delta}\in\utraces{s}\land \project{\sigma'}{L_e^\delta}\in\utraces{e}  \; \land$\\
            $\forall q_s \in Q_s, q_e \in Q_e: s \parcomp e \Trans{\sigma'} q_s \parcomp q_e \implies q_s \parcomp q_e \Trans{\ell}$\\
            \proofstep{\Cref{lem:traces_after_projection_subset}}\\
            $\project{\sigma'}{L_s^\delta}\in\utraces{s}\land \project{\sigma'}{L_e^\delta}\in\utraces{e}  \; \land$\\
            $\forall q_s \in Q_s, q_e \in Q_e: s \parcomp e \Trans{\sigma'} q_s \parcomp q_e \implies q_s \parcomp q_e \Trans{\ell}\;\land$\\
            $s \after (\project{\sigma'}{L_s^\delta}) \subseteq \pi_1(s\parcomp e \after \sigma')\;\land$\\
            $e \after (\project{\sigma'}{L_e^\delta}) \subseteq \pi_2(s\parcomp e \after \sigma)$\\
            \proofstep{\Cref{def:after}: $\after$ and definition $\subseteq$}\\
            $\project{\sigma'}{L_s^\delta}\in\utraces{s}\land \project{\sigma'}{L_e^\delta}\in\utraces{e}  \; \land$\\
            $\forall q_s \in Q_s, q_e \in Q_e: s \parcomp e \Trans{\sigma'} q_s \parcomp q_e \implies q_s \parcomp q_e \Trans{\ell}\;\land$\\
            $\forall q_s' \in Q_s : s \Trans{\project{\sigma'}{L_s^\delta}} q_s' \implies (\exists q_e'\in Q_e: s \parcomp e \Trans{\sigma'} q_s' \parcomp q_e')\;\land$\\
            $\forall q_e'' \in Q_e : e \Trans{\project{\sigma'}{L_e^\delta}} q_e'' \implies (\exists q_s''\in Q_s: s \parcomp e \Trans{\sigma'} q_s'' \parcomp q_e'')$\\
            \proofstep{$\forall$ elimination and $(A \implies B) \land (B \implies C) \implies (A \implies C)$}\\
            $\project{\sigma'}{L_s^\delta}\in\utraces{s}\land \project{\sigma'}{L_e^\delta}\in\utraces{e}  \; \land$\\
            $\forall q_s' \in Q_s:  s \Trans{ \project{\sigma'}{L_s^\delta}} q_s' \implies (\exists q_e' \in Q_e: q_s' \parcomp q_e' \Trans{\ell})\;\land$\\
            $\forall q_e'' \in Q_e:  e \Trans{ \project{\sigma'}{L_s^\delta}} q_e'' \implies (\exists q_s'' \in Q_s: q_s'' \parcomp q_e'' \Trans{\ell})$\\
            \proofstep{\Cref{lem:project_from_parcomp_light} and \cref{def:projection}: $\projectop$}\\
            $\project{\sigma'}{L_s^\delta}\in\utraces{s}\land \project{\sigma'}{L_e^\delta}\in\utraces{e}  \; \land$\\
            $\forall q_s' \in Q_s:  s \Trans{ \project{\sigma'}{L_s^\delta}} q_s' \implies q_s' \Trans{\ell}\;\land$\\
            $\forall q_e'' \in Q_e:  e \Trans{ \project{\sigma'}{L_s^\delta}} q_e'' \implies q_e'' \Trans{\ell})$\\
            \proofstep{\Cref{def:uioco}: $\utraces{}$}\\
            $\project{\sigma'}{L_e^\delta}\cdot\ell \in\utraces{e}  \; \land$\\
            $\project{\sigma'}{L_s^\delta}\cdot\ell \in \utraces{s}$\\
            \proofstep{\Cref{def:projection}: $\projectop$}\\
            $\project{\sigma'\cdot\ell}{L_e^\delta}\in\utraces{e}  \; \land$\\
            $\project{\sigma'\cdot\ell}{L_s^\delta} \in \utraces{s}$\\
        \item[$\ell \in U_s \setminus L_e$:]\ \\
            $i_s \uioco s \;\land$\\
            $\sigma'\cdot\ell\in\utraces{s\parcomp e} \land i_s \parcomp i_e \Trans{\sigma'\cdot\ell}$\\
            \proofstep{\Cref{item:utrace_prefix_closed}}\\
            $i_s \uioco s \;\land$\\
            $\sigma'\in\utraces{s\parcomp e}\land i_s \parcomp i_e \Trans{\sigma'\cdot\ell}$\\
            \proofstep{Apply IH}\\
            $i_s \uioco s \;\land$\\
            $\project{\sigma'}{L_s^\delta}\in\utraces{s}\land \project{\sigma'}{L_e^\delta}\in\utraces{e}  \land i_s \parcomp i_e \Trans{\sigma'\cdot\ell}$\\
            \proofstep{\Cref{lem:project_from_parcomp_IOTS}}\\
            $i_s \uioco s \;\land$\\
            $\project{\sigma'}{L_s^\delta}\in\utraces{s}\land \project{\sigma'}{L_e^\delta}\in\utraces{e}  \; \land$\\
            $i_s \Trans{\project{\sigma'\cdot\ell}{L_s^\delta}}$\\
            \proofstep{\Cref{def:projection}:$\projectop$}\\
            $i_s \uioco s \;\land$\\
            $\project{\sigma'}{L_s^\delta}\in\utraces{s}\land \project{\sigma'}{L_e^\delta}\in\utraces{e}  \; \land$\\
            $i_s \Trans{\project{\sigma'}{L_s^\delta}\cdot\ell}$\\
            \proofstep{\Cref{def:uioco}: $\uioco$}\\
            $\project{\sigma'}{L_s^\delta}\in\utraces{s}\land \project{\sigma'}{L_e^\delta}\in\utraces{e}  \; \land$\\
            $s \Trans{\project{\sigma'}{L_s^\delta}\cdot\ell}$\\
            \proofstep{\Cref{def:uioco}: $\utraces{}$}\\
            $\project{\sigma'}{L_s^\delta}\cdot\ell\in\utraces{s}\land \project{\sigma'}{L_e^\delta}\in\utraces{e}$\\
            \proofstep{\Cref{def:projection}:$\projectop$}\\
            $\project{\sigma'\cdot\ell}{L_s^\delta}\in\utraces{s}\land \project{\sigma'\cdot\ell}{L_e^\delta}\in\utraces{e}$\\
        \item[$\ell \in U_e \setminus L_s$:] Symmetrical to previous case
        \item[$\ell \in U_s \cap I_e$:]\ \\
            $i_s \uioco s \land i_s \mutuallyaccepts{} e\;\land$\\
            $\sigma'\cdot\ell\in\utraces{s\parcomp e} \land i_s \parcomp i_e \Trans{\sigma'\cdot\ell}$\\
            \proofstep{\Cref{item:utrace_prefix_closed}}\\
            $i_s \uioco s \land i_s \mutuallyaccepts{} e\;\land$\\
            $\sigma'\in\utraces{s\parcomp e}\land i_s \parcomp i_e \Trans{\sigma'\cdot\ell}$\\
            \proofstep{Apply IH}\\
            $i_s \uioco s \land i_s \mutuallyaccepts{}  e\;\land$\\
            $\project{\sigma'}{L_s^\delta}\in\utraces{s}\land \project{\sigma'}{L_e^\delta}\in\utraces{e}  \land i_s \parcomp i_e \Trans{\sigma'\cdot\ell}$\\
            \proofstep{\Cref{lem:project_from_parcomp_IOTS}}\\
            $i_s \uioco s \land i_s \mutuallyaccepts{} e\;\land$\\
            $\project{\sigma'}{L_s^\delta}\in\utraces{s}\land \project{\sigma'}{L_e^\delta}\in\utraces{e}  \; \land$\\
            $i_s \Trans{\project{\sigma'\cdot\ell}{L_s^\delta}}$\\
            \proofstep{\Cref{def:projection}:$\projectop$}\\
            $i_s \uioco s \land i_s \mutuallyaccepts{} e\;\land$\\
            $\project{\sigma'}{L_s^\delta}\in\utraces{s}\land \project{\sigma'}{L_e^\delta}\in\utraces{e}  \; \land$\\
            $i_s \Trans{\project{\sigma'}{L_s^\delta}\cdot\ell}$\\
            \proofstep{\Cref{def:uioco}: $\uioco$}\\
            $i_s \mutuallyaccepts{} e\;\land$\\
            $\project{\sigma'}{L_s^\delta}\in\utraces{s}\land \project{\sigma'}{L_e^\delta}\in\utraces{e}  \; \land$\\
            $i_s \Trans{\project{\sigma'}{L_s^\delta}\cdot\ell} \land\; s \Trans{\project{\sigma'}{L_s^\delta}\cdot\ell}$\\
            \proofstep{\Cref{def:uioco}: $\utraces{}$}\\
            $i_s \mutuallyaccepts{} e\;\land$\\
            $\project{\sigma'}{L_s^\delta}\cdot\ell\in\utraces{s}\land \project{\sigma'}{L_e^\delta}\in\utraces{e}  \; \land$\\
            $i_s \Trans{\project{\sigma'}{L_s^\delta}\cdot\ell}$\\
            \proofstep{\Cref{item:trans_transitive}}\\
            $i_s \mutuallyaccepts{} e\;\land$\\
            $\project{\sigma'}{L_s^\delta}\cdot\ell\in\utraces{s}\land \project{\sigma'}{L_e^\delta}\in\utraces{e}  \; \land$\\
            $\exists q_{i_s} \in Q_{i_s}: i_s \Trans{\project{\sigma'}{L_s^\delta}} q_{i_s}\land q_{i_s} \Trans{\ell}$\\
            \proofstep{\Cref{lem:straces_equals_utraces_iots}}\\
            $i_s \mutuallyaccepts{} e\;\land$\\
            $\project{\sigma'}{L_s^\delta}\cdot\ell\in\utraces{s}\land \project{\sigma'}{L_e^\delta}\in\utraces{e} \land \project{\sigma'}{L_s^\delta}\in \utraces{i_s}\;\land $\\
            $\exists q_{i_s} \in Q_{i_s}: i_s \Trans{\project{\sigma'}{L_s^\delta}} q_{i_s}\land q_{i_s} \Trans{\ell}$\\
            \proofstep{\Cref{lem:combine_utrace_imp}}\\
            $i_s \mutuallyaccepts{} e\;\land$\\
            $\project{\sigma'\cdot\ell}{L_s^\delta}\in\utraces{s}\land \project{\sigma'}{L_e^\delta}\in\utraces{e}  \land \sigma'\in \utraces{i_s\parcomp e}\; \land$\\
            $\exists q_{i_s} \in Q_{i_s}: i_s \Trans{\project{\sigma'}{L_s^\delta}} q_{i_s}\land q_{i_s} \Trans{\ell}$\\
            \proofstep{\Cref{def:mutually_accepts,def:accepting}: $\mutuallyaccepts{}$}\\
            $\project{\sigma'\cdot\ell}{L_s^\delta}\in\utraces{s}\land \project{\sigma'}{L_e^\delta}\in\utraces{e}\; \land$\\
            $\exists q_{i_s} \in Q_{i_s}: i_s \Trans{\project{\sigma'}{L_s^\delta}} q_{i_s}\land q_{i_s} \Trans{\ell}$\\
            $\forall q_{i_s}' \in Q_{i_s}, q_e' \in Q_e: i_s \parcomp e \Trans{\sigma'} q_s' \parcomp q_e' \implies \outset{q_{i_s}'}\cap I_e \subseteq \inset{q_e'} \cap U_s$\\
            \proofstep{\Cref{lem:project_from_parcomp_light}}\\
            $\project{\sigma'\cdot\ell}{L_s^\delta}\in\utraces{s}\land \project{\sigma'}{L_e^\delta}\in\utraces{e}\; \land$\\
            $\exists q_{i_s} \in Q_{i_s}: i_s \Trans{\project{\sigma'}{L_s^\delta}} q_{i_s}\land q_{i_s} \Trans{\ell}$\\
            $\forall q_{i_s}' \in Q_{i_s}, q_e' \in Q_e: $\\
            \tab$i_s \Trans{\project{\sigma'}{L_{i_s^\delta}}} q_{i_s}' \land e \Trans{\project{\sigma'}{L_e^\delta}} q_e' \implies \outset{q_{i_s}'}\cap I_e \subseteq \inset{q_e'} \cap U_s$\\
            \proofstep{$\forall$ elimination}\\
            $\project{\sigma'\cdot\ell}{L_s^\delta}\in\utraces{s}\land \project{\sigma'}{L_e^\delta}\in\utraces{e}\; \land$\\
            $\exists q_{i_s} \in Q_{i_s}: i_s \Trans{\project{\sigma'}{L_s^\delta}} q_{i_s}\land q_{i_s} \Trans{\ell}$\\
            $\forall q_e' \in Q_e: $\\
            \tab$i_s \Trans{\project{\sigma'}{L_{i_s}^\delta}} q_{i_s} \land e \Trans{\project{\sigma'}{L_e^\delta}} q_e' \implies \outset{q_{i_s}}\cap I_e \subseteq \inset{q_e'} \cap U_s$\\
            \proofstep{$A\land (A\land B \implies C) \implies (B\implies c)$}\\
            $\project{\sigma'\cdot\ell}{L_s^\delta}\in\utraces{s}\land \project{\sigma'}{L_e^\delta}\in\utraces{e}\; \land$\\
            $\exists q_{i_s} \in Q_{i_s}: q_{i_s} \Trans{\ell}$\\
            $\forall q_e' \in Q_e: e \Trans{\project{\sigma'}{L_e^\delta}} q_e' \implies \outset{q_{i_s}}\cap I_e \subseteq \inset{q_e'} \cap U_s$\\
            \proofstep{\Cref{def:outset}: $\outset{}$ and definition $\subseteq$}\\
            $\project{\sigma'\cdot\ell}{L_s^\delta}\in\utraces{s}\land \project{\sigma'}{L_e^\delta}\in\utraces{e}\; \land$\\
            $\forall q_e' \in Q_e: e \Trans{\project{\sigma'}{L_e^\delta}} q_e' \implies \ell \in \inset{q_e'} \cap U_s$\\
            \proofstep{\Cref{def:inset}: $\inset{}$}\\
            $\project{\sigma'\cdot\ell}{L_s^\delta}\in\utraces{s}\land \project{\sigma'}{L_e^\delta}\in\utraces{e}\; \land$\\
            $\forall q_e' \in Q_e: e \Trans{\project{\sigma'}{L_e^\delta}} q_e' \implies q_e'\Trans{\ell}$\\
            \proofstep{\Cref{def:uioco}: $\utraces{}$}\\
            $\project{\sigma'\cdot\ell}{L_s^\delta}\in\utraces{s}\land \project{\sigma'}{L_e^\delta}\cdot\ell\in\utraces{e}$\\
            \proofstep{\Cref{def:projection}: $\projectop$}\\
            $\project{\sigma'\cdot\ell}{L_s^\delta}\in\utraces{s}\land \project{\sigma'\cdot\ell}{L_e^\delta}\in\utraces{e}$\\

        \item[$\ell \in U_e \cap I_s$:] Symmetrical to previous case
        \item[$\ell = \delta$:]\ \\
            $i_s \uioco s \land i_e \uioco e\;\land$\\
            $\sigma'\cdot\delta\in\utraces{s\parcomp e} \land i_s \parcomp i_e \Trans{\sigma'\cdot\delta}$\\
            \proofstep{\Cref{item:utrace_prefix_closed}}\\
            $i_s \uioco s \land i_e \uioco e\;\land$\\
            $\sigma'\in\utraces{s\parcomp e}\land i_s \parcomp i_e \Trans{\sigma'\cdot\delta}$\\
            \proofstep{Apply IH}\\
            $i_s \uioco s \land i_e \uioco e\;\land$\\
            $\project{\sigma'}{L_s^\delta}\in\utraces{s}\land \project{\sigma'}{L_e^\delta}\in\utraces{e}  \land i_s \parcomp i_e \Trans{\sigma'\cdot\delta}$\\
            \proofstep{\Cref{lem:project_from_parcomp_IOTS}}\\
            $i_s \uioco s \land i_e \uioco e\;\land$\\
            $\project{\sigma'}{L_s^\delta}\in\utraces{s}\land \project{\sigma'}{L_e^\delta}\in\utraces{e}  \; \land$\\
            $i_s \Trans{\project{\sigma'\cdot\delta}{L_s^\delta}} \land\; i_e \Trans{\project{\sigma'\cdot\delta}{L_e^\delta}}$\\
            \proofstep{\Cref{def:projection}:$\projectop$}\\
            $i_s \uioco s \land i_e \uioco e\;\land$\\
            $\project{\sigma'}{L_s^\delta}\in\utraces{s}\land \project{\sigma'}{L_e^\delta}\in\utraces{e}  \; \land$\\
            $i_s \Trans{\project{\sigma'}{L_s^\delta}\cdot\delta} \land\; i_e \Trans{\project{\sigma'}{L_e^\delta}\cdot\delta}$\\            
            \proofstep{\Cref{def:uioco}: $\uioco$}\\
            $\project{\sigma'}{L_s^\delta}\in\utraces{s}\land \project{\sigma'}{L_e^\delta}\in\utraces{e}  \; \land$\\
            $s \Trans{\project{\sigma'}{L_s^\delta}\cdot\delta} \land\; e \Trans{\project{\sigma'}{L_e^\delta}\cdot\delta}$\\
            \proofstep{\Cref{def:uioco}: $\utraces{}$}\\
            $\project{\sigma'}{L_s^\delta}\cdot\delta\in\utraces{s}\land \project{\sigma'}{L_e^\delta}\cdot\delta\in\utraces{e}$\\
            \proofstep{\Cref{def:projection}:$\projectop$}\\
            $\project{\sigma'\cdot\delta}{L_s^\delta}\in\utraces{s}\land \project{\sigma'\cdot\delta}{L_e^\delta}\in\utraces{e}$\\

        \end{case_distinction}

\end{case_distinction}
\end{proof}

\FloatBarrier

\subsection{Algorithm Correctness}
\label{sec:algorithm_correctness}
\newcommand{\GenInputResponseFuncName}{HandleSutOutput}
\SetKwFunction{\GenInputResponseFuncName}{\GenInputResponseFuncName}

This section will discuss the correctness of \Cref{alg:test_eco}. It is difficult to prove the correctness of an imperative and interactive program. We make this simpler by defining a non-interactive variant of the algorithm (\Cref{alg:testcase_alg}) and proving this variant to be correct.

The main difference with \Cref{alg:test_eco} is that instead of interacting with the SUT, it non-deterministically generates a test case. Running this test case with the SUT then corresponds to a possible run of \Cref{alg:test_eco}. \Cref{alg:testcase_alg} covers the same non-deterministic cases as \Cref{alg:test_eco}. A difference is that \Cref{alg:test_eco} implicitly assumes it can switch to handling an output from the $SUT$ if it receives one. \Cref{alg:testcase_alg} makes this explicit by adding transitions for things that could happen outside of the control of the test case to every state, using the $\HandleSutOutput$ function. This essentially moves the non $\delta$ part of case C from \Cref{alg:test_eco} into the $\HandleSutOutput$ function, while case C from \Cref{alg:testcase_alg} specifically waits until quiescence occurs.

By running \Cref{alg:testcase_alg} multiple times, we can generate a test suite. We denote the test suite generated by \Cref{alg:testcase_alg} for a label set $L_i$ and environment specification $e$ as $\TTS[L_i,e]$.

\LinesNotNumbered
\begin{function}[htb]
\DontPrintSemicolon
\functionLabel{GenEcoTest}
\caption{\currentFunName{}(\ProcArgFnt$X_e:\powerset{Q_e}$) }
    \KwData{$e\in\LTS$}
    \KwData{Inputs $I_i$ and outputs $U_i$ of the SUT}
    \KwOut{A partial test case for $i\eco e$, assuming $e$ could be in any state $q\in X_e$}\ \\
    non-deterministically execute one of the following cases:\\
    \textbf{\textit{(A) Stop testing:}}\\
    \Indp
        \nl \Return{$\passState$}\;
    \Indm
    \textbf{\textit{(B) Perform an input $\ell$ on the SUT:}}\\
    \Indp
        \nl choose $\ell \in (\outset{X_e} \cup \inset{X_e})\cap I_i$\;
        \nl \Return \;
        \Indp
            \nl\label{line:sync_out} $\ell\; \Bcomp$ \GenEcoTest{$X_e \after \ell$} $\Bchoice$\;
            \nl\HandleSutOutput{$X_e$}\;
        \Indm
    \Indm
     \textbf{\textit{(C) Observe an output or quiescence from the SUT (Only if $\delta\in\outset{X_e})$:}}\\
    \Indp
    \nl \Return \;
        \Indp
            \nl\label{line:sync_theta}$\theta\; \Bcomp\;$\GenEcoTest{$X_e\after\delta$} $\;\Bchoice$\;
            \nl\HandleSutOutput{$X_e$}\;
            
        \Indm
    \Indm
    \textbf{\textit{(D) Simulate non-interacting behaviour in environment}}\\
    \Indp
        \nl choose $\ell \in (\outset{X_e} \cup \inset{X_e})\setminus L_i^\delta$ \; 
        \nl \Return\;
        \Indp
            \nl\label{line:env_internal_step}$\ell\Bcomp$ \GenEcoTest{$X_e \after \ell$}$\;\Bchoice$\;
            \nl\HandleSutOutput{$X_e$}\;
        \Indm
    \Indm
     \textbf{\textit{(E) Generate non-interacting behaviour of the SUT}}\\
    \Indp
        \nl choose $\ell \in I_i \setminus L_e$ \; 
        \nl \Return\;
        \Indp
            \nl\label{line:sut_internal_step}$\ell\Bcomp$ \GenEcoTest{$X_e$}$\;\Bchoice$\;
            \nl\HandleSutOutput{$X_e$}\;
        \Indm
    \Indm
\end{function}

\begin{function}
\DontPrintSemicolon
\functionLabel{\GenInputResponseFuncName}
\label[FunctionFormat\GenInputResponseFuncName]{fun:GenInputResponse}
\caption{\currentFunName{}(\ProcArgFnt$X_e:\powerset{Q_e}$) }
    \KwData{$e\in\LTS$}
    \KwData{Inputs $I_i$ and outputs $U_i$ of the SUT}
    \KwOut{A partial test case describing how to respond to an output from the $SUT$}
    \Return \;
        \Indp
            \nl\label{line:failtrans}%
            $\Bsum \{\ell\Bcomp \failState \setbar \ell \in  (U_i \cap I_e) \setminus\inset{X_e} \} \Bchoice$\;
            \nl\label{line:sync_input}$\Bsum \{\ell\; \Bcomp\; $\GenEcoTest{$X_e\after \ell$} $\setbar \ell \in U_i\cap\inset{X_e}\}$ \;
        \Indm
\end{function}

\begin{algorithm}
    \caption{$\eco$ test case generation}
    \label{alg:testcase_alg}
    \KwIn{A specification of the environment $e\in \LTS$}
    \KwData{Inputs $I_i$ and outputs $U_i$ of the SUT}
    \KwOut{A test case for $i\eco e$}
    
    \Return\\
    \Indp
        test case $t$ with $I_t=(U_i\cap I_e)\cup \{\theta\}$, $U_t = U_e \cup (I_e\setminus U_i)  \cup (I_i\setminus L_e)$.\\
        $Q_t$ and $T_t$ are given by \GenEcoTest{$e \after \epsilon$}\;
    \Indm
\end{algorithm}
\LinesNumbered

\begin{samepage}
\begin{example}
 \Cref{subfig:testcase_behaviourexpr} shows one possible output of running \Cref{alg:testcase_alg} on \Cref{imp:eco_testing_example} using \Cref{spec:eco_testing_example} as the environment specification. The LTS that is represented by the behaviour expression of \Cref{subfig:testcase_behaviourexpr} is shown in \Cref{subfig:testcase_lts}. This particular example represents the test of waiting for output $\ltslabel{a}$ (Case C), then giving input \ltslabel{r} (Case B), and then waiting for a final output again (Case C). The extra \ltslabel{a} transitions to $\failState$ are created by the \HandleSutOutput function, which adds transitions for every possible output of the SUT to every state of the testcase, in case the SUT gives an output before the testcase can perform the next input. Most testcases could be continued indefinitely, but at some point testing must stop, which is made explicit by taking a transition to $\passState$ (Case A). Each of these passes could be replaced with more test transitions to make the testcase take more possibilities into account. Notably, \Cref{imp:eco_testing_example} is shown here for clarity, but it is not actually used when constructing the testcase in \Cref{subfig:testcase_behaviourexpr}. \Cref{alg:testcase_alg} only takes the labels of \Cref{imp:eco_testing_example} into account, and not the actual transition structure which is considered a black box. This is why there is a $\theta$ transition at the start of the generated testcase, even though \Cref{imp:eco_testing_example} is not quiescent there.
\end{example}
\end{samepage}

\begin{toappendix}
    \begin{lemmarep}
        \label{lem:L_testcase}
        let $t\in \TTS(L_i,e)$\\ 
        $L_t = (U_i\cap I_e)\cup \{\theta\} \cup U_e \cup (I_e\setminus U_i)  \cup (I_i\setminus L_e) = L_e \cup \{\theta\} \cup (I_i\setminus L_e)$
    \end{lemmarep}
    \begin{proof}
        Follows directly from the definition of \Cref{alg:testcase_alg}
    \end{proof}
\end{toappendix}

\equalizeCounters
\begin{figure}[htb]
    \begin{subfigure}[b]{.50\linewidth}
    \centering
    \newimplabel\label{imp:eco_testing_example}
    \begin{tikzpicture}[LTS]
    \node[state,initial above] (1) {1};
    \node[state,below= of 1] (2) {2};

    \path[->] 
        (1) edge [bend left]    node [auto] {\ltslabel{!a}} (2)
            edge [loop right]   node [auto] {\ltslabel{?b}\\\ltslabel{?r}} (1)
        (2) edge [bend left]    node [auto] {\ltslabel{?b}\\\ltslabel{?r}} (1)
        ;
    
    \end{tikzpicture}
    \caption{\currentimp, with I=\{\ltslabel{b},\ltslabel{r}\} and U=\{\ltslabel{a}\}}
    \end{subfigure}
    \qquad
    \begin{subfigure}[b]{.5\linewidth}
    \centering
    \newspeclabel\label{spec:eco_testing_example}
    \begin{tikzpicture}[LTS]
    \node[state,initial] (A) {A};
    \node[state,below=of A] (B) {B};

    \path[->] 
        (A) edge [bend left] node [auto] {\ltslabel{?a}} (B)
        (B) edge [bend left] node [auto] {\ltslabel{!b}} (A)
        ;
    
    \end{tikzpicture}
    \caption{\currentspec, with  I=\{\ltslabel{a}\} and U=\{\ltslabel{b}\}}
    \end{subfigure}
    \\

    \begin{subfigure}[b]{.50\linewidth}
    \[\begin{array}{l}
        \theta \Bcomp \passState \Bchoice
        \Bsum \emptyset \Bchoice
        \Bsum \{\ltslabel{a} \Bcomp 
            (\\\tab
                \Bsum \{\ltslabel{a} \Bcomp \failState\} \Bchoice
                \Bsum \emptyset
                \Bchoice
                \ltslabel{r} \Bcomp 
                (\\\tab\tab
                    \theta \Bcomp \passState \Bchoice
                    \Bsum \emptyset \Bchoice 
                    \Bsum \{\ltslabel{a} \Bcomp \failState\}
                \\\tab)
            \\)
        \}
    \end{array}
    \]
    \caption{Possible output of \Cref{alg:testcase_alg} 
    }
    \label{subfig:testcase_behaviourexpr}
    \end{subfigure}
    \qquad
    \begin{subfigure}[b]{.5\linewidth}
    \begin{tikzpicture}[LTS]
    \setlength{\nodedistance}{0.7\nodedistance}
    \node[state,initial above] (1) {1};
    \node[below left= of 1] (2) {\passState};
    \node[state,below right = of 1] (3) {2};
    \node[below left = of 3] (4) {\failState};
    \node[state,below right = of 3] (5) {3};
    \node[below left = of 5] (6) {\passState};
    \node[below right = of 5] (7) {\failState};

    \path[->] 
        (1) edge []    node [auto] {$\theta$} (2)
            edge []   node [auto] {\ltslabel{?a}} (3)
        (3) edge []    node [auto] {\ltslabel{?a}} (4)
            edge []    node [auto] {\ltslabel{!r}} (5)
        (5) edge []    node [auto] {$\theta$} (6)
            edge []    node [auto] {\ltslabel{?a}} (7)
        ;
    
    \end{tikzpicture}
    
    \caption{LTS representation of \Cref{subfig:testcase_behaviourexpr}}
    \label{subfig:testcase_lts}
    \end{subfigure}
   
    \caption{Example of running \Cref{alg:testcase_alg}}
    \label{fig:eco_testing_example}
\end{figure}

\subsubsection{Proof Sketch}
We prove that the test suite generated by \Cref{alg:testcase_alg} is complete. This means that if an implementation is not $\eco$-conformant to a specification, then \Cref{alg:testcase_alg} can generate a test case that can fail when executed on that implementation. Conversely, none of the test cases generated can fail on an implementation that is $\eco$-conformant.

The core of the proof is to show a correspondence between the lines of $\GenEcoTest$ and \Cref{def:eco_rel}.
This mostly involves rewriting traces between the various transition systems involved. \Cref{lem:project_from_testexec} rewrites between traces through the test execution, the implementation, and the test case. \Cref{lem:project_to_testexec} does the same for traces through parallel composition and test execution. These two lemmas are glued together by \Cref{lem:project_single_trace_from_parcomp}, which uses a more generalised notion of $\mutuallyaccepts{}$, as expressed in \Cref{def:accepting_traces}. Instead of requiring inputs to be defined for all states reachable by $\utraces{}$, \Cref{def:accepting_traces} only requires inputs to be defined for one specific trace. The full proof involves a lot of boilerplate cases, so we only show the core lemmas here and refer to the appendix for the details.

\begin{toappendix}
    \begin{lemmarep}
        \label{lem:after_subset_transitive}
        Let $s\in\LTS$, $\sigma \in L_s^{\delta*}$, $X_s,Y_s\subseteq Q_s$
        \[X_s \subseteq Y_s \implies X_s\after\sigma \subseteq Y_s\after\sigma\]
    \end{lemmarep}
    \begin{proof}
        \ \\
        Take $q\in X_s\after\sigma$. To prove: $q \in Y_s\after\sigma$\\
$q\in X_s\after\sigma \land X_s\subseteq Y_s$\\
\proofstep{\Cref{def:after}: $\after$}\\
$\exists q'\in X_s: q' \Trans{\sigma} q\land X_s\subseteq Y_s$\\
\proofstep{def $\subseteq$}\\
$\exists q'\in Y_s: q' \Trans{\sigma} q$\\
\proofstep{\Cref{def:after}: $\after$}\\
$q\in Y_s\after\sigma$
    \end{proof}
\end{toappendix}

\begin{lemmarep}
    \label{lem:project_from_testexec}
    Let $i\in \IOTS$, $t\in \TTS$, $\sigma\in L_{t\testexec i}^*$, $q_i, q_i'\in Q_i$, $q_t, q_t'\in Q_t$, then
    \[ q_t \testexec q_i \Trans{\sigma} q_t' \testexec q_i' \iff q_t \Trans{\project{\sigma}{L_t}} q_t' \land q_i \Trans{\project{\subst{\sigma}{\theta}{\delta}}{L_i^\delta}} q_i'\]
\end{lemmarep}

\begin{proof}
    \begin{case_distinction}
\item[$\implies:$]
Proof by induction on $\sigma$. The base case is $\sigma=\epsilon$, and the IH is
\[ q_t\testexec q_i \Trans{\sigma'} q_t' \testexec q_i' \implies q_t \Trans{\project{\sigma'}{L_t}} q_t' \land q_i \Trans{\project{\subst{\sigma'}{\theta}{\delta}}{L_i^\delta}} q_i'\]
To prove: the lemma holds for $\sigma=\sigma'\cdot\ell$, $\ell\in L_{q_t \testexec q_i}$. The proof follows from further case distinction on $\ell$ based on \Cref{def:testexec}.\\
\begin{case_distinction}
\item[$\sigma=\epsilon:$]\ \\
    $q_t \testexec q_i \Trans{\epsilon} q_t' \testexec q_i'$\\
    \proofstep{\Cref{def:testexec}: $\testexec$}\\
    $q_t = q_t' \land q_i \Trans{\epsilon} q_i'$\\
    \proofstep{\Cref{def:arrowdefs}: $\Trans{\epsilon}$}\\
    $q_t \Trans{\epsilon} q_t' \land q_i \Trans{\epsilon} q_i'$\\
    \proofstep{\Cref{def:projection,def:substitution}: $\projectop$ and $substitution$}\\
    $q_t \Trans{\project{\epsilon}{L_t}} q_t' \land q_i \Trans{\project{\subst{\epsilon}{\theta}{\delta}}{L_i^\delta}} q_i'$\\
\item[$\sigma=\sigma'\cdot\ell$, $\ell\in L_i\setminus L_t:$]\ \\
    $q_t \testexec q_i \Trans{\sigma'\cdot\ell} q_t' \testexec q_i'$\\
    \proofstep{\Cref{def:arrowdefs}: $\Trans{}$}\\
    $\exists q_t'', q_t'''\in Q_t, q_i'', q_i'''\in Q_i: q_t \testexec q_i \Trans{\sigma'} q_t'' \testexec q_i'' \trans{\ell} q_t'''\testexec q_i'''\Trans{\epsilon} q_t'\testexec q_i'$\\
    \proofstep{Apply IH}\\
    $\exists q_t'', q_t'''\in Q_t, q_i'', q_i'''\in Q_i:q_t'' \testexec q_i'' \trans{\ell} q_t'''\testexec q_i'''\Trans{\epsilon} q_t'\testexec q_i'\;\land$\\
    $q_t \Trans{\project{\sigma'}{L_t}} q_t'' \land q_i \Trans{\project{\subst{\sigma'}{\theta}{\delta}}{L_i^\delta}} q_i''$\\
    \proofstep{\Cref{def:testexec}: $\testexec$. Also means $\ell\in U_i$}\\
    $\exists q_t'', q_t'''\in Q_t, q_i'', q_i'''\in Q_i:q_t'''\testexec q_i'''\Trans{\epsilon} q_t'\testexec q_i'\;\land$\\
    $q_t \Trans{\project{\sigma'}{L_t}} q_t'' \land q_i \Trans{\project{\subst{\sigma'}{\theta}{\delta}}{L_i^\delta}} q_i''\;\land$\\
    $q_t''=q_t'''\land q_i''\trans{\ell}q_i'''$\\
    \proofstep{\Cref{def:arrowdefs}:$\Trans{}$}\\
    $\exists q_t'', q_t'''\in Q_t, q_i'''\in Q_i:q_t'''\testexec q_i'''\Trans{\epsilon} q_t'\testexec q_i'\;\land$\\
    $q_t \Trans{\project{\sigma'}{L_t}} q_t'' \land q_i \Trans{\project{\subst{\sigma'}{\theta}{\delta}}{L_i^\delta}\cdot\ell} q_i'''\;\land$\\
    $q_t''=q_t'''$\\
    \proofstep{Rewrite using $q_t''=q_t'''$}\\
    $\exists q_t'''\in Q_t, q_i'''\in Q_i:q_t'''\testexec q_i'''\Trans{\epsilon} q_t'\testexec q_i'\;\land$\\
    $q_t \Trans{\project{\sigma'}{L_t}} q_t''' \land q_i \Trans{\project{\subst{\sigma'}{\theta}{\delta}}{L_i^\delta}\cdot\ell} q_i'''$\\
    \proofstep{Apply base case}\\
    $\exists q_t'''\in Q_t, q_i'''\in Q_i:q_t''' \Trans{\epsilon} q_t' \land q_i''' \Trans{\epsilon} q_i'\;\land$\\
    $q_t \Trans{\project{\sigma'}{L_t}} q_t''' \land q_i \Trans{\project{\subst{\sigma'}{\theta}{\delta}}{L_i^\delta}\cdot\ell} q_i'''$\\
    \proofstep{\Cref{def:arrowdefs}: $\Trans{}$}\\
    $q_t \Trans{\project{\sigma'}{L_t}} q_t' \land q_i \Trans{\project{\subst{\sigma'}{\theta}{\delta}}{L_i^\delta}\cdot\ell} q_i'$\\
    \proofstep{\Cref{def:projection,def:substitution}: $\projectop$ and $substitution$}\\
    $q_t \Trans{\project{\sigma'\cdot\ell}{L_t}} q_t' \land q_i \Trans{\project{\subst{\sigma'\cdot\ell}{\theta}{\delta}}{L_i^\delta}} q_i'$\\
\item[$\sigma=\sigma'\cdot\ell$, $\ell\in L_t\setminus (L_i\cup \{\theta\}):$]\ \\
    $q_t \testexec q_i \Trans{\sigma'\cdot\ell} q_t' \testexec q_i'$\\
    \proofstep{\Cref{def:arrowdefs}: $\Trans{}$}\\
    $\exists q_t'', q_t'''\in Q_t, q_i'', q_i'''\in Q_i: q_t \testexec q_i \Trans{\sigma'} q_t'' \testexec q_i'' \trans{\ell} q_t'''\testexec q_i'''\Trans{\epsilon} q_t'\testexec q_i'$\\
    \proofstep{Apply IH}\\
    $\exists q_t'', q_t'''\in Q_t, q_i'', q_i'''\in Q_i:q_t'' \testexec q_i'' \trans{\ell} q_t'''\testexec q_i'''\Trans{\epsilon} q_t'\testexec q_i'\;\land$\\
    $q_t \Trans{\project{\sigma'}{L_t}} q_t'' \land q_i \Trans{\project{\subst{\sigma'}{\theta}{\delta}}{L_i^\delta}} q_i''$\\
    \proofstep{\Cref{def:testexec}: $\testexec$}\\
    $\exists q_t'', q_t'''\in Q_t, q_i'', q_i'''\in Q_i:q_t'''\testexec q_i'''\Trans{\epsilon} q_t'\testexec q_i'\;\land$\\
    $q_t \Trans{\project{\sigma'}{L_t}} q_t'' \land q_i \Trans{\project{\subst{\sigma'}{\theta}{\delta}}{L_i^\delta}} q_i''$\\
    $q_i''=q_i'''\land q_t''\trans{\ell}q_t'''\;\land$\\
    \proofstep{\Cref{def:arrowdefs}:$\Trans{}$}\\
    $\exists q_t'''\in Q_t, q_i'', q_i'''\in Q_i:q_t'''\testexec q_i'''\Trans{\epsilon} q_t'\testexec q_i'\;\land$\\
    $q_t \Trans{\project{\sigma'}{L_t}\cdot\ell} q_t''' \land q_i \Trans{\project{\subst{\sigma'}{\theta}{\delta}}{L_i^\delta}} q_i''\;\land$\\
    $q_i''=q_i'''$\\
    \proofstep{Rewrite using $q_i''=q_i'''$}\\
    $\exists q_t'''\in Q_t, q_i'''\in Q_i:q_t'''\testexec q_i'''\Trans{\epsilon} q_t'\testexec q_i'\;\land$\\
    $q_t \Trans{\project{\sigma'}{L_t}\cdot\ell} q_t''' \land q_i \Trans{\project{\subst{\sigma'}{\theta}{\delta}}{L_i^\delta}} q_i'''$\\
    \proofstep{Apply base case}\\
    $\exists q_t''\in Q_t, q_i'''\in Q_i:q_t''' \Trans{\epsilon} q_t' \land q_i''' \Trans{\epsilon} q_i'\;\land$\\
    $q_t \Trans{\project{\sigma'}{L_t}\cdot\ell} q_t''' \land q_i \Trans{\project{\subst{\sigma'}{\theta}{\delta}}{L_i^\delta}} q_i'''$\\
    \proofstep{\Cref{def:arrowdefs}: $\Trans{}$}\\
    $q_t \Trans{\project{\sigma'}{L_t}\cdot\ell} q_t' \land q_i \Trans{\project{\subst{\sigma'}{\theta}{\delta}}{L_i^\delta}} q_i'$\\
    \proofstep{\Cref{def:projection,def:substitution}: $\projectop$ and $substitution$}\\
    $q_t \Trans{\project{\sigma'\cdot\ell}{L_t}} q_t' \land q_i \Trans{\project{\subst{\sigma'\cdot\ell}{\theta}{\delta}}{L_i^\delta}} q_i'$\\
\item[$\sigma=\sigma'\cdot\ell$, $\ell\in L_i\cap L_t:$]\ \\
    $q_t \testexec q_i \Trans{\sigma'\cdot\ell} q_t' \testexec q_i'$\\
    \proofstep{\Cref{def:arrowdefs}: $\Trans{}$}\\
    $\exists q_t'', q_t'''\in Q_t, q_i'', q_i'''\in Q_i: q_t \testexec q_i \Trans{\sigma'} q_t'' \testexec q_i'' \trans{\ell} q_t'''\testexec q_i'''\Trans{\epsilon} q_t'\testexec q_i'$\\
    \proofstep{Apply IH}\\
    $\exists q_t'', q_t'''\in Q_t, q_i'', q_i'''\in Q_i:q_t'' \testexec q_i'' \trans{\ell} q_t'''\testexec q_i'''\Trans{\epsilon} q_t'\testexec q_i'\;\land$\\
    $q_t \Trans{\project{\sigma'}{L_t}} q_t'' \land q_i \Trans{\project{\subst{\sigma'}{\theta}{\delta}}{L_i^\delta}} q_i''$\\
    \proofstep{\Cref{def:testexec}: $\testexec$}\\
    $\exists q_t'', q_t'''\in Q_t, q_i'', q_i'''\in Q_i:q_t'''\testexec q_i'''\Trans{\epsilon} q_t'\testexec q_i'\;\land$\\
    $q_t \Trans{\project{\sigma'}{L_t}} q_t'' \land q_i \Trans{\project{\subst{\sigma'}{\theta}{\delta}}{L_i^\delta}} q_i''\;\land$\\
    $q_i''\trans{\ell}q_i'''\land q_t''\trans{\ell}q_t'''$\\
    \proofstep{\Cref{def:arrowdefs}:$\Trans{}$}\\
    $\exists q_t'''\in Q_t, q_i'''\in Q_i:q_t'''\testexec q_i'''\Trans{\epsilon} q_t'\testexec q_i'\;\land$\\
    $q_t \Trans{\project{\sigma'}{L_t}\cdot\ell} q_t''' \land q_i \Trans{\project{\subst{\sigma'}{\theta}{\delta}}{L_i^\delta}\cdot\ell} q_i'''$\\
    \proofstep{Apply base case}\\
    $\exists q_t''\in Q_t, q_i'''\in Q_i:q_t''' \Trans{\epsilon} q_t' \land q_i''' \Trans{\epsilon} q_i'\;\land$\\
    $q_t \Trans{\project{\sigma'}{L_t}\cdot\ell} q_t''' \land q_i \Trans{\project{\subst{\sigma'}{\theta}{\delta}}{L_i^\delta}\cdot\ell} q_i'''$\\
    \proofstep{\Cref{def:arrowdefs}: $\Trans{}$}\\
    $q_t \Trans{\project{\sigma'}{L_t}\cdot\ell} q_t' \land q_i \Trans{\project{\subst{\sigma'}{\theta}{\delta}}{L_i^\delta}\cdot\ell} q_i'$\\
    \proofstep{\Cref{def:projection,def:substitution}: $\projectop$ and $substitution$}\\
    $q_t \Trans{\project{\sigma'\cdot\ell}{L_t}} q_t' \land q_i \Trans{\project{\subst{\sigma'\cdot\ell}{\theta}{\delta}}{L_i^\delta}} q_i'$\\
\item[$\sigma=\sigma'\cdot\theta:$]\ \\
    $q_t \testexec q_i \Trans{\sigma'\cdot\theta} q_t' \testexec q_i'$\\
    \proofstep{\Cref{def:arrowdefs}: $\Trans{}$}\\
    $\exists q_t'', q_t'''\in Q_t, q_i'', q_i'''\in Q_i: q_t \testexec q_i \Trans{\sigma'} q_t'' \testexec q_i'' \trans{\theta} q_t'''\testexec q_i'''\Trans{\epsilon} q_t'\testexec q_i'$\\
    \proofstep{Apply IH}\\
    $\exists q_t'', q_t'''\in Q_t, q_i'', q_i'''\in Q_i:q_t'' \testexec q_i'' \trans{\theta} q_t'''\testexec q_i'''\Trans{\epsilon} q_t'\testexec q_i'\;\land$\\
    $q_t \Trans{\project{\sigma'}{L_t}} q_t'' \land q_i \Trans{\project{\subst{\sigma'}{\theta}{\delta}}{L_i^\delta}} q_i''$\\
    \proofstep{\Cref{def:testexec}: $\testexec$}\\
    $\exists q_t'', q_t'''\in Q_t, q_i'', q_i'''\in Q_i:q_t'''\testexec q_i'''\Trans{\epsilon} q_t'\testexec q_i'\;\land$\\
    $q_t \Trans{\project{\sigma'}{L_t}} q_t'' \land q_i \Trans{\project{\subst{\sigma'}{\theta}{\delta}}{L_i^\delta}} q_i''\;\land$\\
    $q_i''\trans{\delta}q_i'''\land q_t''\trans{\theta}q_t'''$\\
    \proofstep{\Cref{def:arrowdefs}:$\Trans{}$}\\
    $\exists q_t'''\in Q_t, q_i'''\in Q_i:q_t'''\testexec q_i'''\Trans{\epsilon} q_t'\testexec q_i'\;\land$\\
    $q_t \Trans{\project{\sigma'}{L_t}\cdot\theta} q_t''' \land q_i \Trans{\project{\subst{\sigma'}{\theta}{\delta}}{L_i^\delta}\cdot\delta} q_i'''$\\
    \proofstep{Apply base case}\\
    $\exists q_t''\in Q_t, q_i'''\in Q_i:q_t''' \Trans{\epsilon} q_t' \land q_i''' \Trans{\epsilon} q_i'\;\land$\\
    $q_t \Trans{\project{\sigma'}{L_t}\cdot\theta} q_t''' \land q_i \Trans{\project{\subst{\sigma'}{\theta}{\delta}}{L_i^\delta}\cdot\delta} q_i'''$\\
    \proofstep{\Cref{def:arrowdefs}: $\Trans{}$}\\
    $q_t \Trans{\project{\sigma'}{L_t}\cdot\theta} q_t' \land q_i \Trans{\project{\subst{\sigma'}{\theta}{\delta}}{L_i^\delta}\cdot\delta} q_i'$\\
    \proofstep{\Cref{def:projection,def:substitution}: $\projectop$ and $substitution$}\\
    $q_t \Trans{\project{\sigma'\cdot\theta}{L_t}} q_t' \land q_i \Trans{\project{\subst{\sigma'\cdot\theta}{\theta}{\delta}}{L_i^\delta}} q_i'$\\
\end{case_distinction}

\item[$\impliedby:$]
Proof by induction on $\sigma$. The base case is $\sigma=\epsilon$, and the IH is
\[q_t \Trans{\project{\sigma'}{L_t}} q_t' \land q_i \Trans{\project{\subst{\sigma'}{\theta}{\delta}}{L_i^\delta}} q_i' \implies q_t \testexec q_i \Trans{\sigma'} q_t' \testexec q_i'\]
To prove: the lemma holds for $\sigma=\sigma'\cdot\ell$, $\ell\in L_{q_t \testexec q_i}$. The proof follows from further case distinction on $\ell$ based on \Cref{def:testexec}.\\
\begin{case_distinction}

\item[$\sigma=\epsilon:$]\ \\
    $q_t \Trans{\project{\epsilon}{L_t}} q_t' \land q_i \Trans{\project{\subst{\epsilon}{\theta}{\delta}}{L_i^\delta}} q_i'$\\
    \proofstep{\Cref{def:projection,def:substitution}: $\projectop$ and $substitution$}\\
    $q_t \Trans{\epsilon} q_t' \land q_i \Trans{\epsilon} q_i'$\\
    \proofstep{\Cref{item:testcase_no_tau}: no $\tau$ transitions in $t$}\\
    $q_t = q_t' \land q_i \Trans{\epsilon} q_i'$\\
    \proofstep{\Cref{def:testexec}: $\testexec$}\\
    $q_t \testexec q_i \Trans{\epsilon} q_t' \testexec q_i'$\\

\item[$\sigma=\sigma'\cdot\ell$, $\ell\in I_i\setminus L_t:$] \ \\
    Empty case, as $I_i\setminus L_t \cup L_{i\testexec t} = \emptyset$ (\Cref{lem:L_testexec}).
\item[$\sigma=\sigma'\cdot\ell$, $\ell\in U_i\setminus L_t:$]\ \\
    $q_t \Trans{\project{\sigma'\cdot\ell}{L_t}} q_t' \land q_i \Trans{\project{\subst{\sigma'\cdot\ell}{\theta}{\delta}}{L_i^\delta}} q_i'$\\
    \proofstep{\Cref{def:projection,def:substitution}: $\projectop$ and $substitution$}\\
    $q_t \Trans{\project{\sigma'}{L_t}} q_t' \land q_i \Trans{\project{\subst{\sigma'}{\theta}{\delta}}{L_i^\delta}\cdot\ell} q_i'$\\
    \proofstep{\Cref{def:arrowdefs}: $\Trans{}$}\\
    $\exists q_i'', q_i'''\in Q_i: q_t \Trans{\project{\sigma'}{L_t}} q_t' \land q_i \Trans{\project{\subst{\sigma'}{\theta}{\delta}}{L_i^\delta}} q_i'' \trans{\ell} q_i'''\Trans{\epsilon} q_i'$\\
    \proofstep{Apply IH}\\
    $\exists q_i'', q_i'''\in Q_i: q_i'' \trans{\ell} q_i'''\Trans{\epsilon} q_i'\;\land$\\
    $q_t \testexec q_i \Trans{\sigma'} q_t'\testexec q_i''$\\
    \proofstep{\Cref{def:testexec}: $\testexec$}\\
    $\exists q_i'', q_i'''\in Q_i: q_t' \testexec q_i'' \trans{\ell} q_t' \testexec q_i''' \Trans{\epsilon} q_t' \testexec q_i'\;\land$\\
    $q_t \testexec q_i \Trans{\sigma'} q_t'\testexec q_i''$\\
    \proofstep{\Cref{def:arrowdefs}: $\Trans{}$}\\
    $q_t \testexec q_i \Trans{\sigma'\cdot\ell} q_t'\testexec q_i'$
\item[$\sigma=\sigma'\cdot\ell$, $\ell\in L_t\setminus (L_i\cup \{\theta\}):$]\ \\
    $q_t \Trans{\project{\sigma'\cdot\ell}{L_t}} q_t' \land q_i \Trans{\project{\subst{\sigma'\cdot\ell}{\theta}{\delta}}{L_i^\delta}} q_i'$\\
    \proofstep{\Cref{def:projection,def:substitution}: $\projectop$ and $substitution$}\\
    $q_t \Trans{\project{\sigma'}{L_t}\cdot\ell} q_t' \land q_i \Trans{\project{\subst{\sigma'}{\theta}{\delta}}{L_i^\delta}} q_i'$\\
    \proofstep{\Cref{def:arrowdefs}: $\Trans{}$}\\
    $\exists q_t'', q_t'''\in Q_t: q_t \Trans{\project{\subst{\sigma'}{\theta}{\delta}}{L_i^\delta}} q_t'' \trans{\ell} q_t'''\Trans{\epsilon} q_t' \land  i \Trans{\project{\sigma'}{L_t}} q_i'$\\
    \proofstep{Apply IH}\\
    $\exists q_t'', q_t'''\in Q_t: q_t'' \trans{\ell} q_t'''\Trans{\epsilon} q_t'\;\land$\\
    $q_t \testexec q_i \Trans{\sigma'} q_t''\testexec q_i'$\\
    \proofstep{\Cref{item:testcase_no_tau}: no $\tau$ transitions in $t$}\\
    $\exists q_t'', q_t'''\in Q_t: q_t'' \trans{\ell} q_t'\;\land$\\
    $q_t \testexec q_i \Trans{\sigma'} q_t''\testexec q_i'$\\
    \proofstep{\Cref{def:testexec}: $\testexec$}\\
    $\exists q_t'', q_t'''\in Q_t: q_t'' \testexec q_i' \trans{\ell} q_t' \testexec q_i'\;\land$\\
    $q_t \testexec q_i \Trans{\sigma'} q_t''\testexec q_i'$\\
    \proofstep{\Cref{def:arrowdefs}: $\Trans{}$}\\
    $q_t \testexec q_i \Trans{\sigma'\cdot\ell} q_t'\testexec q_i'$
   
\item[$\sigma=\sigma'\cdot\ell$, $\ell\in L_i\cap L_t:$]\ \\
    $q_t \Trans{\project{\sigma'\cdot\ell}{L_t}} q_t' \land q_i \Trans{\project{\subst{\sigma'\cdot\ell}{\theta}{\delta}}{L_i^\delta}} q_i'$\\
    \proofstep{\Cref{def:projection,def:substitution}: $\projectop$ and $substitution$}\\
    $q_t \Trans{\project{\sigma'}{L_t}\cdot\ell} q_t' \land q_i \Trans{\project{\subst{\sigma'}{\theta}{\delta}}{L_i^\delta}\cdot\ell} q_i'$\\
    \proofstep{\Cref{def:arrowdefs}: $\Trans{}$}\\
    $\exists q_t'', q_t'''\in Q_t, q_i'', q_i'''\in Q_i:$\\
    \tab$q_t \Trans{\project{\sigma'}{L_t}} q_t'' \trans{\ell} q_t''' \Trans{\epsilon} q_t'\land q_i \Trans{\project{\subst{\sigma'}{\theta}{\delta}}{L_i^\delta}} q_i'' \trans{\ell} q_i'''\Trans{\epsilon} q_i'$\\
    \proofstep{Apply IH}\\
    $\exists q_t'', q_t'''\in Q_t, q_i'', q_i'''\in Q_i:q_t'' \trans{\ell} q_t''' \Trans{\epsilon} q_t' \land  q_i'' \trans{\ell} q_i'''\Trans{\epsilon} q_i'\;\land$\\
    $q_t \testexec q_i \Trans{\sigma'} q_t''\testexec q_i''$\\
    \proofstep{\Cref{def:testexec}: $\testexec$}\\
    $\exists q_t'', q_t'''\in Q_t, q_i'', q_i'''\in Q_i: q_t'' \testexec q_i'' \trans{\ell} q_t''' \testexec q_i''' \land  q_t''' \Trans{\epsilon} q_t'  \land q_i''' \Trans{\epsilon} q_i'\;\land$\\
    $q_t \testexec q_i \Trans{\sigma'} q_t''\testexec q_i''$\\
    \proofstep{\Cref{item:testcase_no_tau}: no $\tau$ transitions in $t$}\\
    $\exists q_t''\in Q_t, q_i'', q_i'''\in Q_i: q_t'' \testexec q_i'' \trans{\ell} q_t' \testexec q_i''' \land q_i''' \Trans{\epsilon} q_i'\;\land$\\
    $q_t \testexec q_i \Trans{\sigma'} q_t''\testexec q_i''$\\
    \proofstep{\Cref{def:testexec}: $\testexec$}\\
    $\exists q_t''\in Q_t, q_i'', q_i'''\in Q_i: q_t'' \testexec q_i'' \trans{\ell} q_t' \testexec q_i''' \land q_t' \testexec q_i''' \Trans{\epsilon} q_t' \testexec q_i'\;\land$\\
    $q_t \testexec q_i \Trans{\sigma'} q_t''\testexec q_i''$\\
    \proofstep{\Cref{def:arrowdefs}: $\Trans{}$}\\
    $q_t \testexec q_i \Trans{\sigma'\cdot\ell} q_t'\testexec q_i'$
    
\item[$\sigma=\sigma'\cdot\theta:$]\ \\
    $q_t \Trans{\project{\sigma'\cdot\theta}{L_t}} q_t' \land q_i \Trans{\project{\subst{\sigma'\cdot\theta}{\theta}{\delta}}{L_i^\delta}} q_i'$\\
    \proofstep{\Cref{def:projection,def:substitution}: $\projectop$ and $substitution$}\\
    $q_t \Trans{\project{\sigma'}{L_t}\cdot\theta} q_t' \land q_i \Trans{\project{\subst{\sigma'}{\theta}{\delta}}{L_i^\delta}\cdot\delta} q_i'$\\
    \proofstep{\Cref{def:arrowdefs}: $\Trans{}$}\\
    $\exists q_t'', q_t'''\in Q_t, q_i'', q_i'''\in Q_i:$\\
    \tab$q_t \Trans{\project{\sigma'}{L_t}} q_t'' \trans{\theta} q_t''' \Trans{\epsilon} q_t'\land q_i \Trans{\project{\subst{\sigma'}{\theta}{\delta}}{L_i^\delta}} q_i'' \trans{\delta} q_i'''\Trans{\epsilon} q_i'$\\
    \proofstep{Apply IH}\\
    $\exists q_t'', q_t'''\in Q_t, q_i'', q_i'''\in Q_i:q_t'' \trans{\theta} q_t''' \Trans{\epsilon} q_t' \land  q_i'' \trans{\delta} q_i'''\Trans{\epsilon} q_i'\;\land$\\
    $q_t \testexec q_i \Trans{\sigma'} q_t''\testexec q_i''$\\
    \proofstep{\Cref{def:testexec}: $\testexec$}\\
    $\exists q_t'', q_t'''\in Q_t, q_i'', q_i'''\in Q_i: q_t'' \testexec q_i'' \trans{\theta} q_t''' \testexec q_i''' \land  q_t''' \Trans{\epsilon} q_t'  \land q_i''' \Trans{\epsilon} q_i'\;\land$\\
    $q_t \testexec q_i \Trans{\sigma'} q_t''\testexec q_i''$\\
    \proofstep{\Cref{item:testcase_no_tau}: no $\tau$ transitions in $t$}\\
    $\exists q_t''\in Q_t, q_i'', q_i'''\in Q_i: q_t'' \testexec q_i'' \trans{\theta} q_t' \testexec q_i''' \land q_i''' \Trans{\epsilon} q_i'\;\land$\\
    $q_t \testexec q_i \Trans{\sigma'} q_t''\testexec q_i''$\\
    \proofstep{\Cref{def:testexec}: $\testexec$}\\
    $\exists q_t''\in Q_t, q_i'', q_i'''\in Q_i: q_t'' \testexec q_i'' \trans{\theta} q_t' \testexec q_i''' \land q_t' \testexec q_i''' \Trans{\epsilon} q_t' \testexec q_i'\;\land$\\
    $q_t \testexec q_i \Trans{\sigma'} q_t''\testexec q_i''$\\
    \proofstep{\Cref{def:arrowdefs}: $\Trans{}$}\\
    $q_t \testexec q_i \Trans{\sigma'\cdot\theta} q_t'\testexec q_i'$
    
\end{case_distinction}
\end{case_distinction}
\end{proof}

\begin{toappendix}
\begin{lemmarep}
    \label{lem:map_test_to_utraces}
    Let $i\in \IOTS$, $e\in\LTS$, $t\in \TTS[L_i,e]$,\\
    $\sigma\in L_{t}^*$, $q_t\in (Q_t\setminus\{\failState,\passState\})$:
    \[
        t \Trans{\sigma} q_t \implies \project{\subst{\sigma}{\theta}{\delta}}{L_e^\delta}\in \utraces{e}
    \]
\end{lemmarep}

\begin{proof}
Proof by induction on $\sigma$
\begin{case_distinction}
    \item[Base case $\sigma=\epsilon$:] Trivially true because $epsilon$ is always in the utraces of every $LTS$
    \item[Inductive step $\sigma=\sigma'\cdot\ell$:] \ \\
        IH: \[\forall q_t'\in Q_t: t\Trans{\sigma'} q_t' \implies \project{\subst{\sigma'}{\theta}{\delta}}{L_e^\delta}\in \utraces{e}\]
        Halfway the proof we do further induction on the structure of the transition relation of $t$.

        $t\Trans{\sigma'\cdot\ell} q_t\land q_t\notin\{\passState,\failState\}$\\
        \proofstep{\Cref{item:trans_transitive}}\\
        $\exists q_t'\in Q_t: t\Trans{\sigma'} q_t'\Trans{\ell}q_t\land q_t\notin\{\passState,\failState\}\;\land$\\
        \proofstep{\Cref{def:testcase}: no $\trans{\tau}$ in $t$}\\
        $\exists q_t'\in Q_t: t\Trans{\sigma'} q_t'\trans{\ell}q_t\land q_t\notin\{\passState,\failState\}\;\land$\\
        \proofstep{\Cref{item:pass_fail_state}: $\passState$ and $\failState$ are sink states}\\
        $\exists q_t'\in Q_t: t\Trans{\sigma'} q_t'\trans{\ell}q_t\land q_t'\notin\{\passState,\failState\}\;\land$\\\textbf{}
        \proofstep{Apply IH}\\
        $\exists q_t'\in Q_t: t\Trans{\sigma'} q_t'\trans{\ell}q_t\land q_t'\notin\{\passState,\failState\}\;\land$\\
        $\project{\subst{\sigma'}{\theta}{\delta}}{L_e^\delta}\in \utraces{e}$\\
        \proofstep{\Cref{lem:testcase_eq_genafter}}\\
        $\exists q_t'\in Q_t: q_t'\trans{\ell}q_t \land q_t' = \GenEcoTest\bigl(e \after (\project{\subst{\sigma'}{\theta}{\delta}}{L_e^\delta}) \bigr)\;\land$\\
        $\project{\subst{\sigma'}{\theta}{\delta}}{L_e^\delta}\in \utraces{e}$\\
        \proofstep{Further induction on the structure of $T_t$}\\
        There is a finite number of lines in \cref{alg:testcase_alg} that can result in the transition $q_t'\trans{\ell} q_t$ being added to $t$. We show the goal for each possible line separately.
        \begin{case_distinction}
            \item[\Cref{line:sync_out} of \cref{fun:GenEcoTest}]\ \\
                $\project{\subst{\sigma'}{\theta}{\delta}}{L_e^\delta}\in \utraces{e}\;\land$\\
                $\Bigl(\ell\in \outset[big]{e \after (\project{\subst{\sigma'}{\theta}{\delta}}{L_e^\delta})} \lor \ell\in \inset[big]{e \after (\project{\subst{\sigma'}{\theta}{\delta}}{L_e^\delta})}\Bigr)$\\
                \proofstep{\Cref{def:arrowdefs}:$\Trans{}$}\\
                $\project{\subst{\sigma'}{\theta}{\delta}}{L_e^\delta}\in \utraces{e}\;\land$\\
                $e \Trans{\project{\subst{\sigma'}{\theta}{\delta}}{L_e^\delta}\cdot\ell} \land\;$\\
                $\bigl((\ell\in U_e) \lor (\ell\in I_e \land  \forall q_e\in Q_e: e \Trans{\project{\subst{\sigma'}{\theta}{\delta}}{L_e^\delta}\cdot\ell} q_e \implies q_e \Trans{\ell})\bigr) $\\
                \proofstep{\Cref{def:uioco}: $\utraces{}$}\\
                $\project{\subst{\sigma'}{\theta}{\delta}}{L_e^\delta}\cdot\ell\in \utraces{e}\land\ell\in L_e$\\
                \proofstep{\Cref{def:projection,def:substitution}: $\ell\in L_e^\delta\land \ell\neq \theta$}\\
                $\project{\subst{\sigma'\cdot\ell}{\theta}{\delta}}{L_e^\delta}\in \utraces{e}$
            \item[\Cref{line:sync_theta} of \cref{fun:GenEcoTest}]\ \\
                $\project{\subst{\sigma'}{\theta}{\delta}}{L_e^\delta}\in \utraces{e}\;\land$\\
                $\delta\in \outset[big]{e \after (\project{\subst{\sigma'}{\theta}{\delta}}{L_e^\delta})}\land \ell=\theta$\\
                \proofstep{\Cref{def:arrowdefs}:$\Trans{}$}\\
                $\project{\subst{\sigma'}{\theta}{\delta}}{L_e^\delta}\in \utraces{e}\;\land$\\
                $e \Trans{\project{\subst{\sigma'}{\theta}{\delta}}{L_e^\delta}\cdot\delta} \land\; \ell=\theta $\\
                \proofstep{\Cref{def:uioco}: $\utraces{}$}\\
                $\project{\subst{\sigma'}{\theta}{\delta}}{L_e^\delta}\cdot\delta\in \utraces{e}\land \ell=\theta$\\
                \proofstep{\Cref{def:projection}: $\delta\in L_e^\delta$}\\
                $\project{\subst{\sigma'}{\theta}{\delta}\cdot\delta}{L_e^\delta}\in \utraces{e}\land \ell=\theta$\\
                \proofstep{\Cref{def:projection,def:substitution}: $\ell= \theta$}\\
                $\project{\subst{\sigma'\cdot\ell}{\theta}{\delta}}{L_e^\delta}\in \utraces{e}$
            \item[\Cref{line:env_internal_step} of \cref{fun:GenEcoTest}]\ \\
                $\project{\subst{\sigma'}{\theta}{\delta}}{L_e^\delta}\in \utraces{e}\;\land$\\
                $\Bigl(\ell\in \outset[big]{e \after (\project{\subst{\sigma'}{\theta}{\delta}}{L_e^\delta})}\; \lor$\\
                $\ell \in \inset[big]{e \after (\project{\subst{\sigma'}{\theta}{\delta}}{L_e^\delta})}\Bigr)$\\
                \proofstep{\Cref{def:arrowdefs}:$\Trans{}$}\\
                $\project{\subst{\sigma'}{\theta}{\delta}}{L_e^\delta}\in \utraces{e}\;\land$\\
                $\Bigl((e \Trans{\project{\subst{\sigma'}{\theta}{\delta}}{L_e^\delta}\cdot\ell} \land\; \ell\in U_e)  \;\lor $\\
                $\bigl(e \Trans{\project{\subst{\sigma'}{\theta}{\delta}}{L_e^\delta}\cdot\ell} \land\; \ell\in I_e \land\; (\forall q_e\in Q_e: e \Trans{\project{\subst{\sigma'}{\theta}{\delta}}{L_e^\delta}\cdot\ell} q_e \implies q_e \Trans{\ell})\bigr)\Bigr)$\\
                \proofstep{\Cref{def:uioco}: $\utraces{}$}\\
                $(\project{\subst{\sigma'}{\theta}{\delta}}{L_e^\delta}\cdot\ell\in \utraces{e}\land\ell\in L_e) $\\
                \proofstep{\Cref{def:projection}: $\projectop$}\\
                $\project{\subst{\sigma'}{\theta}{\delta}\cdot\ell}{L_e^\delta}\in \utraces{e} $\\
                \proofstep{\Cref{def:projection,def:substitution}: $\ell\neq \theta$}\\
                $\project{\subst{\sigma'\cdot\ell}{\theta}{\delta}}{L_e^\delta}\in \utraces{e}$
            \item[\Cref{line:sut_internal_step} of \cref{fun:GenEcoTest}]\ \\
                $\project{\subst{\sigma'}{\theta}{\delta}}{L_e^\delta}\in \utraces{e}\;\land \ell\in I_i \setminus L_e$\\
                \proofstep{\Cref{def:projection}: $\projectop$}\\
                $\project{\subst{\sigma'}{\theta}{\delta}\cdot\ell}{L_e^\delta}\in \utraces{e}\land \ell\in I_i \setminus L_e$\\
                \proofstep{\Cref{def:substitution}: $\ell\neq \theta$}\\
                $\project{\subst{\sigma'\cdot\ell}{\theta}{\delta}}{L_e^\delta}\in \utraces{e}$
            \item[\Cref{line:failtrans} of \cref{fun:GenInputResponse}]\ \\
                Contradicts $q_t\notin \{\failState,\passState\}$
            \item[\Cref{line:sync_input} of \cref{fun:GenInputResponse}]\ \\
                $\project{\subst{\sigma'}{\theta}{\delta}}{L_e^\delta}\in \utraces{e}\;\land$\\
                $\ell \in \inset[big]{e \after (\project{\subst{\sigma'}{\theta}{\delta}}{L_e^\delta})} $\\
                \proofstep{\Cref{def:arrowdefs}:$\Trans{}$}\\
                $\project{\subst{\sigma'}{\theta}{\delta}}{L_e^\delta}\in \utraces{e}\;\land$\\
                $e \Trans{\project{\subst{\sigma'}{\theta}{\delta}}{L_e^\delta}\cdot\ell} \;\land $\\
                $\ell\in I_e \land \forall q_e\in Q_e: e \Trans{\project{\subst{\sigma'}{\theta}{\delta}}{L_e^\delta}\cdot\ell} q_e \implies q_e \Trans{\ell}$\\
                \proofstep{\Cref{def:uioco}: $\utraces{}$}\\
                $\project{\subst{\sigma'}{\theta}{\delta}}{L_e^\delta}\cdot\ell\in \utraces{e}\land\ell\in I_e$\\
                \proofstep{\Cref{def:projection,def:substitution}: $\ell\in L_e^\delta\land \ell\neq \theta$}\\
        \end{case_distinction}
\end{case_distinction}
\end{proof}

\begin{lemmarep}
    \label{lem:testcase_eq_genafter}
    Let $i\in \IOTS$, $e\in\LTS$, $t\in \TTS[L_i,e]$,\\
    $\sigma\in L_{t}^*$, $q_t\in (Q_t\setminus\{\failState,\passState\})$:
    \[t \Trans{\sigma } q_t \implies q_t = \GenEcoTest \bigl(e \after \project{(\subst{\sigma}{\theta}{\delta}}{L_e^\delta})\bigr) \]
\end{lemmarep}

\begin{proof}
    \ \\
    \newcommand{\GenInputResponse}{\HandleSutOutput}
    Proof by induction on the length of $\sigma$. 
\begin{case_distinction}
    \item[base case $\sigma=\epsilon$:]\ \\
        Since $t$ does not contain $\tau$ transitions (\itemref{def:testcase}{no_tau}), $t \Trans{\epsilon} q_t$ means $t=q_t$ (\cref{def:arrowdefs}).
        $t=\;$\GenEcoTest{$e\after \epsilon$} (\cref{alg:testcase_alg}), and $\epsilon= \project{\subst{\epsilon}{\theta}{\delta}}{L_e^\delta}$ (\cref{def:projection,def:substitution}).
    \item[Inductive step: $\sigma = \sigma'\cdot\ell$:]
        The IH is
        \[\forall q_t'\in Q_t: t\Trans{\sigma'} q_t' \implies q_t' = \GenEcoTest\bigl(e \after (\project{\subst{\sigma'}{\theta}{\delta}}{L_e^\delta}) \bigr)\]
        
        $t\Trans{\sigma'\cdot\ell} q_t$\\
        \proofstep{\Cref{item:trans_transitive}}\\
        $\exists q_t'\in Q_t: t \Trans{\sigma'} q_t' \Trans{\ell}q_t$\\
        \proofstep{Apply IH}\\
        $\exists q_t'\in Q_t: q_t' \Trans{\ell}q_t\land q_t' = \GenEcoTest\bigl(e \after (\project{\subst{\sigma'}{\theta}{\delta}}{L_e^\delta}) \bigr)$\\
        \proofstep{\Itemref{def:testcase}{no_tau}: no $\tau$ transitions in $t$}\\
        $\exists q_t'\in Q_t: q_t' \trans{\ell}q_t\land q_t' = \GenEcoTest\bigl(e \after (\project{\subst{\sigma'}{\theta}{\delta}}{L_e^\delta}) \bigr)$\\
        \proofstep{Induction on the structure of $T_t$}\\
        The transition function of $t$ is generated by $\GenEcoTest$. From here we see there are 6 ways $q_t'\trans{\ell}$ is possible: lines \ref{line:sync_out}, \ref{line:sync_theta}, \ref{line:env_internal_step} and \ref{line:sut_internal_step} of $\GenEcoTest$, or any of the two lines of $\GenInputResponse$. Lines \ref{line:sync_out}  and \ref{line:env_internal_step} of $\GenEcoTest$ and \cref{line:sync_input} of $\GenInputResponse$ all transition into $\GenEcoTest\bigl(e \after (\project{\subst{\sigma'}{\theta}{\delta}}{L_e^\delta}) \bigr) \after \ell$ from $q_t'$. We also have from these same three possible lines that
        $\ell \in L_e\cap I_i \lor \ell\in L_e\setminus L_i \lor \ell\in U_i\cap I_e$, which means $\ell\in L_e$. 
        Similarly, \cref{line:sync_theta} transitions into  $\GenEcoTest\bigl(e \after (\project{\subst{\sigma'}{\theta}{\delta}}{L_e^\delta}) \bigr) \after \delta$ with $\ell=\theta$, and \cref{line:sut_internal_step} transitions into $\GenEcoTest\bigl(e \after (\project{\subst{\sigma'}{\theta}{\delta}}{L_e^\delta}) \bigr)$ with $\ell\notin L_e$ The last option transitions into $\failState$, but we know $q_t\notin \{\passState,\failState\}$ so this is not possible.
        This gives:\\
        $(q_t = \GenEcoTest\bigl(e \after (\project{\subst{\sigma'}{\theta}{\delta}}{L_e^\delta}) \after\ell \bigr)\land \ell\in L_e)\;\lor$\\
        $(q_t = \GenEcoTest\bigl(e \after (\project{\subst{\sigma'}{\theta}{\delta}}{L_e^\delta}) \bigr)\land \ell\notin L_e)\;\lor$\\
        $(q_t = \GenEcoTest\bigl(e \after (\project{\subst{\sigma'}{\theta}{\delta}}{L_e^\delta}) \after\delta \bigr)\land \ell=\theta)$\\
        \proofstep{\Cref{def:after}: $\after$}\\
        $(q_t = \GenEcoTest\bigl(e \after (\project{\subst{\sigma'}{\theta}{\delta}}{L_e^\delta}\cdot\ell) \bigr)\land \ell\in L_e)\;\lor$\\
        $(q_t = \GenEcoTest\bigl(e \after (\project{\subst{\sigma'}{\theta}{\delta}}{L_e^\delta}) \bigr)\land \ell\notin L_e)\;\lor$\\
        $(q_t = \GenEcoTest\bigl(e \after (\project{\subst{\sigma'}{\theta}{\delta}}{L_e^\delta}\cdot\delta)\bigr)\land \ell=\theta)$\\
        \proofstep{\Cref{def:projection}: $\projectop$}\\
        $(q_t = \GenEcoTest\bigl(e \after (\project{\subst{\sigma'}{\theta}{\delta}\cdot\ell}{L_e^\delta}) \bigr)\land \ell\in L_e)\;\lor$\\
        $(q_t = \GenEcoTest\bigl(e \after (\project{\subst{\sigma'}{\theta}{\delta}\cdot\ell}{L_e^\delta}) \bigr)\land \ell\notin L_e)\;\lor$\\
        $(q_t = \GenEcoTest\bigl(e \after (\project{\subst{\sigma'}{\theta}{\delta}\cdot\delta}{L_e^\delta})\bigr)\land \ell=\theta)$\\
        \proofstep{\Cref{def:substitution}: substitution}\\
        $q_t = \GenEcoTest\bigl(e \after (\project{\subst{\sigma'\cdot\ell}{\theta}{\delta}}{L_e^\delta}) \bigr)$\\
\end{case_distinction}
\end{proof}

\begin{lemma}
    \label{lem:parcomp_test_trace_simplification}
    Let $s\in \LTS$, $i\in \IOTS$, $t\in \TTS[L_i,e]$ $\sigma\in L_{s\parcomp e}^{\delta*}$, $\theta\notin  L_{s\parcomp e}^{\delta}$:
    \[\project{\subst{\project{\subst{\sigma}{\delta}{\theta}}{L_t}}{\theta}{\delta}}{L_e^\delta} \iff  \project{\sigma}{L_e^\delta}\]\\
\end{lemma}
\begin{proof}
       $\project{\subst{\project{\subst{\sigma}{\delta}{\theta}}{L_t}}{\theta}{\delta}}{L_e^\delta}$\\
\proofstep{\Cref{def:projection,def:substitution}: $\delta \notin (\subst{\sigma}{\delta}{\theta}) $}\\
$\project{\subst{\project{\subst{\sigma}{\delta}{\theta}}{L_t^\delta}}{\theta}{\delta}}{L_e^\delta}$\\
\proofstep{\Cref{def:projection,def:substitution}: $\theta\in L_t^\delta\land \delta\in L_t^\delta$}\\
$\project{\project{\subst{\subst{\sigma}{\delta}{\theta}}{\theta}{\delta}}{L_t^\delta}}{L_e^\delta}$\\
\proofstep{\Cref{def:substitution}: $\theta\notin \sigma$}\\
$\project{\project{\sigma}{L_t^\delta}}{L_e^\delta}$\\
\proofstep{\Cref{def:projection,lem:L_testcase}: $L_e^\delta\subseteq L_t^\delta $}\\
$\project{\sigma}{L_e^\delta}$
\end{proof}
\end{toappendix}

\begin{definition}
    \label{def:accepting_traces}
    Let $s,e\in \LTS$ be $\composable$, $\sigma\in L_{s\parcomp e}^{\delta*}$, $q_s\in Q_s$, $q_e\in Q_e$.\\
    Then $q_s \textbf{ accepts } q_e\textbf{ over } \sigma$ ($q_s \accepts{\sigma} q_e$) iff:
    \[\begin{array}{l}
    \forall \sigma',\sigma''\in L_{i\parcomp e}^{\delta*}, q_s'\in Q_s, q_e'\in Q_e:\\ \tab\sigma'\cdot\sigma''=\sigma \land q_s\parcomp q_e \Trans{\sigma'} q_s'\parcomp q_e' \implies \outset{q_e'}\cap I_s\subseteq\inset{q_s'}\cap U_e
    \end{array}\]
\end{definition}

\begin{lemmarep}
\label{lem:project_single_trace_from_parcomp}
let  $s,e\in \LTS$ be $\composable$, $q_s,q_s'\in Q_s$, $q_e,q_e' \in Q_e, \sigma \in  L_{s\parcomp e}^{\delta*}$, then
\[q_s \mutuallyaccepts{\sigma} q_e \implies 
(\ q_s \parcomp q_e \Trans{\sigma} q_s' \parcomp q_e' \ \iff\ 
q_s \Trans{\project{\sigma}{L_s^\delta}} q_s' \:\land\:
q_e \Trans{\project{\sigma}{L_e^\delta}} q_e'\ )\]
\end{lemmarep}

\begin{proof}
    ~\begin{case_distinction}
\item[($\Longrightarrow$):] Proof by induction on $\sigma$. 
    \begin{case_distinction}
    \item[Base case:] covered by \cref{lem:project_from_parcomp_no_delta}
        
    \item[Induction step:] 
        Assume the lemma holds for $\sigma'\in  L_{s \parcomp e}^{\delta*}$.\\
        To proof: the lemma holds for $\sigma =\sigma' \cdot \ell$ with $\ell \in L_{s \parcomp e}^\delta$.  This is divided into two cases based on whether $\ell$ is $\delta$ or not:
        \begin{case_distinction}
            \item[$\ell \neq \delta$:]\ \\
                $q_s \parcomp q_e \Trans{\sigma' \cdot \ell} q_s' \parcomp q_e'$\\
                \proofstep{\Cref{item:trans_transitive}}\\
                $\exists q_s'' \in Q_s, q_e''\in Q_e :$\\
                \tab$q_s \parcomp q_e \Trans{\sigma'} q_s'' \parcomp q_e'' \Trans{\ell} q_s' \parcomp q_e'$\\
                \proofstep{\Cref{lem:project_from_parcomp_no_delta}}\\
                $\exists q_s'' \in Q_s, q_e''\in Q_e :$\\
                \tab$q_s \parcomp q_e \Trans{\sigma'} q_s'' \parcomp q_e'' \land q_s'' \Trans{\project{\ell}{L_s^\delta}} q_s' \land q_e''  \Trans{\project{\ell}{L_e^\delta}} q_e'$\\
                \proofstep{Apply IH}\\
                $\exists q_s'' \in Q_s, q_e'' \in Q_e :$\\
                \tab$q_s \Trans{\project{\sigma'}{L_s^\delta}} q_s'' \land
                q_e \Trans{\project{\sigma'}{L_e^\delta}} q_e'' \land
                q_s'' \Trans{\project{\ell}{L_s^\delta}} q_s' \land q_e''  \Trans{\project{\ell}{L_e^\delta}} q_e'$\\
                \proofstep{\Cref{item:trans_transitive}}\\
                $s \Trans{\project{\sigma'}{L_s^\delta}\cdot (\project{\ell}{L_s^\delta})} q_s' \land
                e \Trans{\project{\sigma'}{L_e^\delta}\cdot(\project{\ell}{L_e^\delta})} q_e'$\\
                \proofstep{\Cref{def:projection}: $\projectop$}\\
                $s \Trans{\project{\sigma'\cdot \ell}{L_s^\delta}} q_s' \land
                e \Trans{\project{\sigma'\cdot \ell}{L_e^\delta}} q_e'$\\
            \item[$\ell = \delta$:]\ \\
                $q_s \parcomp q_e \Trans{\sigma' \cdot \delta} q_s' \parcomp q_e'$\\
                \proofstep{\Cref{def:arrowdefs}: $\Trans{\sigma}$}\\
                $\exists q_s'', q_s''' \in Q_s, q_e'',q_e''' \in Q_e :$\\
                \tab$q_s \parcomp q_e \Trans{\sigma'} q_s'' \parcomp q_e'' \land
                q_s'' \parcomp q_e'' \trans{\delta} q_s''' \parcomp q_e''' \land
                q_s''' \parcomp q_e''' \Trans{\epsilon} q_s' \parcomp q_e' $\\
                \proofstep{\Cref{lem:project_from_parcomp_no_delta}}\\
                $\exists q_s'', q_s''' \in Q_s, q_e'',q_e''' \in Q_e :$\\
                \tab$q_s \parcomp q_e \Trans{\sigma'} q_s'' \parcomp q_e'' \land
                q_s'' \parcomp q_e'' \trans{\delta} q_s''' \parcomp q_e''' \land
                q_s''' \Trans{\epsilon} q_s' \land
                q_e''' \Trans{\epsilon} q_e' $\\
                \proofstep{\Cref{def:delta}: $\delta$}\\
                $\exists q_s'' \in Q_s, q_e''\in Q_e :$\\
                \tab$q_s \parcomp q_e \Trans{\sigma'} q_s'' \parcomp q_e'' \land
                q_s'' \parcomp q_e'' \trans{\delta} q_s'' \parcomp q_e'' \land
                q_s'' \Trans{\epsilon} q_s' \land
                q_e'' \Trans{\epsilon} q_e' $\\
                \proofstep{\Cref{def:delta}: $\delta$}\\
                $\exists q_s'' \in Q_s, q_e''\in Q_e: q_s \parcomp q_e \Trans{\sigma'} q_s'' \parcomp q_e'' \land
                q_s'' \Trans{\epsilon} q_s' \land
                q_e'' \Trans{\epsilon} q_e' $\\
                $\forall \ell \in U_s \cup U_e \cup \{\tau\}: q_s'' \parcomp q_e'' \nottrans{\ell}$\\
                \proofstep{\Cref{def:parcomp}: $\parcomp$}\\
                $\exists q_s'' \in Q_s, q_e''\in Q_e: q_s \parcomp q_e \Trans{\sigma'} q_s'' \parcomp q_e'' \land q_s'' \Trans{\epsilon} q_s' \land q_e'' \Trans{\epsilon} q_e' $\\
                $\forall \ell \in (U_s \setminus L_e) \cup \{\tau\}: q_s'' \nottrans{\ell} \land$\\
                $\forall \ell \in (U_e \setminus L_s) \cup \{\tau\}: q_e'' \nottrans{\ell} \land$\\
                $\forall \ell \in (U_s \cap L_e) \cup (U_e \cap L_s): q_s'' \nottrans{\ell} \lor\; q_e'' \nottrans{\ell}$\\
                \proofstep{\Cref{def:mutually_accepts,def:accepting_traces}: $\mutuallyaccepts{\sigma'\cdot\delta}$ and $\accepts{\sigma'\cdot\delta}$}\\
                $\exists q_s'' \in Q_s, q_e''\in Q_e: q_s \parcomp q_e \Trans{\sigma'} q_s'' \parcomp q_e'' \land q_s'' \Trans{\epsilon} q_s' \land q_e'' \Trans{\epsilon} q_e' $\\
                $\forall \ell \in (U_s \setminus L_e) \cup \{\tau\}: q_s'' \nottrans{\ell} \land$\\
                $\forall \ell \in (U_e \setminus L_s) \cup \{\tau\}: q_e'' \nottrans{\ell} \land$\\
                $\forall \ell \in (U_s \cap L_e) \cup (U_e \cap L_s): q_s'' \nottrans{\ell} \lor\; q_e'' \nottrans{\ell}$\\
                $out(q_s'') \cap I_e \subseteq in(q_e'') \cap U_s\; \land$\\ 
                $out(q_e'') \cap I_s \subseteq in(q_s'') \cap U_e \; \land $\\
                \proofstep{\Cref{def:outset,def:inset}: $out$ and $in$}\\
                $\exists q_s'' \in Q_s, q_e''\in Q_e: q_s \parcomp q_e \Trans{\sigma'} q_s'' \parcomp q_e'' \land q_s'' \Trans{\epsilon} q_s' \land q_e'' \Trans{\epsilon} q_e' $\\
                $\forall \ell \in (U_s \setminus L_e) \cup \{\tau\}: q_s'' \nottrans{\ell} \land$\\
                $\forall \ell \in (U_e \setminus L_s) \cup \{\tau\}: q_e'' \nottrans{\ell} \land$\\
                $\forall \ell \in(U_s \cap L_e) \cup (U_e \cap L_s): q_s'' \nottrans{\ell} \lor\; q_e'' \nottrans{\ell}\land$\\
                $\forall \ell \in U_s \cap I_e: q_s'' \trans{\ell} \implies  q_e'' \Trans{\ell} \land$\\ 
                $\forall \ell \in U_e \cap I_s: q_e'' \trans{\ell} \implies  q_s'' \Trans{\ell}$\\
                \proofstep{$q_s' \nottrans{\tau} \land\; q_e'\nottrans{\tau}$}\\
                $\exists q_s'' \in Q_s, q_e''\in Q_e: q_s \parcomp q_e \Trans{\sigma'} q_s'' \parcomp q_e'' \land q_s'' \Trans{\epsilon} q_s' \land q_e'' \Trans{\epsilon} q_e' $\\
                $\forall \ell \in (U_s \setminus L_e) \cup \{\tau\}: q_s'' \nottrans{\ell} \land$\\
                $\forall \ell \in (U_e \setminus L_s) \cup \{\tau\}: q_e'' \nottrans{\ell} \land$\\
                $\forall \ell \in (U_s \cap L_e) \cup (U_e \cap L_s): q_s'' \nottrans{\ell} \lor\; q_e'' \nottrans{\ell}\land$\\
                $\forall \ell \in U_s \cap I_e : q_s'' \trans{\ell} \implies q_e'' \trans{\ell} \land$\\
                $\forall \ell \in U_e \cap I_s : q_e'' \trans{\ell} \implies q_s'' \trans{\ell}$\\
                \proofstep{$(A \implies B) \iff (\neg B \implies \neg A)$}\\
                $\exists q_s'' \in Q_s, q_e''\in Q_e: q_s \parcomp q_e \Trans{\sigma'} q_s'' \parcomp q_e'' \land q_s'' \Trans{\epsilon} q_s' \land q_e'' \Trans{\epsilon} q_e' $\\
                $\forall \ell \in (U_s \setminus L_e) \cup \{\tau\}: q_s'' \nottrans{\ell} \land$\\
                $\forall \ell \in (U_e \setminus L_s) \cup \{\tau\}: q_e'' \nottrans{\ell} \land$\\
                $\forall \ell \in(U_s \cap L_e) \cup (U_e \cap L_s): q_s'' \nottrans{\ell} \lor\; q_e'' \nottrans{\ell}\land$\\
                $\forall \ell \in U_s \cap I_e : q_e'' \nottrans{\ell} \implies q_s'' \nottrans{\ell}\land$\\
                $\forall \ell \in U_e \cap I_s : q_s'' \nottrans{\ell} \implies q_e'' \nottrans{\ell}$\\
                \proofstep{\Cref{def:composable}: $U_s \cap U_e = \emptyset$}\\
                $\exists q_s'' \in Q_s, q_e''\in Q_e: q_s \parcomp q_e \Trans{\sigma'} q_s'' \parcomp q_e'' \land q_s'' \Trans{\epsilon} q_s' \land q_e'' \Trans{\epsilon} q_e' $\\
                $\forall \ell \in (U_s \setminus L_e) \cup \{\tau\}: q_s'' \nottrans{\ell} \land$\\
                $\forall \ell \in (U_e \setminus L_s) \cup \{\tau\}: q_e'' \nottrans{\ell} \land$\\
                $\forall \ell \in (U_s \cap L_e) \cup (U_e \cap L_s): q_s'' \nottrans{\ell} \lor\; q_e'' \nottrans{\ell}\land$\\
                $\forall \ell \in U_s \cap L_e : q_e'' \nottrans{\ell} \implies q_s'' \nottrans{\ell}\land$\\
                $\forall \ell \in U_e \cap L_s : q_s'' \nottrans{\ell} \implies q_e'' \nottrans{\ell}$\\
                \proofstep{$(A \lor B) \land (A \implies B) \implies B$}\\
                $\exists q_s'' \in Q_s, q_e''\in Q_e: q_s \parcomp q_e \Trans{\sigma'} q_s'' \parcomp q_e'' \land q_s'' \Trans{\epsilon} q_s' \land q_e'' \Trans{\epsilon} q_e' $\\
                $\forall \ell \in (U_s \setminus L_e) \cup \{\tau\}: q_s'' \nottrans{\ell} \land$\\
                $\forall \ell \in (U_e \setminus L_s) \cup \{\tau\}: q_e'' \nottrans{\ell} \land$\\
                $\forall \ell \in U_s \cap L_e :  q_s'' \nottrans{\ell}\land$\\
                $\forall \ell \in U_e \cap L_s :  q_e'' \nottrans{\ell}$\\
                \proofstep{$(X \setminus Y) \cup (X \cap Y) = X$}\\
                $\exists q_s'' \in Q_s, q_e''\in Q_e: q_s \parcomp q_e \Trans{\sigma'} q_s'' \parcomp q_e'' \land q_s'' \Trans{\epsilon} q_s' \land q_e'' \Trans{\epsilon} q_e' $\\
                $\forall \ell \in U_s \cup \{\tau\} : q_s'' \nottrans{\ell} \land$\\ $\forall \ell \in U_e \cup \{\tau\} : q_e'' \nottrans{\ell}$\\
                \proofstep{\Cref{def:delta}: $\delta$}\\
                $\exists q_s'' \in Q_s, q_e''\in Q_e: q_s \parcomp q_e \Trans{\sigma'} q_s'' \parcomp q_e'' \land q_s'' \Trans{\epsilon} q_s' \land q_e'' \Trans{\epsilon} q_e' $\\
                $q_s'' \trans{\delta} q_s'' \land q_e'' \trans{\delta} q_e''$\\
                \proofstep{\Cref{def:arrowdefs}: $\Trans{\delta}$}\\
                $\exists q_s'' \in Q_s, q_e'' \in Q_e :
                q_s \parcomp q_e \Trans{\sigma'} q_s'' \parcomp q_e'' \land
                q_s'' \Trans{\delta} q_s' \land
                q_e'' \Trans{\delta} q_e' $\\
                \proofstep{Apply IH}\\
                $\exists q_s'' \in Q_s, q_e'' \in Q_e :
                s \Trans{\project{\sigma'}{L_s^\delta}} q_s'' \land e \Trans{\project{\sigma'}{L_e^\delta}} q_e'' \land
                q_s'' \Trans{\delta} q_s' \land
                q_e'' \Trans{\delta} q_e' $\\
                \proofstep{\Cref{item:trans_transitive}}\\
                $s \Trans{\project{\sigma'}{L_s^\delta}\cdot \delta} q_s' \land e \Trans{\project{\sigma'}{L_e^\delta} \cdot \delta} q_e' $\\
                \proofstep{\Cref{def:projection}: $\projectop$}\\
                $s \Trans{\project{\sigma'\cdot \delta}{L_s^\delta}} q_s' \land e \Trans{\project{\sigma'\cdot \delta}{L_e^\delta}} q_e' $\\
          
        \end{case_distinction}
    \end{case_distinction}
    \item[($\Longleftarrow$):] Covered by \cref{lem:project_from_parcomp_light}.
\end{case_distinction}
\end{proof}

\begin{toappendix}
\begin{lemmarep}
    \label{lem:project_to_test}
    Let $i\in \IOTS$, $\sigma\in\utraces{i\parcomp e}$, $q_i\in Q_i$, $q_e\in Q_e$:
    \[\begin{array}{l}
        i\parcomp e \Trans{\sigma} q_i \parcomp q_e\;  \land e \mutuallyaccepts{\sigma} i \implies \\
        \tab\exists t \in \TTS[L_i,e], q_t \in Q_t\setminus\{\passState,\failState\}: \\ 
        \tab t \Trans{\project{\subst{\sigma}{\delta}{\theta}}{L_t}} q_t\\ 
    \end{array}\]
\end{lemmarep}
\begin{proof}
    Proof by induction on the length of $\sigma$, with further case distinction based on the last label of $\sigma$. 
\begin{case_distinction}
     \item[Base case $(\sigma=\epsilon)$:]\ \\
        Let $t$ be the test case obtained by choosing case C or B of \cref{fun:GenEcoTest} depending on whether $e\after\epsilon$ is quiescent or not , and then case A in all recursive calls.\\
        Then the proof is trivial, because $t\Trans{\epsilon}t$, and $t$ is not $\passState$ or $\failState$.
    \item[Inductive step $(\sigma=\sigma'\cdot\ell)$]\ \\
        The IH is $\forall q_i'\in Q_i$, $\forall q_e'\in Q_e$:
        \[\begin{array}{ll}
           i\parcomp e \Trans{\sigma'} q_i' \parcomp q_e'\; \implies &\exists t \in \TTS[L_i,e], q_t' \in Q_t\setminus\{\passState,\failState\}: \\ 
           &t \Trans{\project{\subst{\sigma'}{\delta}{\theta}}{L_t}} q_t'\\
        \end{array}\]
        Note that the condition $e \mutuallyaccepts{\sigma'} i$ is dropped, because this is always true for $\sigma'$. ($e \mutuallyaccepts {\sigma'\cdot\ell} i \implies e \mutuallyaccepts{\sigma'} i$), which follows directly from \cref{def:accepting_traces}. Halfway the proof we do a further case distinction on $\ell$ into the following cases, based on the different ways labels are handled by \cref{alg:testcase_alg}: 
        \[\begin{array}{ll@{~~}l}
            (a)& \bigl(\outset{X_e} \cup \inset{X_e}\bigr)\cap I_i & (\textit{Case B})\\
            (b)& \{\delta\} &(\textit{Case C})\\
            (c)& (\outset{X_e}\cup \inset{X_e})\setminus L_i^\delta &(\textit{Case D})\\
            (d)& (U_i\cap I_e)\setminus\inset{X_e} &(\textit{Contradicts } e \mutuallyaccepts{\sigma} i)\\
            (e)&  U_i\cap \inset{X_e}  &(\cref{line:sync_input}\textit{ of }\Cref{fun:GenInputResponse})\\
            (f)& (I_i\cap I_e)\setminus\inset{X_e} &(\textit{Contradicts } 
            \sigma\in \utraces{i\parcomp e})\\
            (g)& U_e\setminus\outset{X_e}&(\textit{Contradicts } 
            \sigma\in \utraces{i\parcomp e})\\
            (h)& I_i\setminus L_e &(\textit{Case E})\\
            (i)& U_i\setminus L_e &(\textit{Simplifies to IH})\\
            (j)& I_e\setminus(\inset{X_e}\cup L_i)&(\textit{Contradicts } 
            \sigma\in \utraces{i\parcomp e})
        \end{array}\]
        where $X_e$ is a set of states of $e$. Cases d, f, g and j are all empty, because they contradict the assumptions. Cases h and i are trivial because they simplify to the IH.
        Together, these cases fully cover all possible labels:
        \begin{enumerate}
            \item[$U_e$:] a + c + g
            \item [$I_e$:] a + c + d + e + f + j
            \item [$U_i$:] d + e + i
            \item[$I_i$:] a + f + g + h
        \end{enumerate}
        $U_i\cap U_e$ is missing because it is empty due to \cref{def:composable}.

        \ \\

        $i\parcomp e \Trans{\sigma'\cdot\ell} q_i\parcomp q_e$\\
        \proofstep{\Cref{item:trans_transitive}}\\
        $\exists q_i' \in Q_i, q_e'\in Q_e: i \parcomp e \Trans{\sigma'} q_i'\parcomp q_e'$\\
        \proofstep{Apply IH}\\
        $\exists t'\in \TTS[L_i,e], q_t'\in Q_{t'}\setminus\{\passState,\failState\}: t'\Trans{\project{\subst{\sigma'}{\delta}{\theta}}{L_t}} q_t'$\\
        \proofstep{\Cref{lem:testcase_eq_genafter}}\\
        $\exists t'\in \TTS[L_i,e], q_t'\in Q_{t'}\setminus\{\passState,\failState\}: t'\Trans{\project{\subst{\sigma'}{\delta}{\theta}}{L_t}} q_t'\;\land$\\
        $X_e = e \after (\project{\subst{\project{\subst{\sigma'}{\delta}{\theta}}{L_t}}{\theta}{\delta}}{L_e^\delta}) \land q_t'= \GenEcoTest(X_e)$\\
        \proofstep{\Cref{lem:parcomp_test_trace_simplification}}\\
        $\exists t'\in \TTS[L_i,e], q_t'\in Q_{t'}\setminus\{\passState,\failState\}: t'\Trans{\project{\subst{\sigma'}{\delta}{\theta}}{L_t}} q_t'\;\land$\\
        $X_e = e \after (\project{\sigma'}{L_e^\delta}) \land q_t'= \GenEcoTest(X_e)$\\
        \proofstep{Case distinction on $\ell$}
        \begin{case_distinction}
        \item[case a, $ \bigl(\outset{X_e} \cup \inset{X_e}\bigr)\cap I_i$:]\ \\ 
            Let $t$ be $t'$, but in $q_t'$ we take choice B of \cref{fun:GenEcoTest}, and then choice A in all recursive calls.\\

            $\exists t\in \TTS[L_i,e], q_t'\in Q_{t}\setminus\{\passState,\failState\}: t\Trans{\project{\subst{\sigma'}{\delta}{\theta}}{L_t}} q_t'\;\land$\\
            $X_e = e \after (\project{\sigma'}{L_e^\delta}) \land q_t'= \GenEcoTest(X_e)$\\
            \proofstep{\Cref{line:sync_out} of \cref{fun:GenEcoTest}}\\
            $\exists t\in \TTS[L_i,e], q_t\in Q_{t}\setminus\{\passState,\failState\}: t\Trans{\project{\subst{\sigma'}{\delta}{\theta}}{L_t}\cdot\ell} q_t\;\land$\\
            $X_e = e \after (\project{\sigma'}{L_e^\delta}) \land q_t= \GenEcoTest(X_e\after\ell)$\\
            \proofstep{\Cref{alg:testcase_alg,def:projection,def:substitution}: $\ell\in L_t$, $\ell\neq\delta$}\\
            $\exists t\in \TTS[L_i,e], q_t\in Q_{t}\setminus\{\passState,\failState\}: t\Trans{\project{\subst{\sigma'\cdot\ell}{\delta}{\theta}}{L_t}} q_t$
        \item[case b, $\ell=\delta$:]\ \\ 
            Let $t$ be $t'$, but in $q_t'$ we take choice C of \cref{fun:GenEcoTest}, and then case A in all recursive calls. Let $\phi = \phi'\cdot\theta$\\

            $\exists t\in \TTS[L_i,e], q_t'\in Q_{t}\setminus\{\passState,\failState\}: t\Trans{\project{\subst{\sigma'}{\delta}{\theta}}{L_t}} q_t'\;\land$\\
            $X_e = e \after (\project{\sigma'}{L_e^\delta}) \land q_t'= \GenEcoTest(X_e)$\\
            \proofstep{\Cref{def:uioco}: $\sigma'\cdot\delta\in\utraces{i\parcomp e}$}\\
            $i\parcomp e \Trans{\sigma'\cdot\delta}\;\land\;\land$\\
            $\exists t\in \TTS[L_i,e], q_t'\in Q_{t}\setminus\{\passState,\failState\}: t\Trans{\project{\subst{\sigma'}{\delta}{\theta}}{L_t}} q_t'\;\land$\\
            $X_e = e \after (\project{\sigma'}{L_e^\delta}) \land q_t'= \GenEcoTest(X_e)$\\
            \proofstep{\Cref{lem:project_single_trace_from_parcomp,def:accepting_traces}: $s \mutuallyaccepts{\sigma'} e$}\\
            $\exists q_i', \in Q_i, q_e'\in Q_e:e \Trans{\project{\sigma'}{L_e^\delta}} q_e' \Trans{\delta} \;\land$\\
            $\exists t\in \TTS[L_i,e], q_t'\in Q_{t}\setminus\{\passState,\failState\}: t\Trans{\project{\subst{\sigma'}{\delta}{\theta}}{L_t}} q_t'\;\land$\\
            $X_e = e \after (\project{\sigma'}{L_e^\delta}) \land q_t'= \GenEcoTest(X_e)$\\
            \proofstep{\Cref{def:outset}:$\outset{}$}\\
            $\delta\in\outset{X_e}\;\land$\\
            $\exists t\in \TTS[L_i,e], q_t'\in Q_{t}\setminus\{\passState,\failState\}: t\Trans{\project{\subst{\sigma'}{\delta}{\theta}}{L_t}} q_t'\;\land$\\
            $X_e = e \after (\project{\sigma'}{L_e^\delta}) \land q_t'= \GenEcoTest(X_e)$\\
            \proofstep{\Cref{line:sync_theta} of \cref{fun:GenEcoTest}}\\
            $\exists t\in \TTS[L_i,e], q_t\in Q_{t}\setminus\{\passState,\failState\}: t\Trans{\project{\subst{\sigma'}{\delta}{\theta}}{L_t}\cdot\theta} q_t\;\land$\\
            $X_e = e \after (\project{\sigma'}{L_e^\delta}) \land q_t= \GenEcoTest(X_e\after\delta)$\\
            \proofstep{\Cref{alg:testcase_alg,def:projection,def:substitution}: $\theta\in L_t$}\\
            $\exists t\in \TTS[L_i,e], q_t\in Q_{t}\setminus\{\passState,\failState\}: t\Trans{\project{\subst{\sigma'\cdot\delta}{\delta}{\theta}}{L_t}} q_t$
        \item[case c, $\ell\in(\outset{X_e}\cup \inset{X_e})\setminus L_i^\delta$:]\ \\
            Let $t$ be $t'$, but in $q_t'$ we take choice D of \cref{fun:GenEcoTest}, and then choice A in all recursive calls.\\

            $\exists t\in \TTS[L_i,e], q_t'\in Q_{t}\setminus\{\passState,\failState\}: t\Trans{\project{\subst{\sigma'}{\delta}{\theta}}{L_t}} q_t'\;\land$\\
            $X_e = e \after (\project{\sigma'}{L_e^\delta}) \land q_t'= \GenEcoTest(X_e)$\\
            \proofstep{\Cref{line:env_internal_step} of \cref{fun:GenEcoTest}}\\
            $\exists t\in \TTS[L_i,e], q_t\in Q_{t}\setminus\{\passState,\failState\}: t\Trans{\project{\subst{\sigma'}{\delta}{\theta}}{L_t}\cdot\ell} q_t\;\land$\\
            $X_e = e \after (\project{\sigma'}{L_e^\delta}) \land q_t= \GenEcoTest(X_e\after\ell)$\\
            \proofstep{\Cref{alg:testcase_alg,def:projection,def:substitution}: $\ell\in L_t$, $\ell\neq\delta$}\\
            $\exists t\in \TTS[L_i,e], q_t\in Q_{t}\setminus\{\passState,\failState\}: t\Trans{\project{\subst{\sigma'\cdot\ell}{\delta}{\theta}}{L_t}} q_t$
        \item[case d, $\ell\in (U_i\cap I_e)\setminus\inset{X_e}$:]\ \\
            $X_e = e \after (\project{\sigma'}{L_e^\delta})$\\
            \proofstep{$\sigma'\cdot\ell\in\utraces{i\parcomp e}$}\\
            $\ell\in\outset{i\parcomp e \after \sigma'}\;\land$\\
            $X_e = e \after (\project{\sigma'}{L_e^\delta})$\\
            \proofstep{\Cref{def:outset,def:after}:$\outset{}$ and $\after$}\\
            $\exists q_i\in Q_i, q_e\in Q_e: i\parcomp e \Trans{\sigma'} q_i\parcomp q_e \land q_i\parcomp q_e \trans{\ell} ;\land$\\
            $X_e = e \after (\project{\sigma'}{L_e^\delta})$\\
            \proofstep{\Cref{def:mutually_accepts,def:accepting_traces}:$\mutuallyaccepts{}$ and $\accepts{\sigma'\cdot\ell}$}\\
            $\exists q_i\in Q_i, q_e\in Q_e: i\parcomp e \Trans{\sigma'} q_i\parcomp q_e \land q_i\parcomp q_e \trans{\ell} ;\land$\\
            $X_e = e \after (\project{\sigma'}{L_e^\delta})\;\land$\\
            $\forall q_e' \in Q_e:  i \parcomp e \Trans{\sigma'} q_i \parcomp q_e' \implies  \outset{q_i} \cap I_e \subseteq \inset{q_e'} \cap U_i$\\
            \proofstep{\Cref{def:outset,def:parcomp}:$\outset{}$ and $T_{\parcomp}$}\\
            $\exists q_i\in Q_i, q_e\in Q_e: i\parcomp e \Trans{\sigma'} q_i\parcomp q_e \land q_i\parcomp q_e \trans{\ell} ;\land$\\
            $\ell\in\outset{q_i}\;\land$\\
            $X_e = e \after (\project{\sigma'}{L_e^\delta})\;\land$\\
            $\forall q_e' \in Q_e:  i \parcomp e \Trans{\sigma'} q_i \parcomp q_e' \implies  \outset{q_i} \cap I_e \subseteq \inset{q_e'} \cap U_i$\\
            \proofstep{Definition $\subseteq$}\\
            $X_e = e \after (\project{\sigma'}{L_e^\delta})\;\land$\\
            $\forall q_e' \in Q_e:  i \parcomp e \Trans{\sigma'} q_i \parcomp q_e' \implies   \ell\in \inset{q_e'}$\\
            \proofstep{\Cref{lem:project_single_trace_from_parcomp}}\\
            $X_e = e \after (\project{\sigma'}{L_e^\delta})\;\land$\\
            $\forall q_e' \in Q_e:  e \Trans{\project{\sigma'}{L_e^\delta}} q_e' \implies   \ell\in \inset{q_e'}$\\
            \proofstep{\Cref{def:outset,def:after}:$\outset{}$ and $\after$}\\
            $\ell\in\inset{X_e}\;\land $
            $X_e = e \after (\project{\sigma'}{L_e^\delta})$\\
            This contradicts the current case of $\ell\in (U_i\cap I_e)\setminus\inset{X_e}$, proving this case is empty.

        \item[case e, $\ell\in  U_i \cap \inset{X_e} $:]\ \\
            Let $t$ be $t'$, but in $q_t'$ we take choice B of \cref{fun:GenEcoTest}, and then choice A in all recursive calls.\\

            $\exists t\in \TTS[L_i,e], q_t'\in Q_{t}\setminus\{\passState,\failState\}: t\Trans{\project{\subst{\sigma'}{\delta}{\theta}}{L_t}} q_t'\;\land$\\
            $X_e = e \after (\project{\sigma'}{L_e^\delta}) \land q_t'= \GenEcoTest(X_e)$\\
            \proofstep{\Cref{line:sync_input} of \cref{fun:GenInputResponse}}\\
            $\exists t\in \TTS[L_i,e], q_t\in Q_{t}\setminus\{\passState,\failState\}: t\Trans{\project{\subst{\sigma'}{\delta}{\theta}}{L_t}\cdot\ell} q_t\;\land$\\
            $X_e = e \after (\project{\sigma'}{L_e^\delta}) \land q_t= \GenEcoTest(X_e\after\ell)$\\
            \proofstep{\Cref{alg:testcase_alg,def:projection,def:substitution}: $\ell\in L_t$, $\ell\neq\delta$}\\
            $\exists t\in \TTS[L_i,e], q_t\in Q_{t}\setminus\{\passState,\failState\}: t\Trans{\project{\subst{\sigma'\cdot\ell}{\delta}{\theta}}{L_t}} q_t$
        \item[case f, $\ell \in (I_i\cap I_e)\setminus\inset{X_e}$:]\ \\ 
            $X_e = e \after (\project{\sigma'}{L_e^\delta})$\\
            \proofstep{$\sigma'\cdot\ell\in\utraces{i\parcomp e}$}\\
            $\ell\in\inset{i\parcomp e \after \sigma'}\;\land$\\
            $X_e = e \after (\project{\sigma'}{L_e^\delta})$\\
            \proofstep{\Cref{lem:traces_after_projection_subset}}\\
            $\ell\in\inset{i\parcomp e \after \sigma'}\;\land$\\
            $X_e\subseteq \pi_2(i\parcomp e \after \sigma')\;\land$\\
            $X_e = e \after (\project{\sigma'}{L_e^\delta})$\\
            \proofstep{\Cref{def:parcomp}:$T_{\parcomp}$}\\
            $\forall q_e \in Q_e, q_i\in Q_i: q_i \parcomp q_e \in i\parcomp e \after \sigma' \implies q_e \Trans{\ell}\;\land$\\
            $X_e\subseteq \pi_2(i\parcomp e \after \sigma')\;\land$\\
            $X_e = e \after (\project{\sigma'}{L_e^\delta})$\\
            \proofstep{\Cref{def:inset}: $\inset{}$ and $\subseteq$ transitive}\\
            $\ell\in \inset{X_e}$\\
            This contradicts the current case of $\ell \in (I_i\cap I_e)\setminus\inset{X_e}$, proving this case is empty.
        \item[case g, $\ell \in U_e\setminus\outset{X_e}$:]\ \\
            $X_e = e \after (\project{\sigma'}{L_e^\delta})$\\
            \proofstep{$\sigma'\cdot\ell\in\utraces{i\parcomp e}$}\\
            $\ell\in\outset{i\parcomp e \after \sigma'}\;\land$\\
            $X_e = e \after (\project{\sigma'}{L_e^\delta})$\\
            \proofstep{\Cref{def:outset,def:after}:$\outset{}$ and $\after$}\\
            $\exists q_i\in Q_i, q_e\in Q_e: i\parcomp e \Trans{\sigma'} q_i\parcomp q_e \land q_i\parcomp q_e \trans{\ell} ;\land$\\
            $X_e = e \after (\project{\sigma'}{L_e^\delta})$\\
            \proofstep{\Cref{lem:project_single_trace_from_parcomp}}\\
            $\exists q_e'\in Q_e:  e \Trans{\project{\sigma'}{L_e^\delta}} q_e' \land q_e' \trans{\ell} ;\land$\\
            $X_e = e \after (\project{\sigma'}{L_e^\delta})$\\
            \proofstep{\Cref{def:outset,def:after}:$\outset{}$ and $\after$}\\
            $\ell\in\outset{X_e}\;\land $
            $X_e = e \after (\project{\sigma'}{L_e^\delta})$\\
            This contradicts the current case of $\ell \in  U_e\setminus\outset{X_e}$, proving this case is empty.

        \item[case h, $\ell\in I_i\setminus L_e$:]\ \\
            Let $t$ be $t'$, but in $q_t'$ we take choice E of \cref{fun:GenEcoTest}, and then choice A in all recursive calls.\\
            
            $\exists t\in \TTS[L_i,e], q_t\in Q_{t}\setminus\{\passState,\failState\}: t\Trans{\project{\subst{\sigma'}{\delta}{\theta}}{L_t}} q_t\;\land$\\
            $X_e = e \after (\project{\project{\sigma'}{L_t^\delta}}{L_e^\delta}) \land q_t= \GenEcoTest(X_e)$\\
            \proofstep{\Cref{line:sut_internal_step} of \cref{fun:GenInputResponse}}\\
            $\exists t\in \TTS[L_i,e], q_t\in Q_{t}\setminus\{\passState,\failState\}: t\Trans{\project{\subst{\sigma'}{\delta}{\theta}}{L_t}\cdot\ell} q_t\;\land$\\
            $X_e = e \after (\project{\sigma'}{L_e^\delta}) \land q_t= \GenEcoTest(X_e)$\\
            \proofstep{\Cref{alg:testcase_alg,def:projection,def:substitution}: $\ell\in L_t$, $\ell\neq\delta$}\\
            $\exists t\in \TTS[L_i,e], q_t\in Q_{t}\setminus\{\passState,\failState\}: t\Trans{\project{\subst{\sigma'\cdot\ell}{\delta}{\theta}}{L_t}} q_t$
        \item[case i, $\ell\in U_i\setminus L_e$:]\ \\
            Take $t = t'$.\\
            $\exists t\in \TTS[L_i,e], q_t\in Q_{t}\setminus\{\passState,\failState\}: t\Trans{\project{\subst{\sigma'}{\delta}{\theta}}{L_t}} q_t\;\land$\\
            $X_e = e \after (\project{\project{\sigma'}{L_t^\delta}}{L_e^\delta}) \land q_t= \GenEcoTest(X_e)$\\
            \proofstep{\Cref{lem:L_testcase,def:projection,def:substitution}: $\ell\notin L_t$}\\
            $\exists t\in \TTS[L_i,e], q_t\in Q_{t}\setminus\{\passState,\failState\}: t\Trans{\project{\subst{\sigma'\cdot\ell}{\delta}{\theta}}{L_t}} q_t\;\land$\\
            $X_e = e \after (\project{\project{\sigma'}{L_t^\delta}}{L_e^\delta}) \land q_t= \GenEcoTest(X_e)$
        \item[case j, $\ell \in I_e\setminus(\inset{X_e}\cup L_i)$:]\ \\ 
            Same as case $\ell \in (I_i\cap I_e)\setminus\inset{X_e}$. I.e. case is empty because it contradicts $\sigma'\cdot\ell\in\utraces{i\parcomp e}$

        \end{case_distinction}
\end{case_distinction}

\end{proof}
\end{toappendix}

\FloatBarrier
\begin{lemmarep}
    \label{lem:project_to_testexec}
    Let $e\in \LTS$, $i\in \IOTS$, $i$ and $e$ be composable, $\sigma\in\utraces{i\parcomp e}$, $q_i\in Q_i$, $q_e\in Q_e$, then
    \[\begin{array}{l}
        i\parcomp e \Trans{\sigma} q_i \parcomp q_e\;  \land e \mutuallyaccepts{\sigma} i \implies \\
        \tab\exists t \in \TTS[L_i,e], q_t \in Q_t\setminus\{\passState,\failState\}: \\ 
        \tab t \testexec i \Trans{\subst{\sigma}{\delta}{\theta}} q_t \testexec q_i
    \end{array}\]
\end{lemmarep}

\begin{proof}
\ \\
    $i\parcomp e \Trans{\sigma} q_i\parcomp q_e \land e \mutuallyaccepts{\sigma} i$\\
\proofstep{\Cref{lem:project_to_test}}\\
$i\parcomp e \Trans{\sigma} q_i\parcomp q_e \land e \mutuallyaccepts{\sigma} i\;\land$\\
$\exists t \in \TTS[L_i,e],q_t\in Q_t\setminus\{\passState,\failState\}: t \Trans{\project{\subst{\sigma}{\delta}{\theta}}{L_t}} q_t $\\
\proofstep{\Cref{lem:project_single_trace_from_parcomp}}\\
$i \Trans{\project{\sigma}{L_i^\delta}} q_i\;\land$\\
$\exists t \in \TTS[L_i,e],q_t\in Q_t\setminus\{\passState,\failState\}: t \Trans{\project{\subst{\sigma}{\delta}{\theta}}{L_t}} q_t $\\
\proofstep{\Cref{lem:L_testexec}:$L_{t\testexec i}$}\\
$i \Trans{\project{\sigma}{L_i^\delta}} q_i \land \subst{\sigma}{\delta}{\theta}\in L_{t\testexec i}^*\;\land$\\
$\exists t \in \TTS[L_i,e],q_t\in Q_t\setminus\{\passState,\failState\}: t \Trans{\project{\subst{\sigma}{\delta}{\theta}}{L_t}} q_t $\\
\proofstep{Reflexivity of =}\\
$i \Trans{\project{\sigma}{L_i^\delta}} q_i \land \subst{\sigma}{\delta}{\theta}\in L_{t\testexec i}^*\;\land$\\
$\project{\subst{\subst{\sigma}{\delta}{\theta}}{\theta}{\delta}}{L_i^\delta} = \project{\subst{\subst{\sigma}{\delta}{\theta}}{\theta}{\delta}}{L_i^\delta}\;\land$\\
$\exists t \in \TTS[L_i,e],q_t\in Q_t\setminus\{\passState,\failState\}: t \Trans{\project{\subst{\sigma}{\delta}{\theta}}{L_t}} q_t $\\
\proofstep{\Cref{def:substitution}: $substitution$, $\theta\notin\sigma$}\\
$i \Trans{\project{\sigma}{L_i^\delta}} q_i \land \subst{\sigma}{\delta}{\theta}\in L_{t\testexec i}^* \;\land$\\
$\project{\subst{\subst{\sigma}{\delta}{\theta}}{\theta}{\delta}}{L_i^\delta} = \project{\sigma}{L_i^\delta}\;\land$\\
$\exists t \in \TTS[L_i,e],q_t\in Q_t\setminus\{\passState,\failState\}: t \Trans{\project{\subst{\sigma}{\delta}{\theta}}{L_t}} q_t $\\
\proofstep{\Cref{lem:project_from_testexec}}\\
$\exists t \in \TTS[L_i,e],q_t\in Q_t\setminus\{\passState,\failState\}: t\testexec i \Trans{\subst{\sigma}{\delta}{\theta}} q_t \testexec q_i $\\
\end{proof}

\begin{toappendix}
    \begin{lemmarep}
        \label{lem:testcase_alg_sound}
        The test suite obtained from all test cases generated by \Cref{alg:testcase_alg} is $\textbf{sound}$.
    \end{lemmarep}
    
    \begin{proof}
        From \Cref{def:exhaustive} we get the goal
\[\forall e \in \LTS, i\in \IOTS: i \eco e \implies i \passes \TTS[L_i,e]\]
We rewrite this to the following goal to prove here:
\[\forall i\in \IOTS:  (\exists t\in \TTS[L_i,e]: i \fails t) \implies i \noteco e\]

We give a proof by contradiction. Let $i\in \IOTS$ be an arbitrary implementation, for which $i \eco e$ holds. Then we show that the existence of a failing test would then imply $i \noteco e$, contradicting the assumption of $i \eco e$ and therefore showing that a failing test case cannot exist. Unpacking the definition of $i\fails t$ (\Cref{def:testrun}) then gives the following:\\
take $i\in \IOTS$, $e\in \LTS$, $t\in \TTS[L_i,e]$, $\sigma\in L_{t\testexec i}^*$, $q_i\in Q_i$ with  $i\eco e$ and $t\testexec i \Trans{\sigma} \failState \testexec q_i$.\\
Without loss of generality we take $\sigma'$ to be the longest prefix of $\sigma$ that does not fail. I.e. $t\testexec i \Trans{\sigma'} q_t' \testexec q_i' \Trans{\ell} \failState \testexec q_i'' \Trans{\sigma''} \failState \testexec q_i$, with $q_t'\neq\failState$, and $\sigma'\cdot\ell\cdot\sigma'' = \sigma$. This trace always exists, because $t$ itself cannot be $\failState$, and always needs at least one label to reach $\failState$ from the initial state as there are no $\tau$ transitions in $t$ (\Cref{alg:testcase_alg}, \Cref{def:testexec}, and \itemref{def:testcase}{no_tau}).\\   
To prove: $i\noteco e$\\

\noindent
$i\eco e \land t\testexec i \Trans{\sigma'} q_t' \testexec q_i' \Trans{\ell} \failState \testexec q_i'' \Trans{\sigma''} \failState \testexec q_i\land q_t'\neq\failState$\\
\proofstep{\Cref{lem:project_from_testexec}}\\
$i\eco e \land t\Trans{\project{\sigma'}{L_t}} q_t'  \Trans{\project{\ell}{L_t}} \failState  \land q_t'\neq\failState \;\land$\\
$i \Trans{\project{\subst{\sigma'}{\theta}{\delta}}{L_i^\delta}}  q_i' \Trans{\project{\subst{\ell}{\theta}{\delta}}{L_i^\delta}}  q_i''\;\land$\\
\proofstep{\Cref{lem:map_test_to_utraces}}\\
$i\eco e \land t\Trans{\project{\sigma'}{L_t}} q_t'  \Trans{\project{\ell}{L_t}} \failState  \land q_t'\neq\failState \;\land$\\
$i \Trans{\project{\subst{\sigma'}{\theta}{\delta}}{L_i^\delta}}  q_i' \Trans{\project{\subst{\ell}{\theta}{\delta}}{L_i^\delta}}  q_i''\;\land$\\
$\project{\subst{\project{\sigma'}{L_t}}{\theta}{\delta}}{L_e^\delta}\in \utraces{e}$\\
\proofstep{\Cref{lem:straces_equals_utraces_iots}}\\
$i\eco e \land t\Trans{\project{\sigma'}{L_t}} q_t'  \Trans{\project{\ell}{L_t}} \failState  \land q_t'\neq\failState \;\land$\\
$i \Trans{\project{\subst{\sigma'}{\theta}{\delta}}{L_i^\delta}}  q_i' \Trans{\project{\subst{\ell}{\theta}{\delta}}{L_i^\delta}}  q_i''\;\land$\\
$\project{\subst{\project{\sigma'}{L_t}}{\theta}{\delta}}{L_e^\delta}\in \utraces{e}\;\land$\\
$\project{\subst{\sigma'}{\theta}{\delta}}{L_i^\delta}\in \utraces{i}$\\
\proofstep{\Cref{lem:testcase_eq_genafter}}\\
$i\eco e \land q_t'  \Trans{\project{\ell}{L_t}} \failState \;\land$\\
$i \Trans{\project{\subst{\sigma'}{\theta}{\delta}}{L_i^\delta}}  q_i' \Trans{\project{\subst{\ell}{\theta}{\delta}}{L_i^\delta}}  q_i''\;\land$\\
$q_t'= \GenEcoTest\bigl(e \after (\project{\subst{\project{\sigma'}{L_t}}{\theta}{\delta}}{L_e^\delta})\bigr)\;\land$\\
$\project{\subst{\project{\sigma'}{L_t}}{\theta}{\delta}}{L_e^\delta}\in \utraces{e}\;\land$\\
$\project{\subst{\sigma'}{\theta}{\delta}}{L_i^\delta}\in \utraces{i}$\\
\proofstep{\Cref{def:projection}: $\sigma'\in L_{t\testexec i}^* \land \delta\notin L_{t\testexec i}^*$}\\
$i\eco e \land q_t'  \Trans{\project{\ell}{L_t}} \failState \;\land$\\
$i \Trans{\project{\subst{\sigma'}{\theta}{\delta}}{L_i^\delta}}  q_i' \Trans{\project{\subst{\ell}{\theta}{\delta}}{L_i^\delta}}  q_i''\;\land$\\
$q_t'= \GenEcoTest\bigl(e \after (\project{\subst{\project{\sigma'}{L_t^\delta}}{\theta}{\delta}}{L_e^\delta})\bigr)\;\land$\\
$\project{\subst{\project{\sigma'}{L_t^\delta}}{\theta}{\delta}}{L_e^\delta}\in \utraces{e}\;\land$\\
$\project{\subst{\sigma'}{\theta}{\delta}}{L_i^\delta}\in \utraces{i}$\\
\proofstep{\Cref{def:projection}: $\theta\in L_t^\delta \land \delta\in L_t^\delta$}\\
$i\eco e \land q_t'  \Trans{\project{\ell}{L_t}} \failState \;\land$\\
$i \Trans{\project{\subst{\sigma'}{\theta}{\delta}}{L_i^\delta}}  q_i' \Trans{\project{\subst{\ell}{\theta}{\delta}}{L_i^\delta}}  q_i''\;\land$\\
$q_t'= \GenEcoTest\bigl(e \after (\project{\project{\subst{\sigma'}{\theta}{\delta}}{L_t^\delta}}{L_e^\delta})\bigr)\;\land$\\
$\project{\project{\subst{\sigma'}{\theta}{\delta}}{L_t^\delta}}{L_e^\delta}\in \utraces{e}\;\land$\\
$\project{\subst{\sigma'}{\theta}{\delta}}{L_i^\delta}\in \utraces{i}$\\
\proofstep{\Cref{lem:L_testcase,def:projection}: $L_e^\delta \subseteq L_t^\delta$}\\
$i\eco e \land q_t'  \Trans{\project{\ell}{L_t}} \failState \;\land$\\
$i \Trans{\project{\subst{\sigma'}{\theta}{\delta}}{L_i^\delta}}  q_i' \Trans{\project{\subst{\ell}{\theta}{\delta}}{L_i^\delta}}  q_i''\;\land$\\
$q_t'= \GenEcoTest\bigl(e \after (\project{\subst{\sigma'}{\theta}{\delta}}{L_e^\delta})\bigr)\;\land$\\
$\project{\subst{\sigma'}{\theta}{\delta}}{L_e^\delta}\in \utraces{e}\;\land$\\
$\project{\subst{\sigma'}{\theta}{\delta}}{L_i^\delta}\in \utraces{i}$\\
\proofstep{\Cref{lem:project_utraces,lem:eco_equals_mutaccepts}}\\
$q_t'  \Trans{\project{\ell}{L_t}} \failState \;\land$\\
$i \Trans{\project{\subst{\sigma'}{\theta}{\delta}}{L_i^\delta}}  q_i' \Trans{\project{\subst{\ell}{\theta}{\delta}}{L_i^\delta}}  q_i''\;\land$\\
$q_t'= \GenEcoTest\bigl(e \after (\project{\subst{\sigma'}{\theta}{\delta}}{L_e^\delta})\bigr)\;\land$\\
$\subst{\sigma'}{\theta}{\delta}\in \utraces{i\parcomp e}$\\
\proofstep{\itemref{def:testcase}{no_tau}}\\
$q_t'  \trans{\project{\ell}{L_t}} \failState\;\land$\\
$i \Trans{\project{\subst{\sigma'}{\theta}{\delta}}{L_i^\delta}}  q_i' \Trans{\project{\subst{\ell}{\theta}{\delta}}{L_i^\delta}}  q_i''\;\land$\\
$q_t'= \GenEcoTest\bigl(e \after (\project{\subst{\sigma'}{\theta}{\delta}}{L_e^\delta})\bigr)\;\land$\\
$\subst{\sigma'}{\theta}{\delta}\in \utraces{i\parcomp e}$\\
\proofstep{\Cref{line:failtrans} of \Cref{fun:GenInputResponse} (Only possible way to get to $\failState$ from non-$\failState$)}\\
$i \Trans{\project{\subst{\sigma'}{\theta}{\delta}}{L_i^\delta}}  q_i' \Trans{\project{\subst{\ell}{\theta}{\delta}}{L_i^\delta}}  q_i''\;\land$\\
$q_t'= \GenEcoTest\bigl(e \after (\project{\subst{\sigma'}{\theta}{\delta}}{L_e^\delta})\bigr)\;\land$\\
$\ell\in U_i\cap I_e \land \ell \notin\inset{\project{\subst{\sigma'}{\theta}{\delta}}{L_e^\delta}}\;\land$\\
$\subst{\sigma'}{\theta}{\delta}\in \utraces{i\parcomp e}$\\
\proofstep{\Cref{def:after,def:inset}: $\after$ and $\inset{}$}\\
$i \Trans{\project{\subst{\sigma'}{\theta}{\delta}}{L_i^\delta}}  q_i' \Trans{\project{\subst{\ell}{\theta}{\delta}}{L_i^\delta}}  q_i''\;\land$\\
$\ell\in U_i\cap I_e \land \exists q_e' \in Q_e: e \Trans{\project{\subst{\sigma'}{\theta}{\delta}}{L_e^\delta}} q_e' \land q_e' \Nottrans{\ell}\;\land$\\
$\subst{\sigma'}{\theta}{\delta}\in \utraces{i\parcomp e}$\\
\proofstep{\Cref{def:projection,def:substitution}: $\ell\in L_i^\delta \land \ell\neq\theta$}\\
$i \Trans{\project{\subst{\sigma'}{\theta}{\delta}}{L_i^\delta}}  q_i' \Trans{\ell}  q_i''\;\land$\\
$\ell\in U_i\cap I_e \land \exists q_e' \in Q_e: e \Trans{\project{\subst{\sigma'}{\theta}{\delta}}{L_e^\delta}} q_e' \land q_e' \Nottrans{\ell}\;\land$\\
$\subst{\sigma'}{\theta}{\delta}\in \utraces{i\parcomp e}$\\
\proofstep{\Cref{def:arrowdefs}:$\Trans{}$}\\
$\exists q_i''' \in Q_i: i \Trans{\project{\subst{\sigma'}{\theta}{\delta}}{L_i^\delta}}  q_i''' \trans{\ell} \;\land$\\
$\ell\in U_i\cap I_e \land \exists q_e' \in Q_e: e \Trans{\project{\subst{\sigma'}{\theta}{\delta}}{L_e^\delta}} q_e' \land q_e' \Nottrans{\ell}\;\land$\\
$\subst{\sigma'}{\theta}{\delta}\in \utraces{i\parcomp e}$\\
\proofstep{\Cref{def:outset,def:inset}: $\outset{}$ and $\inset{}$}\\
$\exists q_i''' \in Q_i: i \Trans{\project{\subst{\sigma'}{\theta}{\delta}}{L_i^\delta}}  q_i'''\land  \ell \in \outset{\ell} \;\land$\\
$\ell\in U_i\cap I_e \land \exists q_e' \in Q_e: e \Trans{\project{\subst{\sigma'}{\theta}{\delta}}{L_e^\delta}} q_e' \land \ell \notin \inset{q_e'} \;\land$\\
$\subst{\sigma'}{\theta}{\delta}\in \utraces{i\parcomp e}$\\
\proofstep{\Cref{lem:project_from_parcomp_light}}\\
$\exists q_i''' \in Q_i, q_e' \in Q_e: i\parcomp e \Trans{\subst{\sigma'}{\theta}{\delta}}  q_i''' \parcomp q_e' \land  \ell \in \outset{\ell} \;\land$\\
$\ell\in U_i\cap I_e \land \ell \notin \inset{q_e'} \;\land$\\
$\subst{\sigma'}{\theta}{\delta}\in \utraces{i\parcomp e}$\\
\proofstep{\Cref{def:mutually_accepts,def:accepting}: $\accepts{}$ and $\mutuallyaccepts{}$}\\
$i\notmutuallyaccepts{} e$\\
\proofstep{\Cref{lem:eco_equals_mutaccepts}}\\
$i\noteco e$

    \end{proof}
    
    \begin{lemmarep}
        \label{lem:testcase_alg_exhaustive}
        The test suite obtained from all test cases generated by \Cref{alg:testcase_alg} is $\textbf{exhaustive}$.
    \end{lemmarep}
    
    \begin{proof}
        From \cref{def:exhaustive} we get the goal
\[\forall i\in \IOTS: i \eco e \impliedby i \passes \TTS[L_i,e]\]
We rewrite this to the following goal to prove here:
\[\forall i\in \IOTS: i \noteco{} e \implies (\exists t\in \TTS[L_i,e]: i \fails t)\]
Let $i\in \IOTS$ be an arbitrary implementation. then
using \cref{lem:eco_equals_mutaccepts,def:mutually_accepts,def:accepting} We rewrite this further to:\\

{
\centering
$(\exists \sigma \in \utraces{i\parcomp e}, q_i\in Q_i, q_e\in Q_e:$\\
$ i\parcomp e \Trans{\sigma} q_i\parcomp q_e \land \outset{q_i}\cap L_e\nsubseteq \inset{q_e}\cap L_i) \implies $\\
$(\exists t\in \TTS[L_i,e]: i \fails t)$\\
}
\noindent
Note that we can drop the option $\outset{q_e}\cap L_i\nsubseteq \inset{q_i} \cap L_e$ because it rewrites to false because $i$ is input enabled (\cref{def:iots}).
We therefore take\\
$\sigma\in \utraces{i\parcomp e}$, $q_i\in Q_i$, $q_e\in Q_e$, $i\parcomp e \Trans{\sigma} q_i\parcomp q_e$, $\ell\in U_i\cap I_e$, $q_i\trans{\ell}$ and $q_e\Nottrans{\ell}$. Without loss of generality, we also assume $i\parcomp e \mutuallyaccepts{\sigma} q_i \parcomp q_e$. This is possible, because if it is not, then by \cref{def:accepting_traces} we have a sub-trace $\sigma'$, also in $\utraces{i\parcomp e}$ by \Cref{item:utrace_prefix_closed}, for which we have exactly the same properties we assumed about $\sigma$. Since $i\parcomp e \mutuallyaccepts{\epsilon} i \parcomp e$ is always true, there is always a shortest sub-trace of $\sigma$ that is accepting and we can still use as a counterexample. This corresponds with the idea of taking the shortest prefix of $\sigma$ that is still a counterexample to $i\eco e$.

To prove: $\exists t \in \TTS[L_i,e]: i \fails t$.\\

\noindent
$\sigma\in \utraces{i\parcomp e}\land i\parcomp e \Trans{\sigma} q_i\parcomp q_e \land i\parcomp e \mutuallyaccepts{\sigma} q_i\parcomp q_e \land \ell\in U_i\cap I_e \land q_i\trans{\ell} \land\; q_e\Nottrans{\ell}$\\
\proofstep{\Cref{lem:project_single_trace_from_parcomp}}\\
$\sigma\in \utraces{i\parcomp e}\land e \Trans{\project{\sigma}{L_e^\delta}} q_e \land i\parcomp e \mutuallyaccepts{\sigma} q_i\parcomp q_e \land \ell\in U_i\cap I_e \land q_i\trans{\ell} \land\; q_e\Nottrans{\ell}$\\
\proofstep{\Cref{lem:project_to_testexec}}\\
$e \Trans{\project{\sigma}{L_e^\delta}} q_e  \land \ell\in U_i\cap I_e \land q_i\trans{\ell} \land\; q_e\Nottrans{\ell}\;\land$\\
$\exists t\in \TTS[L_i,e], q_t\in Q_t\setminus\{\failState,\passState\}: t\testexec i \Trans{\subst{\sigma}{\delta}{\theta}} q_t\testexec q_i$\\
\proofstep{\Cref{lem:project_from_testexec}}\\
$e \Trans{\project{\sigma}{L_e^\delta}} q_e  \land \ell\in U_i\cap I_e \land q_i\trans{\ell} \land\; q_e\Nottrans{\ell}\;\land$\\
$\exists t\in \TTS[L_i,e], q_t\in Q_t\setminus\{\failState,\passState\}: t\testexec i \Trans{\subst{\sigma}{\delta}{\theta}} q_t\testexec q_i\;\land$\\
$t\Trans{\project{\subst{\sigma}{\delta}{\theta}}{L_t}} q_t$\\
\proofstep{\Cref{lem:testcase_eq_genafter}}\\
$e \Trans{\project{\sigma}{L_e^\delta}} q_e  \land \ell\in U_i\cap I_e \land q_i\trans{\ell} \land\; q_e\Nottrans{\ell}\;\land$\\
$\exists t\in \TTS[L_i,e], q_t\in Q_t\setminus\{\failState,\passState\}: t\testexec i \Trans{\subst{\sigma}{\delta}{\theta}} q_t\testexec q_i\;\land$\\
$q_t = \GenEcoTest\bigl(e \after (\project{\subst{\project{\subst{\sigma}{\delta}{\theta}}{L_t}}{\theta}{\delta}}{L_e^\delta}) \bigr)$\\
\proofstep{\Cref{lem:parcomp_test_trace_simplification}}\\
$e \Trans{\project{\sigma}{L_e^\delta}} q_e  \land \ell\in U_i\cap I_e \land q_i\trans{\ell} \land\; q_e\Nottrans{\ell}\;\land$\\
$\exists t\in \TTS[L_i,e], q_t\in Q_t\setminus\{\failState,\passState\}: t\testexec i \Trans{\subst{\sigma}{\delta}{\theta}} q_t\testexec q_i\;\land$\\
$q_t = \GenEcoTest\bigl(e \after (\project{\sigma}{L_e^\delta}) \bigr)$\\
\proofstep{\Cref{def:inset}:$\inset{}$}\\
$\ell\in U_i\cap I_e \land q_i\trans{\ell} \land\; \ell \notin \inset{e \after (\project{\sigma}{L_e^\delta})}\;\land$\\
$\exists t\in \TTS[L_i,e], q_t\in Q_t\setminus\{\failState,\passState\}: t\testexec i \Trans{\subst{\sigma}{\delta}{\theta}} q_t\testexec q_i\;\land$\\
$q_t = \GenEcoTest\bigl(e \after (\project{\sigma}{L_e^\delta}) \bigr)$\\
\proofstep{\Cref{line:failtrans} of \cref{fun:GenInputResponse}}\\
$\ell\in U_i\cap I_e \land q_i\trans{\ell}\land\; q_t\trans{\ell}\failState \;\land$\\
$\exists t\in \TTS[L_i,e], q_t\in Q_t\setminus\{\failState,\passState\}: t\testexec i \Trans{\subst{\sigma}{\delta}{\theta}} q_t\testexec q_i\;\land$\\
\proofstep{\Cref{def:testexec}: $T_{t\testexec i}$}\\
$\exists t\in \TTS[L_i,e], q_t\in Q_t\setminus\{\failState,\passState\} , q_i'\in Q_i:$\\
\tab$t\testexec i \Trans{\subst{\sigma}{\delta}{\theta}} q_t\testexec q_i\land q_t\testexec q_i \trans{\ell} \failState \testexec q_i'$\\
\proofstep{\Cref{def:arrowdefs}: $\Trans{}$}\\
$\exists t\in \TTS[L_i,e], q_t\in Q_t\setminus\{\failState,\passState\} q_i'\in Q_i:$\\
\tab$t\testexec i \Trans{\subst{\sigma}{\delta}{\theta}\cdot\ell} \failState \testexec q_i'$\\
\proofstep{\Cref{def:testrun}}\\
$ i \fails \TTS[L_i,e]$
    \end{proof}
\end{toappendix}

\begin{theoremrep}
    \label{lem:eco_testing_complete}
    The test suite obtained from all test cases generated by \Cref{alg:testcase_alg} is $\textbf{complete}$.
\end{theoremrep}

\begin{proof}
    Follows directly from \Cref{def:complete,lem:testcase_alg_exhaustive,lem:testcase_alg_sound}
\end{proof}

\FloatBarrier
\section{Combining Algorithms}
\label{sec:combining_algorithms}
\declareComposedlabel{spec:combined_eco_uioco_testing_s}{spec:combined_eco_uioco_testing_e}
\declareComposedlabel{imp:combined_eco_uioco_testing_s}{imp:combined_eco_uioco_testing_e}

In previous sections we have introduced algorithms to test whether an implementation is eco conform to the specification of its environment. Previous work has also defined an algorithm for testing whether an implementation $\uioco$ conforms to its own specification \cite{vanderbijl_CompositionalTestingIoco_2004}, which is briefly repeated here as \Cref{alg:test_uioco}. These two algorithms for testing \eco{} and \uioco{} respectively have a lot of overlap. They are both using the same structure of non-deterministically alternating between sending and receiving labels and then giving a pass or fail judgement based on the received labels. The main difference is not in the structure, but in the labels that are chosen to be sent to the SUT, and the judgement passed based on the received labels. This leaves two obvious questions:
\begin{enumerate}
    \item Can we avoid double work by running both algorithms at the same time if the labels chosen by the algorithms are the same?
    \item Can we combine both algorithms into a sensible new one, sending only inputs agreed upon by both algorithms, and giving the verdict fail if either of the two individual algorithms would fail?
\end{enumerate}

The first question is important, as often the majority of testing time is spent waiting for the SUT to either reset or respond to a given input. If the same response can be used for multiple purposes, this should significantly reduce testing time, or allow for running more tests in the same amount of time.
The second question goes one step further, and asks if the inputs that are only part of one algorithm but not of the other are really required. Not giving them would speed up testing even more, but do we lose meaningful coverage this way, or were these inputs not relevant to begin with? We discuss these questions in more detail in the rest of this section. 

We give an algorithm that combines the $\eco$ and $\uioco$ based tests in \Cref{sec:combined on the fly testing}, and then prove this algorithm correct in \Cref{sec:testcase_alogorithm_correctness}. As a trade-off for the optimizations discussed in this section, we have to restrict the class of valid specifications when using \Cref{alg:eco_uioco_combined} to those that do not have non-deterministic underspecification, which will be explained in the proof sketch. In practice this is not a very serious restriction, as this is already a desirable property to have in a model and it does not remove any expressive power.

\subsection{Combined On-the-Fly Testing}
\label{sec:combined on the fly testing}

\Cref{alg:eco_uioco_combined} shows the combination of \Cref{alg:test_eco} and \Cref{alg:test_uioco} according to the ideas outlined above: if either algorithm would report failure (Case C), then the new algorithm also reports failure. Conversely, \Cref{alg:eco_uioco_combined} will only give an input in case B if both of the other algorithms can also give this input in case B. 

This new algorithm immediately shows an advantage of combining the algorithms: Case E of \Cref{alg:test_eco} is now integrated into Case B and no longer needs special treatment. This is because \Cref{alg:test_eco} does not know anything about the non-interacting inputs of the SUT, and therefore needs to overcompensate by assuming any of them can happen at any time. \Cref{alg:eco_uioco_combined} can however extract this information from the specification of the SUT and only needs to cover the inputs that are specified in the current state, instead of the entire input alphabet.

\LinesNotNumbered
\begin{algorithm}
\caption{On-the-Fly Testing for \cioco{}{s}{e}}
\label{alg:eco_uioco_combined}
    \KwIn{$e,s\in\LTS$, connection to the SUT}
    \nl\Assign{$X$}{$s\parcomp e \after \epsilon$}
    non-deterministically execute a finite number of the following cases,\\
    until the test either Passes or Fails:\\
    \textbf{\textit{(A) Stop testing:}}\\
    \Indp
        \nl the test Passes\;
    \Indm
    \textbf{\textit{(B) Perform an input $\ell$ on the SUT:}}\\
    \Indp
        \nl choose $\ell \in \inset[big]{\pi_1(X)} \cap \bigl(\inset{X} \cup \outset{X}\bigr)$\tcp*{Intersection of \Cref{alg:test_uioco,alg:test_eco}}
        \nl send $\ell$ to the SUT\;
        \nl\Assign{$X$}{$X \after \ell$}
    \Indm
    \textbf{\textit{(C) Observe an output or quiescence $\ell\in U_s^\delta$ from the SUT:}}\\
    \algIndentp
    \nl\If{$\ell \in I_e \land \ell\notin \inset[big]{\pi_2(X)} $}
        {
            \nl the test Fails\tcp*{Failure according to \Cref{alg:test_eco}}
        }
    \nl\If{$\ell\notin \outset[big]{\pi_1(X)}$}
        {
            \nl the test Fails\tcp*{Failure according to \Cref{alg:test_uioco}}
        }
    \nl\If{$\ell\neq\delta \lor \delta\in\outset[big]{\pi_2(X)} $}
        {
            \nl\Assign[]{$X$}{$X \after \ell$}\tcp*{Union of \Cref{alg:test_uioco,alg:test_eco}}
        }
    \algIndentm
    \SetInd{\dimexpr\skiprule-\algoskipindent}{\dimexpr\skiptext+\algoskipindent}
    \textbf{\textit{(D) Simulate non-interacting behaviour in the environment}}\\
    \Indp
        \nl choose $\ell \in (\outset{X} \cup \inset{X})\setminus L_s$\tcp*{The same as \Cref{alg:test_eco} } 
        \nl \Assign{$X$}{$X \after \ell$}
    \Indm
\end{algorithm}
\LinesNumbered

We write $\cioco{i}{s}{e}$ iff an implementation $i$ always passes \Cref{alg:eco_uioco_combined} for specification $s$ and environment specification $e$. Testing for $\cioco{i}{s}{e}$ uses less inputs than testing for $\uioco$ and $\eco$ separately. However, we obtain neither the desired $i \eco e$, nor $i\uioco s$. This makes sense, as regular $\uioco$ tests for the correctness of an implementation under any possible environment which correctly interfaces with the specification $s$. When running \Cref{alg:eco_uioco_combined} however, we are instead testing with one specific environment in mind: the part of $e$ that is reachable by the \utraces{} of $s$. The same is true for $i\eco e$, which tests for correct behaviour while allowing any sequence of inputs not coming from $e$. In reality these inputs are however restricted by $s$. This information is not available in \Cref{alg:test_eco} but is taken into account in \Cref{alg:eco_uioco_combined}.

\begin{example}
\label{examp:cioco_notimp_uioco_eco}
In \Cref{fig:combined_eco_uioco_testing_example}, we have $\cref{imp:combined_eco_uioco_testing_s} \notuioco \cref{spec:combined_eco_uioco_testing_s}$: after the utrace $\ltslabel{b}$ we observe quiescence, but the output \ltslabel{x} is expected. Similarly, after the trace $a\cdot r\cdot x$, which is a utrace of $\cref{imp:combined_eco_uioco_testing_s}\parcomp\cref{spec:combined_eco_uioco_testing_e}$, \cref{imp:combined_eco_uioco_testing_s} can produce the output $x$ again which is not accepted at this point in \cref{spec:combined_eco_uioco_testing_e}. This means that we also have \cref{imp:combined_eco_uioco_testing_s} \noteco{} \cref{spec:combined_eco_uioco_testing_e}.

However, we do have $\cioco{\cref{imp:combined_eco_uioco_testing_s}}{\cref{spec:combined_eco_uioco_testing_s}}{\cref{spec:combined_eco_uioco_testing_e}}$. This is because while the behaviour of \cref{imp:combined_eco_uioco_testing_s} after the utrace $b$ is not acceptable, the trace $b$ cannot occur with \cref{spec:combined_eco_uioco_testing_e} as the environment. $b$ is part of the output label set of $\cref{spec:combined_eco_uioco_testing_e}$, but it is not generated while \cref{spec:combined_eco_uioco_testing_s} is in state 1.  Similarly, the \ltslabel{x} transition in state 5 of \cref{imp:combined_eco_uioco_testing_s} is not handled by the environment, but this transition can only be reached by doing an \ltslabel{r} transition in state 2 of \cref{imp:combined_eco_uioco_testing_s} which is not part of any of the \utraces{} of \cref{spec:combined_eco_uioco_testing_s}.
\end{example}

Another argument for why \Cref{alg:eco_uioco_combined} is correct even though it might give the verdict pass in \Cref{examp:cioco_notimp_uioco_eco}, is that after composition we do actually obtain a correct system as shown in \Cref{examp:cioco_fix_postcomp}. 

\begin{example}
\label{examp:cioco_fix_postcomp}
\Cref{fig:combined_eco_uioco_testing_example_composed} shows the composed specifications 
\cref{spec:combined_eco_uioco_testing_composed} and composed implementations 
\cref{imp:combined_eco_uioco_testing_composed} from \Cref{fig:combined_eco_uioco_testing_example}. Here we 
also see that the problematic \ltslabel{?b} transition from state 1 of 
\cref{spec:combined_eco_uioco_testing_s} has been dropped from 
\cref{spec:combined_eco_uioco_testing_composed}. \cref{imp:combined_eco_uioco_testing_composed}  has a couple of
suspicious states hanging below state 2B. However, these extra states are only reachable by an
\ltslabel{?r} transition, which is not part of the \utraces{} of 
\cref{spec:combined_eco_uioco_testing_composed}. Therefore we have 
\cref{imp:combined_eco_uioco_testing_composed} \uioco{} \cref{spec:combined_eco_uioco_testing_composed}, even though \Cref{examp:cioco_notimp_uioco_eco} showed we have both \cref{imp:combined_eco_uioco_testing_s} \notuioco{} \cref{spec:combined_eco_uioco_testing_s} and  \cref{imp:combined_eco_uioco_testing_s} \noteco{} \cref{spec:combined_eco_uioco_testing_e}.
\end{example}

The restrictions imposed by both \cref{spec:combined_eco_uioco_testing_s} and \cref{spec:combined_eco_uioco_testing_e} make some of the transitions and states of \cref{imp:combined_eco_uioco_testing_s} unreachable in practice. 
This means that we don't have to consider these transitions when constructing our test suite. If we also test \cref{imp:combined_eco_uioco_testing_e} in this way, and only use inputs defined by both \cref{spec:combined_eco_uioco_testing_s} and \cref{spec:combined_eco_uioco_testing_e}, then we still obtain correctness of \cref{imp:combined_eco_uioco_testing_s} in a full system ($\cref{imp:combined_eco_uioco_testing_s} \parcomp \cref{imp:combined_eco_uioco_testing_e}$ \uioco{} $\cref{spec:combined_eco_uioco_testing_s} \parcomp \cref{spec:combined_eco_uioco_testing_e}$). This optimisation makes this testing approach faster, but also weaker than the approach introduced in \Cref{sec:testing-accepting-systems}. There, we tested for compositionality between an implementation and a specification of its environment. The results are then transferable if the environment implementation changes. Here, we instead test for the compositionality between two implementations, which means the results are obtained using less time, but have to be retested completely when anything changes.

\equalizeCounters
\begin{figure}[htb]
    \begin{subfigure}[b]{.50\linewidth}
    \setlength{\nodedistance}{0.7\nodedistance}
    \centering
    \newspeclabel\label{spec:combined_eco_uioco_testing_s}
    \begin{tikzpicture}[LTS]
    \node[state,initial above] (1) {1};
    \node[state,below left= of 1] (2) {2};
    \node[state,below right = of 1] (3) {3};
    \node[state,below right= of 2] (4) {4};

    \path[->] 
        (1) edge []    node [inline] {\ltslabel{?a}} (2)
            edge []    node [auto] {\ltslabel{?a}\\\ltslabel{?b}} (3)
        (2) edge [bend left]    node [auto] {\ltslabel{?r}} (1)
            edge []    node [inline] {\ltslabel{!x}} (4)
        (3) edge []    node [inline] {\ltslabel{!x}} (4)
        (4) edge []    node [inline] {\ltslabel{?b}} (1)
        ;
    
    \end{tikzpicture}
    \caption{\currentspec, with I=\{\ltslabel{a},\ltslabel{b},\ltslabel{r}\} and U=\{\ltslabel{x}\}}
    \end{subfigure}
    \qquad
    \begin{subfigure}[b]{.5\linewidth}
    \centering
    \newspeclabel\label{spec:combined_eco_uioco_testing_e}
    \begin{tikzpicture}[LTS]
    \node[state,initial] (A) {A};
    \node[state,below left= \nodedistance and 0.5\nodedistance of A] (B) {B};
    \node[state,right= of B] (C) {C};

    \path[->] 
        (A) edge [] node [inline] {\ltslabel{!a}} (B)
        (B) edge [] node [inline] {\ltslabel{?x}} (C)
        (C) edge [] node [inline] {\ltslabel{!b}} (A)
        ;
    
    \end{tikzpicture}
    \caption{\currentspec, with I=\{\ltslabel{x}\} and U=\{\ltslabel{a},\ltslabel{b}\} }
    \end{subfigure}
    \\

    \begin{subfigure}[b]{.50\linewidth}
    \centering
    \newimplabel\label{imp:combined_eco_uioco_testing_s}
    \begin{tikzpicture}[LTS]
    \setlength{\nodedistance}{0.7\nodedistance}
    \node[state,initial above] (1) {1};
    \node[state,below left= of 1] (2) {2};
    \node[state,below right = of 1] (3) {3};
    \node[state,below right= of 2] (4) {4};
    \node[state,below= of 2] (5) {5};


    \path[->] 
        (1) edge []    node [inline] {\ltslabel{?a}} (3)
            edge []    node [inline] {\ltslabel{?a}} (2)
            edge [loop left] node [auto] {\ltslabel{?r}\\\ltslabel{?b}} (1)
        (2) edge []    node [left] {\ltslabel{?r}} (5)
            edge []    node [inline] {\ltslabel{!x}} (4)
            edge [loop left ]node [auto] {\ltslabel{?a}\\\ltslabel{?b}\\} (2)
        (3) edge []    node [inline] {\ltslabel{!x}} (4)
            edge [loop right ]node [auto] {\ltslabel{?a}\\\ltslabel{?b}\\\ltslabel{?r}} (3)
        (4) edge []    node [inline] {\ltslabel{?b}} (1)
            edge [loop right ]node [auto] {\ltslabel{?a}\\\ltslabel{?r}} (5)
        (5) edge [loop left ]node [auto] {\ltslabel{?a}\\\ltslabel{?b}\\\ltslabel{?r}\\\ltslabel{!x}} (5)

        ;
    
    \end{tikzpicture}
    \caption{\currentimp}
    \end{subfigure}
    \qquad
    \begin{subfigure}[b]{.50\linewidth}
    \centering
    \newimplabel\label{imp:combined_eco_uioco_testing_e}
    \begin{tikzpicture}[LTS]
    \node[state,initial] (A) {A};
    \node[state,below left= \nodedistance and 0.5\nodedistance of A] (B) {B};
    \node[state,right= of B] (C) {C};

    \path[->] 
        (A) edge []             node [inline] {\ltslabel{!a}} (B)
            edge [loop right]   node [auto] {\ltslabel{?x}} (A)
        (B) edge []             node [inline] {\ltslabel{?x}} (C)
        (C) edge [loop right]   node [auto] {\ltslabel{?x}} (C)
            edge []   node [inline] {\ltslabel{!b}} (A)
;    
    \end{tikzpicture}
    \caption{\currentimp}
    \end{subfigure}
    \caption{Examples for \Cref{alg:eco_uioco_combined}}
    \label{fig:combined_eco_uioco_testing_example}
\end{figure}

\begin{figure}[htb]
    \begin{subfigure}[b]{.50\linewidth}
    \centering
    \defineComposedlabel{spec:combined_eco_uioco_testing_s}{spec:combined_eco_uioco_testing_e}\label{spec:combined_eco_uioco_testing_composed}
    \begin{tikzpicture}[LTS]
    \setlength{\nodedistance}{0.7\nodedistance}
    \node[state,initial above] (1A) {1A};
    \node[state,below left= of 1A] (2B) {2B};
    \node[state,below= of 2B] (1B) {1B};
    \node[state,below right= of 1A] (3B) {3B};
    \node[state,below right= of 2B] (4C) {4C};
    
    \path[->] 
        (1A) edge []            node [inline] {\ltslabel{!a}} (2B)
             edge []            node [inline] {\ltslabel{!a}} (3B)
        (2B) edge []            node [inline] {\ltslabel{!x}} (4C)
             edge []            node [left] {\ltslabel{?r}} (1B)
        (3B) edge []            node [inline] {\ltslabel{!x}} (4C)
        (4C) edge []            node [inline] {\ltslabel{!b}} (1A)
             ;
    
    \end{tikzpicture}
    \caption{\cref*{spec:combined_eco_uioco_testing_composed}, with I=\{\ltslabel{r}\} and U=\{\ltslabel{a},\ltslabel{b},\ltslabel{x}\} }
    \end{subfigure}
    \qquad
    \begin{subfigure}[b]{.50\linewidth}
    \centering
    \defineComposedlabel{imp:combined_eco_uioco_testing_s}{imp:combined_eco_uioco_testing_e}\label{imp:combined_eco_uioco_testing_composed}
    \begin{tikzpicture}[LTS]
    \setlength{\nodedistance}{0.7\nodedistance}
    \node[state,initial above] (1A) {1A};
    \node[state,below left= of 1A] (2B) {2B};
    \node[state,below right= of 1A] (3B) {3B};
    \node[state,below = of 2B] (5B) {5B};
    \node[state,below right= of 2B] (4C) {4C};
    \node[state, below = of 5B] (5C) {5C};
    \node[state, below right = of 5B] (5A) {5A};

    \path[->] 
        (1A) edge []            node [inline] {\ltslabel{!a}} (2B)
             edge []            node [inline] {\ltslabel{!a}} (3B)
             edge [loop right]  node [auto] {\ltslabel{?r}} (1A)
        (2B) edge []            node [left] {\ltslabel{?r}} (5B)
             edge []            node [inline] {\ltslabel{!x}} (4C)
        (3B) edge []            node [inline] {\ltslabel{!x}} (4C)
             edge [loop right]  node [auto] {\ltslabel{?r}} (3B)
        (4C) edge []            node [inline] {\ltslabel{!b}} (1A)
             edge [loop right]  node [auto] {\ltslabel{?r}} (4C)
        (5A) edge [loop right]  node [auto] {\ltslabel{?r}\\\ltslabel{!x}} (5A)
             edge []            node [inline] {\ltslabel{!a}} (5B)
        (5B) edge [loop left]   node [auto] {\ltslabel{?r}} (5B)
             edge []            node [left] {\ltslabel{!x}} (5C)
        (5C) edge [loop left]   node [auto] {\ltslabel{?r}\\\ltslabel{!x}} (5C)
             edge []            node [below] {\ltslabel{!b}} (5A)
        
        ;
    
    \end{tikzpicture}
    \caption{\cref*{imp:combined_eco_uioco_testing_composed}}
    \end{subfigure}
    \caption{Composed systems from \Cref{fig:combined_eco_uioco_testing_example}}
    \label{fig:combined_eco_uioco_testing_example_composed}
\end{figure}
\FloatBarrier

\FloatBarrier
\subsection{Algorithm Correctness}
\label{sec:testcase_alogorithm_correctness}
\FloatBarrier
In order to more easily analyse and prove properties about \Cref{alg:eco_uioco_combined}, we follow the same strategy as in \Cref{sec:algorithm_correctness}. \Cref{alg:combined_testcase_alg} is a functional version of \Cref{alg:eco_uioco_combined} that returns a test case instead of directly interacting with the SUT. We will argue the correctness of \Cref{alg:combined_testcase_alg}, and therefore indirectly of \Cref{alg:eco_uioco_combined} by proving properties about these test cases. Since the algorithm is non-deterministic and can therefore return multiple different test cases from the same specifications, we reason about its output as if it is a set of test cases called \TTS[s,e].

\LinesNotNumbered
\begin{function}[htb]
\DontPrintSemicolon
\functionLabel{GenCombinedTest}
\caption{\currentFunName{}(\ProcArgFnt$X:\powerset{Q_{s\parcomp e}}$) }
    \KwData{$s,e\in\LTS$}
    \KwOut{A partial test case for \cioco{}{s}{e}, assuming $s$ and $e$ could be in any state $\pi_1(q)$ and $\pi_2(q)$ respectively, where $q\in X$}
    non-deterministically execute one of the following cases:\\
    \textbf{\textit{(A) Stop testing:}}\\
    \Indp
        \nl\label{line:passtrans}\Return{$\passState$}\;
    \Indm
    \textbf{\textit{(B) Emit a response to the SUT:}}\\
    \algIndentp
        \nl choose $\ell \in \inset[big]{\pi_1(X)} \cap \bigl(\inset{X} \cup \outset{X}\bigr)$\;
        \nl \Return \;
        \Indp
            \nl\label{line:combined_sync_out}$\ell\; \Bcomp$ \GenCombinedTest{$X \after \ell$} $\Bchoice$\;
            \nl\HandleSutOutput{$X$}\;
        \Indm
    \algIndentm
     \textbf{\textit{(C) Observe an output or quiescence from the SUT:}}\\
     
    \algIndentp
    \nl\If{$\delta\notin \outset[big]{\pi_1(X)}$}
        {
            \nl\label{line:theta_uioco_fail}\Assign{$B_\theta$}{$\theta \;\Bcomp\;\failState$}    
        }\nl\ElseIf{$\delta\in \outset{X}$}
        {
            \nl\label{line:theta_env}\Assign[]{$B_\theta$}{$\theta\; \Bcomp\;$\GenCombinedTest{$X \after \delta$}}
        }\nl\Else{
            \nl\label{line:theta_comp}\Assign[]{$B_\theta$}{$\theta_s\; \Bcomp\;$\GenCombinedTest{$X$}}
        }
    \nl\label{line:combined_sync_theta}\Return $B_\theta\Bchoice\;$\HandleSutOutput{$X$}\;
    \algIndentm
    \textbf{\textit{(D) Simulate non-interacting behaviour in environment}}\\
    \Indp
        \nl choose $\ell \in \bigl(\outset{X} \cup \inset{X}\bigr)\setminus L_s$ \; 
        \nl \Return\;
        \Indp
            \nl\label{line:combined_env_internal_step}$\ell\Bcomp$ \GenCombinedTest{$X \after \ell$}$\;\Bchoice$\;
            \nl\HandleSutOutput{$X$}\;
        \Indm
    \Indm
\end{function}

\begin{function}[htb]
\DontPrintSemicolon
\label[FunctionFormat\GenInputResponseFuncName]{fun:\GenInputResponseFuncName_2}%
\label[FunctionFormat\GenInputResponseFuncName]{fun:GenInputResponse_2}
\caption{\GenInputResponseFuncName{}(\ProcArgFnt$X:\powerset{Q_{s\parcomp e}}$) }
    \KwData{$s,e\in\LTS$}
    \KwOut{A partial test case describing how to respond to an output from the $SUT$}
    \Return \;
        \Indp
            \nl\label{line:failtrans_eco}$\Bsum \{\ell\Bcomp \failState \setbar \ell \in  \bigl(U_s \cap I_e\bigr) \setminus\inset[big]{\pi_2(X)} \} \Bchoice$\;
            \nl\label{line:failtrans_uioco}$\Bsum \{\ell\Bcomp \failState \setbar \ell \in  U_s \setminus\outset{X} \} \Bchoice$\;
            \nl\label{line:combined_sync_input}$\Bsum \{\ell\; \Bcomp\; $\GenCombinedTest{$X\after \ell$} $\setbar \ell \in U_s\cap\inset[big]{\pi_2(X)} \cap \outset[big]{X}\}\Bchoice$ \;
            \nl\label{line:combined_unsync_input}$\Bsum \{\ell\; \Bcomp\; $\GenCombinedTest{$X\after \ell$} $\setbar \ell \in (U_s\setminus L_e)\cap \outset{X}\}$ \;
            
        \Indm
\end{function}

\begin{algorithm}[htb]
    \caption{Test case generation for \cioco{}{s}{e}}
    \label{alg:combined_testcase_alg}
    \KwIn{A specification of the SUT $s\in \LTS$}
    \KwIn{A specification of the environment $e\in \LTS$}
    \KwOut{A test case for \cioco{}{s}{e}}
    
    \Return\\
    \Indp
        test case $t$ with $I_t=U_s\cup \{\theta,\theta_s\}$, $U_t = U_e \cup (I_e\setminus U_s)  \cup (I_s\setminus L_e)$.\\
        $Q_t$ and $T_t$ are given by \GenCombinedTest{$s \parcomp e \after \epsilon$}\;
    \Indm
\end{algorithm}

\FloatBarrier

We introduce a new form of listening for quiescence called $\theta_s$. Where regular $\theta$ listens for system wide quiescence, $\theta_s$ only listens for quiescence of $s$. This is required to give a failing verdict if $s$ is quiescent while it is not allowed to be by its own specification. Without this more granular notion, quiescence in one component is sometimes hidden in the composed system when the environment is not quiescent. Listening for $\theta_s$ could give information about the possible states of $s$, which is not available by listening for $\theta$. This extra information is not available when testing for $\uioco{}$-conformance to $s\parcomp e$. To reflect this $\theta_s$ is ignored by not altering the set of possible states $X$ (\cref{line:combined_sync_theta}). It is inserted as an explicit no-op, such that each iteration of the algorithm produces some measurable progress, even if that progress was waiting a bit to make sure the SUT is quiescent. This matches case C of \Cref{alg:eco_uioco_combined} which also does nothing when observing allowed quiescence in $s$ while $e$ is not quiescent.

\begin{toappendix}
\begin{lemma}
    \label{lem:combined_testgen_testcase}
    the output of \Cref{alg:combined_testcase_alg}  is a test case.
\end{lemma}
\begin{proof}
    We need to prove two things according to \Cref{def:testcase}: if $t$ is an output of \Cref{alg:combined_testcase_alg}, then:
\begin{case_distinction}
    \item[$\forall q \in Q_t: q \trans{\ell} $ for exactly one $\ell \in U_t \cup \{\theta\}$]\ \\
        All states of \Cref{fun:GenCombinedTest} are of the form $\ell\Bcomp\GenCombinedTest(\dots)\Bchoice\HandleSutOutput(\dots)$. All transitions in $\HandleSutOutput(\dots)$ are inputs of the testcase, so the only transition that can be an output in a given state is the $\ell$ from $\GenCombinedTest(\dots)$, of which there is just one.

    \item[$\forall q \in Q_t, \ell \in U_s: q \trans{\ell}$]\ \\
        The entirety of $U_s$ is covered by the four transitions of \Cref{fun:GenInputResponse_2}. Technically some labels are covered twice by both \cref{line:failtrans_eco} and \cref{line:failtrans_uioco} ($\ell\in U_s\cap i_e \setminus \Bigl(\outset{X}\cap \inset[big]{\pi_2(X)}\Bigr)$). Since the label, starting state, and target state are all the same we count those as the same transition defined twice.

\end{case_distinction}
    
\end{proof}

\begin{lemmarep}
        \label{lem:L_combined_testcase}
        let $t\in \TTS[s,e]$\\
        $L_t = U_s\cup \{\theta,\theta_s\} \cup U_e \cup (I_e\setminus U_s)  \cup (I_s\setminus L_e)  = L_e \cup \{\theta,\theta_s\} \cup L_s$
    \end{lemmarep}
\begin{proof}
    Follows directly from the definition of \Cref{alg:combined_testcase_alg}
\end{proof}
\end{toappendix}

Notably, \Cref{alg:combined_testcase_alg} is exhaustive, but not sound for system level $\uioco$-correctness. This is by design, as shown in the following example:

\FloatBarrier
\begin{example}
\label{examp:combined_alg_unsound} In \Cref{fig:soundness_counterexamp}
we have $\cref{imp:soundess_counters} \notuioco{} \cref{spec:soundess_counters}$,  $\cref{imp:soundess_countere} \notuioco{} \cref{spec:soundess_countere}$ and $\cref{spec:soundess_counters} \noteco{} \cref{spec:soundess_countere}$. However, this leads to both the composed specification $\cref{spec:soundess_counters}\parcomp\cref{spec:soundess_countere}$ and the composed implementation $\cref{imp:soundess_counters}\parcomp\cref{imp:soundess_countere}$ to drop all meaningful visible behaviour. Since an implementation with no outputs does correctly implement the empty specification, we get $\cref{imp:soundess_counters}\parcomp\cref{imp:soundess_countere} \uioco{} \cref{spec:soundess_counters}\parcomp\cref{spec:soundess_countere}$. This means that in order for \Cref{alg:combined_testcase_alg} to be sound, it would need to pass when running for \cref{imp:soundess_counters} using \cref{spec:soundess_counters} and \cref{spec:soundess_countere} as specifications. Intuitively however, this is not a situation for which you would want to give a verdict pass. The fact that the components are wrong is hidden at the system level because the integration of the specifications is also going wrong. However, these two things should not cancel out the testing verdict, and the test should still fail.
\end{example}

\begin{figure}
    \equalizeCounters
    \captionsetup[subfigure]{justification=centering}
    \begin{subfigure}[b]{.50\linewidth}
    \centering
    \newspeclabel\label{spec:soundess_counters}
    \newimplabel\label{imp:soundess_counters}
    \begin{tikzpicture}[LTS]
    \node[state,initial] (1) {1};
    \node[state,below= of 1] (2) {2};

    \path[->] 
        (1) edge [bend left]    node [auto] {\ltslabel{!a}} (2)
        (2) edge [bend left]    node [auto] {\ltslabel{?b}} (1)
        ;
    
    \end{tikzpicture}\qquad
    \begin{tikzpicture}[LTS]
    \node[state,initial] (1) {1};

    \path[->] 
        (1) edge [loop below]    node [auto] {\ltslabel{?b}\\$\delta$} (1)
        ;
    
    \end{tikzpicture}
    \caption{\currentspec{} (left) and \currentimp{} (right),\\
    with I=\{\ltslabel{b}\} and U=\{\ltslabel{a}\}}
    \end{subfigure}%
    \qquad
    \begin{subfigure}[b]{.50\linewidth}
    \centering
    \newspeclabel\label{spec:soundess_countere}
    \newimplabel\label{imp:soundess_countere}
    \begin{tikzpicture}[LTS]
    \node[state,initial] (A) {A};
    \node[state,below= of A] (B) {B};

    \path[->] 
        (A) edge [bend left]    node [auto] {\ltslabel{!b}} (B)
        (B) edge [bend left]    node [auto] {\ltslabel{?a}} (A)
        ;
    
    \end{tikzpicture}\qquad
    \begin{tikzpicture}[LTS]
    \node[state,initial] (A) {A};

    \path[->] 
        (A) edge [loop below]    node [auto] {\ltslabel{?a}\\$\delta$} (B)
        ;
    
    \end{tikzpicture}
    \caption{\currentspec{} (left) and \currentimp{} (right),\\
    with I=\{\ltslabel{a}\} and U=\{\ltslabel{b}\}}
    \end{subfigure}

    \caption{\Cref{alg:combined_testcase_alg} soundness counter-example}
    \label{fig:soundness_counterexamp}
\end{figure}

\Cref{examp:combined_alg_unsound} suggest that \uioco{}-conformance to the parallel composition of all component specifications might not be the perfect definition of a correct system. If the tests fail there is indeed something wrong with the implementation, but there are still undesirable implementations left that will pass the system level tests this way.
We will however leave refining the definition of what a correct system is as future work for now.

We will prove that \Cref{alg:combined_testcase_alg} is exhaustive, but only for specifications without non-deterministic underspecification. Normal underspecification happens when there are states in a model that do not accept every possible input. What happens if such a state input combination is triggered anyway is thus unspecified. This is a useful modelling tool that allows the models to focus on important behaviour, and still leave room to change less important details in the implementation. Non-deterministic underspecification happens when a specification can be in multiple states after a trace, and only some of these states accept a given input. This is undesirable, as $\uioco$ treats non-deterministic underspecification the same as regular underspecification. 
Adding non-deterministically underspecified inputs does not change the semantics of the model with respect to \uioco{}, but it does make the model harder to analyse. We call the absence of non-deterministically underspecified inputs \utraceClosed{}, as it means that all traces of a system are $\utraces{}$ (\Cref{def:utraceClosed}). This gives the useful property in \Cref{lem:utrace_closed_implies_cartesian_closed}, which means we don't need to keep track of states of the parallel composition ($\powerset{Q_s\times Q_e}$), but instead can keep track of the much smaller set of states of each individual specification ($\powerset{Q_s}\times \powerset{Q_e}$)

\begin{definition}
\label{def:utraceClosed}
$s\in \LTS$ is \utraceClosed{} iff:
\[\forall \sigma \in L_s^{\delta*}: s\Trans{\sigma} \implies \sigma \in \utraces{s} \]
    
\end{definition}

\begin{toappendix} 
\begin{lemmarep}
\label{lem:trans_delta_closed}
Let $s\in \LTS$, $\sigma\in L_s^\delta$
    \[s\Trans{\sigma} q_s \implies s \Trans{\project{\sigma}{L_s}} q_s\]
\end{lemmarep}
\begin{proof}
    Proof by induction on $\sigma$, with further case distinction based on whether the last label of $\sigma$ is $\delta$ or not.
\begin{case_distinction}
    \item[Base case: $\sigma=\epsilon$:]\ \\
        Trivially true due to $\epsilon = \project{\epsilon}{L_s}$ (\cref{def:projection}).
    \item[Inductive step: $\sigma=\sigma'\cdot\ell$:]\ \\
    The IH is:
    \[\forall q_s'\in Q_s: s\Trans{\sigma'} q_s' \implies s \Trans{\project{\sigma}{L_s}} q_s'\]

    \begin{case_distinction}
        \item[Case $\ell\neq\delta$]\ \\
            $s\Trans{\sigma'\cdot\ell} q_s$\\
            \proofstep{\Cref{def:arrowdefs}: $\Trans{\sigma'\cdot\ell}$}\\
            $\exists q_s', q_s''\in Q_s: s\Trans{\sigma'} q_s' \trans{\ell} q_s'' \Trans{\epsilon} q_s$\\
            \proofstep{Apply IH}\\
            $\exists q_s', q_s''\in Q_s: s\Trans{\project{\sigma'}{L_s}} q_s' \trans{\ell} q_s'' \Trans{\epsilon} q_s$\\
            \proofstep{\Cref{def:projection} $\projectop$ and $\ell\neq\delta$}\\
            $\exists q_s', q_s''\in Q_s: s\Trans{\project{\sigma'}{L_s}} q_s' \trans{\project{\ell}{L_s}} q_s'' \Trans{\epsilon} q_s$\\
            \proofstep{\Cref{def:arrowdefs}: $\Trans{}$}\\
            $s\Trans{\project{\sigma'}{L_s}'\cdot\project{\ell}{L_s}} q_s$\\
            \proofstep{\Cref{def:projection} $\projectop$}\\
            $s\Trans{\project{\sigma'\cdot\ell}{L_s}'} q_s$\\
        \item[Case $\ell=\delta$]\ \\
            $s\Trans{\sigma'\cdot\delta} q_s$\\
            \proofstep{\Cref{def:arrowdefs}: $\Trans{\sigma'\cdot\delta}$}\\
            $\exists q_s', q_s''\in Q_s: s\Trans{\sigma'} q_s' \trans{\delta} q_s'' \Trans{\epsilon} q_s$\\
            \proofstep{Apply IH}\\
            $\exists q_s', q_s''\in Q_s: s\Trans{\project{\sigma'}{L_s}} q_s' \trans{\delta} q_s'' \Trans{\epsilon} q_s$\\
            \proofstep{\Cref{def:delta}: $\delta$}\\
            $\exists q_s'\in Q_s: s\Trans{\project{\sigma'}{L_s}} q_s' \trans{\delta} q_s' \Trans{\epsilon} q_s$\\
            \proofstep{\Cref{def:arrowdefs}: $\Trans{}$}\\
            $s\Trans{\project{\sigma'}{L_s}'} q_s$\\
        
    \end{case_distinction}
\end{case_distinction}
\end{proof}
\end{toappendix}

\begin{lemmarep}
\label{lem:utrace_closed_implies_cartesian_closed}
Let $s$ and $e\in \LTS$ be composable and \utraceClosed{}, then\\
    \[s\parcomp e \Trans{\sigma} q_s\parcomp q_e \land s\parcomp e \Trans{\sigma} q_s' \parcomp q_e' \iff s\parcomp e \Trans{\sigma} q_s\parcomp q_e' \land s\parcomp e \Trans{\sigma} q_s' \parcomp q_e \]
    
\end{lemmarep}

\begin{proof}
    Since the bi-implication is symmetrical under alpha renaming, we only proof the left to right side.
The proof follows by induction on $\sigma$, and then further case distinction on the last label of $\sigma$. The interesting case is for $\delta$, as all traces without $\delta$ are trivially handled by \Cref{lem:project_from_parcomp_no_delta}.
\begin{case_distinction}
    \item[Base case $\sigma=\epsilon$] \ \\
            $s\parcomp e \Trans{\epsilon} q_s\parcomp q_e \land s\parcomp e \Trans{\epsilon} q_s' \parcomp q_e'$\\
            \proofstep{\Cref{lem:project_from_parcomp_no_delta}}\\
            $s \Trans{\project{\epsilon}{L_s}} q_s \land e \Trans{\project{\epsilon}{L_e}} q_e \; \land $\\
            $s \Trans{\project{\epsilon}{L_s}} q_s' \land e \Trans{\project{\epsilon}{L_e}} q_e' $\\
            \proofstep{\Cref{lem:project_from_parcomp_no_delta}}\\
            $s\parcomp e \Trans{\epsilon} q_s'\parcomp q_e \land s\parcomp e \Trans{\epsilon} q_s \parcomp q_e'$\\

    \item[Inductive step $\sigma=\sigma'\cdot\ell$]\ \\
    The IH is:  \[\begin{array}{l}
        \forall q_s, q_s'\in Q_s, q_e, q_e'\in Q_e:\\
         \tab s\parcomp e \Trans{\sigma'} q_s\parcomp q_e \land s\parcomp e \Trans{\sigma} q_s' \parcomp q_e' \implies s\parcomp e \Trans{\sigma} q_s\parcomp q_e' \land s\parcomp e \Trans{\sigma} q_s' \parcomp q_e  
    \end{array}\]

    \begin{case_distinction}
        \item[$\ell\neq\delta$:]\ \\
            $s\parcomp e \Trans{\sigma'\cdot\ell} q_s\parcomp q_e \land s\parcomp e \Trans{\sigma'\cdot\ell} q_s' \parcomp q_e'$\\
            \proofstep{\Cref{item:trans_transitive}}\\
            $\exists q_s'', q_s'''\in Q_s, q_e'', q_e'''\in Q_e:$\\
            $s\parcomp e \Trans{\sigma'} q_s''\parcomp q_e'' \Trans{\ell} q_s \parcomp q_e\; \land $\\
            $s\parcomp e \Trans{\sigma'} q_s'''\parcomp q_e''' \Trans{\ell}  q_s' \parcomp q_e' $\\
            \proofstep{Apply IH}\\
            $\exists q_s'', q_s'''\in Q_s, q_e'', q_e'''\in Q_e:$\\
            $s\parcomp e \Trans{\sigma'} q_s''\parcomp q_e''' \land q_s''\parcomp q_e'' \Trans{\ell} q_s \parcomp q_e\; \land $\\
            $s\parcomp e \Trans{\sigma'} q_s'''\parcomp q_e'' \land q_s''' \parcomp q_e''' \Trans{\ell}  q_s' \parcomp q_e' $\\
            \proofstep{\Cref{lem:project_from_parcomp_no_delta}}\\
            $\exists q_s'', q_s'''\in Q_s, q_e'', q_e'''\in Q_e:$\\
            $s\parcomp e \Trans{\sigma'} q_s''\parcomp q_e''' \land q_s'' \Trans{\project{\ell}{L_s}} q_s\land q_e'' \Trans{\project{\ell}{L_e}} q_e \; \land $\\
            $s\parcomp e \Trans{\sigma'} q_s'''\parcomp q_e'' \land q_s''' \Trans{\project{\ell}{L_s}} q_s' \land q_e''' \Trans{\project{\ell}{L_e}} q_e' $\\
            \proofstep{\Cref{lem:project_from_parcomp_no_delta}}\\
             $\exists q_s'', q_s'''\in Q_s, q_e'', q_e'''\in Q_e:$\\
            $s\parcomp e \Trans{\sigma'} q_s''\parcomp q_e''' \land q_s''\parcomp q_e''' \Trans{\ell} q_s \parcomp q_e'\; \land $\\
            $s\parcomp e \Trans{\sigma'} q_s'''\parcomp q_e'' \land q_s''' \parcomp q_e'' \Trans{\ell}  q_s' \parcomp q_e $\\
            \proofstep{\Cref{item:trans_transitive}}\\
            $s\parcomp e \Trans{\sigma'\cdot\ell} q_s\parcomp q_e' \land s\parcomp e \Trans{\sigma'\cdot\delta} q_s' \parcomp q_e$

        \item[$\ell=\delta$]\ \\
            $s\parcomp e \Trans{\sigma'\cdot\delta} q_s\parcomp q_e \land s\parcomp e \Trans{\sigma'\cdot\delta} q_s' \parcomp q_e'$\\
            \proofstep{\Cref{def:arrowdefs,def:delta}:$\Trans{\sigma'\cdot\delta}$}\\
            $\exists q_s'', q_s'''\in Q_s, q_e'', q_e'''\in Q_e:$\\
            $s\parcomp e \Trans{\sigma'} q_s''\parcomp q_e'' \trans{\delta} q_s''\parcomp q_e'' \Trans{\epsilon} q_s \parcomp q_e\; \land $\\
            $s\parcomp e \Trans{\sigma'} q_s'''\parcomp q_e''' \trans{\delta} q_s'''\parcomp q_e''' \Trans{\epsilon} q_s' \parcomp q_e' $\\
            \proofstep{\Cref{lem:trans_delta_closed}}\\
            $\exists q_s'', q_s'''\in Q_s, q_e'', q_e'''\in Q_e:$\\
            $s\parcomp e \Trans{\sigma'} q_s''\parcomp q_e'' \trans{\delta} q_s''\parcomp q_e'' \Trans{\epsilon} q_s \parcomp q_e\; \land $\\
            $s\parcomp e \Trans{\sigma'} q_s'''\parcomp q_e''' \trans{\delta} q_s'''\parcomp q_e''' \Trans{\epsilon} q_s' \parcomp q_e'\; \land $\\
            $s\parcomp e \Trans{\project{\sigma'}{L_{s\parcomp e}}}  q_s''\parcomp q_e''\land s\parcomp e \Trans{\project{\sigma'}{L_{s\parcomp e}}} q_s'''\parcomp q_e'''$\\
            \proofstep{\Cref{lem:project_from_parcomp_no_delta}}\\
            $\exists q_s'', q_s'''\in Q_s, q_e'', q_e'''\in Q_e:$\\
            $s\parcomp e \Trans{\sigma'} q_s''\parcomp q_e'' \trans{\delta} q_s''\parcomp q_e'' \Trans{\epsilon} q_s \parcomp q_e\; \land $\\
            $s\parcomp e \Trans{\sigma'} q_s'''\parcomp q_e''' \trans{\delta} q_s'''\parcomp q_e''' \Trans{\epsilon} q_s' \parcomp q_e'\; \land $\\
            $s\Trans{\project{\sigma'}{L_{s}}}  q_s''\land s\Trans{\project{\sigma'}{L_{s}}} q_s'''\land e \Trans{\project{\sigma'}{L_{e}}}  q_e''\land e \Trans{\project{\sigma'}{L_{e}}} q_e'''$\\
            \proofstep{\Cref{def:utraceClosed,def:uioco}: $\utraces{}$ and $\utraceClosed{}$}\\
            $\exists q_s'', q_s'''\in Q_s, q_e'', q_e'''\in Q_e:$\\
            $s\parcomp e \Trans{\sigma'} q_s''\parcomp q_e'' \trans{\delta} q_s''\parcomp q_e'' \Trans{\epsilon} q_s \parcomp q_e\; \land $\\
            $s\parcomp e \Trans{\sigma'} q_s'''\parcomp q_e''' \trans{\delta} q_s'''\parcomp q_e''' \Trans{\epsilon} q_s' \parcomp q_e'\; \land $\\
            $\forall \ell \in I_s: q_s''\Trans{\ell} \iff q_s''' \Trans{\ell}\;\land$\\
            $\forall \ell \in I_e: q_e''\Trans{\ell} \iff q_e''' \Trans{\ell}$\\
            \proofstep{\Cref{def:parcomp,def:delta}: $T_{\parcomp}$ and $\trans{\delta}$}\\
            $\exists q_s'', q_s'''\in Q_s, q_e'', q_e'''\in Q_e:$\\
            $s\parcomp e \Trans{\sigma'} q_s''\parcomp q_e'' \trans{\delta} q_s''\parcomp q_e'' \Trans{\epsilon} q_s \parcomp q_e\; \land $\\
            $s\parcomp e \Trans{\sigma'} q_s'''\parcomp q_e''' \trans{\delta} q_s'''\parcomp q_e''' \Trans{\epsilon} q_s' \parcomp q_e'\; \land $\\
            $\forall \ell \in I_s: q_s''\Trans{\ell} \iff q_s''' \Trans{\ell}\;\land$\\
            $\forall \ell \in I_e: q_e''\Trans{\ell} \iff q_e''' \Trans{\ell}\;\land $\\
            $\forall \ell \in U_s \cap I_e: (q_s''\trans{\ell} \implies q_e'' \nottrans{\ell}) \land (q_s'''\trans{\ell} \implies q_e''' \nottrans{\ell})\;\land$\\
            $\forall \ell \in U_e\cap I_s: (q_e''\trans{\ell} \implies q_s'' \nottrans{\ell}) \land (q_e'''\trans{\ell} \implies q_s''' \nottrans{\ell})\;\land$\\
            $\forall \ell \in U_s \setminus L_e: q_s''\nottrans{\ell} \land\; q_s''\nottrans{\ell}\;\land$\\
            $\forall \ell \in U_e \setminus L_s: q_e''\nottrans{\ell} \land\; q_e''\nottrans{\ell}\;\land$\\
            $q_s''\nottrans{\tau} \land\; q_s'''\nottrans{\tau} \land\; q_e''\nottrans{\tau} \land\; q_e'''\nottrans{\tau}$\\
            \proofstep{\Cref{def:arrowdefs}: $\Trans{\ell}$}\\
            $\exists q_s'', q_s'''\in Q_s, q_e'', q_e'''\in Q_e:$\\
            $s\parcomp e \Trans{\sigma'} q_s''\parcomp q_e'' \trans{\delta} q_s''\parcomp q_e'' \Trans{\epsilon} q_s \parcomp q_e\; \land $\\
            $s\parcomp e \Trans{\sigma'} q_s'''\parcomp q_e''' \trans{\delta} q_s'''\parcomp q_e''' \Trans{\epsilon} q_s' \parcomp q_e'\; \land $\\
            $\forall \ell \in I_s: q_s''\trans{\ell} \iff q_s''' \trans{\ell}\;\land$\\
            $\forall \ell \in I_e: q_e''\trans{\ell} \iff q_e''' \trans{\ell}\;\land $\\
            $\forall \ell \in U_s \cap I_e: (q_s''\trans{\ell} \implies q_e'' \nottrans{\ell}) \land (q_s'''\trans{\ell} \implies q_e''' \nottrans{\ell})\;\land$\\
            $\forall \ell \in U_e\cap I_s: (q_e''\trans{\ell} \implies  q_s'' \nottrans{\ell}) \land (q_e'''\trans{\ell} \implies q_s''' \nottrans{\ell})\;\land$\\
            $\forall \ell \in U_s \setminus L_e: q_s''\nottrans{\ell} \land\; q_s''\nottrans{\ell}\;\land$\\
            $\forall \ell \in U_e \setminus L_s: q_e''\nottrans{\ell} \land\; q_e''\nottrans{\ell}\;\land$\\
            $q_s''\nottrans{\tau} \land\; q_s'''\nottrans{\tau} \land\; q_e''\nottrans{\tau} \land\; q_e'''\nottrans{\tau}$\\
            \proofstep{Substitution using $q''\trans{\ell} \iff q''' \trans{\ell}$}\\
            $\exists q_s'', q_s'''\in Q_s, q_e'', q_e'''\in Q_e:$\\
            $s\parcomp e \Trans{\sigma'} q_s''\parcomp q_e'' \trans{\delta} q_s''\parcomp q_e'' \Trans{\epsilon} q_s \parcomp q_e\; \land $\\
            $s\parcomp e \Trans{\sigma'} q_s'''\parcomp q_e''' \trans{\delta} q_s'''\parcomp q_e''' \Trans{\epsilon} q_s' \parcomp q_e'\; \land $\\
            $\forall \ell \in U_s \cap I_e: (q_s''\trans{\ell} \implies q_e''' \nottrans{\ell}) \land (q_s'''\trans{\ell} \implies q_e'' \nottrans{\ell})\;\land$\\
            $\forall \ell \in U_e\cap I_s: (q_e''\trans{\ell} \implies  q_s''' \nottrans{\ell}) \land (q_e'''\trans{\ell} \implies q_s'' \nottrans{\ell})\;\land$\\
            $\forall \ell \in U_s \setminus L_e: q_s''\nottrans{\ell} \land\; q_s''\nottrans{\ell}\;\land$\\
            $\forall \ell \in U_e \setminus L_s: q_e''\nottrans{\ell} \land\; q_e''\nottrans{\ell}\;\land$\\
            $q_s''\nottrans{\tau} \land\; q_s'''\nottrans{\tau} \land\; q_e''\nottrans{\tau} \land\; q_e'''\nottrans{\tau}$\\
            \proofstep{\Cref{def:parcomp,def:delta}: $T_{\parcomp}$ and $\trans{\delta}$}\\
            $\exists q_s'', q_s'''\in Q_s, q_e'', q_e'''\in Q_e:$\\
            $s\parcomp e \Trans{\sigma'} q_s''\parcomp q_e'' \trans{\delta} q_s''\parcomp q_e'' \Trans{\epsilon} q_s \parcomp q_e\; \land $\\
            $s\parcomp e \Trans{\sigma'} q_s'''\parcomp q_e''' \trans{\delta} q_s'''\parcomp q_e''' \Trans{\epsilon} q_s' \parcomp q_e'\; \land $\\
            $q_s''\parcomp q_e''' \trans{\delta} q_s''\parcomp q_e''' \land q_s'''\parcomp q_e'' \trans{\delta} q_s'''\parcomp q_e''$\\
            \proofstep{Apply IH}\\
            $\exists q_s'', q_s'''\in Q_s, q_e'', q_e'''\in Q_e:$\\
            $q_s''\parcomp q_e'' \Trans{\epsilon} q_s \parcomp q_e\; \land $\\
            $q_s'''\parcomp q_e''' \Trans{\epsilon} q_s' \parcomp q_e'\; \land $\\
            $s\parcomp e \Trans{\sigma'} q_s''\parcomp q_e''' \trans{\delta} q_s''\parcomp q_e''' \land s\parcomp e \Trans{\sigma'}q_s'''\parcomp q_e'' \trans{\delta} q_s'''\parcomp q_e''$\\
            \proofstep{\Cref{lem:project_from_parcomp_no_delta}}\\
            $\exists q_s'', q_s'''\in Q_s, q_e'', q_e'''\in Q_e:$\\
            $s\parcomp e \Trans{\sigma'} q_s''\parcomp q_e''' \trans{\delta} q_s''\parcomp q_e''' \Trans{\epsilon} q_s\parcomp q_e' \land s\parcomp e \Trans{\sigma'}q_s'''\parcomp q_e'' \trans{\delta} q_s'''\parcomp q_e'' \Trans{\epsilon} q_s'\parcomp q_e'$\\
            \proofstep{\Cref{def:arrowdefs}: $\Trans{\sigma'\cdot\ell}$}\\
            $s\parcomp e \Trans{\sigma'\cdot\delta} q_s\parcomp q_e' \land s\parcomp e \Trans{\sigma'\cdot\delta} q_s' \parcomp q_e$

    \end{case_distinction}
\end{case_distinction}
\end{proof}

\subsubsection{Proof Sketch}

We show that \Cref{alg:combined_testcase_alg} is exhaustive, as expressed in \Cref{lem:combined_testing_exhaustive}. The core of the proof follows the same logic as the proofs for $\Cref{alg:test_uioco}$ and $\Cref{alg:testcase_alg}$. Because of the optimisations present in \Cref{alg:combined_testcase_alg}, the algorithm does not guarantee $i_s \uioco{} s$ or $i_s \eco{} e$. To compensate for this, we prove that if we have $\cioco{i_s}{s}{e}$ but not $i_s \uioco s$ using some counterexample $\sigma$, then either $\sigma$ is not possible under the constraints of $e$, or the test suite for $\cioco{i_e}{e}{s}$ will fail for (a prefix of) $\sigma$.

The high level proof works in two steps: first we link the traces from $s\parcomp e$ to traces trough the generated test cases in \Cref{lem:project_to_combined_test}, and then we link the states of the testcase back to states of $s\parcomp e$ in \Cref{lem:combinedtestcase_eq_genafter}. Because we have constructed the states of the testcase ourselves in \Cref{alg:combined_testcase_alg}, this gives the information required to prove \Cref{alg:combined_testcase_alg} exhaustive in \Cref{lem:combined_testing_exhaustive}. As stated earlier, \Cref{alg:combined_testcase_alg} is not sound, and could therefore also fail for specifications which are technically correct, but practically undesirable. We have to take this into account into the proof here, which is why \Cref{lem:combinedtestcase_eq_genafter,lem:project_to_combined_test} both only make claims about \Cref{alg:combined_testcase_alg} as long as it does not reach $\failState$ for neither $i_s$ nor $i_e$. This then falls away in \Cref{lem:combined_testing_exhaustive}, where we know that all tests passed and therefore cannot reach $\failState$. This is also the point where we require the $\utraceClosed{}$ property for a seemingly minor detail: \Cref{lem:utrace_closed_implies_cartesian_closed} proves that if both test suites reach a point where their implementation was allowed to be quiescent after a trace, then the composed implementation is also allowed to be quiescent after that trace. This property is required for the exhaustiveness of \Cref{alg:combined_testcase_alg}, but is only present for specifications without non-deterministic underspecification.

\begin{lemmarep}
    \label{lem:combinedtestcase_eq_genafter}
    Let $s,e\in \LTS$ be $\composable$, $t\in \TTS[s,e]$,\\
    $\sigma\in L_{t}^*$, $q_t\in (Q_t\setminus\{\passState\})$, then
    \[t \Trans{\sigma } q_t \implies q_t = \GenCombinedTest \bigl(s\parcomp e \after (\project{\subst{\sigma}{\theta}{\delta}}{L_{s\parcomp e}^\delta})\bigr) \lor q_t = \failState\]
\end{lemmarep}

\begin{proof}
    \ \\
    Proof by induction on the length of $\sigma$. 
\begin{case_distinction}
    \item[base case $\sigma=\epsilon$:]\ \\
        Since $t$ does not contain $\tau$ transitions (\itemref{def:testcase}{no_tau}), $t \Trans{\epsilon} q_t$ means $t=q_t$ (\Cref{def:arrowdefs}).
        $t=\;$\GenEcoTest{$s \parcomp e \after \epsilon$} (\Cref{alg:combined_testcase_alg}), and $\subst{\epsilon}{\theta}{\delta} = \epsilon$ (\Cref{def:substitution}).
    \item[Inductive step: $\sigma = \sigma'\cdot\ell$:]
        The IH is:\\
        $\forall q_t'\in Q_t: t\Trans{\sigma'} q_t' \implies q_t' = \GenCombinedTest\bigl(X\bigr) \lor q_t' = \failState$,\\
        Where $X= s \parcomp e \after (\project{\subst{\sigma'}{\theta}{\delta}}{L_{s\parcomp e}^\delta})$\\
        
        $t\Trans{\sigma'\cdot\ell} q_t$\\
        \proofstep{\Cref{item:trans_transitive}}\\
        $\exists q_t'\in Q_t: t \Trans{\sigma'} q_t' \Trans{\ell}q_t$\\
        \proofstep{Apply IH}\\
        $\exists q_t'\in Q_t: q_t' \Trans{\ell}q_t\land q_t' = \GenCombinedTest\bigl(X\bigr) \lor q_t' = \failState$\\
        \proofstep{\Itemref{def:testcase}{no_tau}: no $\tau$ transitions in $t$}\\
        $\exists q_t'\in Q_t: q_t' \trans{\ell}q_t\land q_t' = \GenCombinedTest\bigl(X\bigr) \lor q_t' = \failState$\\
        \proofstep{Induction on the structure of $T_t$}\\
        The transition function of $t$ is generated by \Cref{fun:GenCombinedTest}. From here we see there is a finite amount of ways in which $q_t'\trans{\ell}$ is possible: lines \ref{line:combined_sync_out}, \ref{line:combined_sync_theta} and  \ref{line:combined_env_internal_step} of $\GenEcoTest$, or any of the lines of \Cref{fun:GenInputResponse_2}. We know $q_t'\neq \failState$, because $\failState \nottrans{\ell}$ for any $\ell$, which would contradict the assumptions (\Cref{item:pass_fail_state}). 
        \begin{case_distinction}
            \item[\cref{line:combined_sync_out} of \Cref{fun:GenCombinedTest}:]\ \\
                $q_t = \GenCombinedTest\bigl(X \after \ell\bigr)$\\
                \proofstep{\Cref{def:after}: $\after$}\\
                $q_t = \GenCombinedTest\bigl(s\parcomp e \after (\project{\subst{\sigma'}{\theta}{\delta}}{L_{s\parcomp e}^\delta}\cdot\ell)\bigr)$\\
                \proofstep{\Cref{def:substitution,def:projection}: ($\ell\neq \theta, \ell\in L_{s\parcomp e}$)}\\
                $q_t = \GenCombinedTest\bigl(s\parcomp e \after (\project{\subst{\sigma'\cdot\ell}{\theta}{\delta}}{L_{s\parcomp e}^\delta})\bigr)$
            \item[\cref{line:combined_sync_theta} of \Cref{fun:GenCombinedTest}:]\ \\
                $\Bigl(\delta \notin \outset[big]{\pi_i(X)} \implies \ell=\theta \land q_t = \failState\Bigr)\;\land$\\
                $\Bigl(\delta \in \outset[big]{\pi_1(X)} \land \delta \in \outset[big]{X} \implies $\\
                \tab$\ell=\theta\land q_t = \GenCombinedTest\bigl(s\parcomp e \after (\project{\subst{\sigma'}{\theta}{\delta}}{L_{s\parcomp e}^\delta}\cdot\delta)\bigr)\Bigr)\;\land $\\
                $\Bigl(\delta \in \outset[big]{\pi_1(X)} \land \delta \notin \outset[big]{X} \implies $\\
                \tab$\ell = \theta_s \land q_t = \GenCombinedTest\bigl(s \parcomp e \after (\project{\subst{\sigma'}{\theta}{\delta}}{L_{s\parcomp e}^\delta})\bigr)\Bigr)\;\land $\\
                \proofstep{\Cref{def:substitution,def:projection}: \textbf{substitution} and $\projectop$}\\
                $\Bigl(\delta \notin \outset[big]{\pi_i(X)} \implies q_t = \failState\Bigr)\;\land$\\
                $\Bigl(\delta \in \outset[big]{\pi_1(X)} \implies $\\
                \tab$q_t = \GenCombinedTest\bigl(s \parcomp e \after (\project{\subst{\sigma'\cdot\ell}{\theta}{\delta}}{L_{s\parcomp e}^\delta})\bigr)\Bigr)$\\
                \proofstep{rewrite}\\
                $q_t = \GenCombinedTest\bigl(s \parcomp e \after (\project{\subst{\sigma'\cdot\ell}{\theta}{\delta}}{L_{s\parcomp e}^\delta})\bigr)\Bigr) \lor q_t = \failState$\\
                
            \item[\cref{line:combined_env_internal_step} of \Cref{fun:GenCombinedTest}:]\ \\
                 $q_t = \GenCombinedTest\bigl(X \after \ell\bigr)$\\
                \proofstep{\Cref{def:after}: $\after$}\\
                $q_t = \GenCombinedTest\bigl(s\parcomp e \after (\project{\subst{\sigma'}{\theta}{\delta}}{L_{s\parcomp e}^\delta}\cdot\ell)\bigr)$\\
                \proofstep{\Cref{def:substitution,def:projection}: ($\ell\neq \theta, \ell\in L_{s\parcomp e}$)}\\
                $q_t = \GenCombinedTest\bigl(s\parcomp e \after (\project{\subst{\sigma'\cdot\ell}{\theta}{\delta}}{L_{s\parcomp e}^\delta})\bigr)$
            \item[\cref{line:failtrans_eco} of \Cref{fun:GenInputResponse_2}:]\ \\
                $q_t = \failState$
            \item[\cref{line:failtrans_uioco} of \Cref{fun:GenInputResponse_2}:]\ \\
                 $q_t = \failState$
            \item[\cref{line:combined_sync_input} or \ref{line:combined_unsync_input} of \Cref{fun:GenInputResponse_2}:]\ \\
                 $q_t = \GenCombinedTest\bigl(X \after \ell\bigr)$\\
                \proofstep{\Cref{def:after}: $\after$}\\
                $q_t = \GenCombinedTest\bigl(s\parcomp e \after (\project{\subst{\sigma'}{\theta}{\delta}}{L_{s\parcomp e}^\delta}\cdot\ell)\bigr)$\\
                \proofstep{\Cref{def:substitution,def:projection}: ($\ell\neq \theta, \ell\in L_{s\parcomp e}$)}\\
                $q_t = \GenCombinedTest\bigl(s\parcomp e \after (\project{\subst{\sigma'\cdot\ell}{\theta}{\delta}}{L_{s\parcomp e}^\delta})\bigr)$
            
        \end{case_distinction}
\end{case_distinction}
\end{proof}

\begin{toappendix}
    \begin{lemmarep}
        \label{lem:parcomp_combinedtest_trace_simplification}
        Let $t\in \TTS[s,e]$ $\sigma\in L_{s\parcomp e}^{\delta*}$, $\ell \in L_t \setminus L_{s\parcomp e}^{\delta}$:
        \[\subst{\project{\subst{\sigma}{\delta}{\ell}}{L_t}}{\ell}{\delta} \iff  \sigma\]\\
    \end{lemmarep}
    \begin{proof}
           Direct bi-implication proof. All proofsteps are bi-implifications.\\
$\subst{\project{\subst{\sigma}{\delta}{\ell}}{L_t}}{\ell}{\delta}$\\
\proofstep{\Cref{def:projection,def:substitution}: $\delta \notin (\subst{\sigma}{\delta}{\ell}) $}\\
$\subst{\project{\subst{\sigma}{\delta}{\ell}}{L_t^\delta}}{\ell}{\delta}$\\
\proofstep{\Cref{def:projection,def:substitution}: $\ell\in L_t^\delta\land \delta\in L_t^\delta$}\\
$\project{\subst{\subst{\sigma}{\delta}{\ell}}{\ell}{\delta}}{L_t^\delta}$\\
\proofstep{\Cref{def:substitution}: $\ell\notin \sigma$}\\
$\project{\sigma}{L_t^\delta}$\\
\proofstep{\Cref{def:projection,lem:L_combined_testcase,def:parcomp}: $L_{s\parcomp e}^\delta\subseteq L_t^\delta $}\\
$\sigma$
    \end{proof}
\end{toappendix}

\begin{lemmarep}
    \label{lem:project_to_combined_test}\ \\
    Let $s,e\in \LTS$ be $\composable$, $\sigma\in\utraces{s\parcomp e}$, $q_s\in Q_s$, $q_e\in Q_e$, then
    \[\begin{array}{l}
        s\parcomp e \Trans{\sigma} q_s \parcomp q_e \implies \\
        \tab\Bigl(\exists t_s \in \TTS[s,e], q_{ts} \in Q_{ts}\setminus\{\passState\}: t_s \Trans{\subst{\sigma}{\delta}{\theta}} q_{ts}\Bigr)\; \lor\\
        \tab\Bigl(\exists t_e \in \TTS[e,s]: t_e \Trans{\subst{\sigma}{\delta}{\theta}} \failState\Bigr)
    \end{array}\]
\end{lemmarep}

\begin{proof}
    Proof by induction on $\sigma$, with further case distinctions on the last label of $\sigma$.

    \begin{case_distinction}
        \item[$\sigma = \epsilon$:] \ \\Let $t$ be the test case obtained by choosing case C of \Cref{fun:GenEcoTest}. Then the proof is trivial, because $t\Trans{\epsilon}t$, and $t$ is not $\passState$.
        \item[$\sigma = \sigma'\cdot\ell$:]\ \\
        The IH is $\forall q_s'\in Q_s$, $\forall q_e'\in Q_e$:
            \[\begin{array}{l}
        s\parcomp e \Trans{\sigma'} q_s \parcomp q_e \implies \\
        \tab\bigl(\exists t \in \TTS[s,e], q_t \in Q_t\setminus\{\passState\}: t \Trans{\subst{\sigma'}{\delta}{\theta}} q_t\bigr)
    \end{array}\]
    Note that we exclude the "there is a test that leads to $\failState$" option in the IH. This is because this case would trivially complete the proof ($\failState\Trans{\ell}\failState$ for all $\ell$. (\Cref{def:testcase})).\\
    
        $s\parcomp e \Trans{\sigma'\cdot\ell} q_s\parcomp q_e \land \sigma'\cdot\ell \in \utraces{s\parcomp e}$\\
        \proofstep{\Cref{item:trans_transitive} and apply IH}\\
        $\sigma'\cdot\ell \in \utraces{s\parcomp e}\;\land$\\
        $\exists t'\in \TTS[s,e], q_t'\in Q_{t'}\setminus\{\passState\}: t'\Trans{\subst{\sigma'}{\delta}{\theta}} q_t'$\\
        \proofstep{\Cref{lem:combinedtestcase_eq_genafter}}\\
        $\sigma'\cdot\ell \in \utraces{s\parcomp e}\;\land$\\
        $\exists t'\in \TTS[s,e], q_t'\in Q_{t'}\setminus\{\passState\}: t'\Trans{\subst{\sigma'}{\delta}{\theta}} q_t'\;\land$\\
        $X = s \parcomp e \after (\project{\subst{\subst{\sigma'}{\delta}{\theta}}{\theta}{\delta}}{L_{s\parcomp e}^\delta})\; \land$\\
        $q_t'= \GenCombinedTest(X)$\\
        \proofstep{$\theta\notin L_{s\parcomp e}^\delta$}\\
        $\sigma'\cdot\ell \in \utraces{s\parcomp e}\;\land$\\
        $\exists t'\in \TTS[s,e], q_t'\in Q_{t'}\setminus\{\passState\}: t'\Trans{\subst{\sigma'}{\delta}{\theta}} q_t'\;\land$\\
        $X = s \parcomp e \after \sigma' \land  q_t'= \GenCombinedTest(X)$\\
        \proofstep{Case distinction on $\ell$}
        \begin{case_distinction}
            
            \item[$\ell\in I_s\setminus U_e$:]\ \\
                $\sigma'\cdot\ell \in \utraces{s\parcomp e}\land \ell \in I_s \setminus U_e\;\land$\\
                $\exists t'\in \TTS[s,e], q_t'\in Q_{t'}\setminus\{\passState\}: t'\Trans{\subst{\sigma'}{\delta}{\theta}} q_t'\;\land$\\
                $X = s \parcomp e \after \sigma'\land
                q_t'= \GenCombinedTest(X)$\\
                \proofstep{\Cref{def:uioco}: $\utraces{}$}\\
                $\ell \in \inset{X}\land \ell \in \inset[big]{\pi_1(X)}\;\land$\\
                $\exists t'\in \TTS[s,e], q_t'\in Q_{t'}\setminus\{\passState\}: t'\Trans{\subst{\sigma'}{\delta}{\theta}} q_t'\;\land$\\
                $X = s \parcomp e \after \sigma'\land
                q_t'= \GenCombinedTest(X)$\\
                
                 Let $t$ be $t'$, but in $q_t'$ we take choice B of \Cref{fun:GenCombinedTest}, and then choice A in all recursive calls.\\

                \proofstep{\Cref{line:combined_sync_out} of \Cref{fun:GenCombinedTest}}\\
                $\exists t\in \TTS[s,e], q_t\in Q_{t}\setminus\{\passState\}: t\Trans{\subst{\sigma'}{\delta}{\theta}\cdot\ell} q_t\;\land$\\
                $X = s \parcomp e \after \sigma'\;\land$
                $q_t= \GenCombinedTest(X\after\ell)$\\
                \proofstep{\Cref{alg:combined_testcase_alg,def:projection,def:substitution}: $\ell\in L_t$, $\ell\neq\delta$}\\
                $\exists t\in \TTS[s,e], q_t\in Q_{t}\setminus\{\passState\}: t\Trans{\subst{\sigma'\cdot\ell}{\delta}{\theta}} q_t$
            \item[$\ell\in I_s\cap U_e \cap \protect{\inset[big]{\pi_1(X)}}$:]\ \\
                $\sigma'\cdot\ell \in \utraces{s\parcomp e}\land \ell \in I_s \cap U_e \land \ell \in \inset[big]{\pi_1(X)}\;\land$\\
                $\exists t'\in \TTS[s,e], q_t'\in Q_{t'}\setminus\{\passState\}: t'\Trans{\subst{\sigma'}{\delta}{\theta}} q_t'\;\land$\\
                $X = s \parcomp e \after \sigma'\land
                q_t'= \GenCombinedTest(X)$\\
                \proofstep{\Cref{def:uioco}: $\utraces{}$}\\
                $\ell \in \outset{X}\land \ell \in \inset[big]{\pi_1(X)}\;\land$\\
                $\exists t'\in \TTS[s,e], q_t'\in Q_{t'}\setminus\{\passState\}: t'\Trans{\subst{\sigma'}{\delta}{\theta}} q_t'\;\land$\\
                $X = s \parcomp e \after \sigma'\land
                q_t'= \GenCombinedTest(X)$\\
                
                 Let $t$ be $t'$, but in $q_t'$ we take choice B of \Cref{fun:GenCombinedTest}, and then choice A in all recursive calls.\\

                \proofstep{\Cref{line:combined_sync_out} of \Cref{fun:GenCombinedTest}}\\
                $\exists t\in \TTS[s,e], q_t\in Q_{t}\setminus\{\passState\}: t\Trans{\subst{\sigma'}{\delta}{\theta}\cdot\ell} q_t\;\land$\\
                $X = s \parcomp e \after \sigma'\;\land$
                $q_t= \GenCombinedTest(X\after\ell)$\\
                \proofstep{\Cref{alg:combined_testcase_alg,def:projection,def:substitution}: $\ell\in L_t$, $\ell\neq\delta$}\\
                $\exists t\in \TTS[s,e], q_t\in Q_{t}\setminus\{\passState\}: t\Trans{\subst{\sigma'\cdot\ell}{\delta}{\theta}} q_t$
            \item[$\ell\in (I_s\cap U_e) \setminus\protect{\inset[big]{\pi_1(X)}}$:]\ \\
                $\sigma'\cdot\ell \in \utraces{s\parcomp e}\land \ell \in I_s \cap U_e \land \ell \notin \inset[big]{\pi_1(X)}\;\land$\\
                $\exists t'\in \TTS[s,e], q_t'\in Q_{t'}\setminus\{\passState\}: t'\Trans{\subst{\sigma'}{\delta}{\theta}} q_t'\;\land$\\
                $X = s \parcomp e \after \sigma'\land
                q_t'= \GenCombinedTest(X)$\\
                \proofstep{Apply IH again, with $s$ and $e$ switched, plus the same simplification steps applied before the start of this case distinction.}\\
                $\sigma'\cdot\ell \in \utraces{s\parcomp e}\land \ell \in I_s \cap U_e \land \ell \notin \inset[big]{\pi_1(X)}\;\land$\\
                $\exists t_s'\in \TTS[s,e], q_{ts}'\in Q_{ts'}\setminus\{\passState\}: t_s'\Trans{\subst{\sigma'}{\delta}{\theta}} q_{ts}'\;\land$\\
                $\exists t_e'\in \TTS[e,s], q_{te}'\in Q_{te'}\setminus\{\passState\}: t_e'\Trans{\subst{\sigma'}{\delta}{\theta}} q_{te}'\;\land$\\
                $X = s \parcomp e \after \sigma'\land
                q_{ts}'= \GenCombinedTest(X) \land q_{te}'= \GenCombinedTest(X)$\\
               
                Let $t$ be $t_e'$, but in $q_{te}'$ we take choice C of \Cref{fun:GenCombinedTest}, and then choice A in all recursive calls.\\

                \proofstep{\Cref{line:failtrans_eco} of \Cref{fun:GenCombinedTest}}\\
                $\exists t_e'\in \TTS[e,s], q_{te}'\in Q_{te'}\setminus\{\passState\}: t_e'\Trans{\subst{\sigma'}{\delta}{\theta}\cdot\ell} \failState$\\\proofstep{\Cref{alg:combined_testcase_alg,def:projection,def:substitution}: $\ell\in L_t$, $\ell\neq\delta$}\\
                $\exists t_e'\in \TTS[e,s], q_{te}'\in Q_{te'}\setminus\{\passState\}: t_e'\Trans{\subst{\sigma'\cdot\ell}{\delta}{\theta}} \failState$\\
                
            \item[$\ell \in U_s$:]\ \\
                $\sigma'\cdot\ell \in \utraces{s\parcomp e}\land \ell \in U_s\;\land$\\
                $\exists t'\in \TTS[s,e], q_t'\in Q_{t'}\setminus\{\passState\}: t'\Trans{\subst{\sigma'}{\delta}{\theta}} q_t'\;\land$\\
                $X = s \parcomp e \after \sigma'\land
                q_t'= \GenCombinedTest(X)$\\
                
                Let $t$ be $t'$, but in $q_t'$ we take choice C of \Cref{fun:GenCombinedTest}, and then choice A in all recursive calls.\\

                \proofstep{\Cref{line:combined_sync_input,line:combined_unsync_input,line:failtrans_eco,line:failtrans_uioco} of \Cref{fun:GenInputResponse_2}}\\
                $\exists t\in \TTS[s,e], q_t\in Q_{t}\setminus\{\passState\}: t\Trans{\subst{\sigma'}{\delta}{\theta}\cdot\ell} q_t\;\land$\\
                $X = s \parcomp e \after \sigma'\;\land$\\
                $\biggl(\ell \in \Bigl(I_e\setminus \inset[big]{\pi_2(X)}\Bigr) \cup \Bigl(U_s\setminus \outset{X} \Bigr) \implies  q_t= \failState \biggr)\; \land$\\
                $\biggl(\ell \in \Bigl(\inset[big]{\pi_2(X)}\cup U_s\setminus L_e\Bigr) \cap \outset[Big]{X} \implies$\\
                \tab$ q_t= \GenCombinedTest(X \after \ell)\biggr)$\\
                \proofstep{\Cref{alg:combined_testcase_alg,def:projection,def:substitution}: $\ell\in L_t$, $\ell\neq\delta$}\\
                $\exists t\in \TTS[s,e], q_t\in Q_{t}\setminus\{\passState\}: t\Trans{\subst{\sigma'\cdot\ell}{\delta}{\theta}} q_t\;\land$\\
                $X = s \parcomp e \after \sigma'\;\land$\\
                $\biggl(\ell \in \Bigl(I_e\setminus \inset[big]{\pi_2(X)}\Bigr) \cup \Bigl(U_s\setminus \outset{X} \Bigr) \implies  q_t= \failState \biggr)\; \land$\\
                $\biggl(\ell \in \Bigl(\inset[big]{\pi_2(X)}\cup U_s\setminus L_e\Bigr) \cap \outset[Big]{X} \implies$\\
                \tab$ q_t= \GenCombinedTest(X \after \ell)\biggr)$\\
                \proofstep{$\ell\in U_s$ implies it falls in one of the above two cases.}\\
                $\exists t\in \TTS[s,e], q_t\in Q_{t}\setminus\{\passState\}: t\Trans{\subst{\sigma'\cdot\ell}{\delta}{\theta}} q_t$
            \item[$\ell \in L_e \setminus L_s$:]\ \\
                $\sigma'\cdot\ell \in \utraces{s\parcomp e}\land \ell \in L_e\setminus L_s\;\land$\\
                $\exists t'\in \TTS[s,e], q_t'\in Q_{t'}\setminus\{\passState\}: t'\Trans{\subst{\sigma'}{\delta}{\theta}} q_t'\;\land$\\
                $X = s \parcomp e \after \sigma'\land
                q_t'= \GenCombinedTest(X)$\\
                \proofstep{\Cref{def:uioco}}\\
                $(\ell\in I_e\setminus L_s \implies\ell\in \inset{X} )\;\land$\\
                $(\ell\in U_e \setminus L_s \implies\ell\in \outset{X} )\;\land$\\
                $\exists t'\in \TTS[s,e], q_t'\in Q_{t'}\setminus\{\passState\}: t'\Trans{\subst{\sigma'}{\delta}{\theta}} q_t'\;\land$\\
                $X = s \parcomp e \after \sigma'\land
                q_t'= \GenCombinedTest(X)$\\
               
                Let $t$ be $t'$, but in $q_t'$ we take choice D of \Cref{fun:GenCombinedTest}, and then choice A in all recursive calls.\\

                \proofstep{\Cref{line:combined_env_internal_step} of \Cref{fun:GenCombinedTest}}\\
                $\exists t\in \TTS[s,e], q_t\in Q_{t}\setminus\{\passState\}: t\Trans{\subst{\sigma'}{\delta}{\theta}\cdot\ell} q_t\;\land$\\
                $X = s \parcomp e \after \sigma'\;\land$
                $q_t= \GenCombinedTest(X\after\ell)$\\
                \proofstep{\Cref{alg:combined_testcase_alg,def:projection,def:substitution}: $\ell\in L_t$, $\ell\neq\delta$}\\
                $\exists t\in \TTS[s,e], q_t\in Q_{t}\setminus\{\passState\}: t\Trans{\subst{\sigma'\cdot\ell}{\delta}{\theta}} q_t$

            \item[$\ell = \delta$]\ \\
                $\sigma'\cdot\delta \in \utraces{s\parcomp e}\;\land$\\
                $\exists t'\in \TTS[s,e], q_t'\in Q_{t'}\setminus\{\passState\}: t'\Trans{\subst{\sigma'}{\delta}{\theta}} q_t'\;\land$\\
                $X = s \parcomp e \after \sigma'\land q_t'= \GenCombinedTest(X)$\\
               
                Let $t$ be $t'$, but in $q_t'$ we take choice C of \Cref{fun:GenCombinedTest}, and then choice A in all recursive calls. \\

                \proofstep{\Cref{line:theta_uioco_fail,line:theta_env,line:theta_comp,line:combined_sync_theta} of \Cref{fun:GenCombinedTest}}\\
                $\exists t\in \TTS[s,e], q_t\in Q_{t}\setminus\{\passState\}: t\Trans{\subst{\sigma'}{\delta}{\theta}\cdot\ell} q_t\;\land$\\
                $\Bigl(\delta\notin \outset[big]{\pi_1(X)} \implies q_t = \failState \Bigr)\;\land$\\
                $\Bigl(\delta\in \outset{X} \implies q_t = \GenCombinedTest(X\after \delta)\Bigr)\;\land$\\
                $\biggl(\delta\notin \Bigl(\outset{X} \cup   \outset[big]{\pi_1(X)}\Bigr) \implies q_t = \GenCombinedTest(X)\biggr)\;\land$\\
                $X = s \parcomp e \after \sigma'$\\
                \proofstep{\Cref{alg:combined_testcase_alg,def:projection,def:substitution}: $\theta_\forall \in L_t$}\\
                $\exists t\in \TTS[s,e], q_t\in Q_{t}\setminus\{\passState\}: t\Trans{\subst{\sigma'\cdot\delta}{\delta}{\theta}} q_t$
                   
        \end{case_distinction}
    \end{case_distinction}
\end{proof}

\begin{toappendix}
    
    \begin{lemmarep}
        \label{lem:project_to_combined_testexec}
        Let $\sigma\in\utraces{s\parcomp e}$, $q_s\in Q_s$, $q_e\in Q_e$:
        \[\begin{array}{l}
            s\parcomp e \Trans{\sigma} q_s \parcomp q_e\; \land i_s \parcomp i_e \Trans{\sigma} q_{is} \parcomp q_{i_e} \implies \\
            \tab\Bigl(\exists t_s \in \TTS[s,e], q_{ts} \in Q_{ts}\setminus\{\passState\}: t_s \testexec i_s \Trans{\subst{\sigma}{\delta}{\theta}} q_{ts} \testexec q_{is}\Bigr)\; \lor\\ 
            \tab\Bigl(\exists t_e \in \TTS[e,s]: t_e \testexec i_e \Trans{\subst{\sigma}{\delta}{\theta}} \failState \testexec q_{ie}\Bigr)
        \end{array}\]
    \end{lemmarep}
    
    \begin{proof}
    \ \\
        $ s\parcomp e \Trans{\sigma} q_s \parcomp q_e\; \land i_s \parcomp i_e \Trans{\sigma} q_{is} \parcomp q_{i_e}$\\
\proofstep{\Cref{lem:project_to_combined_test}}\\
$i_s \parcomp i_e \Trans{\sigma} q_{is} \parcomp q_{i_e}\;\land$\\
$\Bigl(\bigl(\exists t_s \in \TTS[s,e], q_{ts} \in Q_{ts}\setminus\{\passState\}: t_s \Trans{\subst{\sigma}{\delta}{\theta}} q_{ts}\bigr)\; \lor$\\
$\bigl(\exists t_e \in \TTS[e,s]: t_e \Trans{\subst{\sigma}{\delta}{\theta}} \failState\bigr)\Bigr)$\\
\proofstep{\Cref{def:testexec,lem:L_combined_testcase}:$L_{t\testexec i}$}\\
$i_s \parcomp i_e \Trans{\sigma} q_{is} \parcomp q_{i_e}  \land\; \subst{\sigma}{\delta}{\theta}\in L_{t\testexec i}^*\;\land$\\
$\Bigl(\bigl(\exists t_s \in \TTS[s,e], q_{ts} \in Q_{ts}\setminus\{\passState\}: t_s \Trans{\subst{\sigma}{\delta}{\theta}} q_{ts}\bigr)\; \lor$\\
$\bigl(\exists t_e \in \TTS[e,s]: t_e \Trans{\subst{\sigma}{\delta}{\theta}} \failState\bigr)\Bigr)$\\
\proofstep{Reflexivity of =}\\
$i_s \parcomp i_e \Trans{\sigma} q_{is} \parcomp q_{i_e}  \land\; \subst{\sigma}{\delta}{\theta}\in L_{t\testexec i}^*\;\land$\\
$\project{\subst{\subst{\sigma}{\delta}{\theta}}{\theta}{\delta}}{L_e^\delta} = \project{\subst{\subst{\sigma}{\delta}{\theta}}{\theta}{\delta}}{L_e^\delta}\;\land$\\
$\project{\subst{\subst{\sigma}{\delta}{\theta}}{\theta}{\delta}}{L_s^\delta} = \project{\subst{\subst{\sigma}{\delta}{\theta}}{\theta}{\delta}}{L_s^\delta}\;\land$\\
$\Bigl(\bigl(\exists t_s \in \TTS[s,e], q_{ts} \in Q_{ts}\setminus\{\passState\}: t_s \Trans{\subst{\sigma}{\delta}{\theta}} q_{ts}\bigr)\; \lor$\\
$\bigl(\exists t_e \in \TTS[e,s]: t_e \Trans{\subst{\sigma}{\delta}{\theta}} \failState\bigr)\Bigr)$\\
\proofstep{\Cref{def:substitution}: $substitution$, $\theta\notin\sigma$}\\
$i_s \parcomp i_e \Trans{\sigma} q_{is} \parcomp q_{i_e}  \land\; \subst{\sigma}{\delta}{\theta}\in L_{t\testexec i}^*\;\land$\\
$\project{\subst{\subst{\sigma}{\delta}{\theta}}{\theta}{\delta}}{L_e^\delta} = \project{\sigma}{L_e^\delta}\;\land$\\
$\project{\subst{\subst{\sigma}{\delta}{\theta}}{\theta}{\delta}}{L_s^\delta} = \project{\sigma}{L_s^\delta}\;\land$\\
$\Bigl(\bigl(\exists t_s \in \TTS[s,e], q_{ts} \in Q_{ts}\setminus\{\passState\}: t_s \Trans{\subst{\sigma}{\delta}{\theta}} q_{ts}\bigr)\; \lor$\\
$\bigl(\exists t_e \in \TTS[e,s]: t_e \Trans{\subst{\sigma}{\delta}{\theta}} \failState\bigr)\Bigr)$\\
\proofstep{\Cref{lem:project_from_parcomp_IOTS}}\\
$i_s\Trans{\project{\sigma}{L_s^\delta}} q_{is} \land i_e\Trans{\project{\sigma}{L_e^\delta}} q_{ie}  \land\; \subst{\sigma}{\delta}{\theta}\in L_{t\testexec i}^*\;\land$\\
$\project{\subst{\subst{\sigma}{\delta}{\theta}}{\theta}{\delta}}{L_e^\delta} = \project{\sigma}{L_e^\delta}\;\land$\\
$\project{\subst{\subst{\sigma}{\delta}{\theta}}{\theta}{\delta}}{L_s^\delta} = \project{\sigma}{L_s^\delta}\;\land$\\
$\Bigl(\bigl(\exists t_s \in \TTS[s,e], q_{ts} \in Q_{ts}\setminus\{\passState\}: t_s \Trans{\subst{\sigma}{\delta}{\theta}} q_{ts}\bigr)\; \lor$\\
$\bigl(\exists t_e \in \TTS[e,s]: t_e \Trans{\subst{\sigma}{\delta}{\theta}} \failState\bigr)\Bigr)$\\
\proofstep{\Cref{lem:project_from_testexec}}\\
$\Bigl(\bigl(\exists t_s \in \TTS[s,e], q_{ts} \in Q_{ts}\setminus\{\passState\}: t_s \testexec i_s \Trans{\subst{\sigma}{\delta}{\theta}} q_{ts} \testexec q_{is}\bigr)\; \lor$\\
$\bigl(\exists t_e \in \TTS[e,s]: t_e \testexec i_e \Trans{\subst{\sigma}{\delta}{\theta}} \failState \testexec q_{ie}\bigr)\Bigr)$\\
    \end{proof}
\end{toappendix}

\begin{toappendix}
\begin{lemmarep}
\label{lem:combinedtest_singlestep_fail}
Let $s$ and $e$ be \utraceClosed{}, $t_s \in \TTS[s,e]$, $t_e \in \TTS[e,s]$\\
$q_{ts}\in Q_{ts}\setminus\passState$, $q_{te}\in Q_{te}\setminus\passState$, $\sigma\in \utraces{s\parcomp e}$, $\ell \in U_{s\parcomp e}^\delta$
    \[q_{ts} = \GenCombinedTest(s\parcomp e \after \sigma) \land q_{te} = \GenCombinedTest(s\parcomp e \after \sigma)\;\land\]
    \[\ell \notin \outset{s\parcomp e \after \sigma} \implies\]
    \[q_{ts} \trans{\subst{\ell}{\delta}{\theta}} \failState \lor  q_{te} \trans{\subst{\ell}{\delta}{\theta}}\failState\]
\end{lemmarep}

\begin{proof}
    Proof by case distinction on $\ell$: 
\begin{enumerate}
    \item $\ell \in U_s$
    \item $\ell \in U_e$
    \item $\ell=\delta \land \pi_1(s\parcomp e \after \sigma) \nottrans{\delta}$
    \item $\ell=\delta \land \pi_2(s\parcomp e \after \sigma) \nottrans{\delta}$
    \item $\ell=\delta \land \pi_1(s\parcomp e \after \sigma) \trans{\delta}\land\; \pi_2(s\parcomp e \after \sigma) \trans{\delta} $
\end{enumerate}

Cases 1 and 2, and cases 3 and 4 are symmetrical, where the error is found by $t_s$ or $t_e$ respectively. The final case is actually not possible, for which we also give a proof.

\begin{case_distinction}
    \item[case $\ell \in U_s$:]\ \\
        $q_{ts} = \GenCombinedTest(s\parcomp e \after \sigma)\land \ell \notin \outset{s \parcomp e \after \sigma}$\\
        \proofstep{\Cref{line:failtrans_uioco}}\\
        $q_{ts} \trans{\ell} \failState$\\
        \proofstep{\Cref{def:projection}: $\projectop$, $\ell\neq \delta$}\\
        $q_{ts} \trans{\subst{\ell}{\delta}{\theta}} \failState$
    \item[case $\ell \in U_e$:] Symmetrical to the previous case.

    \item[case $\ell=\delta \land \pi_1(s\parcomp e \after \sigma) \nottrans{\delta}$:]\ \\
        $q_{ts} = \GenCombinedTest(s\parcomp e \after \sigma) \land \pi_1(s\parcomp e \after \sigma) \nottrans{\delta}$\\
        \proofstep{\Cref{line:theta_uioco_fail}}\\
        $q_{ts} \trans{\theta} \failState$\\
        \proofstep{\Cref{def:projection}: $\projectop$}\\
        $q_{ts} \trans{\subst{\ell}{\delta}{\theta}} \failState$

    \item[case $\ell=\delta \land \pi_2(s\parcomp e \after \sigma) \nottrans{\delta}$:]\ \\
        $q_{te} = \GenCombinedTest(s\parcomp e \after \sigma) \land \pi_2(s\parcomp e \after \sigma) \nottrans{\delta}$\\
        \proofstep{\Cref{def:parcomp}: $s\parcomp e$ = $e\parcomp s$ with projection swapped}\\
        $q_{te} = \GenCombinedTest(e\parcomp s \after \sigma) \land \pi_1(e\parcomp s \after \sigma) \nottrans{\delta}$\\
        \proofstep{\Cref{line:theta_uioco_fail}}\\
        $q_{te} \trans{\theta} \failState$\\
        \proofstep{\Cref{def:projection}: $\projectop$}\\
        $q_{te} \trans{\subst{\ell}{\delta}{\theta}} \failState$

    \item[case $\ell=\delta \land \pi_1(s\parcomp e \after \sigma) \trans{\delta}\land\; \pi_2(s\parcomp e \after \sigma) \trans{\delta} $:]\ \\
        $\pi_1(s\parcomp e \after \sigma) \trans{\delta}\land\; \pi_2(s\parcomp e \after \sigma) \trans{\delta} \land\; \delta \notin \outset{s\parcomp e \after \sigma}$\\
        \proofstep{\Cref{def:after,def:outset}: $\after$, $\pi$ and $\outset{}$}\\
        $\exists q_s\in Q_s, q_e \in Q_e: s\parcomp e \Trans{\sigma} q_s \parcomp q_e \land q_s\trans{\delta}\;\land$\\
        $\exists q_s'\in Q_s, q_e' \in Q_e:s\parcomp e \Trans{\sigma} q_s' \parcomp q_e' \land q_e'\trans{\delta}\;\land$\\
        $\forall q_s''\in Q_s, q_e'' \in Q_e: s\parcomp e \Trans{\sigma} q_s'' \parcomp q_e'' \implies q_s'' \parcomp q_e'' \nottrans{\delta}$\\
        \proofstep{\Cref{lem:utrace_closed_implies_cartesian_closed}}\\
        $\exists q_s\in Q_s, q_e' \in Q_e:s\parcomp e \Trans{\sigma} q_s \parcomp q_e' \land q_e'\trans{\delta}\land\; q_s \trans{\delta}\;\land$\\
        $\forall q_s''\in Q_s, q_e'' \in Q_e: s\parcomp e \Trans{\sigma} q_s'' \parcomp q_e'' \implies q_s'' \parcomp q_e'' \nottrans{\delta}$\\
        \proofstep{\Itemref{lem:parcomp_base_properties}{both_to_parcomp}}\\
        $\exists q_s\in Q_s, q_e' \in Q_e:s\parcomp e \Trans{\sigma} q_s \parcomp q_e' \land q_s \parcomp q_e'\trans{\delta}\;\land$\\
        $\forall q_s''\in Q_s, q_e'' \in Q_e: s\parcomp e \Trans{\sigma} q_s'' \parcomp q_e'' \implies q_s'' \parcomp q_e'' \nottrans{\delta}$\\
        \proofstep{$\bigl(\exists x: P(x) \land \forall y : \neg P(Y)\bigr) \implies False$}\\
        The assumptions contradict, so this case is empty and cannot occur.

\end{case_distinction}
\end{proof}
\end{toappendix}

\begin{theoremrep}
    \label{lem:combined_testing_exhaustive}
    Let $s$ and $e\in \LTS$ be composable and \utraceClosed{}, then 
    \[\cioco{i_s}{s}{e} \land \cioco{i_e}{e}{s} \implies i_s \parcomp i_e \uioco s \parcomp e\]
\end{theoremrep}
\begin{proof}
    We do a proof by negation. We show that under the assumption that $i_s\parcomp i_e \notuioco s\parcomp e$, there must be a failing test case.
From \Cref{def:uioco} we get $\exists \sigma \in \utraces{s\parcomp e}$, $\ell\in U_{s\parcomp e}^\delta$:  $ \ell \in \outset{i_s \parcomp i_e \after \sigma} \land \ell \notin \outset{s \parcomp e \after \sigma}$. 
To Prove:  \\$\exists t_s \in \TTS(s,e), \sigma \in L_{t_s\testexec i_s}^*, q_{is} \in Q_{is}: t_s\testexec i_s \Trans{\sigma} \failState \testexec q_{is}$ or \\
$\exists t_e \in \TTS(e,s), \sigma \in L_{t_e\testexec i_e}^*, q_{ie} \in Q_{ie}: t_e\testexec i_e \Trans{\sigma} \failState \testexec q_{ie}$\\

\noindent
$\sigma\in \utraces{s\parcomp e} \land \ell\in \outset{i_s\parcomp i_e \after \sigma} \land \ell \notin \outset{s\parcomp e \after\sigma}$\\
\proofstep{\Cref{def:outset,def:after}: $\outset{}$ and $\after$}\\
$\sigma\in \utraces{s\parcomp e} \land \ell\in U_{s\parcomp e}^\delta \land \ell \notin \outset{s\parcomp e \after\sigma}\;\land$\\
$\exists q_{ie}\in Q_{ie}, q_{is} \in Q_{is}: i_s \parcomp i_e \Trans{\sigma} q_{is}\parcomp q_{ie} \trans{\ell}\;\land$\\
\proofstep{\Cref{lem:project_to_combined_testexec}}\\
$\sigma\in \utraces{s\parcomp e} \land \ell\in U_{s\parcomp e}^\delta \land \ell \notin \outset{s\parcomp e \after\sigma}\;\land$\\
$\exists q_{ie}\in Q_{ie}, q_{is} \in Q_{is}: q_{is}\parcomp q_{ie} \trans{\ell}\;\land$\\
$\Bigl(\bigl(\exists t_s \in \TTS[s,e], q_{ts} \in Q_{ts}\setminus\{\passState\}: t_s \testexec i_s \Trans{\subst{\sigma}{\delta}{\theta}} q_{ts} \testexec q_{is}\bigr)\; \lor\\ 
\bigl(\exists t_e \in \TTS[e,s]: t_e \testexec i_e \Trans{\subst{\sigma}{\delta}{\theta}} \failState \testexec q_{ie}\bigr)\Bigr)\;\land$\\
$\Bigl(\bigl(\exists t_e \in \TTS[e,s], q_{te} \in Q_{te}\setminus\{\passState\}: t_e \testexec i_e \Trans{\subst{\sigma}{\delta}{\theta}} q_{te} \testexec q_{ie}\bigr)\; \lor\\ 
\bigl(\exists t_s \in \TTS[s,e]: t_s \testexec i_s \Trans{\subst{\sigma}{\delta}{\theta}} \failState \testexec q_{is}\bigr)\Bigr)$\\
\proofstep{\Cref{lem:project_from_testexec}}\\
$\sigma\in \utraces{s\parcomp e} \land \ell\in U_{s\parcomp e}^\delta \land \ell \notin \outset{s\parcomp e \after\sigma}\;\land$\\
$\exists q_{ie}\in Q_{ie}, q_{is} \in Q_{is}: q_{is}\parcomp q_{ie} \trans{\ell}\;\land$\\
$\Bigl(\bigl(\exists t_s \in \TTS[s,e], q_{ts} \in Q_{ts}\setminus\{\passState\}: t_s \testexec i_s \Trans{\subst{\sigma}{\delta}{\theta}} q_{ts} \testexec q_{is}\;\land$\\
\tab$t_s\Trans{\project{\subst{\sigma}{\delta}{\theta}}{L_{ts}}} q_{ts}\bigr)\; \lor\\ 
\bigl(\exists t_e \in \TTS[e,s]: t_e \testexec i_e \Trans{\subst{\sigma}{\delta}{\theta}} \failState \testexec q_{ie}\bigr)\Bigr)\;\land$\\
$\Bigl(\bigl(\exists t_e \in \TTS[e,s], q_{te} \in Q_{te}\setminus\{\passState\}: t_e \testexec i_e \Trans{\subst{\sigma}{\delta}{\theta}} q_{te} \testexec q_{ie}\;\land$\\
\tab$t_e\Trans{\project{\subst{\sigma}{\delta}{\theta}}{L_{te}}} q_{te}\bigr)\; \lor\\ 
\bigl(\exists t_s \in \TTS[s,e]: t_s \testexec i_s \Trans{\subst{\sigma}{\delta}{\theta}} \failState \testexec q_{is}\bigr)\Bigr)$\\
\proofstep{\Cref{lem:combinedtestcase_eq_genafter}}\\
$\sigma\in \utraces{s\parcomp e} \land \ell\in U_{s\parcomp e}^\delta \land \ell \notin \outset{s\parcomp e \after\sigma}\;\land$\\
$\exists q_{ie}\in Q_{ie}, q_{is} \in Q_{is}: q_{is}\parcomp q_{ie} \trans{\ell}\;\land$\\
$\Bigl(\bigl(\exists t_s \in \TTS[s,e], q_{ts} \in Q_{ts}\setminus\{\passState\}: t_s \testexec i_s \Trans{\subst{\sigma}{\delta}{\theta}} q_{ts} \testexec q_{is}\;\land$\\
\tab$q_{ts} = \GenEcoTest(s\parcomp e \after (\project{\subst{\subst{\sigma}{\delta}{\theta}}{\theta}{\delta}}{L_{s\parcomp e}^\delta}))\bigr)\; \lor\\
\bigl(\exists t_e \in \TTS[e,s]: t_e \testexec i_e \Trans{\subst{\sigma}{\delta}{\theta}} \failState \testexec q_{ie}\bigr)\Bigr)\;\land$\\
$\Bigl(\bigl(\exists t_e \in \TTS[e,s], q_{te} \in Q_{te}\setminus\{\passState\}: t_e \testexec i_e \Trans{\subst{\sigma}{\delta}{\theta}} q_{te} \testexec q_{ie}\;\land$\\
\tab$q_{te} = \GenEcoTest(s\parcomp e \after (\project{\subst{\subst{\sigma}{\delta}{\theta}}{\theta}{\delta}}{L_{s\parcomp e}^\delta}))\bigr)\; \lor\\
\bigl(\exists t_s \in \TTS[s,e]: t_s \testexec i_s \Trans{\subst{\sigma}{\delta}{\theta}} \failState \testexec q_{is}\bigr)\Bigr)$\\
\proofstep{\Cref{def:substitution}: $\theta\notin\sigma$, $\sigma \in L_{s\parcomp e}^{\delta*}$}\\
$\sigma\in \utraces{s\parcomp e} \land \ell\in U_{s\parcomp e}^\delta \land \ell \notin \outset{s\parcomp e \after\sigma}\;\land$\\
$\exists q_{ie}\in Q_{ie}, q_{is} \in Q_{is}: q_{is}\parcomp q_{ie} \trans{\ell}\;\land$\\
$\Bigl(\bigl(\exists t_s \in \TTS[s,e], q_{ts} \in Q_{ts}\setminus\{\passState\}: t_s \testexec i_s \Trans{\subst{\sigma}{\delta}{\theta}} q_{ts} \testexec q_{is}\;\land$\\
\tab$q_{ts} = \GenEcoTest(s\parcomp e \after \sigma)\bigr)\; \lor\\
\bigl(\exists t_e \in \TTS[e,s]: t_e \testexec i_e \Trans{\subst{\sigma}{\delta}{\theta}} \failState \testexec q_{ie}\bigr)\Bigr)\;\land$\\
$\Bigl(\bigl(\exists t_e \in \TTS[e,s], q_{te} \in Q_{te}\setminus\{\passState\}: t_e \testexec i_e \Trans{\subst{\sigma}{\delta}{\theta}} q_{te} \testexec q_{ie}\;\land$\\
\tab$q_{te} = \GenEcoTest(s\parcomp e \after \sigma)\bigr)\; \lor\\
\bigl(\exists t_s \in \TTS[s,e]: t_s \testexec i_s \Trans{\subst{\sigma}{\delta}{\theta}} \failState \testexec q_{is}\bigr)\Bigr)$\\
\proofstep{$(A\lor B)\land (C\lor D) \implies B \lor D \lor (A \land B)$}\\
$\sigma\in \utraces{s\parcomp e} \land \ell\in U_{s\parcomp e}^\delta \land \ell \notin \outset{s\parcomp e \after\sigma}\;\land$\\
$\exists q_{ie}\in Q_{ie}, q_{is} \in Q_{is}: q_{is}\parcomp q_{ie} \trans{\ell}\;\land$\\
$\Bigl(\bigl(\exists t_e \in \TTS[e,s]: t_e \testexec i_e \Trans{\subst{\sigma}{\delta}{\theta}} \failState \testexec q_{ie}\bigr)\;\lor$\\
$\bigl(\exists t_s \in \TTS[s,e]: t_s \testexec i_s \Trans{\subst{\sigma}{\delta}{\theta}} \failState \testexec q_{is}\bigr)\;\lor$\\
$\bigl(\exists t_s \in \TTS[s,e], q_{ts} \in Q_{ts}\setminus\{\passState\}: t_s \testexec i_s \Trans{\subst{\sigma}{\delta}{\theta}} q_{ts} \testexec q_{is}\;\land$\\
\tab$q_{ts} = \GenEcoTest(s\parcomp e \after \sigma)\;\land $\\
$\exists t_e \in \TTS[e,s], q_{te} \in Q_{te}\setminus\{\passState\}: t_e \testexec i_e \Trans{\subst{\sigma}{\delta}{\theta}} q_{te} \testexec q_{ie}\;\land$\\
\tab$q_{te} = \GenEcoTest(s\parcomp e \after \sigma)\bigr)\Bigr)$\\
\proofstep{\Cref{lem:combinedtest_singlestep_fail}}\\
$\exists q_{ie}\in Q_{ie}, q_{is} \in Q_{is}: q_{is}\parcomp q_{ie} \trans{\ell}\;\land$\\
$\Bigl(\bigl(\exists t_e \in \TTS[e,s]: t_e \testexec i_e \Trans{\subst{\sigma}{\delta}{\theta}} \failState \testexec q_{ie}\bigr)\;\lor$\\
$\bigl(\exists t_s \in \TTS[s,e]: t_s \testexec i_s \Trans{\subst{\sigma}{\delta}{\theta}} \failState \testexec q_{is}\bigr)\;\lor$\\
$\bigl(\exists t_s \in \TTS[s,e], q_{ts} \in Q_{ts}\setminus\{\passState\}: t_s \testexec i_s \Trans{\subst{\sigma}{\delta}{\theta}} q_{ts} \testexec q_{is}\;\land$\\
$\exists t_e \in \TTS[e,s], q_{te} \in Q_{te}\setminus\{\passState\}: t_e \testexec i_e \Trans{\subst{\sigma}{\delta}{\theta}} q_{te} \testexec q_{ie}\;\land$\\
$(q_{ts}\trans{\subst{\ell}{\delta}{\theta}} \failState \lor q_{te}\trans{\subst{\ell}{\delta}{\theta}} \failState )
\bigr)\Bigr)$\\
\proofstep{\Cref{lem:project_from_parcomp_IOTS,lem:project_from_testexec}}\\
$\exists q_{ie},q_{ie}'\in Q_{ie}, q_{is},q_{is}' \in Q_{is}: $\\
$\Bigl(\bigl(\exists t_e \in \TTS[e,s]: t_e \testexec i_e \Trans{\subst{\sigma}{\delta}{\theta}} \failState \testexec q_{ie}\bigr)\;\lor$\\
$\bigl(\exists t_s \in \TTS[s,e]: t_s \testexec i_s \Trans{\subst{\sigma}{\delta}{\theta}} \failState \testexec q_{is}\bigr)\;\lor$\\
$\bigl(\exists t_s \in \TTS[s,e], q_{ts} \in Q_{ts}\setminus\{\passState\}: t_s \testexec i_s \Trans{\subst{\sigma}{\delta}{\theta}} q_{ts} \testexec q_{is} \trans{\subst{\ell}{\delta}{\theta}} \failState \testexec q_{is}'\;\lor$\\
$\exists t_e \in \TTS[e,s], q_{te} \in Q_{te}\setminus\{\passState\}: t_e \testexec i_e \Trans{\subst{\sigma}{\delta}{\theta}} q_{te} \testexec q_{ie} \trans{\subst{\ell}{\delta}{\theta}} \failState \testexec q_{ie}'\bigr)\Bigr)$\\
\proofstep{\Cref{def:arrowdefs}: $\Trans{\sigma}$}\\
$\exists q_{ie},q_{ie}'\in Q_{ie}, q_{is},q_{is}' \in Q_{is}: $\\
$\Bigl(\bigl(\exists t_e \in \TTS[e,s]: t_e \testexec i_e \Trans{\subst{\sigma}{\delta}{\theta}} \failState \testexec q_{ie}\bigr)\;\lor$\\
$\bigl(\exists t_s \in \TTS[s,e]: t_s \testexec i_s \Trans{\subst{\sigma}{\delta}{\theta}} \failState \testexec q_{is}\bigr)\;\lor$\\
$\bigl(\exists t_s \in \TTS[s,e]: t_s \testexec i_s \Trans{\subst{\sigma}{\delta}{\theta}\cdot \subst{\ell}{\delta}{\theta}} \failState \testexec q_{is}'\;\lor$\\
$\exists t_e \in \TTS[e,s]: t_e \testexec i_e \Trans{\subst{\sigma}{\delta}{\theta}\cdot \subst{\ell}{\delta}{\theta}} \failState \testexec q_{ie}'\bigr)\Bigr)$\\
\end{proof}

\FloatBarrier

\section{Discussion and Impact}
\label{sec:discussion}
In the previous sections we have introduced three ways to do compositional model based testing. In this section we will discuss how compositional testing can be used to deal with several practical testing challenges.

The most immediate possible application is in the field of component based development. During development, there is no full system, but we would still like to test as early as possible on any components that are already finished. After development, the correctness of the composed system is important. Compositional testing provides a way to link the results of component based tests done early during development, to the performance of the full system after composition. This allows for earlier testing, also known as the shift left paradigm \cite{man:smith2001shift}, and effectively pushes tests down to a lower level of the testing pyramid, reducing costs \cite{cohn_SucceedingAgileSoftware_2011}.

Another important application is during system evolution, where parts of a system change over time, but other parts remain unchanged. For instance, components of older systems are often reused when newer systems are developed. Since these components have been successfully deployed in their old system for a while now, testing them again is seen as a lower priority, which is not always without consequences. An infamous example is the Ariane 5 rocket, which crashed after components of the the previously successful Ariane 4 rocket were reused without retesting using the new environment. The root cause was deemed to be an incompatibility in the interface between old and new components, where the newer Ariane 5 components were sending signals not present in the Ariane 4 rocket \cite{lions_Ariane5Flight_1996}. This type of integration problem is hard to find with traditional testing, unless full testing is repeated on every small change in the system.
In this paper we have shown that testing in a component based system can be split into two separate objectives: checking correctness with respect to your own specification ($\uioco$), and checking correctness with respect to your environment ($\mutuallyaccepts{},\eco$). While the former is indeed less important for systems that have been running bug free for a long time, the latter is very important when running a component in a new environment. This also allows focussing the testing effort in the direction where new bugs might be found after an component change: interaction with the environment. 

The component replacement scenario is where the differences between the three testing approaches introduced in this paper become important. Back in \Cref{sec:problemStatement}, we visualized all three of our approaches in \Cref{fig:compMBTsetup}. In this figure, the specifications and implementations are represented as nodes, while the different testing and verification operations are represented as arrows between these nodes. Whenever a node changes, all the arrows touching that node have to re-run. When using the $\mutuallyaccepts{}$ based approach, we verified that all possible component implementations of all component specifications are compositional. This means that on changing a component implementation, only that component needs to be retested with respect to its own specification. If a specification changes, then the compositionality verification also needs to be redone. With the $\eco$ based approach, half of this versatility is dropped: one implementation is tested to be compositional with all valid implementations of its environment specification. In this approach, any change always results in having to redo some $\uioco$ and $\eco$ based testing, while other testing results remain valid. This is indicated by the arrows in \Cref{fig:compMBTsetup}.
Our third approach, testing for $\cioco{}{s}{e}$, is optimized for systems in which the specifications have settled, but some of the implementations might still get updates. In this scenario, changing an implementation only requires rerunning the tests for that implementation, but changing a specification requires rerunning everything.

Finally, another practical problem to be solved when diagnosing the root cause of a failing test, is fault localization. Once a test fails, this usually produces a high level trace trough the system, along with a message that an unexpected outcome was observed. The process of finding the precise location in the system where the failure originated based on the information provided by the failed test is called fault localization, and it can be very time consuming. One common way to deal with this is trace minimization, where heuristic search tries to find a smaller trace that also leads to the same failing test. This is very effective, but suffers from the drawback that all traces still have to start from the entry-point of the system, go trough all involved components, and only then reach the error. By restricting tests to one component at a time the search space for the problem becomes much smaller, which can be achieved by using the contrapositive of \Cref{lem:compositional_testing_mutaccepts}. For example, if the checkout system crashes when you have more than 100 items in your cart, a system wide (model based) test would first need to fill a cart with 100 items, producing a very long trace. A compositional testing approach could instead test the cart and the checkout system separately and generate possible carts for the checkout to use, without having to go trough the cart API to fill them. This approach is orthogonal to traditional trace minimization, and could be combined for an even greater effect, but this is outside the scope of this paper.
\section{Related Work}
\label{sec:related-work}

The work in this paper is based on earlier work done on compositional model-based testing \cite{frantzen_ModelBasedTestingEnvironmental_2007,vanderbijl_CompositionalTestingIoco_2004,noroozi_ImprovingInputOutputConformance_2014}, and  specifically on the theory of environmental conformance ($\eco$) \cite{frantzen_ModelBasedTestingEnvironmental_2007}. This last work also describes an approach for testing compositionality, similar to our approach in \Cref{sec:testing-accepting-systems}. They however do not link to any existing form of correctness such as $\uioco{}$ and simply set their relation $\eco$ to be the definition of correctness. By linking the definition to $\uioco$, we provide a stronger argumentation of the benefits of testing compositionality. The scope considered by \cite{frantzen_ModelBasedTestingEnvironmental_2007} is also narrower than ours. They assume that the two components form a closed system, where the only observable behaviour is the communication between these components. We lift this assumption to make it easier to compose more than two systems while maintaining associativity.

Interface automata \cite{dealfaro_InterfaceAutomata_2001,dealfaro_InterfaceTheoriesComponentBased_2001,dealfaro_InterfaceBasedDesign_2005}, which are a form of contract based design \cite{benveniste_ContractsSystemDesign_2018,incer_InterfaceAutomataHypercontracts_2022},  model both the behaviour of a component and the constraints it puts on its environment. Similar to our model of LTS, interface automata also model restrictions on the environment: if a model does not define an input transition, then the environment is expected not to give that input. In the theory of interface automata an optimistic approach is taken, which means two specifications are compositional if there is a valid environment that avoids all communication errors. In contrast, compositionality for \uioco{} requires a pessimistic approach, where two specifications are compositional if all possible valid environments avoid communication errors. It is however likely that these two viewpoints can be combined to create a more permissible definition of compositionality, by differentiating between the final composition and intermediate steps. The final composition needs to be free of errors, and should therefore take a pessimistic approach. Errors in intermediate steps could however still be avoided by composition with further specifications, which could make these errors unreachable. This could be represented by a more optimistic approach to composition. 

Earlier work on combining interface automata with $\ioco{}$ was done by \textcite{daca_CompositionalSpecificationsIoco_2014}. They define ambiguous states in the context of parallel composition which correspond to our notion of communication errors. $\mutuallyaccepts{}$ then coincides with the absence of ambiguous states reachable by $\utraces{}$. \citeauthor{daca_CompositionalSpecificationsIoco_2014} also define friendly composition, a variant on parallel composition which removes transitions from the specifications until all ambiguous states are unreachable, and give pseudocode for an algorithm to compute this. This partly corresponds to our verification based approach in \Cref{sec:eco-verificaction}, minus the pruning. Our algorithm also takes non-deterministic underspecification into account by being based on $\uioco$ instead of $\ioco{}$, and we give a formal proof of correctness. \citeauthor{daca_CompositionalSpecificationsIoco_2014} do not give any proofs for their algorithm, but do run some practical experiments. We believe these two results compliment each other and show the promising nature of compositional testing.

Apart from labeled transition systems, there are many other formalisms that can be used to describe component specifications, such as CSP \cite{roscoe_UnderstandingConcurrentSystems_2010}, LOTOS \cite{bolognesi_IntroductionISOSpecification_1987}, (E)FSM \cite{andrews_TestingWebApplications_2005,cheng_AutomaticFunctionalTest_1993}, UML \cite{_UnifiedModelingLanguage_2017,rathod_StructuralBehavioralModeling_2013} or BPMN \cite{silver_BPMNMethodStyle_2011,lubke_BPMNBasedModelDrivenTesting_2017}, etc. A good formalism has clear semantics, but is also easy to learn and use for practitioners. We find that LTSes have a good balance between ease of use and formal semantics, while being slightly more expressive than FSMs. Process algebra such as CSP and LOTOS are often very hard to use for non experts, while UML and BPMN already have broad support in industry but lack a unified semantics. Our theory is however likely to be transferable to any other formalism, as the core problem of making sure that send and receive operations match, is universal.

Component based design \cite{mahapatro_ReviewingLandscapeComponentBased_2024} is a popular approach to designing software systems. This naturally leads to also designing tests in a component based way \cite{kuliamin_ComponentArchitectureModelbased_2010,beydeda_StateArtTesting_2003,deussen_OnlineTestPlatform_2002,schatz_IntegratingComponentTests_2010}. A challenge here is to relate the coverage and guarantees of the component based tests to those offered by system level tests, to reason about when to stop testing or where to develop more tests. The theory developed in this paper can help to make this connection, by pointing out where the interactions between components are not covered by the component level tests.

\FloatBarrier
\section{Concluding Remarks}
\label{sec:conclusion}

In this paper we have introduced three approaches to do compositional model based testing. We can check compositionality directly on the specification level, which means that all \uioco{}-conformant implementations are also compositional. If formal verification is not feasible, we can instead test if an implementation is compositional with respect to the specification of its environment. Finally, we give a third approach where we optimize for one specific set of specifications. This gives a smaller test suite, but puts some additional restrictions on the specifications, and requires much more retesting if any specification ever changes. We then introduced algorithms for all three of these testing approaches. All three algorithms allow compositional testing: correctness of all components implies correctness of the system. This allows picking the best approach for a given situation. Small, simple models work well with formal verification, while a still changing system with models containing data would work better with the testing approach. If the system is older and a bit more stable, optimizing for the current set of specifications can bring testing time down even further.

The use of compositional model based testing means testing can focus on the component level, giving several advantages: components have smaller models than the whole system, making them easier to create, understand and maintain; retesting when replacing components becomes easier; and smaller models lead to smaller counter-examples which makes using the results of failing tests easier. These advantages should lower the barrier to entry for model-based testing, as now a couple of component models can get one started. Later on, this still allows for the possibility to expand with more models if desired, without losing completeness or having to throw away models.

\subsection*{Future Work}
\label{sec:future-work}
As shown in \Cref{sec:combining_algorithms}, taking the parallel composition of several component specifications to be the specification for the whole system has some issues. The composed specification correctly forbids illegal behaviour, but it does not require all, or even any, allowed behaviour to be present in the system implementation. This causes seemingly nonsensical implementations to still pass \uioco{} based testing at the system level. One way to deal with this is to check if the generated system level specification has certain desirable behaviour. The component based testing done so far can then be used to show that the implementation also has this desired behaviour. Defining this desirable behaviour could for instance be done by linking the \uioco{} based testing to another often used notion of correctness: system requirements or features. These often already exist in the form of short natural language descriptions of things a correct system should be able to do, such as those used with the popular behaviour driven development (BDD) paradigm \cite{rose_BDDBooksFormulation_2021}. Whether these informal behaviour descriptions can be integrated with a formal technique such as MBT is still an open question. 

A closely related problem is that of adding or removing labels when working with multiple levels of abstraction. When combining low level components, the set of observable behaviour might be different then when taking a system level view. To translate between different abstraction levels, action hiding or action refinement can be used. These often have complicated interactions with the concept of quiescence, which is central to the $\uioco$ testing theory. It is still unclear how best to integrate viewpoints from different abstraction levels into model based testing, while maintaining compositionality as much as possible.

We use a definition of parallel composition common to several other papers. This definition assumes synchronous communication between components, which means that one component can block communication attempts from another component. In settings where this is not realistic, another form of composition might make more sense, for instance using buffers. Some of our results might be transferable to other forms of composition, depending on how these composition operators deal with inputs, outputs and quiescence. Asynchronous communication could for instance be handled by also explicitly modelling the communication medium \cite{noroozi_SynchronyAsynchronyConformance_2015}. 

When composing systems with more than two components, we need to check if an already composed system mutually accepts a new component: $(s_1\parcomp s_2) \mutuallyaccepts{} s_3$. This is possible with the given algorithms, but does not scale well as the state space of the composed system grows exponentially. It would be much easier to check $s_1 \mutuallyaccepts{} s_3$ and $s_2 \mutuallyaccepts{} s_3$ directly, however, it is still unclear whether $(s_1\parcomp s_2) \mutuallyaccepts{} s_3 \iff s_1 \mutuallyaccepts{} s_3\land s_2 \mutuallyaccepts{} s_3$. An alternative approach could also be to redefine both composition and mutual acceptance as operations on sets of specifications, instead of just two at a time. Similarly, a more refined notion of quiescence might simplify some of the notation and definitions used in our paper. Applying partial quiescence over a subset of all labels to $\ioco{}$ is explored in \cite{noroozi_DecomposabilityInputOutput_2013}.

While we believe we have given a solid theoretical foundation for both the correctness and usefulness of compositional model based testing using $\uioco$, it is still important to verify this using practical experiments. This has fallen out of scope for this paper, but some relevant experiments on the use compositional testing with $\ioco{}$ can be found in \cite{daca_CompositionalSpecificationsIoco_2014}.

Finally, we currently only give an extensional definition for \cioco{i}{s}{e}, by testing using \Cref{alg:eco_uioco_combined}. An alternative intentional definition is also possible, which could be used to further validate \Cref{alg:eco_uioco_combined}. One possible road for this is a definition of the form $\cioco{i}{s}{e} \iff i \uioco f(s,e) \land i \eco f'(s,e)$, where $f$ and $f'$ represent the reachable part of $s\parcomp e$ where some labels are hidden. Formalizing this precisely is tricky however, because of the interactions between hiding labels and quiescence.

\newpage
\printbibliography

\end{document}